\documentclass[12pt,a4paper]{report}
\usepackage{bm}
\usepackage{epsfig}
\usepackage{graphics}

\usepackage{graphicx}

\begin{document}
\ifx\href\undefined\else\hypersetup{linktocpage=true}\fi 

\renewcommand\theequation{\ifnum\value{chapter}>0{\thechapter.\arabic{equation}}\fi}

\newcommand{\lsim}{\mbox{\raisebox{-.9ex}{~$\stackrel{\mbox{$<$}}{\sim}$~}}}

\newcommand{\gsim}{\mbox{\raisebox{-.9ex}{~$\stackrel{\mbox{$>$}}{\sim}$~}}}

\newcommand\vev[1]{{\langle {#1} \rangle}}

\renewcommand\({\left(}
\renewcommand\){\right)}
\renewcommand\[{\left[}
\renewcommand\]{\right]}

\newcommand\del{{\mbox {\boldmath $\nabla$}}}

\newcommand\eq[1]{Eq.~(\ref{#1})}
\newcommand\eqs[2]{Eqs.~(\ref{#1}) and (\ref{#2})}
\newcommand\eqss[3]{Eqs.~(\ref{#1}), (\ref{#2}), and (\ref{#3})}
\newcommand\eqsss[4]{Eqs.~(\ref{#1}), (\ref{#2}), (\ref{#3})
and (\ref{#4})}
\newcommand\eqssss[5]{Eqs.~(\ref{#1}), (\ref{#2}), (\ref{#3}),
(\ref{#4}) and (\ref{#5})}
\newcommand\eqst[2]{Eqs.~(\ref{#1})--(\ref{#2})}
\newcommand\eqref[1]{(\ref{#1})}
\newcommand\eqsref[2]{(\ref{#1}) and (\ref{#2})}
\newcommand\eqssref[3]{(\ref{#1}), (\ref{#2}), and (\ref{#3})}
\newcommand\eqsssref[4]{(\ref{#1}), (\ref{#2}), (\ref{#3})
and (\ref{#4})}
\newcommand\eqssssref[5]{(\ref{#1}), (\ref{#2}), (\ref{#3}),
(\ref{#4}) and (\ref{#5})}
\newcommand\eqstref[2]{(\ref{#1})--(\ref{#2})}

\newcommand\pa{\partial}
\newcommand\pdif[2]{\frac{\pa #1}{\pa #2}}

\newcommand\ee{\end{equation}}
\newcommand\be{\begin{equation}}
\newcommand\eea{\end{eqnarray}}
\newcommand\bea{\begin{eqnarray}}

\newcommand\mpl{M_{\rm P}}

\newcommand\dbibitem[1]{\bibitem{#1}\hspace{1cm}#1\hspace{1cm}}
\newcommand{\dlabel}[1]{\label{#1} \ \ \ \ \ \ \ \ #1\ \ \ \ \ \ \ \ }
\def\dcite#1{[#1]}

\def\calf{{\cal F}}
\def\calh{{\cal H}}
\def\call{{\cal L}}
\def\calm{{\cal M}}
\def\caln{{\cal N}}
\def\calp{{\cal P}}
\def\calr{{\cal R}}
\def\calpr{{\calp_\calr}}

\newcommand\bfa{{\bf a}}
\newcommand\bfb{{\bf b}}
\newcommand\bfc{{\bf c}}
\newcommand\bfd{{\bf d}}
\newcommand\bfe{{\bf e}}
\newcommand\bff{{\bf f}}
\newcommand\bfg{{\bf g}}
\newcommand\bfh{{\bf h}}
\newcommand\bfi{{\bf i}}
\newcommand\bfj{{\bf j}}
\newcommand\bfk{{\bf k}}
\newcommand\bfl{{\bf l}}
\newcommand\bfm{{\bf m}}
\newcommand\bfn{{\bf n}}
\newcommand\bfo{{\bf o}}
\newcommand\bfp{{\bf p}}
\newcommand\bfq{{\bf q}}
\newcommand\bfr{{\bf r}}
\newcommand\bfs{{\bf s}}
\newcommand\bft{{\bf t}}
\newcommand\bfu{{\bf u}}
\newcommand\bfv{{\bf v}}
\newcommand\bfw{{\bf w}}
\newcommand\bfx{{\bf x}}
\newcommand\bfy{{\bf y}}
\newcommand\bfz{{\bf z}}

\newcommand\yr{\,\mbox{yr}}
\newcommand\sunit{\,\mbox{s}}
\newcommand\munit{\,\mbox{m}}
\newcommand\wunit{\,\mbox{W}}
\newcommand\Kunit{\,\mbox{K}}
\newcommand\muK{\,\mu\mbox{K}}

\newcommand\metres{\,\mbox{meters}}
\newcommand\mm{\,\mbox{mm}}
\newcommand\cm{\,\mbox{cm}}
\newcommand\km{\,\mbox{km}}
\newcommand\kg{\,\mbox{kg}}
\newcommand\TeV{\,\mbox{TeV}}
\newcommand\GeV{\,\mbox{GeV}}
\newcommand\MeV{\,\mbox{MeV}}
\newcommand\keV{\,\mbox{keV}}
\newcommand\eV{\,\mbox{eV}}
\newcommand\Mpc{\,\mbox{Mpc}}

\newcommand\msun{M_\odot}

\newcommand\sub[1]{_{\rm #1}}
\newcommand\su[1]{^{\rm #1}}

\newcommand{\one}{_1}
\newcommand{\two}{_2}

\newcommand\mone{^{-1}}
\newcommand\mtwo{^{-2}}
\newcommand\mthree{^{-3}}
\newcommand\mfour{^{-4}}
\newcommand\mhalf{^{-1/2}}
\newcommand\half{^{1/2}}
\newcommand\threehalf{^{3/2}}
\newcommand\mthreehalf{^{-3/2}}

\renewcommand{\P}{{\cal P}}
\newcommand{\f}{f_{\rm NL}}
\newcommand{\mk}{{\mathbf k}}
\newcommand{\mq}{{\mathbf q}}

\newcommand{\zetag}{{\zeta\sub g}}
\newcommand\sigmas{{\sigma^2}}

\newcommand{\fnl}{{f\sub{NL}}}
\newcommand{\fnltilde}{{\tilde f\sub{NL}}}
\newcommand{\fnli}{ {f_{ {\rm NL}i } } }

\newcommand\kmax{{k\sub{max}}}
\newcommand\bfkp{{{\bfk}'}}
\newcommand\bfkpp{{{\bfk}''}}
\newcommand\bfpp{{{\bfp}'}}

\newcommand\tp{{(2\pi)}}
\newcommand\tpq{{(2\pi)^3}}
\newcommand\tps{{(2\pi)^6}}

\renewcommand\ni{{N_{,i}}}
\newcommand\nj{{N_{,j}}}
\newcommand\nij{{N_{,ij}}}
\newcommand\nii{{N_{,ii}}}

\newcommand{\la}{\label}
\newcommand{\ci}{\cite}

\pagestyle{empty}%
\begin{titlepage}
\begin{center}
{\LARGE{\bf The Origin of the Large-Scale Structure}}\\
\medskip
{\LARGE{\bf in the Universe:}}\\
\bigskip
{\LARGE{\bf Theoretical and Statistical Aspects}}\\
\vspace{2.8cm}
{\bf by\\}
\vspace{2.8cm}
{\bf YEINZON RODR\'IGUEZ GARC\'IA\\
Physicist, MSc\\}
\vspace{3.5cm}
{\bf A thesis submmited in partial fulfillment\\
of the requirements for the degree of\\
Doctor of Philosophy\\
in\\
Physics}\\
\vspace{3.8cm}
{\bf LANCASTER UK\\
LANCASTER UNIVERSITY\\
FACULTY OF ENVIRONMENTAL AND NATURAL SCIENCES\\
DEPARTMENT OF PHYSICS\\
2005\\}

\end{center}
\end{titlepage}

\newpage

\begin{center}

\vspace*{5cm}

{\bf Thesis directed by} \\
{\bf Dr. David H. Lyth (Lancaster University)} \\
{\bf and approved by}




\vspace{3cm}

$\_\_\_\_\_\_\_\_\_\_\_\_\_\_\_\_\_\_\_\_\_\_\_\_\_\_\_\_\_\_\_\_\_\_\_\_\_\_\_\_\_\_\_\_\_\_\_\_\_\_\_\_\_\_\_\_\_$

{\bf Internal Examiner:  Dr. Konstantinos Dimopoulos} \\
\hspace{2.5cm}{\bf (Lancaster University)}

\vspace{2cm}

$\_\_\_\_\_\_\_\_\_\_\_\_\_\_\_\_\_\_\_\_\_\_\_\_\_\_\_\_\_\_\_\_\_\_\_\_\_\_\_\_\_\_\_\_\_\_\_\_\_\_\_\_\_\_\_\_\_$

{\bf External Examiner:  Dr. Edmund J. Copeland} \\
\hspace{4cm}{\bf (Nottingham University)}

\end{center}

\newpage
\begin{titlepage}

\vspace*{0cm}\hspace{10cm}\textit{To Luz Yasmid.} \\
\hspace*{9.3cm}\textit{Three years = one life.}

\vspace{1.3cm}

\hspace*{0.8cm}\textit{Men in their arrogance claim to understand the nature of creation,} \\
\hspace*{1.5cm}\textit{and devise elaborate theories to describe its behaviour. But always} \\
\hspace*{1.5cm}\textit{they discover in the end that God was quite a bit more clever than} \\
\hspace*{11.2cm}\textit{they thought.} \\
\hspace*{6cm}\textit{Sister Miriam Godwinson,} \\
\hspace*{6cm}\textit{``We Must Dissent''}

\vspace{1.3cm}

\hspace*{2cm}\textit{Begin with a function of arbitrary complexity. Feed it values,} \\
\hspace*{2cm}\textit{``sense data''. Then, take your result, square it, and feed it back} \\
\hspace*{2.8cm}\textit{into your original function, adding a new set of sense data.} \\
\hspace*{2.1cm}\textit{Continue to feed your results back into the original function ad} \\
\hspace*{1.3cm}\textit{infinitum. What do you have? The fundamental principle of human} \\
\hspace*{11cm}\textit{consciousness.} \\
\hspace*{6cm}\textit{Academician Prokhor Zakharov,} \\
\hspace*{6cm}\textit{``The Feedback Principle''}

\vspace{1.3cm}

\hspace*{6cm}\textit{Einstein would turn over in his grave!} \\
\hspace*{4.5cm}\textit{Not only does God play dice... the dice are loaded.} \\
\hspace*{6cm}\textit{Chairman Sheng-ji Yang,} \\
\hspace*{6cm}\textit{``Looking God in the Eye''}

\vspace{1.3cm}

\hspace*{3.1cm}\textit{The supreme task of the physicist is to arrive at those} \\
\hspace*{3.3cm}\textit{universal elementary laws from which the cosmos can be} \\
\hspace*{8.7cm}\textit{built up by pure deduction.} \\
\hspace*{6cm}\textit{Albert Einstein}

\vspace{1.3cm}

\hspace*{2.1cm}\textit{Beware, you who seek first and final principles, for you are} \\
\hspace*{2.3cm}\textit{trampling the garden of an angry God and he awaits you just} \\
\hspace*{9.1cm}\textit{beyond the last theorem.} \\
\hspace*{6cm}\textit{Sister Miriam Godwinson,} \\
\hspace*{6cm}\textit{``But for the Grace of God''}

\end{titlepage}

\chapter*{Acknowledgments}

\pagestyle{plain}

There are so many people to acknowledge, so many reasons to do it, and it is so easy to let me take by the excitement of the moment, that I will try to be as short and objective as I can.
\begin{itemize}
\item To my wife Luz Yasmid, who has shared with me all those beautiful moments in England, and all her efforts, constancy, and dreams. She is an essential part of this success.
\item To all my relatives, specially my fathers Ram\'on and Rosalba, my siblings Yohany and Yamile, my grandmother Leonilde, and my little niece Jinneth. Without them, no PhD, no thesis, no anything. I love them all.
\item To my friend David, who has guided me through the physics paths. In absence of my father during these three years, David has reminded me many times of him.
\item To all our friends throughout these three years, specially to Margareth, Rebecca, Mati, Mila, Flora, Terry, Tessa, Julieta, Juan, Jos\'e Roberto, Patricia, Jos\'e Roberto Jr., Francis, Jos\'e Daniel, Santiago, Laura, Carlos Miguel, Mar\'{\i}a Isabel, Ang\'elica, Jos\'e, Karla, Jeannette, Juan Carlos, Mauricio, Pablo, Pili, Luis, John Jairo, Marcos, Simone, Isadora, Carlos, Julia, Abel, Margarita, Sof\'{\i}a, Daniel, Gerardo, Ana Cecilia, Iv\'an, Rosario, Vladimir, Ra\'ul, Oliverio, Jos\'e Juan, Dulce, Federico, Florencia, Agust\'{\i}n, Santiago, Ignacio, Ricardo, Alejandra, Mar\'{\i}a Francisca, Manuel, and Maribel.
\item To Elspeth and Andy, and the Across Cultures group. They have made a great effort to make all the international students and their families feel like at home. We will never forget such a praiseworthy action.
\item To the fathers Paul and Hugh, and the sister Ella, for being such excellent friends, and for helping my wife and creating such a wonderful environment for her.
\item To the members of the Cosmology and Astroparticle Physics Group: David, Kostas, John, Lotfi, Karim, Ignacio, Juan Carlos, Leila, and Chia-Min.
\item To the United Kingdom, our second fatherland.
\item To my late uncle Julio, who always supported me and who died days before my travel to the UK.
\item To my fathers and to my cousin Julio, who have served as debtors of my loan-scholarships.
\item To Carlos, who I began my physics career with.
\item To the Centro de Investigaciones at the Universidad Antonio Nari\~no, which supported me in my application to the COLCIENCIAS loan-scholarship.
\item To the Colombian Institute for the Science and Technology Development ``Francisco Jos\'e de Caldas'' COLCIENCIAS, for its full postgraduate loan-scholarship.
\item To the Foundation for the Future of Colombia COLFUTURO, Universities UK, and the Department of Physics of Lancaster University, for their partial financial support.
\end{itemize}

\vspace{1cm}

\textit{Lancaster UK, June 28th 2005.}

\chapter*{Abstract}

We review some theoretical and statistical aspects of the origin of the 
large-scale structure in the Universe, in view of the two most widely known and
accepted scenarios: the {\it inflaton} scenario (primordial curvature perturbation $\zeta$
generated by the quantum fluctuations of the light scalar field $\varphi$ that drives inflation, named the {\it inflaton}),
and the {\it curvaton} scenario
($\zeta$ generated by the quantum
fluctuations of a weakly coupled light scalar field $\sigma$ that does not drive inflation, named the {\it curvaton}). Among the theoretical
aspects, we
point out the impossibility of having a low inflationary energy scale in the simplest curvaton model.
A couple of modifications to the
simplest setup are explored, corresponding
to the implementation of a second (thermal)
inflationary period whose end makes the curvaton field `heavy', triggering either its oscillations
or immediate decay. Low scale inflation is then possible to attain with $H_\ast$ (the
Hubble parameter a few Hubble times after horizon exit) being as low as $1$ TeV.
Among the statistical aspects, we study the bispectrum $B_\zeta(k_1,k_2,k_3)$ of $\zeta$ whose normalisation $\fnl$ gives information about the level of non-gaussianity in the
primordial curvature perturbation.
In connection with $\fnl$, several conserved and/or gauge invariant quantities described
as the second-order curvature perturbation have been given in the literature. We review each of these
quantities showing how to interpret one in terms of the others, and analyze the respective expected
$\fnl$ in
both the inflaton and the curvaton scenarios as well as in other less known models for the generation of primordial perturbations and/or non-gaussianities.
The $\delta N$ formalism
turns out to be a powerful technique to compute $\fnl$ in multi-component slow-roll inflation, as the knowledge of the evolution of some family of unperturbed universes is the only requirement.
We present for the first time this formalism and apply it to selected examples.

\tableofcontents

\chapter{Introduction\label{introduccion}}

The Friedmann-Robertson-Walker (FRW) cosmological model (known also as the standard
or Big-Bang cosmological model) \cite{friedman,robertson1,robertson2,robertson3,walker} is the successful framework that describes the observed properties of the Universe: homogeneity and isotropy at large scales, Hubble expansion, and almost 14 billion years of evolution in agreement with globular clusters and radioactive isotopes dating. The additional predictions of the cosmic background radiation, confirmed by Penzias and Wilson's discovery in 1965 \cite{dicke,penziaswilson}, and the relative abundances of light elements \cite{alpher1,alpher2,gamow,hoyle,olive,wagoner,walkeretal} in full agreement with observation, have 
established a solid base for the study of the Universe throughout its history, turning the old speculative cosmology into well established and experimentally supported science \cite{garcia}.

The introduction of a period of exponential expansion (called inflationary) \cite{albste,guth,linde82}, prior to the Big-Bang, brought an elegant solution to the horizon, flatness, and unwanted relics problems that were present in the original standard cosmological model \cite{albste,guth,kt,linde82,riotto}. The horizon problem, or {\it why is the Cosmic Microwave Background radiation {\rm (CMB)} temperature highly isotropic?}, was solved as the accelerated expansion blows up a region initially in thermal equilibrium to a much bigger size, making the horizon at the end of the matter dominated era be still inside that region; as a result all regions in the sky appear today at the same background temperature. The same accelerated expansion makes the comoving horizon shrink so rapidly that our local patch of the universe becomes extremely flat despite its actual topology, solving this way the flatness problem, or {\it why is our observable Universe almost perfectly flat?}.
The huge dilution of the abundances of unwanted relics (e.g. topological defects: magnetic monopoles, cosmic strings, and domain walls) caused by the exponential grow of the size of the Universe during inflation and the huge production of entropy by the decay of the scalar field that drives inflation, gave solution to the unwanted relics problem, or {\it where are the topological defects (and some other troublesome stuff) predicted by the standard cosmology?} \footnote{Be aware however that a minimum of inflationary expansion (at least 70 e-folds which correspond to a minimum of $10^{-36}$ seconds of inflation) is required to successfully solve the horizon, flatness, and unwanted relics problems, assuming standard evolution.}. 

In spite of its success at solving the above mentioned problems, the inflationary period became perhaps more important because of its
ability to stretch the quantum fluctuations of the fields living in the FRW spacetime \cite{bst,guthpi82,hawking,linde82a,mukhanov,mukhanovrep,riotto,starobinsky1}, making them classical \cite{albrecht,grishchuk,guthpi,lombardo,lyth84} and almost constant soon after horizon exit. They correspond to small inhomogeneities in the energy density and are responsible, via gravitational attraction, of the large-scale structure seen today in the Universe (see Figs. \ref{2dfgs} and \ref{growing}). If this scenario turned to be correct, the energy density inhomogeneities should have left their trace in the CMB released at the time of recombination. Indeed, the Cosmic Background Explorer (COBE) in 1992 \cite{cobe} found small anisotropies in the CMB temperature of the order of 1 part in $10^5$ (with average temperature \mbox{$T_0 = 2.725 \pm 0.002$ K} \cite{bennett}), on scales of order thousands of Megaparsecs. With 30 times better angular resolution and sensitivity than COBE, the Wilkinson Microwave Anisotropy Probe (WMAP) \cite{wmap} confirmed this picture in 2003 (see \hbox{Fig. \ref{wmapsky}}), measuring in turn the cosmological parameters with a $1\%$ order precision \cite{observation} on scales of order tens of Megaparsecs. The PLANCK satellite \cite{planck}, due to be launched in 2007, will be able to refine these observations (see Fig. \ref{plancksky}). With 10 times better angular resolution and sensitivity than WMAP, PLANCK promises to determine the temperature anisotropies with a resolution of the order of 1 part in $10^6$, and the cosmological parameters with a $0.1\%$ order precision.

\begin{figure}[t]
\begin{center}
\includegraphics[height=9cm]{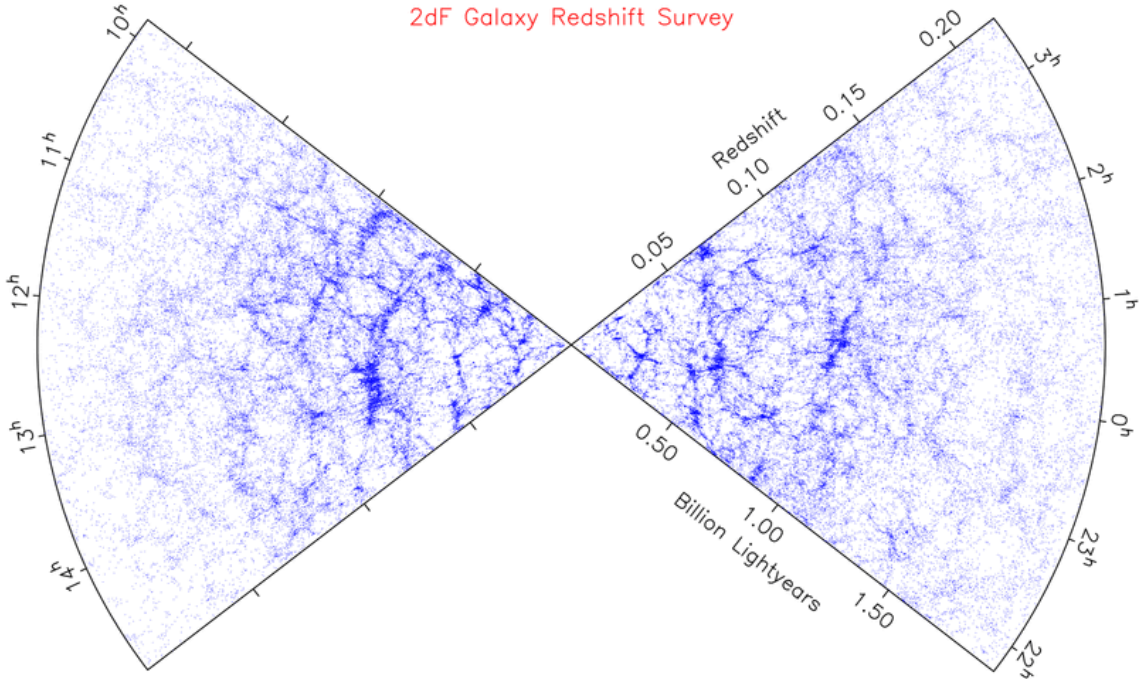}
\caption{The 2-degree Field Galaxy Redshift Survey (2dFGRS) map \cite{colless}. The 2dFGRS obtained spectra for 245591 objects, mainly galaxies. Reliable redshifts were obtained for 221414 galaxies. The galaxies cover an area of approximately 1500 square degrees selected from the extended APM Galaxy Survey in three regions: a north galactic pole strip, a south galactic pole strip and random fields scattered around the south galactic pole strip. The figure shows the map of the galaxy distribution produced from the completed survey. The filamentary structure can be qualitatively compared with that predicted by the Cold Dark Matter model with dark energy ($\Lambda$CDM model) (see Fig. \ref{growing}) (Courtesy of the Anglo-Australian Observatory's 2dFGRS group \cite{2dFGRS}). \label{2dfgs}}
\end{center}
\end{figure}

\begin{figure}
\begin{center}
\leavevmode
\hbox{
\epsfxsize=1.85in
\epsffile{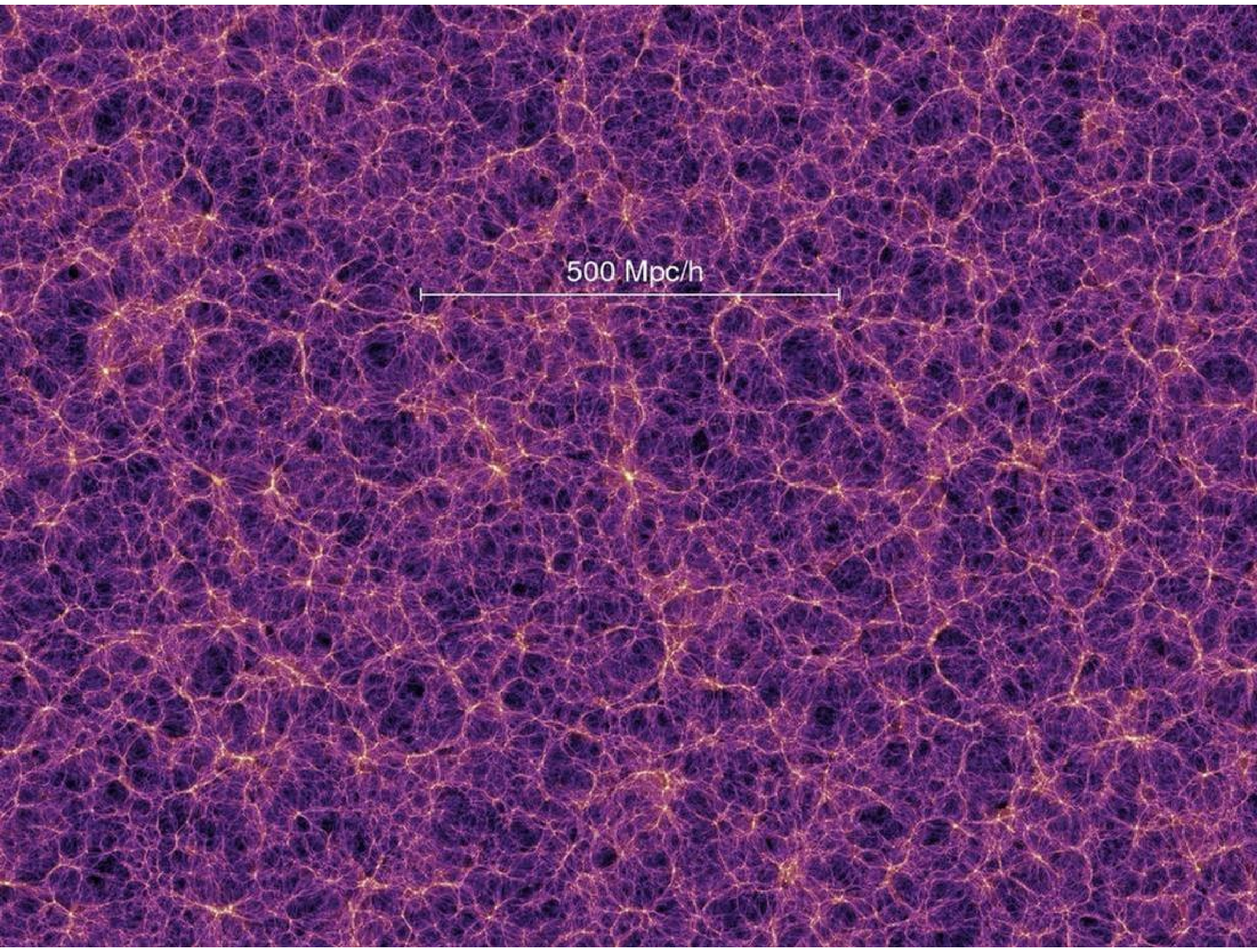}}
\hbox{\hspace{1mm}%
\epsfxsize=1.85in
\epsffile{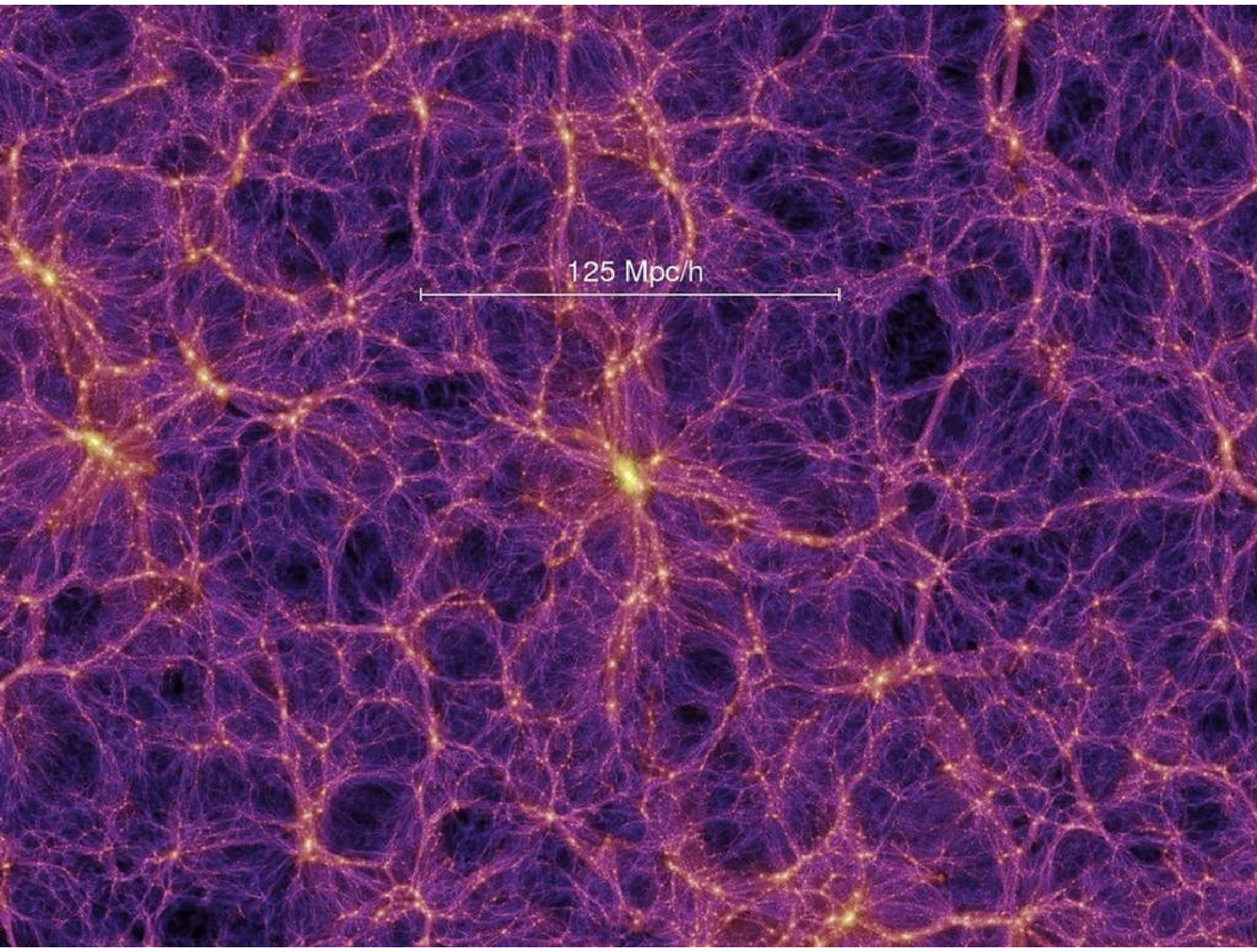}}
\hbox{\hspace{1mm}%
\epsfxsize=1.85in
\epsffile{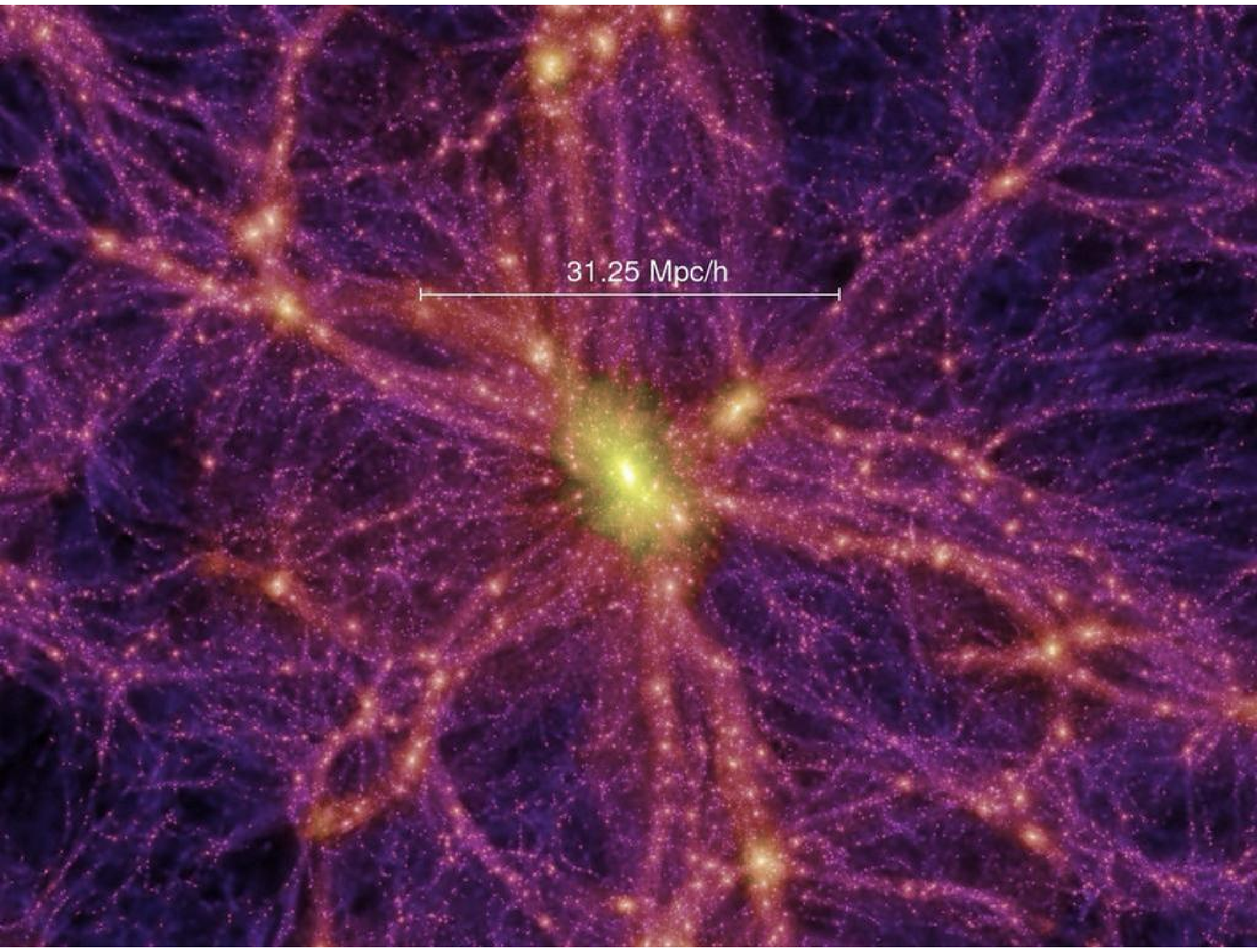}}
\hbox{
\epsfxsize=1.85in
\epsffile{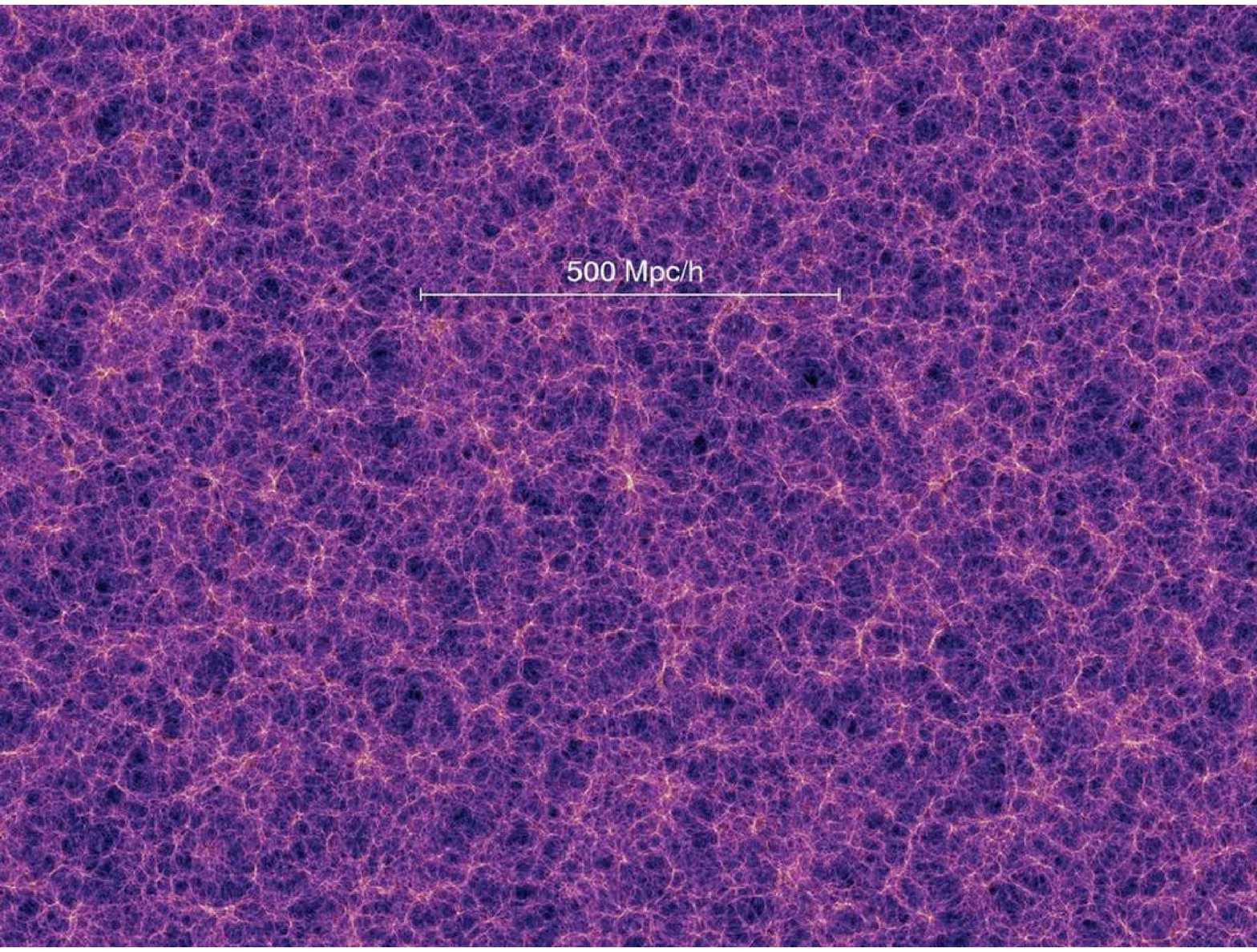}}
\hbox{\hspace{1mm}%
\epsfxsize=1.85in
\epsffile{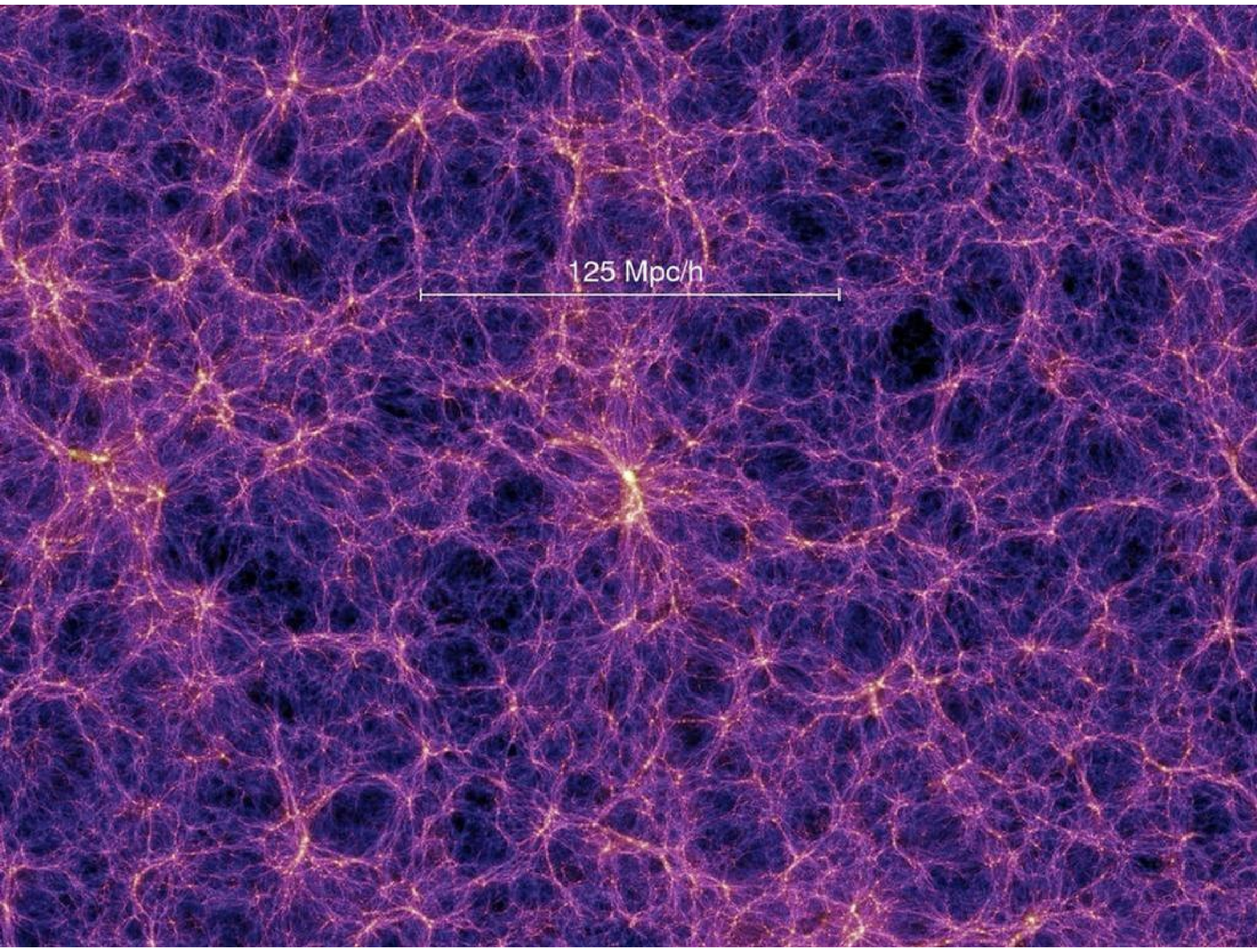}}
\hbox{\hspace{1mm}%
\epsfxsize=1.85in
\epsffile{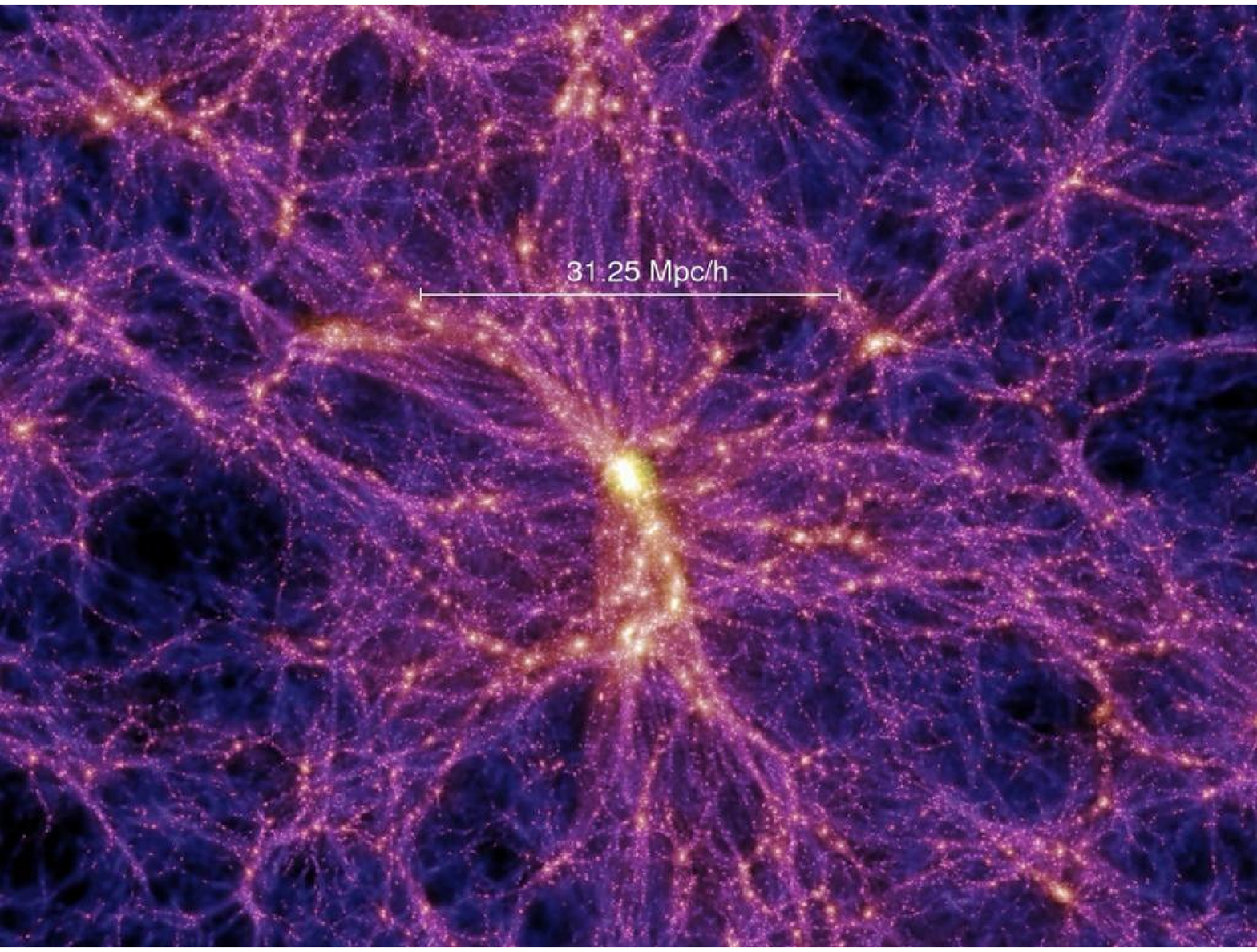}
\put(-413,108){Redshift $z=0$ ($t = 13.6$ Gyr)}}
\hbox{
\epsfxsize=1.85in
\epsffile{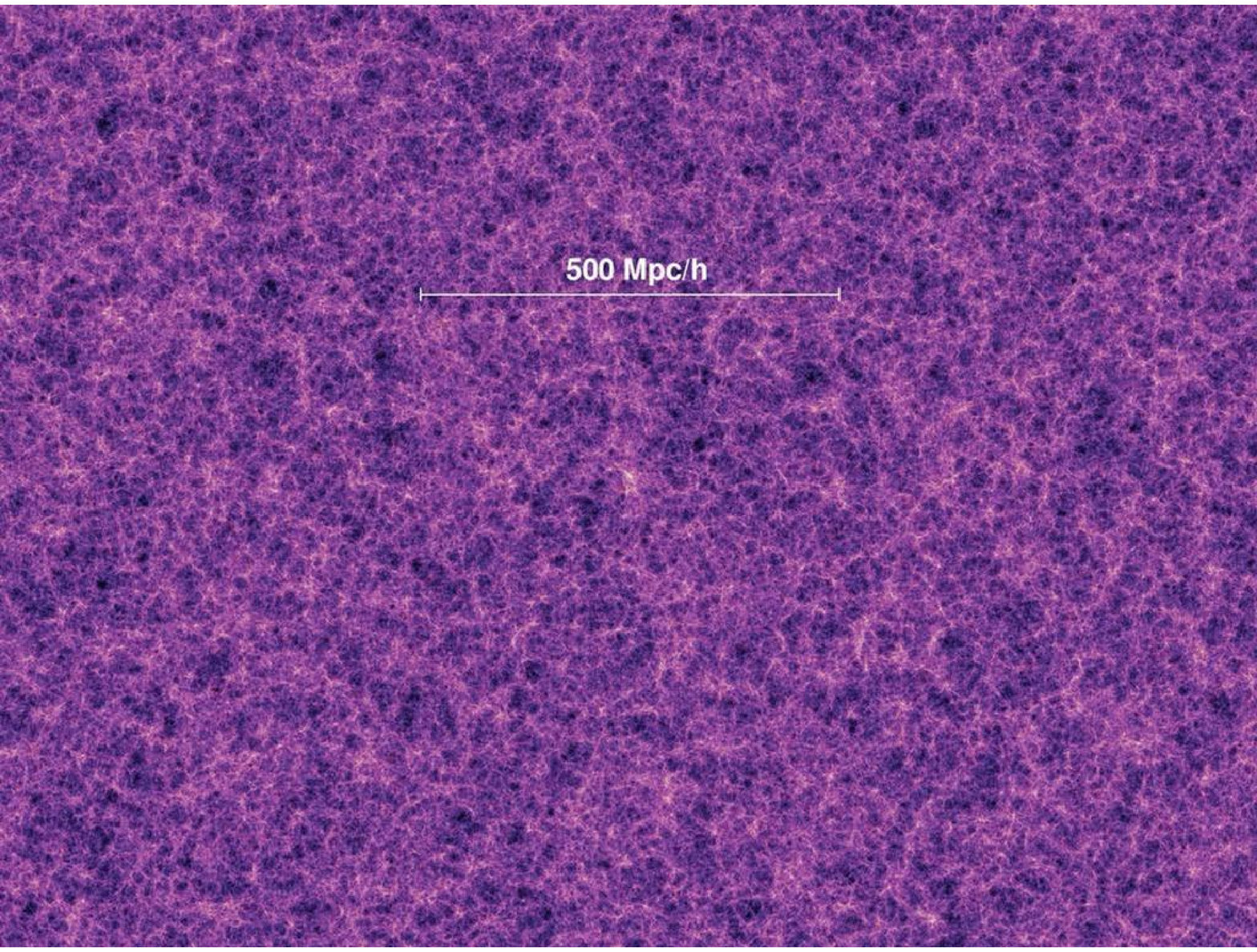}}
\hbox{\hspace{1mm}%
\epsfxsize=1.85in
\epsffile{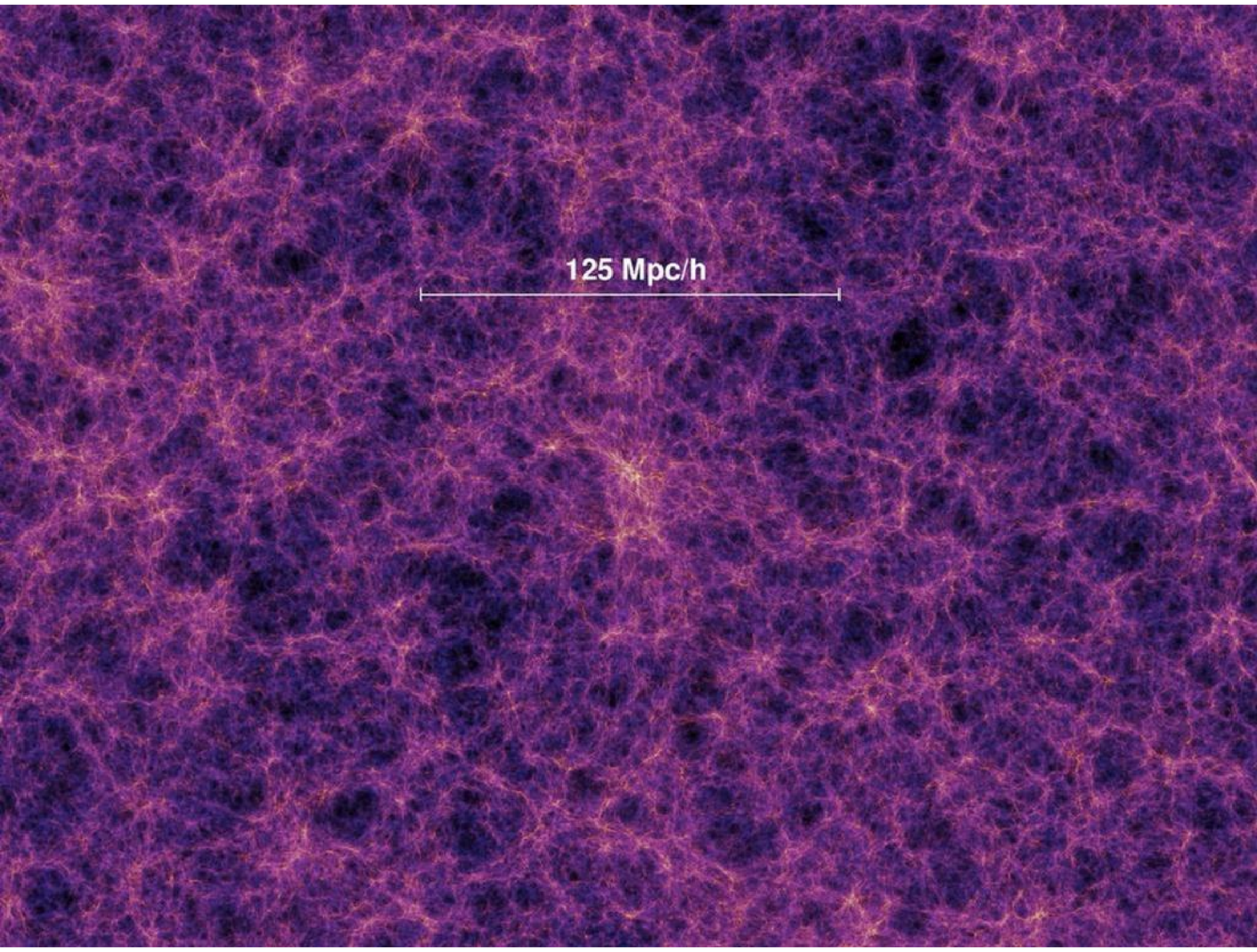}}
\hbox{\hspace{1mm}%
\epsfxsize=1.85in
\epsffile{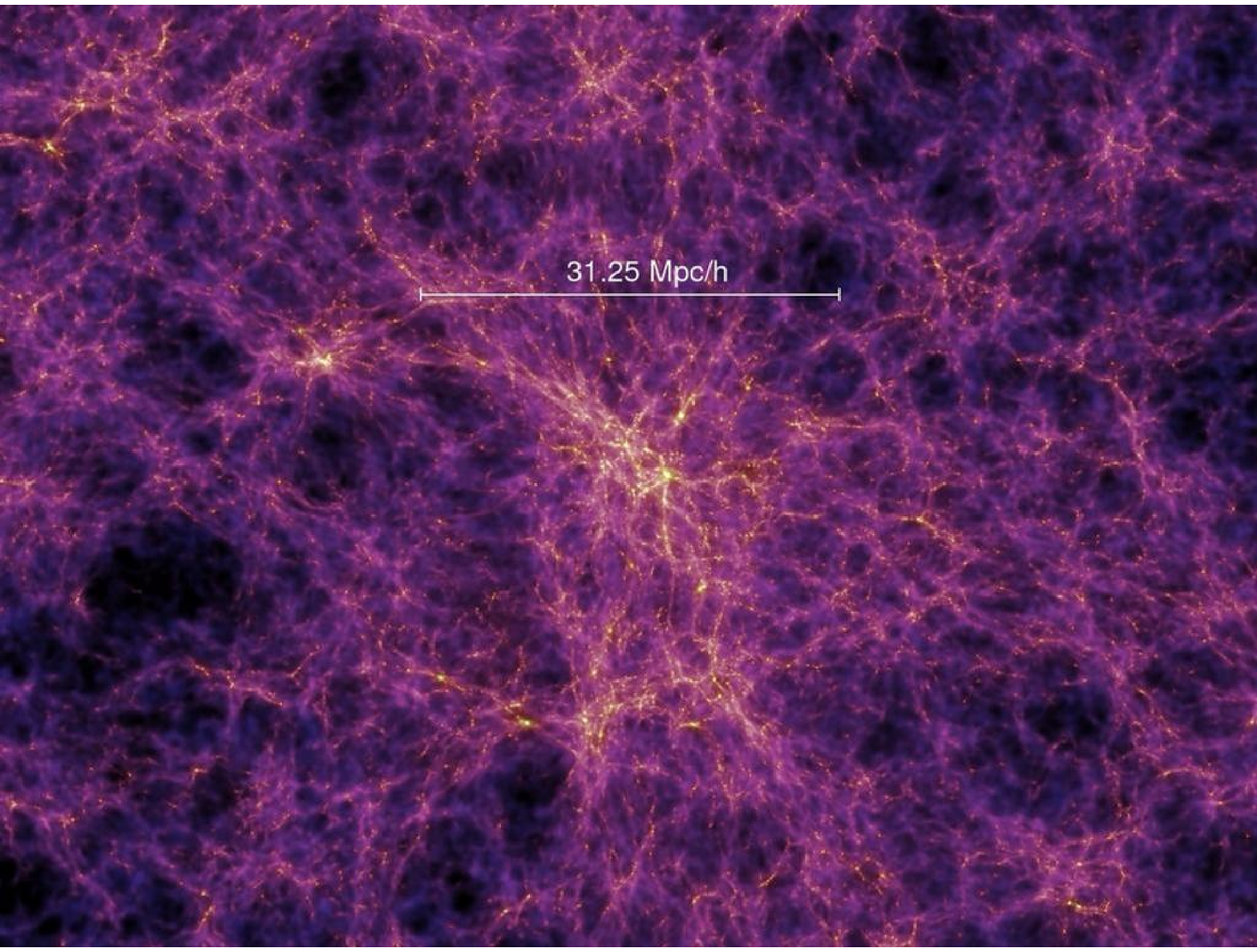}
\put(-413,108){Redshift $z=1.4$ ($t = 4.7$ Gyr)}}
\hbox{
\epsfxsize=1.85in
\epsffile{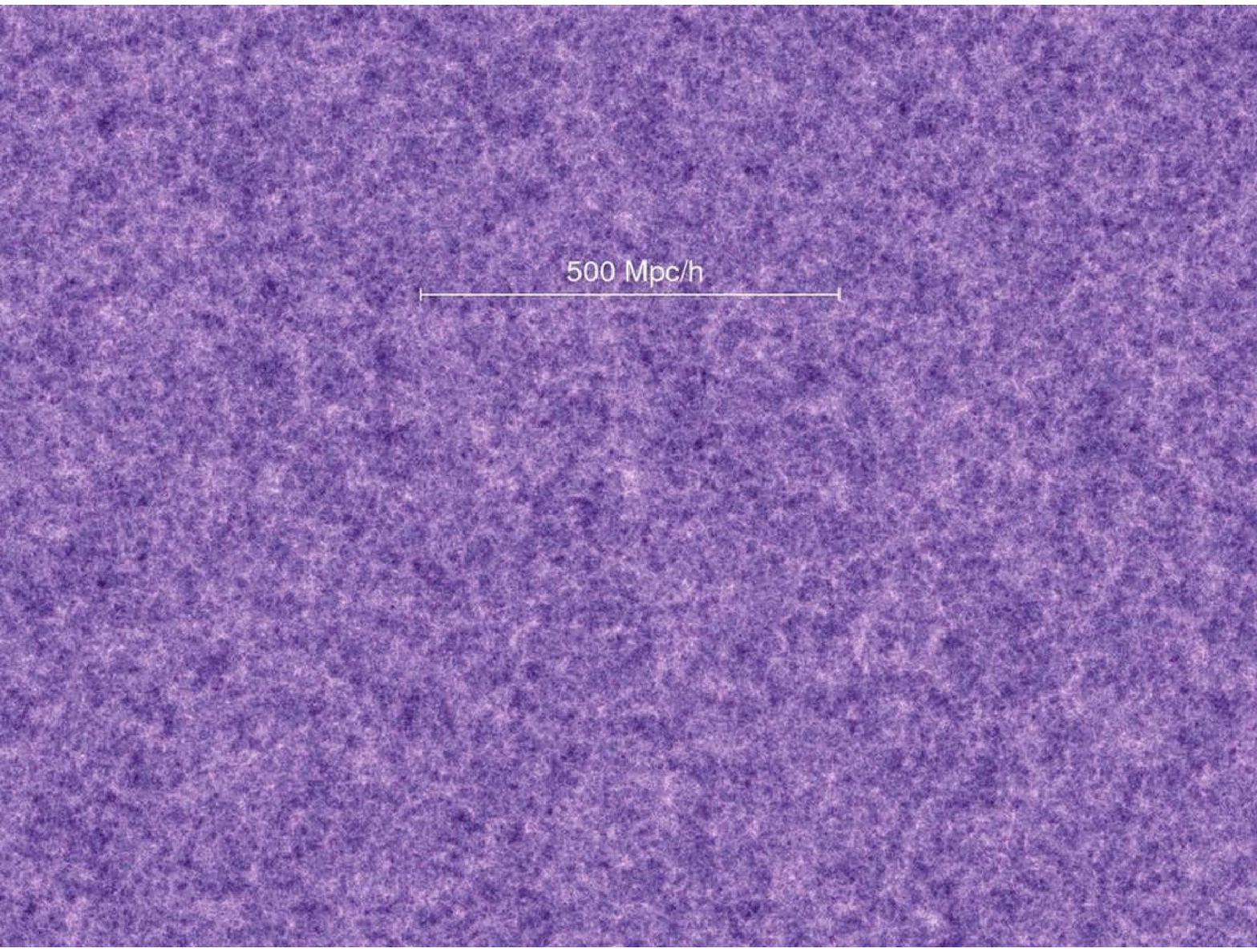}}
\hbox{\hspace{1mm}%
\epsfxsize=1.85in
\epsffile{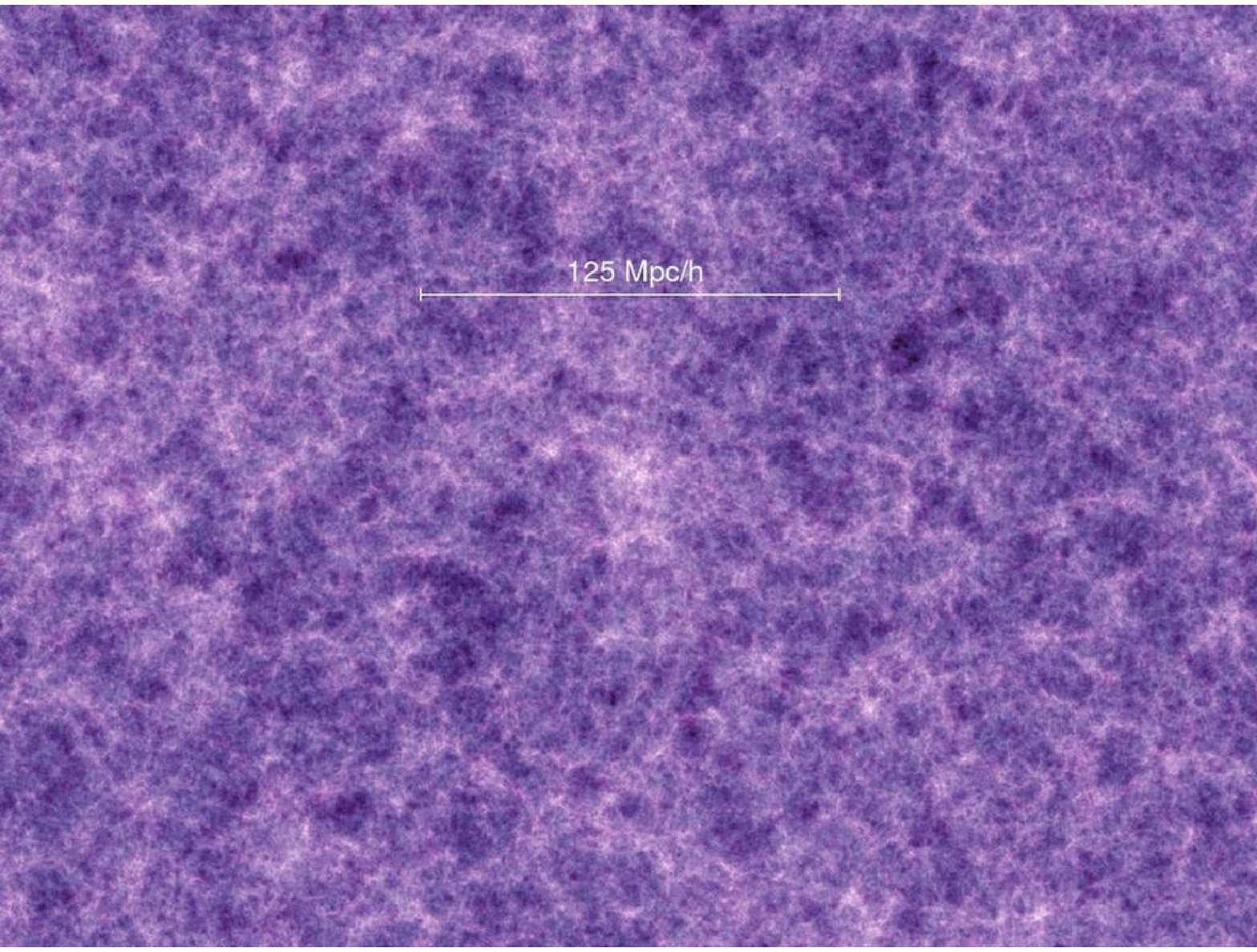}}
\hbox{\hspace{1mm}%
\epsfxsize=1.85in
\epsffile{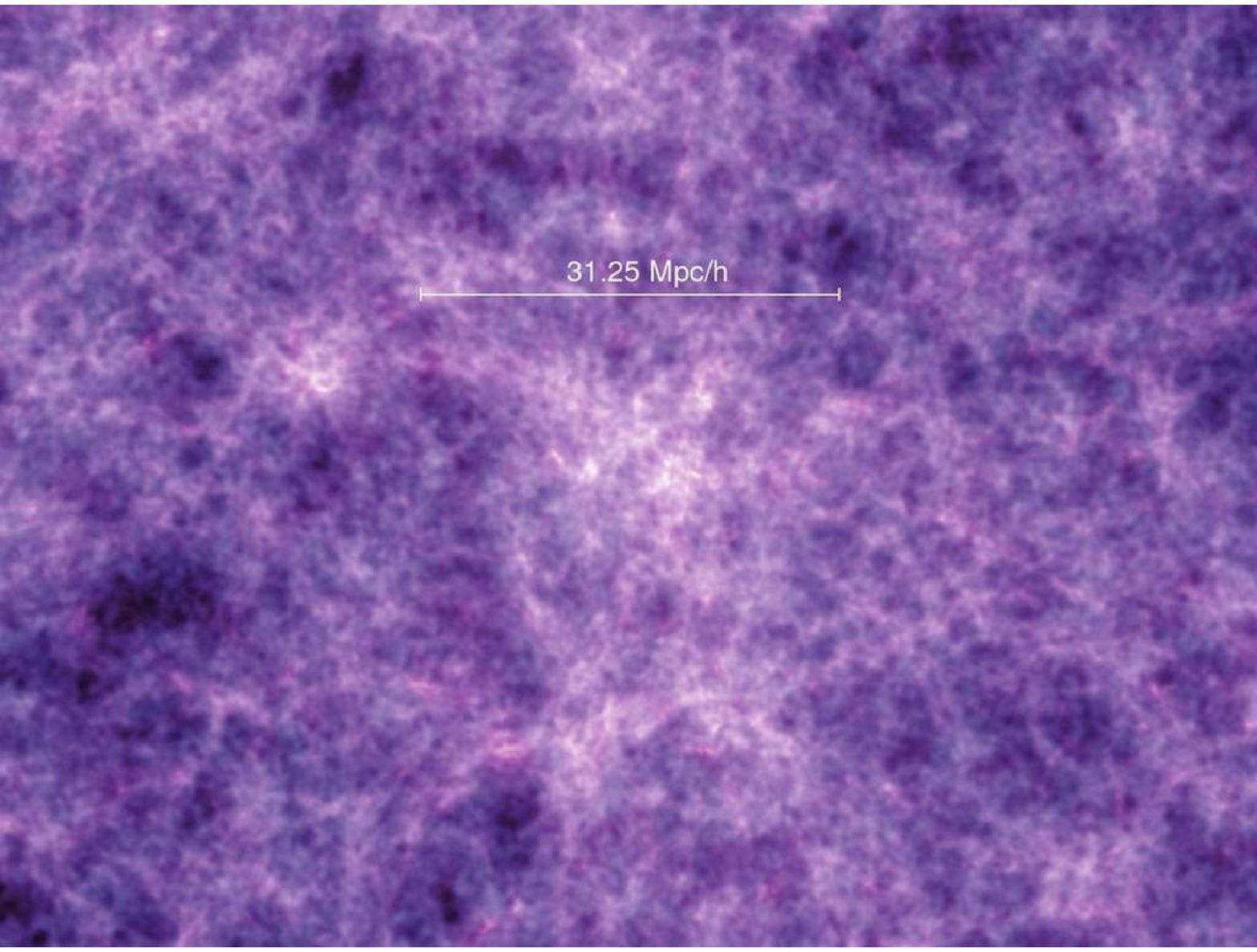}
\put(-413,108){Redshift $z=5.7$ ($t = 1.0$ Gyr)}
\put(-413,-10){Redshift $z=18.3$ ($t = 0.21$ Gyr)}}
\caption{These slices show the dark matter distribution in the Universe, based on the Millennium Simulation \cite{springel}: the largest N-body simulation carried out thus far (more than $10^{10}$ particles). By zooming in on a massive cluster of galaxies, the slices highlight the morphology of the structure on different scales, and the large dynamic range of the simulation ($10^5$ per dimension in 3D). The zoom extends from scales of a few Gpc down to resolved substructures as small as ~100 Mpc. The slices through the density field are all 15 Mpc/$h$ thick, with $h = 0.71 \pm 0.03$ \cite{observation}. For each redshift, three panels are displayed. Subsequent panels zoom in by a factor of four with respect to the previous ones. This filamentary structure can be qualitatively compared with that observed by the 2-degree Field Galaxy Redshift Survey (\hbox{see Fig. \ref{2dfgs}}) (Courtesy of the Virgo Consortium for cosmological supercomputer simulations \cite{filaments}). \label{growing}}
\end{center}
\end{figure}

\begin{figure}[t]
\begin{center}
\includegraphics[width=15cm]{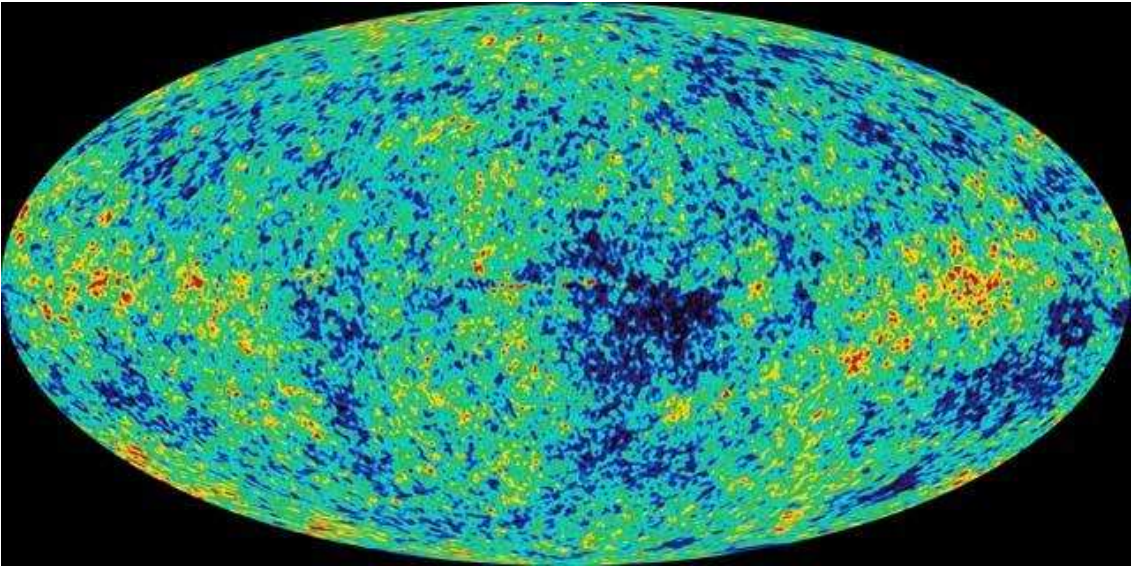}
\caption{CMB temperature anisotropies as seen by the WMAP satellite \cite{bennett}. The oval shape is a projection to display the whole sky. The temperature anisotropies are found to be of the order of 1 part in $10^5$. The background temperature is $T_0 = 2.725 \pm 0.002$ K; regions at that temperature are in very light blue. The hottest regions (in red) correspond to $\Delta T = 200 \mu{\rm K}$. The coldest regions (in very dark blue) correspond to $\Delta T = -200 \mu{\rm K}$ (Courtesy of the NASA/WMAP Science Team \cite{wmap}). \label{wmapsky}}
\end{center}
\end{figure}

\begin{figure}[t]
\begin{center}
\includegraphics[width=15cm]{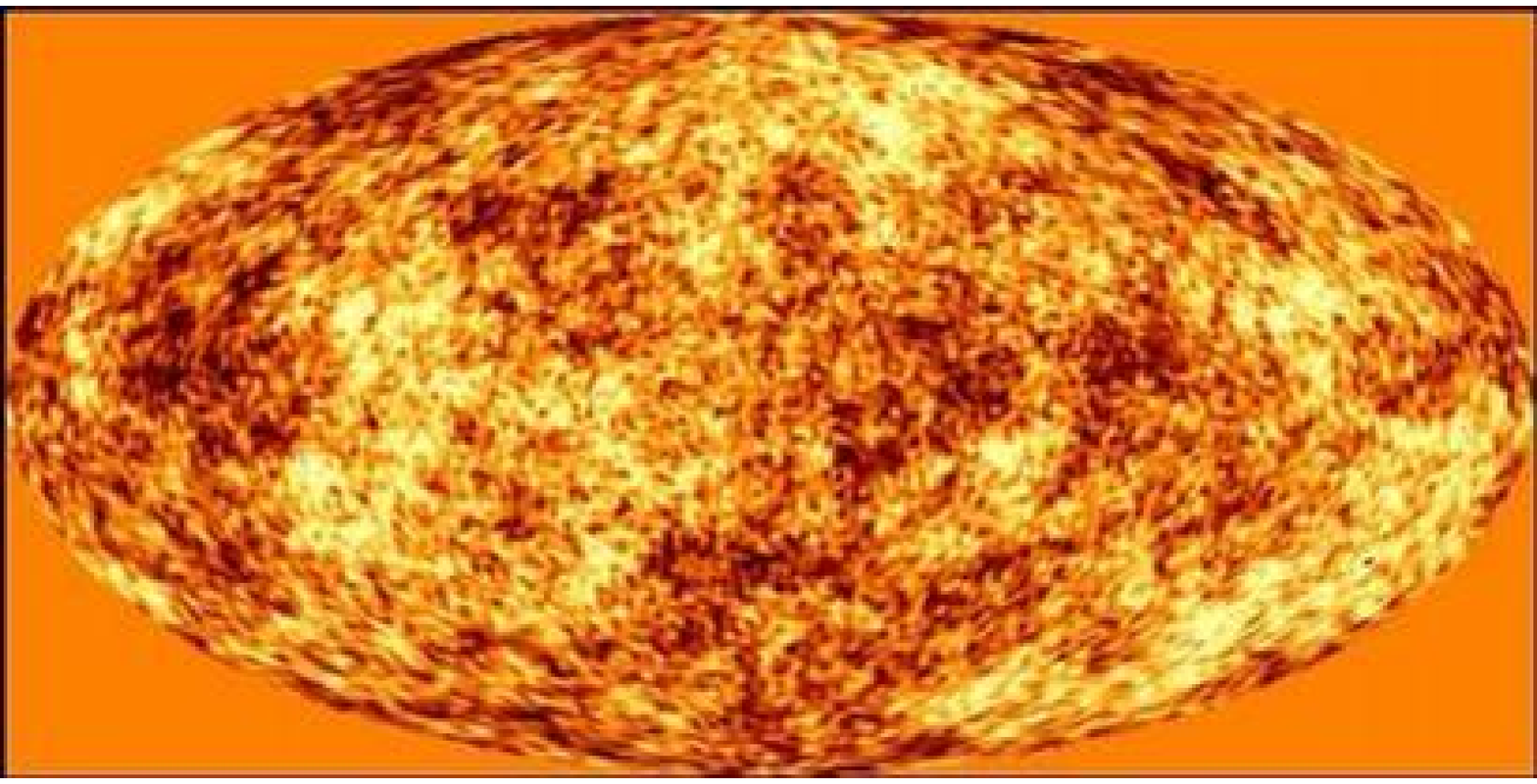}
\caption{Simulation of the CMB temperature anisotropies as seen by the PLANCK satellite. PLANCK will provide a map of the CMB field at all angular resolutions greater than 10 arcminutes and with a temperature resolution of the order of 1 part in $10^6$ (ten times better than WMAP) (Courtesy of ESA's PLANCK mission \cite{planck}). \label{plancksky}}
\end{center}
\end{figure}

The anisotropies in the CMB temperature\footnote{From now on, and unless otherwise stated, the perturbation $\delta y$ in any quantity $y$ will be regarded as first-order in cosmological perturbation theory. Unperturbed quantities will be denoted by a subscript 0 unless otherwise stated.} $\delta T/T_0$ are directly related to the perturbations in the energy density $\delta \rho/\rho_0$ at the time of recombination (Sachs-Wolfe effect), whose primarily origin is the stretched quantum fluctuations of one or several scalar fields $\phi_i$ that fill the Universe during inflation \cite{book,sachs}\footnote{In Eqs. (\ref{connections1}) and (\ref{connections2}) all the quantities are evaluated at time of last scattering, being $a$ the global expansion parameter and $H \equiv \dot{a}/a$ the global Hubble parameter. A dot means derivative with respect to the cosmic time. The subscripts $k$ stand for the fourier modes with comoving wavenumber $k$.}:
\be
\left(\frac{\delta T}{T_0}\right)_k = -\frac{1}{2}\left(\frac{aH}{k}\right)^2 \left(\frac{\delta \rho}{\rho_0}\right)_k \,. \label{connections1}
\ee
The perturbations in the energy density at the time of recombination can in turn be quantified by the
gauge-invariant primordial curvature perturbation $\zeta$ \cite{book,sachs}:
\be
\left(\frac{\delta \rho}{\rho_0}\right)_k = \frac{2}{5}\left(\frac{k}{aH}\right)^2 \zeta_k \,, \label{connections2}
\ee
which, on flat slices\footnote{A choice of coordinates defines a {\it threading} of spacetime into lines of fixed spatial coordinates $x_i$, and a {\it slicing} into constant time $t$ hypersurfaces. The flat slices are defined such that the intrinsic spatial curvature $^{(3)}R$ vanishes in those hypersurfaces.}, can be expressed in terms of only the fluctuations in the fields $\phi_i$. For instance, in the case of only one scalar field $\varphi$ present during inflation, \hbox{$\zeta$ \cite{bardeen,bst,mukhanov,mukhanovrep}} is given by
\be
\zeta = -H_{\rm inf} \frac{\delta \varphi}{\dot{\varphi}_0} \,,
\ee
where $H_{\rm inf}$ is the global Hubble parameter during inflation.
The curvature perturbation $\zeta$ is a convenient quantity to describe the primordial perturbations since it is conserved on superhorizon scales ($k \ll aH_{\rm inf}$), as long as the pressure is a unique function of the energy density \cite{sasaki1,rs1,llmw}, and it is well defined even after the scalar fields $\phi_i$ have decayed. We will define $\zeta$ in a rigorous way in Chapter \ref{mechanismsiandc}.

The scalar fields $\phi_i$ we have been talking about might or might not have dominated the energy density during inflation and, therefore, driven the inflationary stage prior to the Big-Bang\footnote{Although the waterfall field in hybrid inflation dominates the energy density, it does not drive inflation. This is not in contradiction with our previous statement since, in the strict sense, the waterfall field fluctuations are suppressed as the waterfall mass is much bigger than $H_*$
(the star `$\ast$' denoting the global Hubble parameter evaluated a few Hubble times after horizon exit) \cite{book,linde90,linde91,linde94,treview}.}.
Our adopted definition for the {\it inflaton} field in this thesis will be the light field $\varphi$ that dominates the energy density during inflation\footnote{By light field we mean a field $\phi$ whose mass $m_\phi$ is much less than $H_*$.} and drives the exponential expansion. This field in most cases parameterises the distance along the inflationary trajectories. Until recently, the most widely known and accepted scenario for the origin of the density perturbations identified the inflaton with the scalar field whose fluctuations were responsible for the primordial density perturbations. This scenario, called {\it the inflaton scenario}\footnote{For simplicity we will just consider single-component inflationary models of the slow-roll variety in the present introduction and in the following chapter. The multi-component case, which contains one inflaton and one or more `light non-inflaton fields', will be considered in Chapters \ref{fourthpaper} and \ref{fifthpaper}.} \cite{albste,book,linde82,treview}, describes very well the properties of $\zeta$, leading to an almost scale invariant power spectrum
\be
\mathcal{P}_\zeta (k) \equiv A_\zeta^2 \left(\frac{k}{aH_{\rm inf}}\right)^{n_\zeta} \,, \label{spectrumzeta}
\ee
which is defined by the statistical average
\be
\langle \zeta_{{\bf k}_1} \zeta_{{\bf k}_2} \rangle \equiv \frac{2\pi^2}{k^3} \delta^3({\bf k}_1 + {\bf k}_2) \mathcal{P}_\zeta (k) \,,
\ee
calculated over an ensemble of universes.
The amplitude $A_\zeta$ and spectral index $n_\zeta$ in Eq. (\ref{spectrumzeta}) \footnote{In this thesis we will use natural units such that $c = \hbar = k_B =1$, and Newton's gravitational constant given by $8\pi G \equiv m_P^{-2}$, with $m_P = 2.436 \times 10^{18}$ GeV being the reduced Planck mass.}:
\bea
A_\zeta &\simeq& - \frac{H_*}{\sqrt{8 \varepsilon} \pi m_P} \,, \label{AmInflation} \\
n_\zeta &=& 2\eta_\varphi - 6\varepsilon \,,
\eea
are functions of the slow-roll parameters
\bea
\varepsilon &\equiv& - \left(\frac{\dot{H}_{\rm inf}}{H_{\rm inf}^2}\right)_\ast \,, \label{srepsilon} \\
\eta_\varphi &\equiv& \varepsilon - \frac{\ddot{\varphi}_0}{H_{\rm inf}\dot{\varphi}_0} \label{sreta}\,,
\eea
that characterize the inflationary behaviour and satisfy the slow-roll conditions \mbox{$\varepsilon \ll 1$} and $|\eta_\varphi| \ll 1$
\cite{liddle,book,treview,riotto}.
The scale of inflation, given by $H_*$, is not completely determined in the inflaton scenario by the already measured amplitude $|A_\zeta|$ \cite{observation}, since the slow-roll parameter $\varepsilon$ is unknown (except for the upper bound $\varepsilon \lsim 0.01$ \cite{observation}).

The next statistical significant quantity after $\mathcal{P}_\zeta(k)$, the bispectrum $B_\zeta(k_1,k_2,k_3)$ defined by the statistical average
\be
\langle \zeta_{{\bf k}_1} \zeta_{{\bf k}_2} \zeta_{{\bf k}_3} \rangle \equiv (2\pi)^{-3/2} B_\zeta(k_1,k_2,k_3) \delta^3({\bf k}_1 + {\bf k}_2 + {\bf k}_3) \,,
\ee
is also well described in the inflaton scenario as its normalisation $\fnl$, defined by
\be
B_\zeta(k_1,k_2,k_3) \equiv  -\frac{6}{5} \fnl \left[\frac{2\pi^2}{k_1^3} \mathcal{P}_\zeta(k_1) \frac{2\pi^2}{k_2^3} \mathcal{P}_\zeta(k_2) + {\rm cyclic \ permutations}\right] \,,
\ee
is suppressed by $\varepsilon$ and $\eta_\varphi$ \cite{maldacena} describing a highly gaussian set of perturbations:
\be
\fnl = \frac{5}{12} [2\eta_\varphi - 6\varepsilon - 2\epsilon f(k_1,k_2,k_3)] \,. \label{fnlinflaton}
\ee
The value of the scale dependent function $f$ in the previous expression lies in the range $0 \leq f \leq 5/6$, being precisely determined by the respective wavevector configuration \cite{maldacena}.

We make a couple of observations \cite{abbott,fabbri,liddle,book,rubakov} which are valid not only for the inflaton scenario but also for the curvaton one discussed below. First, the amplitude $A_T$ of the power spectrum of gravitational waves
\be
\mathcal{P}_T (k) \equiv A_T^2 \left(\frac{k}{aH_{\rm inf}}\right)^{n_T} \,,
\ee
which is defined by the statistical average
\be
\sum_{ij} \langle h^{ij}_{{\bf k}_1} h^{ij}_{{\bf k}_2} \rangle \equiv \frac{2\pi^2}{k^3} \delta^3({\bf k}_1 + {\bf k}_2) \mathcal{P}_T (k) \,,
\ee
where $h_{ij}$ is the tensor perturbation in the perturbed metric tensor,
depends only on the scale of inflation (given by $H_\ast$):
\be
A_T \simeq \frac{\sqrt{2}H_*}{\pi m_P} \,.
\ee
Second, its tilt $n_T$ is a function only of $\varepsilon$:
\be
n_T = -2\varepsilon \,.
\ee
In view of our previous discussion about $\mathcal{P}_\zeta (k)$, we can write a  consistency relation involving $A_\zeta$, $A_T$, and $n_T$, which presents itself as a nice prediction of the inflaton scenario:
\be
r_{T\zeta} \equiv \frac{A_T^2}{A_\zeta^2} = -8n_T \,. \label{consistency}
\ee
Its confirmation, as well as
a negative detection of non-gaussianity
by both WMAP \cite{komatsu03,spergel,wmap} and PLANCK \cite{spergel,planck} satellites, would give strong support to the inflaton scenario as the correct framework to understand the origin of the large-scale structure in the Universe. We will discuss in more detail the inflaton scenario in Chapter \ref{mechanismsiandc}.

An alternative to the inflaton scenario is when
the weakly coupled light field $\sigma$, whose fluctuations are responsible for the primordial density perturbations, does {\it not} dominate the energy density, and therefore does not drive inflation. This scenario is called {\it the curvaton scenario} \cite{lyth03c,lyth02,moroi01b} (see also Refs. \cite{moroi01a,earlier1,earlier2}), and the field $\sigma$ receives the name {\it curvaton}. Introduced in 2002, this scenario describes also very well the properties of $\zeta$, with an almost scale invariant power spectrum \mbox{$\mathcal{P}_\zeta(k) \equiv A_\zeta^2 (k/aH_{\rm inf})^{n_\zeta}$} whose spectral index $n_\zeta$ written as
\be
n_\zeta = 2\eta_\sigma - 2\varepsilon \,,
\ee
is function of $\varepsilon$ and
\be
\eta_\sigma \equiv \frac{m_P^2}{V} \frac{\partial^2 V}{\partial \sigma^2_0} \approx \frac{m_\sigma^2}{3H_\ast^2} \ll 1 \,,
\ee
being $V$ the scalar potential,
and whose amplitude $A_\zeta$ is given by $H_\ast$, the unperturbed component of the curvaton field during inflation $\sigma_\ast$, and the fractional global curvaton energy density \mbox{$\Omega_{\rm dec} \equiv \rho_{\sigma_0} / \rho_{\rm total_0}$} just before the curvaton decay:
\be
A_\zeta \simeq \frac{H_\ast \Omega_{\rm dec}}{3\pi \sigma_\ast} \,.
\ee
The scale of inflation in the curvaton scenario, given by $H_\ast$, does not depend on $\varepsilon$ nor on $\eta_\sigma$, instead it depends on the free parameters $\sigma_*$ and $\Omega_{\rm dec}$. A consequence from the previous expression is that there is no analogous to the consistency relation [\hbox{c.f. Eq. (\ref{consistency})}] for the curvaton scenario. However, there are distinctive non-gaussian signatures that can allow us to distinguish this model from other scenarios for the origin of the large-scale structure in the Universe (see below) \cite{review,curvaton,lyth04,lr,lyth03c}.
The curvaton scenario will be studied in detail in Chapter \ref{mechanismsiandc}.

Perhaps the main motivation for the introduction of the curvaton scenario is that it liberates the inflaton field from the generation of the primordial perturbations \cite{liber,mota,motayo}. This is particularly good from the particle physics point of view since the intrinsic difficulty at embedding the inflaton scenario in a particle physics model is greatly alleviated. Indeed, in the inflaton scenario, the energy scale of inflation is likely to be quite high\footnote{The upper bound on $H_*$ can be obtained from Eq. (\ref{AmInflation}) and the current bound \cite{observation} on the slow-roll parameter
$\varepsilon \lsim 0.01$.} ($H_* \lsim 10^{14}$ GeV) \cite{lyth84a} in order to produce the required level of primordial perturbations $|A_\zeta| \approx 5 \times 10^{-5}$ \cite{observation}\footnote{Nevertheless there are examples where an $\varepsilon$ parameter of order $\sim 10^{-24}$ is naturally obtained, so that the right amount of primordial perturbations is generated for $H_\ast \sim 10^3$ GeV (see e.g. Ref.~\cite{snat}).}. This makes very difficult the identification of the inflaton field with one of the scalar fields present in the supersymmetry (SUSY) breaking sector or with one of the Minimal Supersymmetric Standard Model (MSSM) flat directions. This is because the characteristic SUSY energy scale, which depends on the symmetry breaking scheme, goes typically from \mbox{$H_* \sim 10^{-2}$ GeV} to $H_* \sim 10^5$ GeV \cite{lythlsi,lyth04}. In contrast, in the curvaton scenario, the energy scale has to be much lower, satisfying \mbox{$H_* \lsim 10^{12}$ GeV} \cite{lyth03c,lyth02,moroi01b}, so that the inflaton field does not contribute to the curvature perturbation [\hbox{c.f. Eq. (\ref{AmInflation})}]\footnote{An interesting scenario is when both the inflaton and the curvaton fields contribute to $\zeta$ \cite{ferrer,langlois}. This is the case where the curvaton starts oscillating around the minimum of its potential when it already contributes significantly to the total energy density $\rho_{\rm total}$. The upper bound \mbox{$H_* \lsim 10^{12}$ GeV} is, in this case, therefore relaxed. In this thesis we will consider only the standard curvaton scenario, where the inflaton field does not contribute to $\zeta$ and the curvaton oscillations begin when the curvaton energy density $\rho_\sigma$ is still subdominant.}. The last bound can be taken as an anti-smoking gun for the curvaton scenario because a positive gravitational wave signal would require $H_* \gsim 10^{12}$ GeV \cite{CMB1,CMB2,CMB3,seto,sigurdson,smith}, ruling out the curvaton as the source of primordial density fluctuations. Some people can see this as a bad feature of the scenario in question as it sends an unpromising message to all those who are making big efforts to detect gravitational waves. However, as described in Ref. \cite{pilo}, the curvaton scenario may well be consistent with detectable gravitational waves as far as the inflaton is a `heavy' field ($m_\varphi \gsim H_\ast$), violating this way the slow-roll condition $|\eta_\varphi| \ll 1 \ \footnote{Of course, inflation in this case is not of the slow-roll variety. Some possibilities are fast-roll \cite{fastroll} or hilltop inflation \cite{lotfi}.}$, and suppressing the amplitude of the curvature perturbation produced by the inflaton itself.

One of our main concerns in this thesis is the possibility to accommodate low-scale inflation in the curvaton scenario. Unfortunately, even when the energy scale may in principle be greatly reduced compared to the inflaton scenario, the simplest curvaton model still requires a quite high inflationary energy scale satisfying \hbox{$H_\ast > 10^7$ GeV} \cite{lyth04}. This of course makes impossible the embedding of the inflaton field within the framework of a SUSY particle physics model. In Chapters \ref{firstpaper} and \ref{secondpaper} we explore two modifications to the curvaton model which can instead allow inflation at a low scale \cite{yeinzon,matsuda03,postma04,yeinzon04}. In the first modification \cite{yeinzon} the end of a second (thermal) inflationary stage \cite{lazarides86,lyth95,lyth96}, driven by the rolling of a flaton field $\chi$ coupled to the curvaton, makes the curvaton mass $m_\sigma$ increase suddenly at some moment after the end of inflation but before the onset of the curvaton oscillations. This proposal can work but not in a completely natural way. Nevertheless, we show that inflation with $H_\ast$ as low as \mbox{1 TeV} or lower is possible to attain. In the second modification \cite{postma04,yeinzon04} the increment in $m_\sigma$ at the end of the thermal inflation era is so huge that the decay rate overtakes the Hubble parameter and the curvaton field decays immediately. The advantage of this second modification is that low scale inflation is achieved for more natural values in the relevant parameter space. 

Aside from the previously introduced theoretical aspects of the origin of the large-scale structure in the Universe, we also study some of its statistical aspects. Conversely to the inflaton scenario [c.f. Eq. (\ref{fnlinflaton})], $\zeta$ in the simplest curvaton model may present a sizable non-gaussian component if the curvaton does not dominate the energy density before decaying \cite{review,curvaton,lr,lyth03c}. More specifically, according to the expression
\be
\fnl = \frac{5}{3} + \frac{5}{6} r - \frac{5}{4r} \,, \label{fnlcurvatone}
\ee
which is valid in the curvaton scenario, $|\fnl| \gg 1$ is obtained if $r$ is very small. In the previous expression $r$ is defined by
\be
r \equiv \frac{3\rho_{\sigma_0}}{4\rho_{r_0} + 3\rho_{\sigma_0}} \,, \label{fnlf}
\ee
and is
evaluated just before the curvaton decay (being $\rho_r$ the global radiation energy density).
This is of extreme importance since the next WMAP data release \cite{komatsu03,spergel,wmap}, or in its defect the future PLANCK satellite data \cite{spergel,planck}, will either detect non-gaussianity or put strong constraints on $\fnl$, offering the possibility of successfully discriminating among the different inflaton and curvaton models. The current constraint on $\fnl$, according to the first-year WMAP data, is $|\fnl| \lsim 10^2$ \cite{komatsu03}. The next data release is expected to lower this bound by one order of magnitude or so \cite{spergel}.

The non-linearity parameter $\fnl$, if independent of position, is closely related to the second-order curvature perturbation $\zeta_2$ defined by
\bea
\zeta({\bf x}) &\equiv& \zeta_1({\bf x}) + \frac{1}{2} \zeta_2({\bf x}) \nonumber \\
&\equiv& \zeta_g({\bf x}) - \frac{3}{5} \fnl (\zeta_g^2({\bf x}) - \langle \zeta_g^2 \rangle) \,, \label{zeta2fnl}
\eea
where $\zeta_g$ is gaussian with $\langle \zeta_g \rangle = 0$. 
In connection with this issue we point out that several conserved and/or gauge invariant quantities described as the second-order curvature perturbation have been given in the literature \cite{acquaviva,lyth,maldacena,malik}. In Chapter \ref{fourthpaper} we revisit various scenarios for the generation of second-order non-gaussianity in $\zeta$, employing for the first time a unified notation and focusing on $f\sub{NL}$ \cite{lr}. When $\zeta$ first appears a few Hubble times after horizon exit, $|f\sub{NL}|$ is much less than $1$ and is, therefore, negligible. Thereafter $\zeta$ (and hence $f\sub{NL}$) is conserved as long as the pressure is a unique function of the energy density (adiabatic pressure) \cite{sasaki1,rs1,llmw}. Non-adiabatic pressure comes presumably only from the effect of fields,  other than  the one pointing along the inflationary trajectory, which are light during inflation (light non-inflaton fields) \cite{wandsreview,gordon}. Our expectation is that, although during single-component inflation $\fnl$ is constant, multi-component inflation might generate $|f\sub{NL}|\sim 1$ or bigger. We mention some recent proposals where non-gaussianity can be generated during the preheating stage following inflation \cite{ev1,ev2,enqvistnew}, and conjecture that preheating  can  affect $\fnl$ only in atypical scenarios where it involves light non-inflaton fields \cite{steve2,bdck,steve1,enqvistnew,kr}. We also study the curvaton scenario and derive Eq. (\ref{fnlcurvatone}), showing that the simplest model typically gives $\fnl\ll -1$ or $\fnl=+5/4$. The inhomogeneous reheating scenario \cite{inhomog1,inhomog3,modreh} (see also Refs. \cite{emp,matarrese,maz,mazpo03,inhomog4,inhomog2}), where $\zeta$ is generated by the inhomogeneities in the inflaton decay rate during reheating, is quickly reviewed showing that it can give a wide range of values for $\fnl$ \cite{inhomog4,inhomog2}. One important conclusion from this chapter is that it will be crucial to calculate  the precise observational limit on $\fnl$ using second order theory in case that, unless there is a detection, observation could eventually provide a limit $|\fnl|\lsim 1$ \cite{spergel}.

A new and interesting proposal is the extension to second order of the $\delta N$ formalism \cite{ss} (see also Refs. \cite{starobinsky1,starobinsky2}), used initially to calculate $\zeta$ at first order and the spectral index in multi-component slow-roll models of inflation \cite{ss}\footnote{For an extension to multi-component non slow-roll models see Ref. \cite{lee}.}. In this formalism $\zeta(t,{\bf x})$ is expressed as the perturbation in the amount of expansion
\be
N(t,{\bf x}) \equiv \ln \left[\frac{\tilde{a}(t,{\bf x})}{a(t_{\rm in})}\right] \,,
\ee
from an initially flat slice at time $t_{in}$ to a final slice of uniform energy density at time $t$:
\be
\zeta(t,{\bf x}) = \delta N \equiv N(t,{\bf x}) - N_0(t) \,.
\ee
Here
\be
N_0(t) \equiv \ln\left[\frac{a(t)}{a(t_{\rm in})}\right] \,,
\ee
is the unperturbed amount of expansion, and $\tilde{a} (t, {\bf x})$ is the local expansion parameter.
Thus, $\zeta$ up to second order is given by
\be
\zeta(t,{\bf x}) = \sum_i N_{,i}(t) \delta \phi_i + \frac{1}{2} \sum_{ij} N_{,ij}(t) \delta \phi_i \delta \phi_j \,,
\ee
where 
\bea
N_{,i} &=& \frac{\partial N}{\partial \phi_{i_0}} \,, \\
N_{,ij} &=& \frac{\partial^2 N}{\partial \phi_{i_0} \partial \phi_{j_0}} \,,
\eea
and the fields $\phi_i$, evaluated a few Hubble times after horizon exit, are those relevant for the generation of $\zeta$. In view of Eq. (\ref{zeta2fnl}), this presents as a powerful method to calculate $\fnl$ in any multi-component slow-roll inflationary model for the generation of $\zeta$ \cite{lr1}. In Chapter \ref{fifthpaper} we give for the first time this formalism, which allows us to extract all the stochastic  properties of $\zeta$ if the initial field perturbations are gaussian\footnote{In the case that the initial field perturbations are non-gaussian, there is an additional contribution to Eq. (\ref{magicf}) which is strongly wavevector dependent \cite{seery2}. This contribution is in any case very small \cite{ignacio}, being $\lsim 10^{-2}$.}. The elegance and power of this method lies in the fact that the calculation requires only the knowledge of the evolution of some family of unperturbed universes. The following formula is given for $\fnl$ in terms of $N(t,\phi_{i_0})$ and its derivatives \cite{bl,lr1}:
\be
-\frac35 \fnl = \frac{\sum_{ij}\ni\nj\nij}{2\[ \sum_i \ni^2 \]^2 } + \ln(kL) \frac{A_\zeta^2}  2 \frac{\sum_{ijk} \nij N_{,jk} N_{,ki}}{\[ \sum_i \ni^2  \]^3} \,, \label{magicf}
\ee
where $k^{-1}$ is a typical cosmological scale and $L$ is the size of the region within which the stochastic properties are specified, so that $\ln(kL) \sim 1$.
We apply the above formula to the Kadota and Stewart modular inflation model \cite{ks1,ks2}, the curvaton scenario \cite{lyth03c,lyth02,moroi01b}, and the multi-component `hybrid' inflation model of Enqvist and V$\ddot{{\rm a}}$ihk$\ddot{{\rm o}}$nen \cite{ev}. The relation of this formula to cosmological perturbation theory is also explained.

The conclusions of this thesis are drawn in Chapter \ref{conclusiones}.

\chapter{Two mechanisms for the origin of the large-scale structure} \label{mechanismsiandc}

\section{Introduction}

In the standard inflationary scenario \cite{albste,book,linde82,treview} a single scalar field, named the inflaton, is responsible
for the solution of the horizon, flatness, and unwanted relics problems, as well as for the origin of the large
scale structure seen in the observable Universe. This double mission for the inflaton field imposes strong constraints on the parameters of the inflationary models, leading to big intrinsic difficulties at building successful and realistic models of inflation \cite{book,treview}. To rescue the well motivated inflationary models that fail at generating the required level of primordial perturbations \cite{liber}, the inflaton field is left in charge of driving inflation {\it only}. The other task, the generation of the primordial perturbations, is assigned to a weakly coupled light field $\sigma$ {\it different} from the inflaton. This is the curvaton scenario \cite{lyth03c,lyth02,moroi01b} (see also Refs. \cite{moroi01a,earlier1,earlier2}), where the original curvature perturbation $\zeta_\sigma$, associated to and produced by $\sigma$ during inflation, is gradually transformed into the total curvature perturbation $\zeta$ \footnote{During inflation and until the start of the radiation dominated epoch just after reheating, $\zeta_\sigma$ is actually an {\it isocurvature} perturbation. The reason of the name {\it isocurvature} is because $\zeta_\sigma$ during that time does not contribute at all to the total curvature perturbation $\zeta$.}. The conversion process starts during the radiation dominated epoch that follows the reheating stage produced by the inflaton decay\footnote{The cause of the conversion process is the relative redshifting between the radiation and the curvaton fluid energy densities. The curvaton at this stage is considered a matter fluid since the period of its oscillations around the minimum of its potential is much less than the characteristic expansion time scale \cite{turner}. This makes the average curvaton pressure essentially zero.}.

This chapter will give some preliminary definitions and  basic facts about the inflaton and the curvaton scenarios, such as the first-order perturbations in the metric, the precise definition of $\zeta$, the slow-roll conditions, the characteristics of the spectrum of $\zeta$ in both the inflaton and the curvaton scenarios, and the spectrum of gravitational waves and the consequences derived from its possible detection. Having this information at hand, it will be easier to follow the main discussions of this thesis that are exposed in Chapters \ref{firstpaper} to \ref{fifthpaper}.

\section{Metric perturbations and the primordial curvature perturbation $\zeta$} \label{metricandzeta}

Our observable expanding Universe is well described as homogeneous and isotropic, down to scales of order of tens of Megaparsecs. The FRW line element
\be
ds^2 = -dt^2 + a^2(t) \left(\frac{dr^2}{1 - \kappa r^2} + r^2 d\theta^2 + r^2 \sin^2 \theta d\phi^2 \right) \,,
\ee
describes such an Universe \cite{friedman,robertson1,robertson2,robertson3,walker}, assuming that the galaxies are always at the same comoving spherical coordinates $r$, $\theta$, and $\phi$, and that the intergalactic space is continually increasing with time due to the expansion parameter (scale factor) $a(t)$. The proper time, once synchronized in all the galaxies, runs at the same rate everywhere, so the metric is well described by a global \hbox{{\it cosmic time} $t$}. Finally, the topology of the Universe can be classified by the parameter $\kappa$ as closed ($\kappa > 0$), open ($\kappa = 0$), or flat ($\kappa < 0$), according to whether the space is finite but unbounded, infinite with certain curvature, or infinite but strictly flat {\it a la Minkowski} \footnote{The actual value of $\kappa$ is conventional since it depends on the {\it chosen} value for the present expansion parameter $a_{\rm today}$.}.
The inflationary stage however shrinks the comoving horizon so rapidly that the present Universe looks so extremely flat. This makes the parameter $\kappa$ be completely irrelevant when going back in time, so that we can safely discard it when studying the inflationary and early Universe processes. Thus, and switching to cartesian coordinates $x^i$ and {\it conformal time} $\eta$ defined by $d\eta \equiv dt/a$, the FRW line element looks
\be
ds^2 = a^2(\eta) \left(-d\eta^2 + \delta_{ij} dx^i dx^j \right) \,. \label{unperturbedle}
\ee

\subsection{First-order perturbations in the FRW line element: classification and number of degrees of freedom} \label{freecounting}

The homogeneity and isotropy assumptions make a good description of the Universe, but they are not in any case perfect conditions.
The observed galaxy distribution and the temperature fluctuations in the CMB (see Figs. \ref{2dfgs} - \ref{plancksky})
reveal the importance of modifying the FRW line element so that it accounts for the small deviations from homogeneity and isotropy required to correctly describe such large-scale structures. The introduction of small inhomogeneities in the FRW line element, and the truncation at first-order of the perturbed Einstein equations, are well justified since the galaxy density contrast $\delta \rho/\rho_0$ and the CMB temperature anisotropies $\delta T/T_0$ are observed to be of order $10^{-5}$ \cite{bennett,lahav,percival}. In consequence, considering only scalar degrees of freedom, the most generic first-order perturbed line element reads \cite{bardeen,kodama,malikthesis,mukhanov,mukhanovrep,riotto,sasaki}
\be
ds^2 = a^2(\eta) \left[-(1 + 2\phi_G)d\eta^2 + 2 \, \partial_iB \, d\eta dx^i + ((1 - 2\psi)\delta_{ij} + D_{ij}E) dx^i dx^j \right] \,, \label{perturbedline}
\ee
where $D_{ij} \equiv \partial_i\partial_j - (1/3)\delta_{ij} \nabla^2$, and $\phi_G$, $B$, $\psi$, and $E$ are scalar quantities.

The peculiar form of the perturbed line element in Eq. (\ref{perturbedline}), as well as the right number of scalar degrees of freedom in that expression, are justified as follows:
\begin{itemize}
\item The full perturbed metric tensor is generically described by {\it scalar}, {\it vector}, and {\it tensor} perturbations.
\item The total number of degrees of freedom for a symmetric tensor in an $(n+1)$-dimensional spacetime is $(n+2)(n+1)/2$.
\item The $g_{00}$ entry can be written as $-(1 + 2\phi_G)$, where $\phi_G$ corresponds to a purely scalar perturbation (1 scalar degree of freedom).
\item The $g_{0i}$ entries can be parameterised by $B_i \equiv \partial_i B + \upsilon_i$, where $\vec{\nabla} \cdot \vec{\upsilon} = 0$. The term $\partial_i B$ corresponds to a purely scalar perturbation (1 scalar degree of freedom). The term $\upsilon_i$ corresponds to a purely vector perturbation ($n-1$ vector degrees of freedom because of the $\vec{\nabla} \cdot \vec{\upsilon} = 0$ constraint).
\item The $g_{ij}$ entries can be written as $(1 - 2\psi) \delta_{ij} + \Pi_{ij}$, where $\psi$ is a purely scalar perturbation (1 scalar degree of freedom) that accounts for the trace of $g_{ij}$, and $\Pi_{ij}$ is a symmetric traceless tensor.
\item $\Pi_{ij} = \Pi_{ij}^S + \Pi_{ij}^V + \Pi_{ij}^T$ is expressed in terms of a purely scalar perturbation (encoded in $\Pi_{ij}^S$), a purely vector perturbation (encoded in $\Pi_{ij}^V$), and a purely tensor perturbation (encoded in $\Pi_{ij}^T$).
\item $\Pi_{ij}^S$ can be written as $D_{ij}E$ because it is a symmetric traceless tensor corresponding to the purely scalar perturbation $E$ (1 scalar degree of freedom). 
\item $\Pi_{ij}^V$ can be written as $(\partial_i \Pi_j + \partial_j \Pi_i)/2$, where $\Pi_i$ is a purely vector perturbation satisfying $\vec{\nabla} \cdot \vec{\Pi} = 0$ ($n-1$ vector degrees of freedom).
\item $\Pi_{ij}^T$ corresponds to a purely tensor perturbation such that $\partial_i \Pi_{ij}^T = 0$. Considering the previous three items, and the fact that the number of degrees of freedom coming from a traceless symmetric tensor in an $n$-dimensional space is $(n+1)n/2 -1$, the number of tensor degrees of freedom is $(n-2)(n+1)/2$.
\item The total number of scalar degrees of freedom (4), vector degrees of freedom ($2(n-1)$), and tensor degrees of freedom ($(n-2)(n+1)/2$), add up to reproduce the total number of degrees of freedom of the metric tensor $g_{\mu \nu}$ given in the second item: $(n+2)(n+1)/2$.
\end{itemize}
Two very important facts follow from the first-order perturbed Einstein equations \cite{bardeen,kodama,malikthesis,mukhanovrep}. First, at first order the vector perturbations are decoupled from the scalar perturbations being anyway usually neglected due to their rapid decrease with time. Second, the tensor perturbations at first order also decouple from their scalar counterparts. These two facts, along with the parameterisation of the metric tensor discussed in the above items, justify the use of the \hbox{Eq. (\ref{perturbedline})} as the most generic first-order metric perturbed line element that describes the energy density (scalar) perturbations in the FRW spacetime.

\subsection{The curvature perturbation $\psi$ and its non-invariance under infinitesimal coordinate transformations} \label{zetaninvariant}

In the perturbed line element of Eq. (\ref{perturbedline}), $\psi$ represents the intrinsic spatial curvature $^{(3)}R$ on hypersurfaces of constant conformal time $\eta$ \cite{bardeen}:
\be
^{(3)}R = \frac{4}{a^2}\nabla^2 \psi \,,
\ee
where the operator $\nabla^2$ is the comoving Laplacian operator.
For this reason the quantity $\psi$ is usually referred to as the {\it curvature perturbation}. The curvature perturbation $\psi$, as well as the other scalar perturbations $\phi_G$, $B$, and $E$, are however not invariant under a coordinate transformation. Indeed, from the most general infinitesimal coordinate transformation\footnote{The vector component that should appear in the transformation of $x_i$ has been discarded. This is because the infinitesimal first-order vector shifts
contribute to the transformation rules of the vector perturbations {\it only}.}
\bea
\eta &\rightarrow& \eta + \xi(x^\mu) \,, \label{transtau} \\
x_i &\rightarrow& x_i + \partial_i \beta (x^\mu) \,, \label{transx}
\eea
the scalar perturbations $\phi_G$, $B$, $\psi$, and $E$ transform as \cite{bardeen}
\bea
\phi_G &\rightarrow& \phi_G - \xi' - \mathcal{H} \xi \,, \label{transphi} \\
B &\rightarrow& B + \xi - \beta' \,, \label{transB} \\
\psi &\rightarrow& \psi + \frac{1}{3}\nabla^2 \beta + \mathcal{H} \xi \,, \label{transpsi} \\
E &\rightarrow& E - 2\beta \,, \label{transE}
\eea
where a prime means derivation with respect to the conformal time $\eta$, and $\mathcal{H}$ is the global conformal Hubble parameter defined by $\mathcal{H} \equiv a'/a$.
The above transformation rules are under the physical proviso that the perturbed line element $ds^2$ in Eq. (\ref{perturbedline}) should remain invariant.

The parameters $\xi$ and $\beta$ that give account of the infinitesimal coordinate transformation in Eqs. (\ref{transtau}) and (\ref{transx}) can be adjusted (fixing the slicing and the threading) so that two of the four scalar perturbations in Eqs. (\ref{transphi}) to (\ref{transE}) vanish. The longitudinal gauge, for instance, corresponds to choose $B$ and $E$ equal to zero.
In this specific gauge the gravitational potential $\phi_G$ becomes equal to the curvature perturbation $\psi$ up to first order \cite{kodama,malikthesis,mukhanov,mukhanovrep,riotto} as long as the considered fluid is described by a perfect isotropic stress, examples of such a fluid being the inflaton and/or the curvaton fields. All of this leads us to say with confidence that the curvature perturbation $\psi$ really represents the effect of the inhomogeneities in the FRW spacetime and it is, therefore, the quantity to study.

\subsection{The gauge-invariant curvature perturbation $\zeta$} \label{acgauge}

To parameterise adequately the inhomogeneities in the FRW spacetime, we need a quantity invariant under the coordinate (gauge) transformations in Eqs. (\ref{transtau}) and (\ref{transx}). This is not completely possible to do, but we may define a quantity which is invariant under transformations in time only. This is done taking into account that, for any scalar quantity $f(x^\mu)$ different to $\phi_G$, $B$, $\psi$, and $E$, the transformation law in the associated perturbation $\delta f(x^\mu)$ is given by
\be
\delta f \rightarrow \delta f - f_0' \xi \,.
\ee

The first gauge invariant quantity that we may define is that which represents the curvature perturbation $\psi$ in the comoving slices, defined them as the constant time hypersurfaces where there is no flux of energy. Considering the inflaton field $\varphi$, the comoving slices coincide with those where $\varphi$ is uniform ($\delta \varphi_{\rm com} = 0$) \cite{kodama,sasaki1,mukhanovrep,llmw} so that the time translation $\xi_{\rm com}(x^\mu)$ required to go from a generic slice to the comoving slice is:
\be
\xi_{\rm com} = \frac{1}{\varphi_0'}[\delta \varphi - \delta\varphi_{\rm com}] = \frac{\delta \varphi}{\varphi_0'} \,.
\ee
Therefore the comoving curvature perturbation $\psi_{\rm com}$, denoted from now on as $-\zeta$, is written in terms of $\psi$ and $\varphi$ in the generic slice as
\bea
\zeta &=& -\psi - \mathcal{H}\xi_{\rm com} \nonumber \\
&=& -\psi - \mathcal{H} \frac{\delta \varphi}{\varphi_0'} \,. \label{zeta_def1}
\eea
The overall minus sign is just a convention, chosen in this thesis to match the agreed definition of $\zeta$ by most of the authors.

The second gauge invariant quantity that we may define represents the curvature perturbation $\psi$ in the slices of uniform energy density. Again, and following the same steps as before, we have to consider the infinitesimal time translation $\xi_{\rm uni}(x^\mu)$ required to go from a generic slice to the uniform energy density slice where $\delta \rho_{\rm uni} = 0$:
\be
\xi_{\rm uni} = \frac{1}{\rho_0'}[\delta \rho - \delta\rho_{\rm uni}] = \frac{\delta \rho}{\rho_0'} \,.
\ee
Thus, the curvature perturbation in the uniform density slice $\psi_{\rm uni}$, denoted from now on as $-\zeta$, is given by
\bea
\zeta &=& -\psi - \mathcal{H}\xi_{\rm uni} \nonumber \\
&=& -\psi - \mathcal{H} \frac{\delta \rho}{\rho_0'} \,. \label{zeta_def2}
\eea
We have chosen to denote both the comoving and the uniform density slice curvature perturbations with the same letter $\zeta$ because they are equivalent on superhorizon scales ($k \ll aH_{\rm inf}$) \cite{bardeen}. 
The superhorizon scale region is of great importance because $\zeta$ is conserved in that region if the adiabatic condition, described below, is satisfied. In addition, at superhorizon scales $\zeta$ becomes truly gauge-invariant because the contribution of $\beta$ in Eq. (\ref{transpsi}) gets suppressed by the spatial derivatives, so that $\psi$ does not depend on the changes in the threading\footnote{
Notice that if $\varphi_0' \rightarrow 0$ or $\rho_0' \rightarrow 0$ the curvature perturbations in Eqs. (\ref{zeta_def1}) and (\ref{zeta_def2}) blow up. To avoid such a disaster we need very small values for $\calh$ so as to generate the observed value for $\zeta$.  If for some reason $\varphi_0' = 0$ or $\rho_0' = 0$, $\zeta$ as given in Eq. (\ref{zeta_def1}) or (\ref{zeta_def2}) becomes ill defined and we would need to look for a better well defined gauge invariant quantity that represents the intrinsic curvature perturbation $\psi$.}.

The conservation of $\zeta$ is guaranteed as long as the pressure $P$ is a unique function of the energy density $\rho$ (the adiabatic condition). The last statement follows from the local energy conservation equation in the uniform density slicing at large scales \cite{sasaki1,rs1,llmw}:
\be
\dot{\rho}(t) = -3(H + \dot{\zeta})[\rho(t) + P(t,{\bf x})] \,. \label{localece}
\ee
If $P$ satisfies the adiabatic condition then it becomes spatially uniform ($\zeta = \langle \zeta \rangle$), and so does $\dot{\zeta}$. As a result
\be
\dot{\zeta} = 0 \,,
\ee
because
\be
\dot{\zeta} = \langle \dot{\zeta} \rangle = \frac{d}{dt} \langle \zeta \rangle = 0 \,.
\ee
If $P$ does not satisfy the adiabatic condition, i.e. if it has a non-adiabatic component $P_{\rm nad}$, \hbox{Eq. (\ref{localece})} implies
\be
\dot{\zeta} = - \frac{H}{\rho_0 + P_0} \delta P_{\rm nad} \,.
\ee
The curvaton scenario is one example where the existence of a non-adiabatic pressure perturbation makes $\zeta$ evolve from the negligible curvature perturbation produced by the inflaton to the right value observed today. The non-adiabatic pressure perturbation is, in this case, the result of the presence of two weakly interacting fluids, the curvaton matter fluid and the radiation fluid, in the period that follows the inflaton decay and reheating.

\section{Inflation and its effect on the spectrum of perturbations of a non-dominating massless scalar field during a de Sitter stage} \label{inflation}

Any period in the history of the Universe during which the expansion is accelerated is denominated as {\it inflationary} \cite{albste,guth,book,linde82,treview}. The inflationary stage prior to the Hot Big-Bang has the nice property to stretch the quantum fluctuations of the scalar fields living in the FRW spacetime \cite{bst,guthpi82,hawking,linde82a,mukhanov,mukhanovrep,riotto,starobinsky1},  so that they become classical \cite{albrecht,grishchuk,guthpi,lombardo,lyth84} and almost constant, sourcing the primordial density inhomogeneities responsible for the presently observed large-scale structure. The amplitude of the spectrum of the classical field perturbations (for light fields) is generically the same for all kinds of quasi de Sitter models. However, the spectral index is written down in a certain way for fields that dominate the energy density and in another way for fields that do not.
In this section we will describe inflation, paying special attention to the constraint imposed by it on the slow-roll parameter $\varepsilon$, and review the properties of the power spectrum of perturbations of a massless scalar field during a de Sitter stage, characterized the latter by a constant Hubble parameter $H_{\rm inf} = H_\ast$ during inflation.

\subsection{Inflation} \label{epless1}

Inflation can be rigorously defined as the period when the global expansion parameter $a$ satisfies the condition
\be
\ddot{a} > 0 \,. \label{doubledota}
\ee
Such a condition translates into a definite requirement on the global Hubble parameter during inflation $H_{\rm inf}$. From the definition of $H_{\rm inf}$ in terms of $a$, which can be written alternatively as the following evolution equation:
\be
a(t) = a_{\rm ini} \exp\left(\int_{t_{\rm ini}}^t H_{\rm inf} dt\right) \,, \label{aintegral}
\ee
being $a_{\rm ini}$ the expansion parameter at some time $t_{\rm ini}$, the inflationary condition in Eq. (\ref{doubledota}) is satisfied while
\be
\dot{H}_{\rm inf} > - H^2_{\rm inf} \,.
\ee
The last expression reduces to an upper bound on the slow-roll parameter $\varepsilon$ defined in \mbox{Eq. (\ref{srepsilon})}:
\be
\varepsilon \equiv -\left(\frac{\dot{H}_{\rm inf}}{H^2_{\rm inf}}\right)_\ast < 1 \,.
\ee

Inflation is held as long as $\varepsilon < 1$, being one possibility when $H_{\rm inf}$ is constant in time. This scenario is commonly recognized as the de Sitter stage, and it will be useful at studying the spectrum of perturbations of a massless scalar field. Other possibilities correspond to nonvanishing values for $\dot{H}_{\rm inf}$, being the quasi de Sitter case ($\varepsilon \ll 1$) the most popular. Indeed, the quasi de Sitter stage supplemented by the condition $|\eta_\varphi| \ll 1$, being $\eta_\varphi$ the slow-roll parameter defined in Eq. (\ref{sreta}), is what is known as {\it slow-roll inflation}. This variety of inflation will be discussed in \hbox{Subsection \ref{slow_roll}}.  Meanwhile in the following subsection, we will study the fluctuations of a non-dominating massless scalar field during a de Sitter stage. Later on we will generalise these results to the fluctuations of non-dominating and dominating massive scalar fields in a quasi de Sitter stage.

\subsection{Spectrum of perturbations of a non-dominating massless scalar field during a de Sitter stage} \label{spectrum1}

In this subsection we will consider the effects of a de Sitter inflationary stage on the fluctuations of a massless scalar field that does not dominate the energy density. Let's call this field $\phi$. During inflation, and before horizon exit, the fluctuations in $\phi$ can still be regarded as quantum operators. If we further assume that $\phi$ is almost a non-interacting field, we can write down the field perturbation operator $\delta \hat{\phi} ({\bf x}, t)$ in terms of the usual creation and annihilation operators $\hat{a}^\dagger_{\bf k}$ and $\hat{a}_{\bf k}$:
\be
\delta \hat{\phi}({\bf x},t) = \int \frac{d^3k}{(2\pi)^{3/2}} \exp(i {\bf k \cdot x}) \delta \hat{\phi}_{\bf k} (t) \,,
\ee
where
\be
\delta \hat{\phi}_{\bf k} (t) \equiv \omega_k (t) \hat{a}_{\bf k} + \omega_k^\ast (t) \hat{a}^\dagger_{\bf -k} \,.
\ee

As it was discussed in the introduction of this thesis, the properties of the primordial curvature perturbation $\zeta$ are specified by the spectrum $\mathcal{P}_\zeta(k)$, which is defined by the statistical average $\langle \zeta_{{\bf k}_1} \zeta_{{\bf k}_2} \rangle$ over an ensemble of universes. The same definition can be applied for the perturbations in $\phi$, so that the only information we need to know to calculate $\mathcal{P}_{\delta \phi}(k)$ is the quantum state of the Universe during inflation, being the most reasonable choice the vacuum state\footnote{The vacuum state guarantees the homogeneity and isotropy in the whole space.}.
Since the universes in the ensemble are all in the vacuum state during inflation, the statistical average $\langle \delta \phi_{{\bf k}_1} \delta \phi_{{\bf k}_2} \rangle$ is now very easy to calculate corresponding to the expectation value $\langle 0 | \delta \hat{\phi}_{{\bf k}_1} \delta \hat{\phi}_{{\bf k}_2} | 0 \rangle$. It is straightforward then to recognize that the spectrum $\mathcal{P}_{\delta \phi} (k)$ of the $\delta \phi$ perturbations, defined by
\be
\langle \delta \phi_{{\bf k}_1} \delta \phi_{{\bf k}_2} \rangle \equiv \frac{2\pi^2}{k^3} \delta^3 ({\bf k}_1 + {\bf k}_2) \mathcal{P}_{\delta \phi} (k) \,,
\ee
is given by the simple formula
\be
\mathcal{P}_{\delta \phi} (k) = \frac{k^3}{2\pi^2} |\omega_k|^2 \,. \label{easyp}
\ee

To calculate $|\omega_k|$ we need to solve the Klein-Gordon equation for the fluctuations in $\phi$. The usual Klein-Gordon equation
\be
\ddot{\omega}_k + 3H_\ast \dot{\omega}_k + \frac{k^2}{a^2} \omega_k = 0 \,, \label{perturbedkge}
\ee
is only applicable if the inflaton field (that which dominates the energy density) is unable to generate significant primordial perturbations. That guarantees that the line element in Eq. (\ref{unperturbedle}) is {\it not} modified by any scalar perturbation as $\phi$ is assumed not to dominate the energy density during inflation.
As a consequence, the usual structure of the Klein-Gordon equation is unmodified.

Eq. (\ref{perturbedkge}) is easily solved by making the following change of variables:
\be
\omega_k \equiv \frac{\lambda_k}{a} \,, \label{omegalambda}
\ee
and going to conformal time where the expansion parameter in a de Sitter stage ($H_{\rm inf}$ being constant)
\be
a(t) = a_{\rm ini} \exp[H_\ast (t - t_{\rm ini})] \,,
\ee
is given by
\be
a(\eta) = - \frac{1}{H_\ast \eta} \,, \label{atau}
\ee
with conformal time $\eta$ taking negative values.
Thus, the Klein-Gordon equation in Eq. (\ref{perturbedkge}) reduces to
\be
\lambda_k'' + \left(k^2 - \frac{2}{\eta^2}\right) \lambda_k = 0 \,, \label{modperturbedkge}
\ee
whose exact solution is
\be
\lambda_k = \frac{\exp(-ik\eta)}{\sqrt{2k}} \left(1 - \frac{i}{k\eta}\right) \,. \label{lambdasol}
\ee
The reader might worry about the fact that Eq. (\ref{lambdasol}) is valid up to a multiplicative constant. To solve this ambiguity we note that in the subhorizon regime \mbox{($k \gg aH_{\rm inf}$)} Eq. (\ref{modperturbedkge}) reproduces the Klein-Gordon equation for a massless scalar field in Minkowski spacetime. The solution for $\lambda_k$ in Eq. (\ref{lambdasol}) satisfies the Bunch and Davies normalisation \cite{bdbook,bd} on subhorizon scales, so that no integration constant should amplify it.

The subhorizon regime is useful in the sense that we can find the adequate solution and normalisation for $\lambda_k$. However, the superhorizon regime ($k \ll aH_{\rm inf}$) is much more interesting because $\zeta$ defined in that regime is truly gauge invariant and conserved as long as the pressure satisfies the adiabatic condition. From Eqs.~(\ref{omegalambda}), (\ref{atau}), and (\ref{lambdasol}), the magnitude of the mode function $\omega_k$ for superhorizon scales turns out to be constant and it is given by
\be
|\omega_k| \approx \frac{H_\ast}{\sqrt{2k^3}} \,.
\ee
The latter expression allows us to obtain the spectrum of perturbations in $\phi$, given by Eq. (\ref{easyp}):
\be
\mathcal{P}_{\delta \phi}(k) \approx \left(\frac{H_\ast}{2\pi}\right)^2 \,. \label{finalspec1}
\ee
As we can see, the spectrum $\mathcal{P}_{\delta \phi}(k)$ of the perturbations of a massless scalar field $\phi$ in a de Sitter stage is constant and scale invariant. We will see later on that the previous amplitude is generically the same for any kind of quasi de Sitter model as far as we deal with light fields. A possible scale dependence will arise if the field under consideration has a finite mass and/or if the inflationary stage is not completely de Sitter ($\dot{H}_{\rm inf} \neq 0$).

\section{The curvaton scenario: an example of a non-dominating light scalar field during a quasi de Sitter stage} 

In the previous section we described the spectrum of perturbations of a non-domina\-ting massless scalar field during an inflationary period with $H_{\rm inf}$ being constant (de Sitter stage). Now we move a step ahead considering a non-dominating scalar field $\sigma$ whose mass $m_\sigma$ satisfies $m_\sigma \ll H_\ast$ (light field), during an inflationary period where the Hubble parameter $H_{\rm inf}$ is not constant but evolves slowly satisfying the condition $-\dot{H}_{\rm inf}/H^2_{\rm inf} \ll 1$ so that $\varepsilon \ll 1$ (quasi de Sitter stage). We will see that the introduction of a slowly varying $H_{\rm inf}$ and a small mass $m_\sigma$ for the field $\sigma$ results in a small scale dependence for $\mathcal{P}_{\delta \sigma} (k)$ which is parameterised by $\varepsilon$ and $\eta_\sigma \equiv m_\sigma^2 / 3H_\ast^2$. The amplitude of $\mathcal{P}_{\delta \sigma} (k)$ will turn out to be the same as that for a non-dominating massless scalar field in a de Sitter stage, which was the case described in Subsection \ref{spectrum1}. This example will help us to
understand the properties of the curvature perturbation produced in
one of the most satisfying
models proposed
to explain the origin of the large-scale structure in the Universe: the curvaton scenario.

\subsection{Spectrum of perturbations of a non-dominating light scalar field during a quasi de Sitter stage} \label{spectrum2}

The calculation of the spectrum of perturbations of a non-dominating light scalar field $\sigma$ during a quasi de Sitter stage closely resembles that done in Subsection \ref{spectrum1}. We will now have to consider the mass $m_\sigma$ of $\sigma$ and the running of $H_{\rm inf}$ given by the slow-roll parameter $\varepsilon$. Since the inflaton field is supposed to produce negligible curvature perturbation, and $\sigma$ does not dominate the energy density during inflation, the line element is still given by Eq. (\ref{unperturbedle}). This means that the Klein-Gordon equation for the mode functions $\omega_k$ does not change compared with that for the mode functions of the background component of $\sigma$:
\be
\ddot{\omega}_k + 3 H_{\rm inf} \dot{\omega}_k + \left(\frac{k^2}{a^2} + m_\sigma^2\right) \omega_k = 0 \,.
\ee

In the quasi de Sitter stage the expansion is almost exponential; nevertheless it is better described by the following evolution equation:
\be
a(t) = a_{\rm ini} [1 + H_{\rm inf}(t_{\rm ini}) \varepsilon (t - t_{\rm ini})]^{1/\varepsilon} \,,
\ee
where $H_{\rm inf}(t_{\rm ini})$ is the Hubble parameter at the time $t_{\rm ini}$. The last expression can be written down using the conformal time $\eta$ in an easier way:
\be
a(\eta) = -\frac{1}{H_{\rm inf}(\eta) \eta (1 - \varepsilon)} \,,
\ee
having in mind that $\eta$ takes negative values. Going to conformal time and making the change of variables
\be
\omega_k \equiv \frac{\lambda_k}{a} \,,
\ee
the following equation of motion for $\lambda_k$ is obtained:
\be
\lambda_k'' + \left[k^2 - \frac{1}{\eta^2}\left(\upsilon_\sigma^2 - \frac{1}{4}\right)\right] \lambda_k = 0 \,, \label{kgmodified}
\ee
where 
\bea
\upsilon_\sigma &\equiv& \left[\frac{1}{4} - \frac{3\eta_\sigma - 2 + \varepsilon}{(1 - \varepsilon)^2}\right]^{1/2} \nonumber \\
&\approx& \frac{3}{2} + \varepsilon - \eta_\sigma \,. \label{up_var_eta}
\eea
The solution for such an equation is given in terms of the Hankel's functions of the first and second kind $H_{\upsilon_\sigma}^{(1)}$ and $H_{\upsilon_\sigma}^{(2)}$ \cite{brandenberger,riotto,sasaki}:
\be
\lambda_k = \sqrt{-\eta} [c_1(k) H_{\upsilon_\sigma}^{(1)}(-k\eta) + c_2(k) H_{\upsilon_\sigma}^{(2)}(-k\eta)] \,, \label{hankel}
\ee
where $c_1(k)$ and $c_2(k)$ are integration constants that are determined going to the subhorizon regime ($k \gg aH_{\rm inf}$), which corresponds to $-k\eta \gg 1$, and normalising the solution according to Bunch and Davies \cite{bdbook,bd}\footnote{The requirement of a solution for $\lambda_k$ normalised a la Bunch and Davis is justified because Eq. (\ref{kgmodified}) on subhorizon scales reduces to the Klein-Gordon equation for a massless scalar field in Minkowski spacetime.}. Indeed, taking into account that in the subhorizon regime the Hankel's functions are well approximated by \cite{brandenberger,riotto,sasaki}
\bea
H_{\upsilon_\sigma}^{(1)} (-k\eta \gg 1) &\approx& \sqrt{\frac{2}{\pi k \eta}} \exp\left[-i\left(k\eta + \frac{\pi}{2} \upsilon_\sigma +\frac{3\pi}{4}\right)\right] \,, \\
H_{\upsilon_\sigma}^{(2)} (-k\eta \gg 1) &\approx& \sqrt{\frac{2}{\pi k \eta}} \exp\left[i\left(k\eta + \frac{\pi}{2} \upsilon_\sigma +\frac{3\pi}{4}\right)\right] \,,
\eea
the Bunch and Davies normalisation
\be
\lambda_k = \frac{\exp(-ik\eta)}{\sqrt{2k}} \,,
\ee
is obtained in the subhorizon regime by choosing the following values for the integration constants:
\bea
c_1(k) &=& \frac{\sqrt{\pi}}{2} \exp \left[i\left(\upsilon_\sigma + \frac{1}{2}\right)\frac{\pi}{2}\right] \,, \\
c_2(k) &=& 0 \,.
\eea
Thus, the exact solution in Eq. (\ref{hankel}) is rewritten as
\be
\lambda_k = \frac{\sqrt{\pi}}{2} \exp \left[i\left(\upsilon_\sigma + \frac{1}{2}\right)\frac{\pi}{2}\right] \sqrt{-\eta} H_{\upsilon_\sigma}^{(1)} (-k\eta) \,.
\ee

To find out the spectrum of $\delta \sigma$ on superhorizon scales ($k \ll aH_{\rm inf}$), corresponding to $-k\eta \ll 1$, we need to consider the behaviour of $\lambda_k$ in that regime. This is given by the following approximation for the Hankel's function of the first kind on superhorizon scales \cite{brandenberger,riotto,sasaki}
\be
H_{\upsilon_\sigma}^{(1)} (-k\eta \ll 1) \approx \sqrt{\frac{2}{\pi}} \exp \left(-i\frac{\pi}{2}\right) 2^{\upsilon_\sigma - \frac{3}{2}} \frac{\Gamma(\upsilon_\sigma)}{\Gamma(3/2)} (-k\eta)^{-\upsilon_\sigma} \,.
\ee
The magnitude of the mode function $\omega_k$ on superhorizon scales is then almost constant and approximately given by
\bea
|\omega_k| &\approx& [2(1 - \varepsilon)]^{\upsilon_\sigma - \frac{3}{2}} (1 - \varepsilon) \frac{\Gamma(\upsilon_\sigma)}{\Gamma(3/2)} \frac{H_{\rm inf}}{\sqrt{2k^3}} \left(\frac{k}{aH_{\rm inf}}\right)^{\frac{3}{2} - \upsilon_\sigma} \nonumber \\
&\approx& \frac{H_\ast}{\sqrt{2k^3}} \left(\frac{k}{aH_{\rm inf}}\right)^{\eta_\sigma - \varepsilon} \,. \label{presolution}
\eea
where we have used the relation involving $\upsilon_\sigma$, $\varepsilon$, and $\eta_\sigma$, established in Eq. (\ref{up_var_eta}).

Making use of the expression in Eq. (\ref{easyp}), the spectrum of perturbations in $\sigma$ is finally written down as
\be
\mathcal{P}_{\delta \sigma} (k) \approx \left(\frac{H_\ast}{2\pi}\right)^2 \left(\frac{k}{aH_{\rm inf}}\right)^{n_{\delta \sigma}} \,, \label{amplitudePsigma}
\ee
whose spectral index $n_{\delta \sigma}$ is given by
\be
n_{\delta \sigma} = 2\eta_\sigma - 2\varepsilon \,. \label{indexPsigma}
\ee
Comparing the previous result with that found for the case of a non-dominating massless scalar field during a de Sitter stage [c.f. Eq. (\ref{finalspec1})], we see that the amplitude of $\mathcal{P}_{\delta \sigma} (k)$ is exactly the same for both cases, but now we have a small scale dependence parameterised by the slow evolution of $H_{\rm inf}$ and the mass $m_\sigma$ of the light field $\sigma$, characterized by the smallness of the parameters $\varepsilon$ and $\eta_\sigma$.

\subsection{The curvaton scenario} \label{zetaandcurvaton}

In the basic curvaton setup the Hubble parameter during inflation $H_{\rm inf}$ is assumed to be slowly varying ($\varepsilon \ll 1$), the curvaton
energy density $\rho_\sigma$ during inflation is assumed to be negligible, and the inflaton field is supposed to produce a negligible curvature perturbation $\zeta_r$ \cite{lyth03c,lyth02,moroi01b} which is imprinted to the radiation fluid while the inflaton decays. After the end of inflation the Universe is composed by the almost unperturbed radiation fluid that originates from the reheating process following the inflaton decay, and the weakly coupled \cite{CD} light curvaton field $\sigma$ \footnote{The curvaton field must be weakly coupled to avoid premature thermalisation, which would erase all the information about the generatated curvature perturbation \cite{CD}.} whose unperturbed component is kept frozen at a value $\sigma_\ast$ until the Hubble parameter $H$ becomes of the order of the curvaton mass $m_{\sigma}$ during the radiation dominated epoch. Once $\sigma$ is unfrozen, it begins oscillating around the minimum of its potential, which is taken to be quadratic, with an oscillation period which rapidly becomes much less than the characteristic expansion time scale. This ensures that the average curvaton pressure vanishes and, therefore, $\sigma$ may be considered as a matter fluid \cite{turner}. During the oscillatory period, the curvaton energy density $\rho_\sigma$ decreases with time according to $\rho_\sigma \propto a^{-3}$, while the radiation energy density $\rho_r$ decreases with time faster than $\rho_\sigma$ according to $\rho_r \propto a^{-4}$. Eventually the curvaton will decay, but by that time the contribution of $\rho_\sigma$ to the total energy density will be big enough for the original isocurvature perturbation $\zeta_\sigma$, generated by $\sigma$ during inflation and which is not negligible, to become the {\it total} curvature perturbation $\zeta$.

Since the oscillations of $\sigma$ around the minimum of its potential are so fast, we can approximate its energy density by
\begin{equation}
\rho_\sigma(t,{\bf x}) \approx \frac{1}{2} m_\sigma^2 \sigma_a^2(t,{\bf x}) \,, \label{sigma_ed}
\end{equation}
where $\sigma_a(t,{\bf x})$ is the amplitude of the oscillations. Notice that, under these circumstances, the expression for the curvaton energy density in Eq. (\ref{sigma_ed}) corresponds also to the expression for the curvaton potential $V(\sigma)$. The no appearance of quartic or higher order terms in the potential is essential for the success of the model because, otherwise, the density ratio $\rho_\sigma/\rho_r$ would not increase with time \cite{CD}.

Making use of the curvature perturbation definition in Eq. (\ref{zeta_def2}), we can write down the total $\zeta$ in the curvaton scenario as:
\begin{equation}
\zeta \equiv -\psi -H \left(\frac{\delta \rho}{\dot{\rho}_0}\right)_{\rm total} \,, \label{totalcp}
\end{equation}
where the total energy density $\rho_{\rm total}$ is simply the addition of the curvaton and radiation energy densities $\rho_\sigma$ and $\rho_r$,
that define the conserved curvaton and radiation curvature perturbations $\zeta_\sigma$ and $\zeta_r$ \footnote{The curvature perturbations $\zeta_\sigma$ and $\zeta_r$, associated to the curvaton and the radiation fluids respectively, are conserved since the adiabatic condition is satisfied separately being both fluids non-interacting.}:
\begin{eqnarray}
\zeta_\sigma &\equiv& -\psi - H \frac{\delta \rho_\sigma}{ \dot \rho_{\sigma_0}} = -\psi + \frac 13 \frac{\delta \rho_\sigma}{\rho_{\sigma_0}} \,, \label{curvcp} \\
\zeta_r &\equiv& -\psi - H \frac{\delta \rho_r}{\dot \rho_{r_0}} = -\psi + \frac 14 \frac{\delta \rho_r}{\rho_{r_0}} \,. \label{radcp}
\end{eqnarray}
In the above expressions we have employed the background continuity equation \cite{foster,kt,book}
\be
\dot{\rho}_0 + 3H(\rho_0 + P_0) = 0 \,,
\ee
where $P_0=0$ for a matter fluid, and $P_0 = \rho_0/3$ for a radiation fluid.
Combining \hbox{Eqs. (\ref{totalcp}), (\ref{curvcp}), and (\ref{radcp})}, the total curvature perturbation $\zeta$ can then be written down as the weighted sum
\begin{equation}
\zeta = (1-r)\zeta_r + r\zeta_\sigma \,, \label{zetaandf}
\end{equation}
with modulation factor
\begin{equation}
r \equiv \frac{3\rho_{\sigma_0}}{4\rho_{r_0} + 3\rho_{\sigma_0}} \,. \label{fmodulation}
\end{equation}

Notice that just at the beginning of the radiation dominated epoch that follows the reheating stage produced by the inflaton decay, $r$ is almost zero since $\rho_\sigma$ is negligible by that time; therefore $\zeta \approx \zeta_r$ which is negligible. However, $r$ grows in time due to the relative redshifting between $\rho_\sigma$ and $\rho_r$ until when eventually $\sigma$ decays. In view of Eqs. (\ref{zetaandf}) and (\ref{fmodulation}), the total $\zeta$ grows then in time approaching more and more to the curvaton curvature perturbation $\zeta_\sigma$. One extreme example is when $\sigma$ has dominated the energy density before decaying; in that case $r \approx 1$ and therefore \mbox{$\zeta \approx \zeta_\sigma$}. When $\sigma$ decays, $\zeta$ is imprinted in remaining radiation fluid starting this way the gravitational instability process that ends up with the presently observed large-scale structure. 

As we have already pointed out, one of the requirements of the curvaton scenario is that the curvature perturbation produced by the inflaton during inflation $\zeta_r$ is completely negligible compared with that produced by the curvaton $\zeta_\sigma$ during the same period: $\zeta_r \ll \zeta_\sigma$. Under that assumption, the expression for the total $\zeta$ after $\sigma$ decays comes from Eq. (\ref{zetaandf}) as
\be
\zeta \approx r \zeta_\sigma \,,
\ee
in the sudden decay approximation \cite{lyth02}. For a model that goes beyond 
this approximation
the expression for the total $\zeta$ in terms of $\zeta_\sigma$ is only obtained by means of numerical calculations, the result being in that case \cite{lyth03c,mwu}
\be
\zeta \approx \Omega_{\rm dec} \zeta_\sigma \,, \label{beyondsdecay}
\ee
where $\Omega_{\rm dec}$ is the fractional global curvaton energy density just before the curvaton decay:
\begin{equation}
\Omega_{\rm dec} = \left(\frac{\rho_{\sigma_0}}{\rho_{\rm total_0}}\right)_{\rm dec} \,. \label{fomegadef}
\end{equation}
As we mentioned in Chapter \ref{introduccion}, and will explain in Chapter \ref{fourthpaper}, the normalisation $\fnl$ of the bispectrum $B_\zeta (k_1,k_2,k_3)$ of $\zeta$ in the curvaton scenario is directly related to $\Omega_{\rm dec}$ if the latter is not so close to 1 \cite{review,curvaton,lr,lyth03c}:
\be
\fnl \approx -\frac{5}{4\Omega_{\rm dec}} \,. \label{firstfnldef}
\ee
The parameter $\fnl$ gives information about the level of non-gaussianity present in $\zeta$, and the actual bound on it, coming from WMAP data \cite{komatsu03}, is $|\fnl| \lsim 10^2$. This bound translates into a lower bound for $\Omega_{\rm dec}$, that combined with the obvious energy density condition $\Omega_{\rm dec} \leq 1$, gives the allowed range
\begin{equation}
0.01 \lsim \Omega_{\rm dec} \leq 1 \,. \label{fWMAPr}
\end{equation}
The present lower bound on $\Omega_{\rm dec}$ is likely to be increased \cite{spergel} by the next WMAP data release or the future PLANCK satellite data if non-gaussianity effects are not detected.

Once we have studied how the curvature perturbation is produced in the curvaton scenario, we proceed now to study the spectrum $\mathcal{P}_\zeta (k)$ of $\zeta$. 
In view of Eq. (\ref{sigma_ed}), and having in mind that the equation of motion for $\delta \sigma_a$ is the same as that for the background field $\sigma_{a_0}$, throughout inflation and during the post-inflationary period, as long as the non-gauge invariant curvature perturbation $\psi$ is fixed to be zero\footnote{That is indeed the case while the curvature perturbation $\zeta_r$ in the radiation fluid is taken to be negligible \cite{sloth}, which is one of the key assumptions in the curvaton scenario.}, we can relate the contrast in the energy density of $\sigma$ at any time $t$ with the contrast in $\sigma$ some time after horizon exit but before the onset of the curvaton oscillations:
\begin{equation}
\frac{\delta \rho_\sigma }{\rho_{\sigma_0}} \approx 2 \left(\frac{\delta \sigma_a}{\sigma_{a_0}}\right) \simeq 2\frac%
{\delta \sigma}{\sigma_\ast} \,. \label{contrastes}
\end{equation}
From Eqs. (\ref{curvcp}), (\ref{beyondsdecay}), and (\ref{contrastes}), $\zeta$ is expressed in terms of the perturbations in $\sigma$ a few Hubble times after horizon exit:
\be
\zeta \approx \frac{2}{3}\Omega_{\rm dec}\frac{\delta \sigma}{\sigma_\ast} \,, 
\ee
and in consequence the spectrum $\mathcal{P}_\zeta(k)$ is given by
\be
\mathcal{P}_\zeta (k) \approx \frac{4}{9} \Omega_{\rm dec}^2 \frac{\mathcal{P}_{\delta\sigma}(k)}{\sigma_\ast^2} \,.
\ee
The curvaton field $\sigma$ is a light field whose energy density is negligible during inflation; therefore the discussion and results of Subsection \ref{spectrum2} apply [c.f. Eqs. (\ref{amplitudePsigma}) and (\ref{indexPsigma})], giving as a result
\be
\mathcal{P}_\zeta (k) \equiv A_\zeta^2 \left(\frac{k}{a H_{\rm inf}}\right)^{n_\zeta} \simeq \left[\frac{H_\ast \Omega_{\rm dec}}{3\pi \sigma_\ast}\right]^2 \left(\frac{k}{a H_{\rm inf}}\right)^{2\eta_\sigma - 2\varepsilon} \,. \label{Pkcurvaton}
\ee
The spectral index $n_\zeta$ is in good agreement with observation, which requires an almost-scale invariant power spectrum \cite{observation}:
\be
-0.048 < n_\zeta < 0.016 \,.
\ee

Unfortunately the Hubble parameter a few Hubble times after horizon exit $H_\ast$, which gives information about the inflationary energy scale, is not predicted by the amplitude of $\mathcal{P}_\zeta(k)$ since $\sigma_\ast$ is an unknown and unbounded parameter. Nevertheless, a lower bound for $H_\ast$ will be found in Chapter \ref{firstpaper} by taking into consideration other effects 
that give a different relation between $H_\ast$ and $\sigma_\ast$, and the amplitude $A_\zeta$ in \hbox{Eq. (\ref{Pkcurvaton})}
once the WMAP normalisation ($|A_\zeta| \approx 5 \times 10^{-5}$) \cite{observation} is taken into account:
\begin{equation}
\sigma_\ast \approx (1.5 \pi \times 10^{-4})^{-1} \Omega_{\rm dec} H_\ast.
\end{equation}
What is important however to emphasise at this point is that the biggest possible value for $H_\ast$ in the curvaton scenario is for sure $10^{12}$ GeV. Otherwise the curvature perturbation $\zeta_r$, produced by the inflaton field during inflation, would contribute significantly to the total $\zeta$, spoiling the main motivation for the proposal of the curvaton scenario\footnote{The only way to have $H_\ast > 10^{12}$ GeV in the curvaton scenario while making $\zeta_r$ negligible is by requiring the inflaton field not to be light during inflation ($m_\varphi \geq H_\ast$) \cite{pilo}. A non slow-roll inflationary model is in that case compulsory.}. The justification of this assertion will be given in the following section.

\section{The inflaton scenario} \label{section_inflaton}

In this section we will discuss the main facts about the inflaton scenario where inflation is assumed to be of the slow-roll variety \cite{albste,book,linde82,treview}. Slow-roll inflation corresponds to the case where the inflaton field $\varphi$ slowly-roll down towards the minimum of its potential. We will specify the slow-roll conditions and see what their consequences are on the shape of the inflaton potential as well as on the value and structural form of the slow-roll parameters $\varepsilon$ and $\eta_\varphi$. Being $\varphi$ in the inflaton scenario the responsible of driving inflation and also of generating the curvature perturbation $\zeta$, the power spectrum $\mathcal{P}_\zeta(k)$ of $\zeta$ presents definite signatures that are expressed in terms of $\varepsilon$ and $\eta_\varphi$. We will calculate $\mathcal{P}_\zeta(k)$ and see what the constraints on the inflaton potential are in order to produce enough primordial perturbations. The Hubble parameter during inflation $H_\ast$ will turn out to be likely quite high \mbox{($H_\ast \lsim 10^{14}$ GeV)} for the inflaton scenario to be consistent with the amplitude of perturbations observed by WMAP. The scale of inflation is, therefore, likely high enough to
impose severe constraints on concrete inflation models \cite{peiris}.

\subsection{The slow-roll conditions} \label{slow_roll}

We begin by considering the Friedmann and continuity equations, derived from the background Einstein equations for the FRW cosmological model \cite{foster,kt,book}, that relate the Hubble parameter at any time with the global energy density and pressure of the fluid that fills the Universe:
\bea
H^2 &=& \frac{\rho_0}{3m_P^2} \,, \\
\dot{\rho}_0 &=& - 3H(\rho_0 + P_0) \,.
\eea
A direct consequence of both equations is that the second derivative of the global expansion parameter $a$ with respect to the cosmic time is given by a simple relation involving $\rho_0$ and $P_0$:
\be
\frac{\ddot{a}}{a} = - \frac{\rho_0 + 3P_0}{6m_P^2} \,.
\ee
This expression tells us that, to satisfy the inflationary condition $\ddot{a} > 0$, the pressure of the fluid that fills the Universe must be negative satisfying
\be
\rho_0 + 3P_0 < 0 \,. \label{icrhop}
\ee
As an application of the above formula we may study the dynamics of the inflaton field $\varphi$ knowing that, from the energy momentum tensor for a homogeneous scalar field \cite{book}, the unperturbed energy density and pressure of $\varphi$ are given by
\bea
\rho_{\varphi_0} &=& \frac{1}{2}\dot{\varphi}_0^2 + V(\varphi_0) \,, \\
P_{\varphi_0} &=& \frac{1}{2}\dot{\varphi}_0^2 - V(\varphi_0) \,.
\eea
The inflationary condition in Eq. (\ref{icrhop}) is then satisfied provided that
\be
\dot{\varphi}_0^2 < V(\varphi_0) \,.
\ee

The most popular type of inflationary models assume that the kinetic energy density of the inflaton field is much less than the potential energy density:
\be
\frac{1}{2} \dot{\varphi}_0^2 \ll V(\varphi_0) \,, \label{firstslowroll}
\ee
which corresponds intuitively to a very flat potential along which the inflaton field $\varphi$ slowly-roll down during inflation towards the minimum of its potential \cite{book,treview}. If the expression in Eq. (\ref{firstslowroll}) is supplemented by the condition
\be
|\ddot{\varphi}_0| \ll |3H_{\rm inf} \dot{\varphi}_0| \,, \label{secondslowroll}
\ee
the inflaton field $\varphi$ satisfies what is known as the {\it slow-roll} conditions \cite{liddle}. As we will see, these conditions can be expressed in terms of the parameters $\varepsilon$ and $\eta_\varphi$ that parameterise the spectral index and amplitude of $\mathcal{P}_\zeta(k)$ in the inflaton scenario. Notice that, under the slow-roll conditions, the background field $\varphi_0$ follows the slow-roll equation of motion
\be
3H_{\rm inf} \dot{\varphi_0} \approx - \frac{\partial V}{\partial \varphi_0} \,,
\ee
which corresponds to the background Klein-Gordon equation under the condition given by Eq. (\ref{secondslowroll}).

As discussed in Subsection \ref{epless1}, the requirement to have a period of accelerated expansion is easily expressed as an upper bound on the slow-roll parameter $\varepsilon$ that describes the rate of change of the Hubble parameter a few Hubble times after horizon exit:
\be
\varepsilon \equiv -\left(\frac{\dot{H}_{\rm inf}}{H_{\rm inf}^2}\right)_\ast < 1 \,.
\ee
The true reason why $\varepsilon$ is called a slow-roll parameter is because it is constrained to be much less than 1 under the slow-roll conditions in Eqs. (\ref{firstslowroll}) and (\ref{secondslowroll}), being easily expressed in terms of the unperturbed inflaton potential $V(\varphi_0)$ and its derivative with respect to $\varphi$ \cite{liddle}:
\be
\varepsilon \approx \frac{m_P^2}{2V^2} \left(\frac{\partial V}{\partial \varphi_0}\right)^2 \ll 1 \,. \label{vandep}
\ee
The {\it flatness} condition on the potential $V(\varphi)$ required for $\varphi$ to slowly-roll during inflation is evident from the above expression.

Two slow-roll conditions (Eqs. (\ref{firstslowroll}) and (\ref{secondslowroll})) require constraints on two slow-roll parameters. One of them is that given in Eq. (\ref{vandep}) in terms of $\varepsilon$; the other one is given in terms of the parameter $\eta_\varphi$ already defined in Eq. (\ref{sreta}):
\be
\eta_\varphi \equiv \varepsilon - \frac{\ddot{\varphi}_0}{H_{\rm inf} \dot{\varphi}_0} \,.
\ee
The respective constraint on $\eta_\varphi$ and its relation with $V(\varphi)$ are obtained once we take into consideration the slow-roll conditions in Eqs. (\ref{firstslowroll}) and (\ref{secondslowroll}) \cite{liddle}:
\be
|\eta_\varphi| \approx \left|\frac{m_P^2}{V} \frac{\partial^2V}{\partial \varphi_0^2}\right| \ll 1 \,. \label{eandv}
\ee
This relation again shows how flat the potential of the inflaton field ought to be to drive inflation. This is particularly good in order to generate enough inflation as $\varphi$ spends a lot of time rolling along the flat part of its potential, which is in turn perhaps the main motivation to have an inflationary slow-roll model. 

From the practical point of view, inflation is said to start when $V(\varphi)$ satisfies both Eqs. (\ref{vandep}) and (\ref{eandv}), and ends when any of them is violated. Let's however remember that, in any case, the slow-roll conditions
are sufficient but not necessary
to drive inflation. Strictly speaking inflation may end some time after the slow-roll conditions are violated, but this time is very small compared with the 70 e-folds or so of inflation required to solve the horizon, flatness, and unwanted relics problems, under standard evolution.

\subsection{Spectrum of perturbations of a dominating light scalar field during a quasi de Sitter stage} \label{dSpec}

When we consider a scalar field $\varphi$ that dominates the energy density during inflation and whose perturbations are sizable enough, the spacetime stops being perfectly smooth so that we have to leave the comfortable unperturbed metric line element in Eq. (\ref{unperturbedle}) to adopt the perturbed line element described in Eq. (\ref{perturbedline}). As a result the Klein-Gordon equation for the Fourier modes $\omega_k$ of the perturbations in $\varphi$ is modified to take into account the backreaction of the metric \cite{kodama,mukhanov,mukhanovrep,riotto}:
\be
\ddot{\omega}_k + 3H_{\rm inf} \dot{\omega}_k + \left( \frac{k^2}{a^2} + \frac{\partial^2 V}{\partial \varphi_0^2}\right) \omega_k  =  - 2\phi_{G_k} \frac{\partial V}{\partial \varphi_0} + \dot{\phi}_{G_k} \dot{\varphi}_0 + 3\dot{\psi}_k \dot{\varphi}_0 - k^2 B_k \dot{\varphi}_0 \,. 
\ee
The above equation looks quite difficult to manage but, fortunately, we can eliminate some of the scalar degrees of freedom by fixing the gauge and using the perturbed Einstein equations for the inflaton field $\varphi$ \cite{kodama,mukhanov,mukhanovrep,riotto}. For instance, going to the longitudinal gauge, we can fix the scalar perturbations $B$ and $E$ to be zero in the metric line element of Eq. (\ref{perturbedline}), whereas the non-diagonal part of the $ij$ component of the perturbed Einstein equations requires $\phi_G = \psi$ being the stress associated to $\varphi_0$ completely isotropic. The modified Klein-Gordon equation reduces in this case to
\be
\ddot{\omega}_k + 3H_{\rm inf} \dot{\omega}_k + \left(\frac{k^2}{a^2} + \frac{\partial^2 V}{\partial \varphi_0^2}\right) \omega_k = - 2\psi_k \frac{\partial V}{\partial \varphi_0} + 4 \dot{\psi}_k \dot{\varphi}_0 \,. 
\ee

To solve the previous equation we still require to know the behaviour of $\psi$. To that aim we take advantage of the $00$, $0i$, and the diagonal part of the $ij$ components of the perturbed Einstein equations in the longitudinal gauge \cite{kodama,mukhanov,mukhanovrep,riotto}:
\bea
-3H_{\rm inf} (\dot{\psi} + H_{\rm inf} \psi) + \frac{\nabla^2 \psi}{a^2} &=& \frac{1}{2m_P^2} \left(\dot{\varphi}_0 \delta \dot{\varphi} - \dot{\varphi}_0^2 \psi + \frac{\partial V}{\partial \varphi_0} \delta \varphi \right) \,, \\
\dot{\psi} + H_{\rm inf} \psi &=& \frac{1}{2m_P^2} \dot{\varphi}_0 \delta \varphi \,, \label{0icomponent} \\
\left(2 \frac{\ddot{a}}{a} + H_{\rm inf}^2 \right)\psi + 4H_{\rm inf}\dot{\psi} + \ddot{\psi} &=& \frac{1}{2m_P^2} \left(\dot{\varphi}_0 \delta \dot{\varphi} - \dot{\varphi}_0^2 \psi - \frac{\partial V}{\partial \varphi_0} \delta \varphi \right) \,,
\eea
which combined give the following equation for $\psi_k$ in terms of the slow-roll parameters $\varepsilon$ and $\eta$:
\be
\ddot{\psi}_k + H_{\rm inf} (1 + 2\eta - 2\varepsilon) \dot{\psi}_k + 2H_{\rm inf}^2 (\eta - 2\varepsilon) \psi_k + \frac{k^2}{a^2} \psi_k = 0 \,.
\ee
A quick look at the previous expression reveals that, on superhorizon scales, $\psi$ behaves as $\dot{\psi}_k \approx 2(2\varepsilon - \eta) H_{\rm inf} \psi_k$ so that $|4\dot{\psi}_k\dot{\varphi}_0| \ll |2\psi_k \partial V/\partial\varphi_0|$, whereas on subhorizon scales $\psi_k \approx 0$. In view of this, and by making use of the $0i$ component of the perturbed Einstein equations [\hbox{c.f. Eq. (\ref{0icomponent})}], which may also be written down as
\be
\dot{\psi} + H_{\rm inf} \psi = \varepsilon H_{\rm inf}^2 \frac{\delta \varphi}{\dot{\varphi}_0} \,,
\ee
we conclude that, on superhorizon scales,
\be
\psi_k \approx \frac{\varepsilon H_{\rm inf} \omega_k}{\dot{\varphi}_0} \,, \label{psilongitudinal}
\ee
so that the equation of motion for $\omega_k$ is finally given by
\bea
\ddot{\omega}_k + 3H_{\rm inf}\dot{\omega}_k + \left(\frac{k^2}{a^2} + \frac{\partial^2 V}{\partial \varphi_0^2} \right) \omega_k = \left\{
\begin{array}{c}
0 \,, \hspace{1.65cm} {\rm for} \ k \gg aH_{\rm inf} \,, \\
6\varepsilon H_{\rm inf}^2 \omega_k \,, \ \ {\rm for} \ k \ll aH_{\rm inf} \,.
\end{array}
\right\}
\eea

This kind of differential equation is much more familiar to us, and we know that it can be solved 
going to conformal time and making the usual change of variables
\be
\lambda_k \equiv \frac{\omega_k}{a} \,.
\ee
The resultant equation of motion for $\lambda_k$ is then
\be
\lambda_k'' + \left[k^2 - \frac{1}{\eta^2}\left(\upsilon_\varphi^2 - \frac{1}{4}\right)\right] \lambda_k = 0 \,,
\ee
with $\upsilon_\varphi$ defined by
\bea
\upsilon_\varphi &\equiv& \left[\frac{1}{4} - \frac{3\eta_\varphi - 2 - 5\varepsilon}{(1 - \varepsilon)^2}\right]^{1/2} \nonumber \\
&\approx& \frac{3}{2} + 3\varepsilon  - \eta_\varphi \,.
\eea
The solution for this equation is immediate, based on the results found in Subsection~\ref{spectrum2}. The magnitude of the mode function $\omega_k$ on superhorizon scales is then almost constant and given by
\bea
|\omega_k| &\approx& [2(1 - \varepsilon)]^{\upsilon_\varphi - \frac{3}{2}} (1 - \varepsilon) \frac{\Gamma(\upsilon_\varphi)}{\Gamma(3/2)} \frac{H_{\rm inf}}{\sqrt{2k^3}} \left(\frac{k}{aH_{\rm inf}}\right)^{\frac{3}{2} - \upsilon_\varphi} \nonumber \\
&\approx& \frac{H_\ast}{\sqrt{2k^3}} \left(\frac{k}{aH_{\rm inf}}\right)^{\eta_\varphi - 3\varepsilon} \,,
\eea
which is used to calculate the spectrum $\mathcal{P}_{\delta \varphi}(k)$ of perturbations in the inflaton field $\varphi$ by means of Eq. (\ref{easyp}):
\be
\mathcal{P}_{\delta \varphi} (k) \approx \left(\frac{H_\ast}{2\pi}\right)^2 \left(\frac{k}{aH_{\rm inf}}\right)^{n_{\delta \varphi}} \,. \label{domspectrum}
\ee
The spectrum of perturbations in $\varphi$ is almost scale-invariant with spectral index $n_{\delta \varphi}$ given in terms of the slow-roll parameters $\varepsilon$ and $\eta_\varphi$:
\be
n_{\delta \varphi} = 2\eta_\varphi - 6\varepsilon \,. \label{secondindex}
\ee

Comparing the spectrum obtained [c.f. Eqs. (\ref{domspectrum}) and (\ref{secondindex})] with that for a non-dominating scalar field [c.f. Eqs. (\ref{amplitudePsigma}) and (\ref{indexPsigma})], we see that the backreaction of the metric only affects the spectral index of the spectrum of perturbations. The amplitude remains the same either the respective field dominates the energy density or not.

\subsection{The spectrum of $\zeta$ in the inflaton scenario} \label{pzetainflation}

Now we are in position to calculate the spectrum of the curvature perturbation $\zeta$ in the inflaton scenario, based on the results found in the previous subsection. We first begin by invoking the definition of $\zeta$ given in Eq. (\ref{zeta_def1}) in terms of the $\varphi$ field:
\be
\zeta = -\psi - H_{\rm inf} \frac{\delta \varphi}{\dot{\varphi}_0} \,,
\ee
which, on superhorizon scales, reduces to
\be
\zeta_k = - (1 + \varepsilon) H_{\rm inf} \frac{\omega_k}{\dot{\varphi}_0} \simeq - H_{\rm inf} \frac{\omega_k}{\dot{\varphi}_0} \,,
\ee
where the expression in Eq. (\ref{psilongitudinal}) has been used.

The spectrum $\mathcal{P}_\zeta(k)$ of $\zeta$ is, in view of the latter, given in terms of the spectrum $\mathcal{P}_{\delta \varphi}(k)$ of the perturbations in $\varphi$:
\be
\mathcal{P}_\zeta (k) \approx \left(\frac{H_\ast}{\dot{\varphi}_0}\right)^2 \mathcal{P}_{\delta \varphi} (k) \,,
\ee
which, according to Eq. (\ref{domspectrum}), gives the final expression
\bea
\mathcal{P}_\zeta (k) \equiv A_\zeta^2 \left(\frac{k}{aH_{\rm inf}}\right)^{n_\zeta} &\approx& \left[\frac{H_\ast^2}{2\pi \dot{\varphi}_0} \right]^2 \left(\frac{k}{aH_{\rm inf}}\right)^{2\eta_\varphi - 6\varepsilon} \nonumber \\
&=& \left[\frac{H_\ast}{\sqrt{8\varepsilon}\pi m_P}\right]^2 \left(\frac{k}{aH_{\rm inf}}\right)^{2\eta_\varphi - 6\varepsilon} \,, \label{Aspectruminf}
\eea
where, in the second line, the amplitude $A_\zeta$ is written down in terms of $\varepsilon$.

In the inflaton scenario $\varphi$ is responsible of driving inflation and also of generating the required level of primordial perturbations measured by WMAP \mbox{($|A_\zeta| \approx 5 \times 10^{-5}$} \cite{observation}). That imposes the following constraint on the Hubble parameter during inflation $H_\ast$ in terms of $\varepsilon$:
\be
H_\ast \approx 10^{15} \sqrt{\varepsilon} \ {\rm GeV} \,,
\ee
that combined with the present bound $\varepsilon \lsim 0.01$ coming from spectral index and gravitational waves constraints \cite{observation} requires $H_\ast \lsim 10^{14}$ GeV \cite{lyth84a}. Despite the fact that the inflationary energy scale, directly related to the value of $H_\ast$, is in the inflaton scenario regulated by the parameter $\varepsilon$, low-scale inflation may well be obtained but only at the expense of a very small $\varepsilon$, which in turn requires a high level of fine-tuning \cite{peiris} (see however Ref. \cite{snat}). As a consequence serious problems appear when trying to build successful particle physics inflationary models \cite{book,treview}. The relevance of finding such a kind of low-scale inflation models is evident since the inflaton field could be identified with one of the MSSM flat directions or one of the scalar fields in the SUSY breaking sector (see for example Refs. \cite{berkooz,dine,enqvistrep,kasuya,lythlsi,lyth04}).

Going back to the curvaton scenario, we said that one of the assumptions of the model was a negligible curvature perturbation generated by the inflaton. This assumption may be quantified by requiring $\zeta_r$ to be, say, at most $1\%$ of the total $\zeta$, which means, from Eq. (\ref{Aspectruminf}), that $H_\ast$ in the curvaton scenario must satisfy
\be
H_\ast \lsim 10^{13} \sqrt{\varepsilon} \ {\rm GeV} \lsim 10^{12} \ {\rm GeV} \,.
\ee
In the following section we will show how such an upper bound on $H_\ast$ makes the detection of gravitational waves an anti-smoking gun for the curvaton scenario.

\section{Gravitational waves}

Primordial tensor-type perturbations in spacetime are regarded as gravitational waves, being unsourced
during inflation and susceptible to be decomposed in a polarisation tensor basis. They also propagate in a way that each subhorizon mode function follows a harmonic wave equation of motion, i.e. the Klein-Gordon equation associated to a massless scalar field in Minkowski spacetime.

The relevant quantity to discuss now is $\Pi^T_{ij}$, which is the tensor component of the full perturbed metric tensor, and that, as shown in Subsection \ref{freecounting}, is characterized by two degrees of freedom in our three dimensional space as well as by satisfying the transversality condition $\partial_i \Pi^T_{ij} = 0$. Hereafter we will write $\Pi_{ij}^T$ as $2h_{ij}$ as it is done in most of the literature. Displaying only the tensor perturbation, the metric line element in conformal time reads then
\be
ds^2 = a^2(\eta) [-d\eta^2 + (\delta_{ij} + 2 h_{ij}(\eta,{\bf x}))dx^i dx^j] \,. \label{pmetrich}
\ee
Tensor perturbations are decoupled from their scalar and vector counterparts. The line element in Eq. (\ref{pmetrich}) shows then that $h_{ij}$ is a gauge invariant quantity.

The Einstein-Hilbert action involving $h_{ij}$, and given in a general way by 
\be
S_E \equiv - \frac{m_P^2}{2} \int d^4x (-g)^{1/2} R \,, \label{actiong}
\ee
being $g$ the determinant of the $g_{\mu\nu}$ metric tensor and $R$ the Ricci scalar,
is given as a function of the kinetic term associated to $h_{ij}$ as obtained from Eq. (\ref{pmetrich}):
\be
S_E = -\frac{m_P^2}{2} \int d^4x (-g)^{1/2} \frac{1}{2} \partial_\mu h_{ij} \partial^\mu h_{ij} \,. \label{actionh}
\ee
Notice that no more terms have been added to Eqs. (\ref{actiong}) and (\ref{actionh}) because no tensor-type contributions to the energy-momentum tensor exist during the inflationary period. The primordial tensor perturbations $h_{ij}$ are in consequence unsourced so that they propagate freely throughout space following a harmonic (on subhorizon scales) wave propagation pattern.

To clearly show how the $h_{ij}$ perturbations propagate, we apply to them the same kind of treatment we do to the scalar perturbations in previous sections. We begin by decomposing $h_{ij}$ in canonically normalised Fourier modes $h_k^p \,$:
\be
h_{ij} (\eta, {\bf x}) = \frac{\sqrt{2}}{m_P} \int \frac{d^3k}{(2\pi)^{3/2}} \exp(i{\bf k} \cdot {\bf x}) \sum_p \varepsilon_{ij}(p,{\bf k}) h^p_k (\eta) + h.c. \,, \label{classgravamp}
\ee
where $p = +,\times$ are the two degrees of freedom (polarisation states), and the factors $\varepsilon_{ij} (p,{\bf k})$ are the polarisation tensors that satisfy
\bea
\sum_i k_i \varepsilon_{ij} (p,{\bf k}) &=& 0 \,, \\
\sum_{ij} \varepsilon_{ij}^\ast (p, {\bf k}) \varepsilon_{ij} (p', {\bf k}) &=& 2\delta_{pp'} \,, \label{secondpgw} \\
\sum_{ijl} \varepsilon^{ilk} \varepsilon_{ij}^{\ast}(+, {\bf
k}) \varepsilon_{jl}(\times, {\bf k}) &=& -\sum_{ijl} \varepsilon^{ilk}
\varepsilon_{ij}^{\ast}(\times, {\bf k}) \varepsilon_{jl}(+, {\bf
k}) = 2\frac{k_k}{\left| {\bf k} \right|} \,, \label{thirdpgw} \\
\sum_{ijl} \varepsilon^{ilk} \varepsilon_{ij}^{\ast}(+, {\bf k}) \varepsilon_{jl}(+,
{\bf k}) &=& -\sum_{ijl}\varepsilon^{ilk} \varepsilon_{ij}^{\ast}(\times, {\bf k})
\varepsilon_{jl}(\times, {\bf k}) = 0 \,, \label{fourthpgw}
\eea
according to the transversality condition and the properties of the rotational transformations \cite{LRQ1,LRQ}\footnote{In the expressions of Eqs. (\ref{thirdpgw}) and (\ref{fourthpgw}), $\varepsilon^{ijk}$ is the totally antisymmetric Levi-Civita tensor.}. Next, we recognize that 
the mode functions $h_k^p$, which satisfy the Klein-Gordon equation of motion derived from the Einstein action in \hbox{Eq. (\ref{actionh})} \cite{lifshitz,starob}:
\be
\ddot{h}_k^p +3H_{\rm inf} \dot{h}_k^p + \frac{k^2}{a^2} h_k^p = 0 \,,
\ee
are better handled if we rescale them as
\be
h_k^p \equiv \frac{z_k^p}{a} \,.
\ee
Thus, the equation of motion for $z_k^p$ is the same as that for a massless scalar field\footnote{That property reflects the masslessness of the {\it graviton} (the gravity messenger particle).}:
\be
z_k'' + \left[k^2 - \frac{1}{\eta^2}\left(\upsilon_h^2 - \frac{1}{4}\right)\right] z_k = 0 \,, \label{eqforgw}
\ee
and reduces in the subhorizon limit ($-k\eta \gg 1$) to the Klein-Gordon equation in Minkowski spacetime \cite{hwang}.
We point out that, in deriving the previous expression, we have worked in conformal time during a quasi de Sitter stage. The expansion parameter $a(\eta)$ is in this case given by
\be
a(\eta) = -\frac{1}{H_{\rm inf}(\eta) \eta (1-\varepsilon)} \,,
\ee
where the conformal time $\eta$ takes negative values, and the parameter $\upsilon_h$ is
\bea
\upsilon_h &=& \left[\frac{1}{4} + \frac{2 - \varepsilon}{(1 - \varepsilon)^2}\right]^{1/2} \nonumber \\
&\approx& \frac{3}{2} + \varepsilon \,.
\eea
The solution to Eq. (\ref{eqforgw}) is well known from previous sections (see specifically Eq. (\ref{presolution}) in Subsection \ref{spectrum2}), and reduces to the almost time-independent value
\bea
|h_k^p| &\approx& [2(1 - \varepsilon)]^{\upsilon_h - \frac{3}{2}} (1 - \varepsilon) \frac{\Gamma(\upsilon_h)}{\Gamma(3/2)} \frac{H_{\rm inf}}{\sqrt{2k^3}} \left(\frac{k}{aH_{\rm inf}}\right)^{\frac{3}{2}-\upsilon_h} \nonumber \\
&\approx& \frac{H_\ast}{\sqrt{2k^3}} \left(\frac{k}{aH_{\rm inf}}\right)^{-\varepsilon} \,, \label{hkp}
\eea
for the magnitude of the mode function $h_k^p$ in the superhorizon regime.

We are interested in the statistical properties of the gravitational waves which are well described by the spectrum $\mathcal{P}_T(k)$ defined by the statistical average
\be
\sum_{ij} \langle h^{ij}_{{\bf k}_1} h^{ij}_{{\bf k}_2} \rangle \equiv \frac{2\pi^2}{k^3} \delta^3 ({\bf k}_1 + {\bf k}_2) \mathcal{P}_T(k) \,, \label{defspecgrav}
\ee
over an ensemble of universes. Here $h^{ij}_{{\bf k}}$ stands for
\be
h^{ij}_{{\bf k}} \equiv \frac{\sqrt{2}}{m_P} \sum_p \varepsilon_{ij}(p,{\bf k}) h^p_k + h.c. \,.
\ee
To calculate the statistical average during inflation, we must promote the gravitational wave amplitude to an operator $\hat{h}_{ij}$ by introducing the creation and annihilation operators $\hat{a}^{p\dagger}_{\bf k}$ and $\hat{a}^p_{\bf k}$ that depend on the polarisation $p$ and wave vector ${\bf k}$, and satisfy the commutation relation \cite{hwang,LRQ1}
\be
[\hat{a}^p_{\bf k},\hat{a}^{p'\dagger}_{\bf k'}] = \delta^3 ({\bf k} - {\bf k'}) \delta_{pp'} \,.
\ee
The gravitational amplitude operator $\hat{h}_{ij}(\eta, {\bf x})$ that generalizes Eq. (\ref{classgravamp}) is then given by
\be
\hat{h}_{ij} (\eta, {\bf x}) = \int \frac{d^3k}{(2\pi)^{3/2}} \exp(i{\bf k} \cdot {\bf x}) \hat{h}^{ij}_{\bf k} (\eta) \,,
\ee
with
\be
\hat{h}^{ij}_{\bf k} (\eta) \equiv \frac{\sqrt{2}}{m_P} \sum_p \left[ \varepsilon_{ij}(p,{\bf k}) h^p_k (\eta) \hat{a}_{\bf k}^p + \varepsilon_{ij}^\ast (p,-{\bf k}) h^{\ast p}_k (\eta) \hat{a}_{-{\bf k}}^{p \dagger} \right] \,.
\ee

Being all the universes in the ensemble in the vacuum state during inflation, the statistical average $\langle h^{ij}_{{\bf k}_1} h^{ij}_{{\bf k}_2} \rangle$ is easily identified with the expectation value $\langle 0 | \hat{h}^{ij}_{{\bf k}_1} \hat{h}^{ij}_{{\bf k}_2} | 0 \rangle$. The spectrum of gravitational perturbations $\mathcal{P}_T(k)$, defined by Eq. (\ref{defspecgrav}), is then
\bea
\mathcal{P}_T(k) &=& \frac{k^3}{\pi^2 m_P^2}  \sum_{ij} \sum_{pp'} \varepsilon_{ij} (p, {\bf k}) \varepsilon^\ast_{ij} (p',{\bf k}) h_k^p (\eta) h_k^{\ast p'} (\eta) \nonumber \\
&=& \frac{4k^3}{\pi^2 m_P^2} |h_k^p|^2 \,,
\eea
where one of the properties of the polarisation tensors [c.f. Eq. (\ref{secondpgw})] has been used. Now we can make use of the result in Eq. (\ref{hkp}) to finally arrive to a definite expression for $\mathcal{P}_T(k)$ on superhorizon scales in terms of $H_\ast$ and $\varepsilon$ \cite{abbott,fabbri,liddle,book,rubakov}:
\be
\mathcal{P}_T(k) \equiv A_T^2 \left(\frac{k}{aH_{\rm inf}}\right)^{n_T} \approx \left [\frac{\sqrt{2}H_\ast}{\pi m_P}\right]^2 \left(\frac{k}{aH_{\rm inf}}\right)^{-2\varepsilon} \,. \label{finalgwspectrum}
\ee

This nice result shows that the inflationary energy scale, given by $H_\ast$, can be known from a direct measurement of the amplitude $A_T$. Unfortunately, at the moment all the efforts to detect gravity waves have been fruitless, leaving only the upper bound \hbox{$H_\ast \lsim 10^{14}$ GeV} \cite{observation}. In addition, technological restrictions impose the lower bound $H_\ast \gsim 10^{12}$ GeV if gravity waves may some day be detected \cite{CMB1,CMB2,CMB3,seto,sigurdson,smith}. A positive detection would kill then the curvaton scenario because, as we had discussed in Subsection \ref{pzetainflation}, the inflationary energy scale in this scenario is required to satisfy $H_\ast \lsim 10^{12}$ GeV to make the inflaton field $\varphi$ not to generate enough curvature perturbation. Only in non-slow roll inflationary models \cite{pilo} (specifically if $\eta_\varphi \equiv m_\varphi^2/3H_\ast^2> 1$), the energy scale during inflation could be high enough to allow the detection of gravity waves consistent with a negligible contribution of $\varphi$ to $\zeta$.

We end up this section by reporting the existence of a consistency relation between the curvature perturbation spectrum $\mathcal{P}_\zeta (k)$ and the gravitational waves one $\mathcal{P}_T (k)$ in the inflaton scenario \cite{abbott,fabbri,book,rubakov}. The ratio between the amplitudes of both spectra [c.f. \hbox{Eqs. (\ref{Aspectruminf}) and (\ref{finalgwspectrum})}] is given by the slow-roll parameter $\varepsilon$:
\be
r_{T \zeta} \equiv \frac{A_T^2}{A_\zeta^2} = 16 \varepsilon \,, \label{randvar}
\ee
which in turn gives information about the spectral index of $\mathcal{P}_T (k) \,$:
\be
n_T = -2\varepsilon \,.
\ee
The ratio $r_{T \zeta}$ is then consistently related to the spectral index $n_T$, this relation being given by the expression
\be
r_{T \zeta} = -8 n_T \,. \label{constzeta}
\ee
No consistency relation of this type is encountered in the curvaton scenario or in other scenarios for the origin of the large-scale structure in the Universe, although it is true that $r_{T\zeta}$ is always smaller in the multi-component inflationary case \cite{treview}, so that its future confirmation would mean good news for the inflaton scenario. Nevertheless, if the consistency relation turns out to be experimentally wrong, that does not mean necessarily that the inflationary paradigm is wrong, just that the single-field variant is not nature's choosing. Anyway, the non-gaussianity signatures associated to each model would add up to the consistency relation in Eq. (\ref{constzeta}), to act as powerful discriminators for models that give origin to the primordial energy density perturbations (see Chapters \ref{fourthpaper} and \ref{fifthpaper}).

\section{Conclusions}

The curvature perturbation $\zeta$ is a well defined quantity, gauge-invariant and conserved on large scales (if the adiabatic condition is satisfied), that allows us to quantify the primordial energy density inhomogeneities produced during inflation. The statistical properties of $\zeta$ are given by the spectrum $\mathcal{P}_\zeta (k)$ whose amplitude and spectral index strongly depend on the specific mechanism of production of density inhomogeneities. This makes $\mathcal{P}_\zeta (k)$ act as a discriminator for the different production mechanisms, at least as the spectral index $n_\zeta$ and the possible relation of $A_\zeta$ with the gravitational waves spectrum $\mathcal{P}_T (k)$ are concerned \cite{book,treview}. The inflationary energy scale, given by $H_\ast$, is well determined by the amplitude of $\mathcal{P}_T (k)$ so that the current upper bound on $A_T$ leads to $H_\ast \lsim 10^{14}$ GeV \cite{observation}. The energy scale may be well below $10^{14}$ GeV, but only at the expense of a high level of fine tuning to make the slow-roll parameter $\varepsilon$ be extremely below 1. This reflects how constrained is the inflaton potential in the inflaton scenario, in order to produce enough curvature perturbation while driving inflation. As a result the particle physics motivated inflationary models are quite unrealistic if we insist that the inflaton field $\varphi$ has to produce the energy density inhomogeneities \cite{peiris}. It is here where the curvaton scenario comes to the rescue: by requiring $\varphi$ just to drive inflation, the weakly coupled curvaton field $\sigma$ is in charge of giving origin to $\zeta$ \cite{lyth03c,lyth02,moroi01b}. The inflationary energy scale is in this case easily lowered so as to possibly associate $\varphi$ with one of the fields appearing in supersymmetric extensions of the Standard Model of particle physics. Gravitational waves are in this case so tiny to ever be detected, since the current detection technology restricts $H_\ast$ to be above $10^{12}$ GeV \cite{CMB1,CMB2,CMB3,seto,sigurdson,smith}. Which scenario for the generation of $\zeta$ is correct will be determined by future observations. At the moment, we will just try to do our best to successfully integrate cosmology and particle physics, being the first step the determination of the lower bound for $H_\ast$ in the simplest curvaton model \cite{lyth04}. As we will see in the next chapter, $H_\ast$ in such a model is still quite high, being $H_\ast > 10^7$ GeV, so that a modification to the basic setup is urgently needed. Two modifications to the simplest curvaton model will be explored in Chapters \ref{firstpaper} and \ref{secondpaper} \cite{yeinzon,yeinzon04}, showing that low scale inflation with $H_\ast \sim 1$ TeV or lower is possible to be obtained.

\chapter{Low scale inflation and the curvaton mechanism} \label{firstpaper}

\section{Introduction}

The primordial curvature perturbation $\zeta$ is generated presumably from the 
perturbation of some scalar field, which in turn is generated from the 
vacuum fluctuation during inflation. The scalar field responsible for the
primordial curvature perturbation is traditionally supposed to be the 
inflaton field $\varphi$, i.e. the field responsible for the dynamics and, in particular,
the end of inflation \cite{book}. 
This `inflaton hypothesis' is economical, but it is 
quite difficult to implement and, if many scalar fields exist, it presumably 
is not particularly likely. An alternative is that the  curvature perturbation 
is generated by the weakly coupled curvaton field $\sigma$, which could dominate (though not necessarily) 
the energy density before it decays \cite{lyth03c,lyth02,moroi01b} (see also \hbox{Refs. \cite{moroi01a,earlier1,earlier2}}).
According to this `curvaton hypothesis', the contribution of the inflaton
to the curvature perturbation is negligible. This is especially true if 
the energy scale of inflation is much lower than the scale of grand 
unification, the latter scale being the typical requirement of the traditional inflaton 
hypothesis\footnote{Although some exceptions exist (see for example Ref. \cite{snat}).}.
In fact, one of the advantages of the curvaton scenario is the relaxation
of the constraints on the inflationary energy scale, which alleviates many 
tuning problems in inflation model--building and allows for the construction 
of realistic, theoretically well-motivated inflation models \cite{liber,mota,motayo}.

In the simplest version of the curvaton model though,
the scale of inflation is still required to be quite high corresponding to 
Hubble parameter \mbox{$H_\ast>10^7$\,GeV} \cite{lyth04}. 
The purpose of this chapter is to systematically
explore a modification of the curvaton model which can instead allow
inflation at an even lower scale \cite{yeinzon}\footnote{Low scale inflation has also been
studied \cite{postma04inh} in the context of the `inhomogeneous reheating scenario'
\cite{inhomog1,inhomog3,modreh} (see also Refs. \cite{emp,matarrese,maz,mazpo03,inhomog4,inhomog2}),
where the inhomogeneities in the inflaton decay rate during inflation give origin to
$\zeta$.}. To be specific, we aim for 
 $H_\ast\sim 10^3$\,GeV, which holds if the inflationary potential is 
generated by some mechanism of  gravity-mediated supersymmetry 
breaking which holds also in the vacuum.

We begin by presenting some known bounds in a unified notation.
Then we consider the possibility that the  curvaton mass increases 
suddenly at some moment after the end of inflation but before the onset of 
the curvaton oscillations \cite{yeinzon}.

\section{The bounds on the scale of inflation}

In this section we present four bounds on the scale of inflation, in terms
of three parameters which encode possible modifications of the simplest 
curvaton scenario. These bounds have been presented at least implicitly in
earlier works \cite{lyth04,matsuda03,postma04} but not in the unified notation 
that we 
employ. The advantage of this notation is that it will allow us to compare the 
bounds in various situations, establishing with ease which is  the most 
crucial. The three parameters are
\begin{itemize}
\item
The  ratio $\epsilon\equiv\sigma_*/\sigma_{\rm osc}$, where
$\sigma_*$ is the global value of the curvaton field at horizon exit and 
$\sigma_{\rm osc}$ is its global value when it starts to oscillate.
\item
The ratio $f\equiv H_{\rm osc}/\tilde m_\sigma$, where $H_{\rm osc}$ is the 
Hubble parameter at the start of the oscillations and $\tilde m_\sigma$
is the effective curvaton mass after the onset of the oscillations.
\item
The ratio $\delta\equiv\sqrt{H_{\rm osc}/H_*}$ where $H_*$ is the Hubble 
parameter a few Hubble times after horizon exit.
\end{itemize}

\subsection{Curvaton physics considerations}

The  observed value of the nearly scale invariant spectrum of curvature 
perturbations, parameterised by the amplitude $A_\zeta$, is 
$|A_\zeta| \approx 5\times 10^{-5}$ \cite{observation}.
In the curvaton scenario $\zeta$ is given by \cite{lyth03c,lyth02,moroi01b} [c.f. Eq. (\ref{fomegadef})]
\begin{equation}
\zeta \approx \Omega_{\rm dec}\zeta_\sigma \,,
\label{zeta}
\end{equation}
where \mbox{$\Omega_{\rm dec}\leq 1$} is the density fraction of the global curvaton energy
density $\rho_{\sigma_0}$ over the global total energy density of the Universe $\rho_{\rm {total_0}}$ at the time
of the decay of the curvaton:
\begin{equation}
\Omega_{\rm dec}\equiv\left(\frac{\rho_{\sigma_0}}{\rho_{\rm {total_0}}}\right)_{\rm dec} \,,
\end{equation}
and $\zeta_\sigma$ is the curvature perturbation of the curvaton field
$\sigma$, which is \cite{CD} [c.f. Eq.~(\ref{curvcp})]
\begin{equation}
\zeta_\sigma\sim\left(\frac{\delta\sigma}{\sigma}\right)_{\rm dec}\approx
\left(\frac{\delta\sigma}{\sigma}\right)_{\rm osc} \,,
\label{zs1}
\end{equation}
where `osc' denotes the time when the curvaton oscillations begin and `dec'
denotes the time of curvaton decay.

In all the cases which we consider,
\begin{equation}
\left(\frac{\delta\sigma}{\sigma}\right)_*\simeq
\left(\frac{\delta\sigma}{\sigma}\right)_{\rm osc} \,,
\label{ds/s}
\end{equation}
where `*' denotes the epoch when the cosmological scales exit the horizon 
during inflation. The above typically holds true because the curvaton (being a 
light field) is frozen during and after inflation until the onset of its 
oscillations. However, this does not mean that 
\mbox{$\sigma_*\simeq\sigma_{\rm osc}$} 
necessarily. Indeed, in the case of a pseudo Nambu-Goldstone boson (PNGB) curvaton with a 
varying order parameter $v$, the curvaton field is associated with the angular 
displacement $\theta$ from the minimum of its potential as \cite{dlnr,yeinzon}
\begin{equation}
\sigma\equiv\sqrt 2\,v\theta \,.
\label{theta}
\end{equation}
Therefore, even though after the end of inflation, $\theta$ remains 
approximately frozen (the angular motion is over damped), we may have 
\mbox{$\epsilon\ll 1$}, where
\begin{equation}
\epsilon\equiv\frac{\sigma_*}{\sigma_{\rm osc}}\,,
\label{eps}
\end{equation} 
because [cf. Eq.~(\ref{theta})] 
\mbox{$v_*=\epsilon v_{\rm osc}\ll v_{\rm osc}$}. However, in this 
case too, for the curvaton fractional perturbation we find
\begin{equation}
\left(\frac{\delta\sigma}{\sigma}\right)_*
=\left(\frac{\delta\theta}{\theta}\,\right)_*\simeq
\left(\frac{\delta\sigma}{\sigma}\right)_{\rm osc} \,,
\label{fractional}
\end{equation}
which agrees nicely with Eq.~(\ref{ds/s}). 

Now, for the perturbation of the curvaton we have the following value for the
amplitude $A_{\delta \sigma_\ast}$ of the spectrum of perturbations [c.f. Eq. (\ref{amplitudePsigma})]
\begin{equation}
A_{\delta\sigma_*} \simeq \frac{H_*}{2\pi}\,.
\label{dsH}
\end{equation}
Combining Eqs.~(\ref{eps}) and (\ref{dsH}) we find
\begin{equation}
A_{\delta\sigma_{\rm osc}}\simeq\frac{H_*}{2\pi\epsilon}\,,
\label{dsosc}
\end{equation}
which means that, if the order parameter of a PNGB curvaton grows, the curvaton
perturbation is amplified by a factor $\epsilon^{-1}$ \cite{yeinzon}.

From Eqs.~(\ref{zeta}) and (\ref{zs1}) we have
\begin{equation}
\sigma_{\rm osc}\sim \Omega_{\rm dec}\,\frac{\delta\sigma_{\rm osc}}{\zeta} = \Omega_{\rm dec} \frac{A_{\delta \sigma_{\rm osc}}}{A_\zeta} \,.
\end{equation}
Using Eq.~(\ref{dsosc}), we can recast the above as
\begin{equation}
\sigma_{\rm osc}\sim\frac{H_*\Omega_{\rm dec}}{\pi\epsilon A_\zeta} \,.
\label{sosc}
\end{equation}

\subsection{The main bound on the scale of inflation}

For the density fraction at the onset of the curvaton oscillations we have:
\begin{equation}
\left(\frac{\rho_{\sigma_0}}{\rho_{\rm {total_0}}}\right)_{\rm osc}\sim f^{-2}
\left(\frac{\sigma_{\rm osc}}{m_P}\right)^2 \,,
\label{rhofracosc}
\end{equation}
where 
\begin{equation}
f\equiv\frac{H_{\rm osc}}{\tilde m_\sigma} \,,
\label{fff}
\end{equation}
and
we used that 
\mbox{$(\rho_{\sigma_0})_{\rm osc}\simeq
\frac{1}{2}\tilde m_\sigma^2\sigma_{\rm osc}^2$} 
and \mbox{$(\rho_{\rm {total_0}})_{\rm osc}=3H_{\rm osc}^2m_P^2$}. Here, $\tilde m_\sigma$
denotes the effective mass of the curvaton {\em after} the onset of its 
oscillations. In the basic setup of the curvaton hypothesis this effective mass
 is the bare mass $m_\sigma$. If this is the case then 
\mbox{$\tilde m_\sigma=m_\sigma\simeq H_{\rm osc}$} (i.e. \mbox{$f\simeq 1$}).
However, in the heavy curvaton scenario, the mass of the curvaton
is supposed to be suddenly incremented at some time after the end of the 
inflationary epoch due to a coupling of the form 
$\lambda \chi^2 \sigma^2$ with a field $\chi$ which acquires a large vacuum
 expectation value (VEV) at some time after the end of inflation \cite{yeinzon,lyth04,yeinzon04}. In this case
\mbox{$\tilde{m}_\sigma^2=m_\sigma^2+\lambda\langle\chi\rangle^2\approx 
\lambda\langle\chi\rangle^2\gg H_{\rm osc}^2$} (\hbox{i.e. \mbox{$f\ll 1$}}).

Now, we need to consider separately the cases when the curvaton decays before
it dominates the Universe (\mbox{$\Omega_{\rm dec}\ll 1$}) or after it does
so (\mbox{$\Omega_{\rm dec}\sim 1$}). Note, that the WMAP constraints 
on non-gaussianity in the CMB \cite{komatsu03} impose a lower bound on $\Omega_{\rm dec}$, which allows the
range \cite{lyth03c} [c.f. Eqs. (\ref{firstfnldef}) and (\ref{fWMAPr})]
\begin{equation}
0.01 \lsim \Omega_{\rm{dec}} \leq 1 \,. 
\label{WMAPr}
\end{equation} 
Because of the above bound we might require that the density ratio $\rho_\sigma/\rho_{\rm total}$
grows substantially after the end of inflation. Typically, in
the curvaton scenario this does indeed take place after the curvaton begins
oscillating, but only if the curvaton oscillates in a quadratic potential 
during the radiation era. As it was shown in Ref.~\cite{CD}, if the curvaton 
oscillates in a quartic or even higher order potential, its density ratio does 
not increase with time (it may well decrease instead) and satisfying the bound 
in Eq.~(\ref{WMAPr}) might be very hard. Due to this fact, in the following,
we assume that
the period of oscillations 
occurs in the radiation era with a quadratic potential. Hence, we consider that
\mbox{$H_{\rm osc}\leq\Gamma_{\rm inf}$}, being $\Gamma_{\rm inf}$ the inflaton
decay rate.

Suppose, at first, that the curvaton decays before dominating the density of
the Universe so that \mbox{$\Omega_{\rm dec}\ll 1$}.
Assuming that the curvaton oscillates in a quadratic potential, during the 
radiation epoch, its density fraction grows as 
\mbox{$\rho_{\sigma_0}/\rho_{\rm {total_0}}\propto H^{-1/2}$}. 
Therefore, at curvaton decay we have
\begin{equation}
\Omega_{\rm{dec}}\sim
\frac{\tilde m_\sigma^2\sigma_{\rm{osc}}^2}{T_{\rm{dec}}
H_{\rm{osc}}^{3/2}m_P^{3/2}} \,, 
\label{Tdec}
\end{equation}
where we used Eq.~(\ref{rhofracosc}) and also that 
\mbox{$(\rho_{\rm total_0})_{\rm dec}\sim T_{\rm dec}^4$}, with $T_{\rm dec}$ being the radiation temperature just after the curvaton decay.
Using Eq.~(\ref{sosc}) the above can be recast as
\begin{equation}
H_*\sim\pi\epsilon A_\zeta f
\frac{m_P}{\sqrt{\Omega_{\rm dec}}}
\left(\frac{H_{\rm dec}}{H_{\rm osc}}\right)^{1/4} \,,
\label{H*1}
\end{equation}
where we used that \mbox{$T_{\rm dec}^2\sim H_{\rm dec}m_P$}.

Now, suppose that the curvaton decays after it dominates the Universe so that
\mbox{$\Omega_{\rm dec}\sim 1$}. Since 
\mbox{$(\rho_\sigma/\rho_{\rm total})_{\rm dom}\simeq 1$}
by definition, using again that, during the radiation epoch, 
\mbox{$\rho_{\sigma_0}/\rho_{\rm total_0}\propto H^{-1/2}$} and in view of 
Eq.~(\ref{rhofracosc}), we obtain
\begin{equation}
H_{\rm dom}\sim H_{\rm osc} f^{-4}
\left(\frac{\sigma_{\rm osc}}{m_P}\right)^4 \,,
\label{Hdom}
\end{equation}
where `dom' denotes the time of curvaton domination\footnote{Here we define $H_{\rm dom}$ by $H_{\rm dom} = H_{\rm eq}$, where $H_{\rm eq}$ is the Hubble parameter at the time when the global curvaton energy density $\rho_{\sigma_0}$ makes equal to the global radiation energy density $\rho_{r_0}$.}. Employing again 
Eq.~(\ref{sosc}), the above can be written as
\begin{equation}
H_*\sim\pi\epsilon A_\zeta
f m_P
\left(\frac{H_{\rm dom}}{H_{\rm osc}}\right)^{1/4} \,.
\label{H*2}
\end{equation}

Combining Eqs.~(\ref{H*1}) and (\ref{H*2}) we find that, in all cases
\begin{equation}
H_* \sim \pi\epsilon A_\zeta f\frac{m_P}{\sqrt{\Omega_{\rm dec}}}
\left(\frac{\max\{H_{\rm dom}, H_{\rm dec}\}}{H_{\rm osc}}\right)^{1/4} \,.
\label{H*0}
\end{equation}
This can be rewritten as 
\begin{equation}
H_* \sim \Omega_{\rm dec}^{-2/5}
\left(\frac{H_*}{H_{\rm osc}}\right)^{1/5}
\left(\frac{\max\{H_{\rm dom}, H_{\rm dec}\}}{H_{\rm BBN}}\right)^{1/5}
(\pi\epsilon A_\zeta f)^{4/5}(T_{\rm BBN}^2m_P^3)^{1/5} \,,
\label{H*}
\end{equation}
or equivalently (using \mbox{$V_*^{1/4}\sim\sqrt{H_*m_P}$})
\begin{equation}
V_*^{1/4} \sim \Omega_{\rm dec}^{-1/5}
\left(\frac{H_*}{H_{\rm osc}}\right)^{1/10}
\left(\frac{\max\{H_{\rm dom}, H_{\rm dec}\}}{H_{\rm BBN}}\right)^{1/10}
(\pi\epsilon A_\zeta f)^{2/5}(T_{\rm BBN}m_P^4)^{1/5} \,,
\label{V*}
\end{equation}
where `BBN' denotes the epoch of Big-Bang Nucleosynthesis (BBN)
(\mbox{$T_{\rm BBN}\sim 1$ MeV}).
Now, according to Eq.~(\ref{WMAPr}) we have \mbox{$\Omega_{\rm dec}\leq 1$}.
Also, we require that the curvaton decays before BBN, i.e. 
\mbox{$H_{\rm dec}>H_{\rm BBN}$}.
Hence, the above provides the following bounds
\begin{equation}
\mbox{\framebox{%
\begin{tabular}{c}\\
$H_*>(\pi\epsilon A_\zeta f)^{4/5}(T_{\rm BBN}^2m_P^3)^{1/5}\sim
(\epsilon f)^{4/5}\times 10^7 \ {\rm GeV}$,\\
\\
$V_*^{1/4}>(\pi\epsilon A_\zeta f)^{2/5}(T_{\rm BBN}m_P^4)^{1/5}\sim
(\epsilon f)^{2/5}\times 10^{12} \ {\rm GeV}$.\\
\\
\end{tabular}}}
\label{H*bound}
\label{V*bound}
\end{equation}
In the standard setup of the curvaton scenario \mbox{$\epsilon=f=1$} and
the above bounds do not allow inflation at low energy scales to take place
\cite{lyth04}. However, we see that if either $\epsilon$ or $f$ are much
smaller than unity the lower bound on the inflationary scale can be 
substantially relaxed and low scale inflation can be accommodated\footnote{The relevance of a low $\epsilon$ makes evident in Ref. \cite{yeinzon} where the scenario of the curvaton as a PNGB is studied. The specific explored model refers to a PNGB whose order parameter $v$ is increased after the cosmological scales exit the horizon during inflation, but before the onset of the curvaton oscillations. That makes $\epsilon$ very small.}. Still, 
though, there are more bounds to be considered.

\subsection{Other bounds related to curvaton decay}

Firstly, let us consider the bound coming from the fact that the decay rate
of the curvaton field cannot be arbitrarily small. Indeed, in view of 
the fact that the curvaton interactions are at least of gravitational strength,
we find the following decay rate for the curvaton
\begin{equation}
\Gamma_\sigma \approx 
\gamma_\sigma 
\frac{\tilde{m}_\sigma^3}{m_P^2}\leq\tilde m_\sigma \,, 
\label{decay_rate}
\label{Gs}
\end{equation}
where \mbox{$\gamma_\sigma \gsim 1$}. 

Suppose, at first, that the curvaton decays after the onset of its 
oscillations, as in the basic setup of the curvaton scenario. In this case, 
\mbox{$\Gamma_\sigma\leq H_{\rm osc}$} and \mbox{$H_{\rm dec}=\Gamma_\sigma$}.
Using the fact that 
\mbox{max$\{H_{\rm dom}, \Gamma_\sigma\}\geq\Gamma_\sigma$},
Eq.~(\ref{decay_rate}) suggests
\begin{equation}
\frac{\max\{H_{\rm dom}, H_{\rm dec}\}}{H_{\rm osc}}\geq 
\gamma_\sigma 
f^{-1}
\left(\frac{\tilde m_\sigma}{m_P}\right)^2 \,.
\end{equation}
Including the above into Eq.~(\ref{H*0}) the latter becomes
\begin{equation}
H_* \geq 
\sqrt{\gamma_\sigma}(\pi\epsilon A_\zeta)^2\sqrt{f}\,
\frac{m_P}{\Omega_{\rm dec}}\left(\frac{H_{\rm osc}}{H_*}\right) \,,
\label{Hbound0}
\end{equation}
which results in the bounds
\begin{equation}
\mbox{\framebox{%
\begin{tabular}{c}\\
$H_*\geq 
(\pi\epsilon A_\zeta)^2\sqrt{f}\;\delta^2\,m_P
\sim
\epsilon^2\sqrt f\;\delta^2
\times 10^{11} \ {\rm GeV}$, \\
\\
$V_*^{1/4}\geq\pi\epsilon A_\zeta f^{1/4}\delta\,m_P
\sim
\epsilon f^{1/4}\delta
\times 10^{14} \ {\rm GeV}$, \\
\\
\end{tabular}}}
\label{Hbound}
\label{Vbound}
\end{equation}
where we have defined 
\begin{equation}
\delta\equiv
\sqrt{\frac{H_{\rm{osc}}}{H_\ast}} \,,
\label{dratio}
\label{fdelta}
\end{equation}
which must be really small in order to reduce the bounds in Eq. (\ref{H*bound}) to satisfactory levels.
In the case of a PNGB curvaton we see that the bounds in 
Eq.~(\ref{Hbound}) are drastically reduced with $\epsilon$, compared with
the bounds in Eq.~(\ref{H*bound}).

Now, provided we demand that the curvaton field does not itself result in a
period of inflation, we see that the curvaton cannot dominate the Universe 
before the onset of its oscillations. This results into the constraint
\begin{equation}
\left(\frac{\rho_{\sigma_0}}{\rho_{\rm total_0}}\right)_{\rm osc} \leq 1
\;\Leftrightarrow\;
\tilde m_\sigma\leq\pi\epsilon A_\zeta\,\delta^2\frac{m_P}{\Omega_{\rm dec}}
\;\Leftrightarrow\;
f\geq\frac{\Omega_{\rm dec}H_*}{(\pi\epsilon A_\zeta)m_P} \,,
\label{cons_m_n}
\label{mfbound}
\end{equation}
where we used Eqs.~(\ref{sosc}), (\ref{rhofracosc}), (\ref{fff}) and 
(\ref{fdelta}). Inserting the above into Eq.~(\ref{Hbound0}) we obtain
\begin{equation}
H_*\geq
\gamma_\sigma(\pi\epsilon A_\zeta)^3\delta^4\,
\frac{m_P}{\Omega_{\rm dec}} \,,
\label{Hbound1}
\end{equation}
which results in the bounds
\begin{equation}
\mbox{\framebox{%
\begin{tabular}{c}\\
$H_*\geq (\pi\epsilon A_\zeta)^3\delta^4m_P\sim
\epsilon^3\delta^4\times 10^7 \ {\rm GeV}$, \\
\\
$V_*^{1/4}\geq 
(\pi\epsilon A_\zeta)^{3/2}\delta^2m_P\sim
\epsilon^{3/2}\delta^2\times 10^{12} \ {\rm GeV}$. \\
\\
\end{tabular}}}
\label{Hbound-0}
\label{first}
\label{Vbound-0}
\end{equation}
A similar bound is reached with the use of the upper bound on 
$\tilde{m}_\sigma$
\begin{equation}
\tilde m_\sigma\leq\gamma_\sigma^{-1/3}(H_{\rm osc}m_P^2)^{1/3} \,,
\label{msbound}
\label{m_dr}
\end{equation}
which comes from $\Gamma_\sigma\leq H_{\rm osc}$ and
Eq.~(\ref{decay_rate}), 
instead of the bound in Eq. (\ref{cons_m_n}). Inserting the above into 
Eq.~(\ref{Hbound0}) one finds [cf. Eq.~(\ref{Hbound1})]
\begin{equation}
H_*\geq
\gamma_\sigma(\pi\epsilon A_\zeta)^3\delta^4\,
\frac{m_P}{\Omega_{\rm dec}^{3/2}} \,,
\label{Hbound2}
\end{equation}
which, again, results in the bound in Eq.~(\ref{Hbound-0}), as it was 
suggested in Ref.~\cite{postma04}.

In the heavy curvaton scenario we have \mbox{$\epsilon=1$} and
also \mbox{$H_{\rm osc}\simeq\min\{H_{\rm pt}, \tilde m_\sigma\}$}, where 
$H_{\rm pt}$ corresponds to the phase transition which increases the effective
mass of the curvaton. Then, if \mbox{$\delta\rightarrow 1$}, the bounds in 
Eq.~(\ref{Hbound-0})
are not possible to be relaxed below the standard case discussed in 
Ref.~\cite{lyth04} despite the fact that we may have \mbox{$f\ll 1$} in 
Eqs.~(\ref{H*bound}) and (\ref{Hbound}). 
Therefore, in the heavy curvaton 
scenario we require \mbox{$\delta\ll 1$}, i.e. {\em the onset of the curvaton
oscillations has to occur much later than the end of inflation} so that 
\mbox{$H_*\gg H_{\rm osc}\geq\Gamma_\sigma$} \cite{matsuda03}. In this case, as can be seen in
Eq.~(\ref{Hbound-0}), 
it is easy to lower the bound
on the inflationary scale even for a not-so-small value of $\delta$.
This is a very nice feature of this scenario. Note also, that in the case of a 
PNGB curvaton \mbox{$H_{\rm osc}\sim m_\sigma\ll H_*$} and $\delta$
is very small necessarily. Because, in this case, \mbox{$\epsilon\ll 1$},
it is straightforward to see that the bounds in Eq.~(\ref{Hbound-0}) are much
weaker than the bounds in Eq.~(\ref{H*bound}).

As it was pointed out in Ref.~\cite{postma04}, the sudden increment in the 
curvaton mass might lead to a growth in the curvaton decay rate enough for 
\mbox{$\Gamma_\sigma>H_{\rm pt}$}. This would force the curvaton to decay 
immediately and we can write 
\mbox{$H_{\rm osc}\sim H_{\rm pt}\sim H_{\rm dec}$}.
Obviously, in this case we cannot have \mbox{$H_{\rm dec}<H_{\rm dom}$} and
there is no period when \mbox{$\rho_{\sigma_0}/\rho_{\rm total_0}\propto H^{-1/2}$}. This 
means that \mbox{$(\rho_{\sigma_0}/\rho_{\rm total_0})_{\rm osc}\sim\Omega_{\rm dec}$}. Using
Eqs.~(\ref{sosc}) and (\ref{rhofracosc}) it is easy to find
\begin{equation}
H_*\sim\pi\epsilon A_\zeta f
\frac{m_P}{\sqrt{\Omega_{\rm dec}}} \,,
\label{H*00}
\end{equation}
which results in the following bounds
\begin{equation}
\mbox{\framebox{%
\begin{tabular}{c}\\
$H_*\geq\pi\epsilon A_\zeta f\,m_P\sim\epsilon f\times 10^{14} 
\ {\rm GeV}$, \\
\\
$V_*^{1/4}\geq\sqrt{\pi\epsilon A_\zeta f}\,m_P\sim
(\epsilon f)^{1/2}\times 10^{16} \ {\rm GeV}$. \\
\\
\end{tabular}}}
\label{Hbound-00}
\label{second}
\label{Vbound-00}
\end{equation}
It is evident that the above bounds may challenge the WMAP constraint for the
curvaton scenario \cite{liber} leading to excessive curvature 
perturbations from the inflaton field if $\varepsilon$ and/or $f$ are not much 
smaller than unity. 

The bounds in Eqs. (\ref{H*bound}), (\ref{Hbound}), and  (\ref{Hbound-0}) 
 provide the basis for our investigation \cite{yeinzon}, leaving the fourth bound in Eq.
(\ref{Hbound-00}) to be considered in the next chapter \cite{yeinzon04}.
As a matter of completeness we have considered all the other 
possible bounds coming from the requirements that 
\mbox{$\Gamma_\sigma < \tilde{m}_\sigma$} and 
\mbox{$H_{\rm dec} \geq H_{\rm BBN}$}. We have found that these bounds lead
to consistent and/or weaker constraints than the above four.

\section{The case of a heavy curvaton}

In this section we are going to consider the so called `heavy curvaton 
scenario'
 where an increment in the curvaton mass, at some moment after the end of inflation
 but before the onset of the curvaton oscillations, leads to a huge decrease
 of the inflationary scale through the attainment of a very small parameter
 $\delta$ [cf. Eq. (\ref{first})]. We will do so by the implementation of a
 second inflationary period following the idea first presented in Ref. \cite{matsuda03}\footnote{Note
 however that any post-inflationary phase transition could serve the purpose of giving an effective mass
 to the curvaton field.}.
 We identify this second inflationary period as the thermal inflation one which
 triggers the increment in the curvaton mass when the flaton field, that responsible
 for the generation of the thermal inflation era, rolls down towards the minimum
 of the potential.

\subsection{The thermal inflation model} \label{thermal}
Thermal inflation was introduced as a very nice mechanism to get rid of some unwanted relics that the main inflationary epoch is not able to dilute, without affecting the density perturbations generated during ordinary inflation. As its name suggests, thermal inflation relies on the finite-temperature effects on the {\it flaton} scalar potential. A flaton field $\chi$ could be defined as a field with mass 
$m_\chi$ and vacuum expectation value $M \gg m_\chi$ 
\cite{lyth95,lyth96}. 
More specifically, a flaton field is a MSSM flat direction lifted by non-renormalisable terms. SUSY breaking provides soft terms which create a large vacuum expectation value because the absence of quartic terms in the potential. 
The possible candidates for a flaton field within particle physics are either 
one of the many expected gauge singlets in string theory \cite{polchinski} or the GUT Higgs (which is a scalar field charged under the GUT gauge symmetry but neutral under the Standard Model one) with $m_\chi \sim 10^3$ GeV and $M \sim 10^{16}$ GeV \cite{lyth95}\footnote{Note, though, that in some GUT models there are additional Higgs fields with much smaller vevs \cite{guts4,guts2,guts3,guts1}.}.
After the period of reheating following the main inflationary epoch, the thermal background modifies the flaton potential $V$ trapping the flaton field at the origin and preventing it to roll-down towards $M$ \cite{barreiro96,lazarides86}. At this stage the total energy density $\rho_{\rm total}$ and pressure $P_{\rm total}$ are
\begin{eqnarray}
\rho_{\rm total} &=& V + \rho_r \,, \nonumber \\
P_{\rm total} &=& -V + \frac13 \rho_r \,,
\end{eqnarray}
making the condition for thermal inflation, $\rho_{\rm total_0} + 3P_{\rm total_0} < 0$, valid when the global thermal energy density $\rho_{r_0}$ falls below the height of the potential $V_h$,
which corresponds to a temperature of roughly $V_h^{1/4}$. Thermal inflation ends when the finite temperature becomes
ineffective at confining the field, at a temperature of order $m_\chi$, so the number of e-folds this inflationary period lasts is
\begin{equation}
N = \ln\left(\frac{a_{\rm{end}}}{a_{\rm{start}}}\right) 
=\ln \left(\frac{T_{\rm{start}}}{T_{\rm{end}}}\right) \sim \ln\left(\frac{V_h^{1/4}}{m_\chi}\right) 
\sim\frac 12\ln\left(\frac{M}{m_\chi}\right) \sim 10 \,.
\end{equation}
Here we have used the fact that, in a flaton potential of the form
\begin{equation}
V = V_h - (m_\chi^2 - gT^2) |\chi|^2 + \sum_{n=1}^{\infty} \lambda_n
m_P^{-2n} |\chi|^{2n+4} \,,
\end{equation}
where the $n$th term dominates:
\begin{eqnarray}
\tilde{m}_\chi^2 &=& 2 (n+1) m_\chi^2 \,, \\
M^{2n+2} m_P^{-2n} &=& [2(n+1)(n+2) \lambda_n]^{-1} \tilde{m}_\chi^2 \,, \\
V_h &=& [2(n+2)]^{-1} \tilde{m}_\chi^2 M^2 \,.
\end{eqnarray}
Note that the $gT^2$ contribution to the effective mass of the flaton field stands for the effect of the thermal background, which changes the slope of the potential in the $\chi$ direction and traps the flaton field at the origin of the potential \cite{barreiro96,lazarides86}.
It is worthwhile to mention that the potential is stabilized by non-renormalisable terms, with dimensionless couplings $\lambda_n \sim 1$ to make the theory valid up to the Planck scale.  Notice also that the $\lambda_4 |\chi|^4$ term is absent in the potential; otherwise, the vacuum
expectation value $M$ would not be much bigger than $\tilde{m}_\chi$, spoiling the suppression of the unwanted relics.

Before embedding the thermal inflation epoch and the curvaton mechanism into a single model, we want to clarify some issues about the nature of the interactions that produce the thermal background.
If the flaton is a GUT Higgs, it is coupled with those fields charged under the GUT gauge symmetry, in particular with those the inflaton field decays into. That collection of particles makes the thermal background, and its interaction with the flaton field produces the thermal correction. If the flaton field is a gauge singlet it still can be coupled, via Yukawa coupling terms, with some other fields, possibly in a hidden sector, that the inflaton field decays into. Again, a thermal correction is generated. The actual interactions and decay rate are not important as the main objective of this chapter and the next one is to obtain some particle physics model-independent information about the possibility of reconciling low scale inflation with the curvaton mechanism, in a scenario that involves a second period of inflation (thermal inflation), without going into the details of the identification of all the relevant fields (inflaton, flaton, and curvaton) in the framework of a particle physics model (GUT theories, MSSM, etc.), which would make the results highly particle physics model-dependent. The flaton could be either a gauge singlet or the GUT Higgs.  In the former case the flaton can be coupled with some other fields that the inflaton field decays into, via Yukawa coupling terms, and the specific interactions would be known once we choose what of the many gauge singlets expected in string theory is the flaton.
In the latter case the interactions in the GUT models are already known. The specific interactions are important of course, but there are so many possibilities that the general result would be hidden behind the characteristics associated to any definite particle physics model.

Having discussed the nature of the flaton interactions, and guided by the result in Ref. \cite{matsuda03}, we proceed to implement a second inflationary stage into the curvaton
scenario in order to lower the main inflationary energy scale.  If this second epoch of inflation is the thermal inflation one devised in Refs. \cite{lazarides86,lyth95,lyth96} we would be solving not only the issue of the ordinary inflation energy scale but also the moduli problem present in the standard cosmology \cite{banks,carlos,coughlan,ellislina,ellisnaq}.

In the curvaton model supplemented by a thermal inflation epoch two fields $\chi$ and $\sigma$, which we identify as the flaton and the curvaton fields respectively, are embedded into the radiation background left by the inflaton decay. It is assumed that the curvaton field could be either a gauge singlet \cite{polchinski}, the Peccei-Quinn field \cite{PQ}, a PNGB \cite{chun,dlnr,hofmann}, or a MSSM flat direction \cite{enqvistmpl,enqvist,ekm,enqvistrep,martin,hamaguchi,kasuya2,mcdonald2,mcdonald,mcdonald1,murayama,postma2,pomaz},
and has just a quadratic interaction with the flaton one so that the unperturbed component is frozen at some value $\sigma_\ast$ until the time when the flaton field is released from the origin and rolls down towards the minimum of the potential. This in turn signals the end of the thermal inflation era and the beginning of the oscillations of the curvaton field around the minimum of its quadratic potential \cite{CD}. The flaton field, in addition to the non renormalisable terms with $\lambda_n \sim 1$ that stabilize the potential and make its slope in the $\chi$ direction be very flat, presents a quadratic interaction with the curvaton field. The complete expression for the potential is
\begin{equation}
V(\chi, \sigma) = V_h - (m_\chi^2 - g T^2) |\chi|^2 + m_\sigma^2 |\sigma|^2 + \lambda |\chi|^2 |\sigma|^2 + \sum_{n=1}^\infty \lambda_n m_P^{-2n} |\chi|^{2n+4} \,, \label{mypotential}
\end{equation}
where $m_\chi \sim 10^3 \ {\rm GeV}$ due to the soft SUSY contributions in a gravity mediated SUSY breaking scheme. Under these circumstances the condition for an inflationary period, $\rho_{\rm total_0} + 3P_{\rm total_0} < 0$, is satisfied when the global thermal energy density $\rho_{r_0}$ falls below $V_h$. Of course, this period of thermal inflation ends when the effect of the thermal background becomes unimportant, at a temperature $T \sim m_\chi$, liberating the flaton field to roll down towards the minimum of the potential and letting it acquire a large vacuum expectation value $M$ given by:
\begin{equation}
M \simeq \frac{V_h^{1/2}}{m_\chi} \,.
\end{equation}
The evolution of the energy densities associated to the different fluids in this case are sketched in Fig. \ref{figstandard}.

\begin{figure}
\begin{center}
\includegraphics[width=15cm,height=20cm]{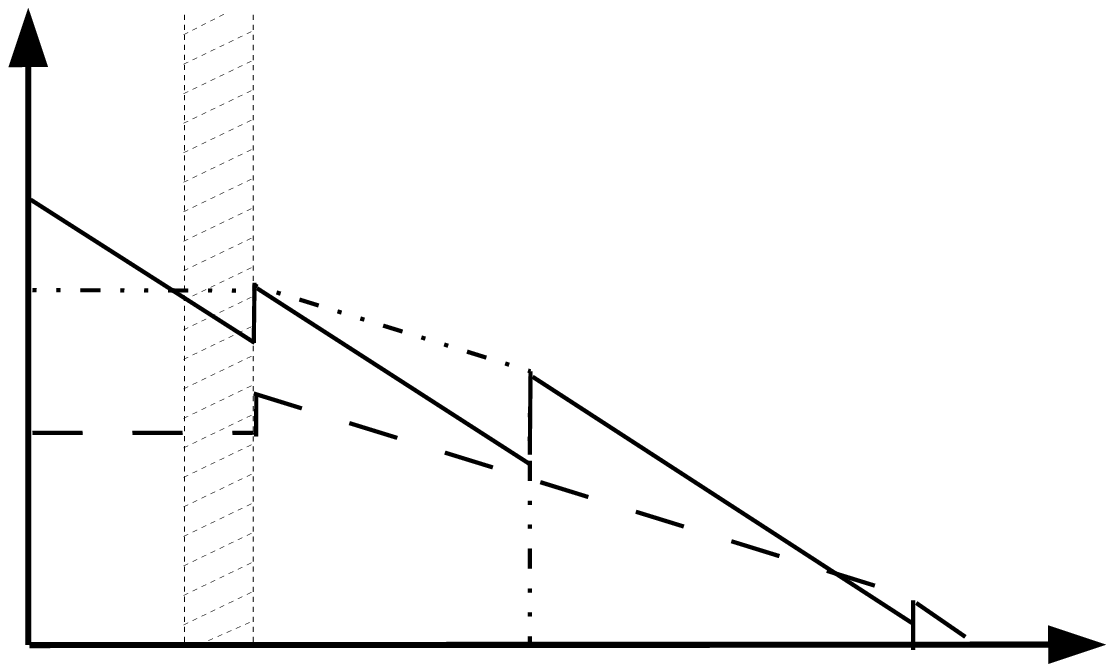}
\put(-340,520){$\log \rho$}
\put(-323,468){$V_h$}
\put(-280,385){$A$}
\put(-265,385){$B$}
\put(-204,385){$C$}
\put(-123,385){$D$}
\put(-80,385){$\log a$}
\vspace*{-13.5cm}
\caption{Evolution of the energy densities in the thermal inflation model where the curvaton field $\sigma$ has some time to oscillate before decaying \cite{yeinzon}. The continuous line corresponds to the global radiation energy density $\rho_{r_0}$, the dashed dotted line corresponds to the global flaton energy density $\rho_{\chi_0}$, and the dashed line corresponds to the global curvaton energy density $\rho_{\sigma_0}$. The horizontal axis represents the expansion parameter $a$. From the left to $A$ radiation dominates the energy density, although it decreases following \mbox{$\rho_{r_0} \propto a^{-4}$}. At this stage the flaton and curvaton fields $\chi$ and $\sigma$ are frozen at \mbox{$\chi_0 = 0$} and $\sigma_0 = \sigma_\ast$ making their energy densities constants. When $\rho_{r_0}$ reaches $V_h$ at $A$, thermal inflation begins. The thermal inflation period lasts until $B$ when the temperature $T$ becomes of the order of the flaton mass $m_\chi$. Thermal inflation stage is portrade by the dashed region. After thermal inflation ends, the parametric resonance process transforms a substantial fraction of $\rho_\chi$ into $\rho_r$ \cite{wandsreview,kofman,kls,preheat1}. The flaton field is liberated by this time and begins oscillating around the minimum of its potential, behaving then as a matter fluid with $\rho_{\chi_0} \propto a^{-3}$. The curvaton field increments suddenly its mass $m_\sigma$ at $B$ as a result of the oscillations of $\chi$ around the vacuum expectation value $M$, which yields to a much bigger, but still subdominant, $\rho_\sigma$. The increment is enough for the effective curvaton mass $\tilde{m}_\sigma$ to overtake $H_{\rm pt}$ (the Hubble parameter at $B$) so that $\sigma_0$ gets unfrozen and starts oscillating around $\sigma_0 = 0$. The curvaton field behaves then as a matter fluid so that $\rho_{\sigma_0} \propto a^{-3}$. By the time $C$, $\chi$ already dominates the energy density before decaying into radiation. The curvaton field continues to oscillate until $D$ when it decays into radiation after having come to dominate (though not necessarily) the total energy density. The curvature perturbation is transfered to the radiation at this moment due to the decay of $\sigma$.
\label{figstandard}}
\end{center}
\end{figure}

Let's assume that the usual inflation and its corresponding reheating have already
 happened, so that the flaton and
the curvaton fields are embedded into a radiation bath. Therefore, even when
 the minimum of the potential is located at
$\chi_0 = M_\chi(\sigma_*) \neq 0$ and $\sigma_0 = 0$, $\chi$ is trapped at the
 origin because of the finite-temperature effects and $|\sigma_0| = \sigma_\ast
 \neq 0$ because $m_\sigma < H < H_\ast$. Thus,
 the value of the scalar potential at this stage
is:
\begin{equation}
V(\chi_0 = 0, \sigma_0 = \sigma_\ast) = V_h + m_\sigma^2 \sigma_\ast^2 \,,
\end{equation}
with
\begin{eqnarray}
\tilde{m}_\chi^2 &=& 2 (n+1) (m_\chi^2 - \lambda |\sigma_0|^2) \,, \\
M_\chi^{2n+2} m_P^{-2n} &=& [(n+2) \lambda_n ]^{-1} (m_\chi^2 - \lambda |\sigma_0|^2) \,, \\
V_h &=& [2 (n + 2)]^{-1} (\tilde{m}_\chi^2 M_\chi^2) \mid _{\sigma_0 = 0} \,. 
\end{eqnarray}

When the thermal energy density falls below $V_h + m_\sigma^2 \sigma_\ast^2$
 thermal inflation begins. This period
lasts until the temperature is of the order the effective mass of the flaton
 field which is $\tilde{m}_\chi = %
(m_\chi^2 - \lambda \sigma_\ast^2)^{1/2}$. Note that $\lambda\sigma_*^2<m_\chi^2$
 because otherwise there is no thermal inflation. Then, we obtain a first constraint
 on the value of the parameter $\lambda$:
\begin{equation}
\lambda < \frac{m_\chi^2}{\sigma_\ast^2} \sim \frac{10^{-2} \,
 {\rm GeV}^2}{H_\ast^2 \Omega_{\rm dec}^2} \,, \label{const_lambda}
\end{equation}
where we have used the Eq. (\ref{sosc}) and focused on $m_\chi \sim 10^3$ GeV
 which comes from
the gravity-mediated SUSY breaking contributions.

When thermal inflation ends the thermal energy density is no longer dominant.
 The Hubble parameter at the end of thermal
inflation is then associated to the energy density coming from the curvaton
 and the flaton fields:
\begin{equation}
H_{\rm{osc}}^2=\frac{\rho_T+V(\chi_0=0,\sigma_0=\sigma_\ast)}{3m_P^2} 
\sim\frac{m_\chi^2M^2}{3m_P^2} \,,
\end{equation}
so that
\begin{equation}
H_{\rm{osc}} \sim 10^{-16} M \,,
\end{equation}
and therefore the parameter $f$ [cf. Eq. (\ref{fff})] is
\begin{equation}
f \equiv \frac{H_{\rm osc}}{\tilde{m}_\sigma} \sim 10^{-16} \frac{M}{\tilde{m}_\sigma} \,, \label{fagain}
\end{equation}
where $M \equiv M_\chi \mid_{\sigma_0 = 0}$ is somewhere in the range $10^{3}
 \hspace{1mm} {\rm GeV} \ll M \lsim 10^{18}%
\hspace{1mm} {\rm GeV}$.

With this so-low value for the Hubble parameter at the end of thermal inflation, the parameter $\delta$ [cf. Eq. (\ref{fdelta})] is
\begin{equation}
\delta
\sim 10^{-8}\sqrt{\frac{M}{H_\ast}} \,,
\end{equation}
so that the bounds in Eqs. (\ref{H*bound}) and (\ref{Hbound}) become\footnote{The bound in Eq. (\ref{Hbound-0}) is consistent with low scale inflation in view of $M \lsim 10^{18}$ GeV. Notice also that, in the heavy curvaton mechanism, $\epsilon = 1$ because there is no
amplification of the curvaton perturbations.}:
\begin{eqnarray}
H_\ast &>& 10^{-6} \, {\rm GeV}\;
\frac{M^{4/5}}{\tilde{m}_\sigma^{4/5}\Omega_{\rm dec}^{2/5}} \,, \label{H_I_1} \\
H_\ast &>& 10^{-7} \, {\rm GeV}^{1/2}\; 
\frac{M^{3/4}}{\tilde{m}_\sigma^{1/4}\Omega_{\rm dec}^{1/2}} \,. \label{H_I_2}
\end{eqnarray}
The effective mass of the curvaton
field after the end of thermal inflation, i.e., when $\bar \chi = M_\chi$ and
 $\bar \sigma = 0$ are the average over the oscillations of the
flaton and the curvaton fields, is
\begin{equation}
\tilde{m}_\sigma = (m_\sigma^2 + \lambda M^2)^{1/2} \,. \label{fcef}
\end{equation}
Note that we are focusing in the case of a final
curvaton decay rate $\Gamma_\sigma$ smaller than the Hubble parameter at the
 beginning of the oscillations $H_{\rm{osc}}$.
This is to allow the curvaton field to decay after the flaton field so that we
 can keep working in the simplest curvaton scenario where the curvaton field
 oscillates in a radiation background \cite{lyth03c,lyth02,moroi01b} (see Fig. \ref{figstandard}).

Making use of the constraint in Eq. (\ref{const_lambda}) and the expression
 in Eq. (\ref{fcef}), and taking into account that the bare curvaton mass $m_\sigma$
 is
smaller than the Hubble parameter $H_{\rm{osc}}$ at the end of thermal inflation,
 we obtain an upper bound on the effective
mass of the curvaton field:
\begin{equation}
\tilde{m}_\sigma < 10^{-1} \, {\rm GeV} \frac{M}{H_\ast \Omega_{\rm dec}} \,. \label{mboundE1}
\end{equation}
This bound, when applied to Eq. (\ref{H_I_1}), is consistent with low scale inflation.
When Eq. (\ref{mboundE1}) is applied to \hbox{Eq. (\ref{H_I_2})}, we obtain
 a lower bound for $H_\ast$ which is consistent too with low-energy scale inflation since $M \lsim 10^{18}$ GeV:
\begin{equation}
H_\ast > 10^{-9} \, {\rm GeV}^{1/3} M^{2/3} \Omega_{\rm dec}^{-1/3} \,. \label{morebounds1}
\end{equation}
The last inequality is stronger than that of Eq. (\ref{H_I_1}) only while the
 effective mass of the curvaton field is
\begin{equation}
\tilde{m}_\sigma > 10^2 \ {\rm GeV}^{10/11} M^{1/11} \Omega_{\rm dec}^{2/11} \,.
 \label{cem}
\end{equation}
Otherwise, we still need to consider the expression in Eq. (\ref{H_I_1}).

\subsection{Required parameter space} \label{rpsk}

Once we have checked the viability of a low-energy scale inflation we proceed
 to investigate the required range of values
for the parameters of the Lagrangian. Remember that we are going to focus on the
 gravity-mediated SUSY breaking
scheme where the Hubble parameter during inflation is $H_\ast \sim m_{3/2} \sim
 10^3$ GeV. After thermal inflation has ended, the flaton and curvaton fields
 start to oscillate,
eventually decaying into thermalised radiation (see Fig. \ref{figstandard}). The decay process is distinguished
 by the decay rate. The field with the
biggest decay rate will decay first. The flaton and curvaton decay rates are
 given by
\begin{eqnarray}
\Gamma_\chi\approx\gamma_\chi\frac{m_\chi^3}{M^2} \, & \quad{\rm and}\quad &
\Gamma_\sigma\approx\gamma_\sigma\frac{\tilde{m}_\sigma^3}{m_P^2} \,, 
\end{eqnarray}
with $\gamma_\chi \lsim 1$ \cite{lazarides86,lyth95,lyth96} and $\gamma_\sigma \gsim 1$. Since we like the curvaton mechanism not to be modified, the flaton field must
decay well before the curvaton decay. This requires
\begin{equation}
\tilde{m}_\sigma^3 \ll m_\chi^3 \frac{m_P^2}{M^2} \sim \frac{10^{46} \ {\rm GeV}^5}{M^2} \,. \label{mboundA}
\end{equation}
Now, using the expression in Eq. (\ref{H_I_1}), which is relevant for $\tilde{m}_\sigma
 \leq 10^2 \ {\rm GeV}^{10/11} M^{1/11} \Omega_{\rm dec}^{2/11}$ [cf. Eq. (\ref{cem})],
 we require
\begin{equation}
\tilde{m}_\sigma > 10^{-11} M \Omega_{\rm dec}^{-1/2} \,, \label{mboundB}
\end{equation}
in order to obtain low-energy scale inflation. 
Note that, combining the above with Eq.~(\ref{fagain}), we find
\begin{equation}
f<10^{-5}\sqrt{\Omega_{\rm dec}}\ll 1 \,,
\end{equation}
as required by the heavy curvaton scenario.
Similarly to the above, 
using the expression in Eq. (\ref{H_I_2}), which is relevant for $\tilde{m}_\sigma
 > 10^2 \ {\rm GeV}^{10/11} M^{1/11} \Omega_{\rm dec}^{2/11}$ \hbox{[cf. Eq. (\ref{cem})]},
 we require
\begin{equation}
\tilde{m}_\sigma > 10^{-40} {\rm GeV}^{-2} M^3 \Omega_{\rm dec}^{-2} \,. \label{mboundC}
\end{equation}

Thus, for values of $\tilde{m}_\sigma$ less than $10^2 \ {\rm GeV}^{10/11} M^{1/11}
 \Omega_{\rm dec}^{2/11}$ the required range of values for $\tilde{m}_\sigma$
 is\footnote{The bound in Eq. (\ref{mboundA}) is weaker than $\tilde{m}_\sigma
 < 10^2 \ {\rm GeV}^{10/11} M^{1/11}$ within the allowed range for $M$ (see Fig. \ref{lambda_g}).}:
\begin{equation}
10^{-11} M < \tilde{m}_\sigma < 10^2 \, {\rm GeV}^{10/11} M^{1/11} \,,
 \label{first_region}
\end{equation}
where the lower bound comes from Eq. (\ref{mboundB}). The vacuum expectation
 value $M$ is in the range
\begin{equation}
10^{12} \, {\rm GeV} \lsim M \lsim 10^{14} \ {\rm GeV} \,,
\end{equation}
where the lower bound comes from the solution to the moduli problem as we will
 see later, and the upper
bound comes from Eq. (\ref{first_region}).
On the other hand, for values of $\tilde{m}_\sigma$ bigger than $10^2 \ {\rm
 GeV}^{10/11} M^{1/11} \Omega_{\rm dec}^{2/11}$ the required range of values
 for $\tilde{m}_\sigma$ is:
\begin{equation}
\max\{10^2 \, {\rm GeV}^{10/11} M^{1/11}, 10^{-40} \,
 {\rm GeV}^{-2} M^3 \} <\tilde{m}_\sigma < 10^{15} \, {\rm
 GeV}^{5/3} / M^{2/3} \,, \label{moreconstraints4}
\end{equation}
where we have used Eqs. (\ref{mboundA}) and (\ref{mboundC}), and $M$ can
 be, a priori, up to $m_P$.
We have considered all the other possible constraints on $\tilde{m}_\sigma$
 and found they are irrelevant compared with those in Eq. (\ref{first_region})
 and Eq. (\ref{moreconstraints4}).

\begin{figure}[t]
\begin{center}
\leavevmode
\hbox{\hspace{-2.5cm}%
\epsfxsize=5.5in
\epsffile{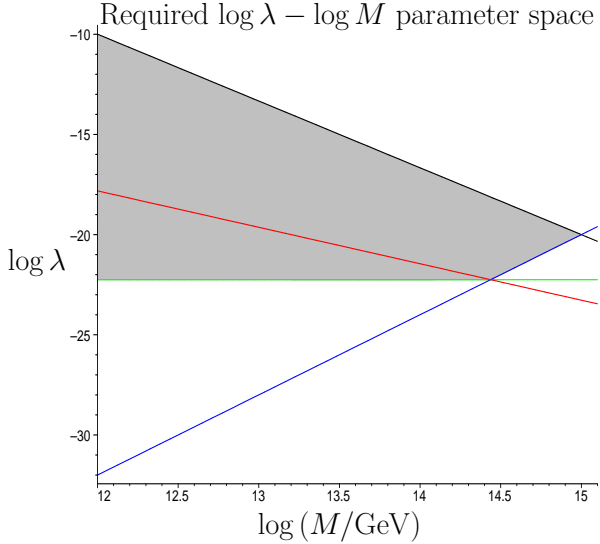}}
\vspace{-11cm}
\caption{
Required $\lambda - M$ parameter space (grey region) as a logarithmic plot.
 The two lines in the middle of the graph correspond to the limits in Eq. (\ref{first_region})
 which are valid up to the meeting point of the three lowest lines. The slanting
 lines correspond to the limits in Eq. (\ref{moreconstraints4}). Note that it
 is impossible to satisfy the conditions in Eq. (\ref{moreconstraints4}) beyond
 the meeting point of the highest and the lowest lines. Such a range of values for $M$ suggests
 that the flaton field cannot be the GUT Higgs field studied in Ref. \cite{lyth95}. The flaton
 field should be, therefore, one of the gauge singlets present in string theory \cite{polchinski}. \label{lambda_g}}
 \end{center}
\end{figure}

Fig. \ref{lambda_g} shows the required parameter space $\lambda$ vs $M$ (grey
 region) as a logarithmic plot. We have made use of the definition of the curvaton
 effective mass $\tilde{m}_\sigma$ in terms of the coupling constant $\lambda$
 and the vacuum expectation value $M$ [c.f. Eq. (\ref{fcef})]:
\begin{equation}
\tilde{m}_\sigma^2 \approx \lambda M^2 \,,
\end{equation}
and the required parameter space $\tilde{m}_\sigma$ vs $M$ studied before. Note
 that for values of $M$ higher than $\sim 10^{15}$ GeV it is impossible to satisfy
 Eq. (\ref{moreconstraints4}), so our final range for $M$ is
\begin{equation}
10^{12} \, {\rm GeV} \lsim M \lsim 10^{15} \, {\rm GeV} \,. \label{whitinM}
\end{equation}

The required values for $\lambda$, according to Fig. \ref{lambda_g}:
\be
10^{-22} \lsim \lambda \lsim 10^{-10} \,,
\ee
 are in agreement with the upper bound in the
 \hbox{Eq. (\ref{const_lambda})}:
\begin{equation}
\lambda < \frac{10^2 \, {\rm GeV}^2}{H_\ast^2} \sim 10^{-4} \,, \label{morebounds2}
\end{equation}
and with the lower bound
\begin{equation}
\lambda > \frac{H_{\rm{osc}}^2}{M^2}%
\approx \frac{m_\chi^2}{3 m_P^2} \sim 10^{-31} \,, \label{morebounds3}
\end{equation} 
which follows from $\tilde{m}_\sigma > H_{\rm{osc}}$.

In view of the allowed range of values for $M$ [c.f. Eq. (\ref{whitinM})], we conclude that our flaton
field cannot be the GUT Higgs field investigated in Ref. \cite{lyth95}. We must remember however that in some other
GUT models there are additional Higgs fields with much smaller vevs \cite{guts4,guts2,guts3,guts1} so they
are still good flaton candidates. The flaton field as a gauge singlet in string theory \cite{polchinski} remains as a
viable option.

Once we have found the required parameter space for $\lambda$ we must do the
 same for the other relevant parameter of the
Lagrangian: the bare mass of the curvaton $m_\sigma$.
The only
bound
on $m_\sigma$ is
\begin{equation}
m_\sigma < H_{\rm{osc}} \sim 10^{-16} M \,, \label{morebounds5}
\end{equation}
which is related to the fact that the oscillations of the curvaton around the
 minimum begin due to the sudden increment
in the curvaton mass at the end of thermal inflation.
That means, in view of Eq. (\ref{whitinM}), that
\be
m_\sigma \lsim 10^{-1} {\rm GeV} \,. \label{whitinbarem}
\ee
Such a small value for $m_\sigma$, taking into account the soft supersymmetric contributions of
order the gravitino mass for any scalar field which is not protected by a global symmetry, leads us
to point a PNGB as a viable curvaton candidate \cite{chun,dlnr,hofmann}.  

Finally, we still need 
to understand the lower bound $M
 \gsim 10^{12}$ GeV. To do that,
we must study the solution to the moduli problem\footnote{In the following subsection we correct one mistake in Ref. \cite{yeinzon} which led to a reduced parameter space for $m_\sigma$. Conclusions are different of course, but they are now more positive than before.}.

\subsection{Solution to the moduli problem}

Among the unwanted relics that the inflationary epoch is not able to dilute
 are the moduli \cite{banks,carlos,coughlan,ellislina,ellisnaq}. Moduli fields are flaton
fields with a vacuum expectation value of order the Planck mass. The decays
 of the flaton and the curvaton fields
increment the entropy, so that the big-bang moduli abundance, defined as that
 produced before thermal inflation and given by
\cite{lyth96}
\begin{equation}
\frac{n_\Phi}{s} \sim \frac{\Phi^2}{10 m_P^{3/2} m_\Phi^{1/2}} \,,
\end{equation}
where $\Phi$ is the vacuum expectation value of the moduli fields, gets suppressed
 by three factors. One is
\begin{equation}
\Delta_{PR} \simeq \frac{g_\ast(T_{PR})}{g_\ast(T_C)} \frac{T_{PR}^3}{T_C^3} \,,
\end{equation} 
due to the parametric resonance process \cite{wandsreview,kofman,kls,preheat1} following the end of the thermal inflation
 era, where the $g_\ast$ are the total internal particle degrees of freedom, $T_{PR}$ is the
temperature just after the
period of preheating, and $T_C$ is the temperature at the end of thermal inflation; another is
\begin{equation}
\Delta_\chi \simeq \frac{4 \beta V_h / 3 T_\chi}{(2 \pi^2 / 45) g_\ast(T_{PR})
 T_{PR}^3} \,,
\end{equation} 
due to the flaton decay, where $T_\chi$ is the temperature just after the decay\footnote{This is assuming that the flaton has come to dominate the energy
 density just before decaying (see Fig. \ref{figstandard}).}, and $\beta$ is the fraction of the total
energy density left in the flatons by the parametric resonance process and the increment in the energy density of the curvaton ($\beta \lsim 1$); the other is
\begin{equation}
\Delta_\sigma \simeq \frac{4 \tilde{m}_\sigma^2 \sigma_{\rm{osc}}^2 / 3 \Omega_{\rm{dec}} T_{\rm dec}}{(2 \pi^2 / 45) g_\ast(T_\chi) T_\chi^3} \,,%
\label{ds}
\end{equation}
due to the curvaton decay, where $T_{\rm dec}$
is the associated reheating temperature which must be bigger than $1$ MeV
not to disturb the nucleosynthesis process\footnote{We have assumed that $\rho_\sigma$ does not change appreciably from the time when $T = T_C$ to the time when $T=T_\chi$. This is a good approximation since $\Gamma_\chi\gg\Gamma_\sigma$.}.
 This enhancement in the entropy depends on the temperature just after the flaton decay
\begin{equation}
T_\chi \simeq \frac{10^{13} \hspace{1mm} {\rm GeV}^2}{M} \gamma_\chi^{1/2} \,,
 \label{temp_flat_decay}
\end{equation}
which is obtained by setting $\Gamma_\chi \sim H$ and assuming that the flaton
 decay products thermalise promptly. 
Thus, the abundance of the big-bang moduli after thermal inflation is:
\begin{eqnarray}
\frac{n_\Phi}{s} &\sim& \frac{\Phi^2}{10 m_P^{3/2} m_\Phi^{1/2} \Delta_{PR} \Delta_\chi \Delta_\sigma} \sim \frac{\Phi^2 T_\chi^4 T_{\rm dec} T_C^3}{10^5 \beta V_h m_\Phi^{1/2} \tilde{m}_\sigma^2 \Omega_{\rm{dec}} m_P^{3/2} H_\ast^2} \nonumber \\
&\gsim& 10^{48} \ {\rm GeV}^8 \,\lambda^{-1}M^{-8} \gamma_\chi^2 \left(\frac{\Phi}{m_P}\right)^2 \left(\frac{T_{\rm dec}}{1 \, {\rm MeV}}\right) \left(\frac{T_C}{m_\Phi}\right)^3 \times \nonumber \\
&& \times \left(\frac{m_\Phi}{10^3 \, {\rm GeV}}\right)^{1/2}\frac{1}{\beta} \left(\frac{m_\Phi^2 M^2}{V_h}\right) 
\frac{1}{\Omega_{\rm{dec}}} \left(\frac{10^3 \, {\rm GeV}}{H_\ast}\right)^2 \,. \label{fabundance}
\end{eqnarray}
The lower bound
\begin{equation}
\lambda \gsim \frac{10^{60} \, {\rm GeV}^8}{M^8} \gamma_\chi^2 \,, \label{entropyb1}
\end{equation}
is obtained when taking into account the restriction $n_\Phi / s \lsim 10^{-12}$ coming from nucleosynthesis \cite{ellis92}. This is a weaker bound on $\lambda$ than those presented in Fig. \ref{lambda_g}.

Let's have a look at the thermal inflation moduli abundance defined as that
 produced during the preheating stage following the end of the thermal inflation era 
\begin{eqnarray}
\frac{n_{\Phi_T}}{s} &\sim&
\frac{\Phi^2_T V_h^2 / 10 m_\Phi^3 m_P^4}{(2 \pi^2 / 45) g_\ast(T_{PR}) T_{PR}^3 \Delta_\chi \Delta_\sigma} \sim \frac{\Phi^2_T V_h T_\chi^4 T_{\rm dec}}{ 10^7 \beta m_\Phi^3 \tilde{m}_\sigma^2 \Omega_{\rm{dec}} m_P^4 H_\ast^2} \nonumber \\
&\gsim& 10^{-4} \, {\rm GeV}^4 \lambda^{-1}M^{-4} \gamma_\chi^2 \left(\frac{\Phi_T}{m_P}\right)^2 \left(\frac{T_{\rm dec}}{1 \, {\rm MeV}} \right) \frac{1}{\beta} \;\times \nonumber \\
&& \times \left(\frac{10^3 \, {\rm GeV}}{m_\Phi}\right) \left(\frac{V_h}{m_\Phi^2 M^2}\right)\frac{1}{\Omega_{\rm{dec}}}
\left(\frac{10^3 \, {\rm GeV}}{H_\ast}\right)^2 \,. \label{fsupp_abun_2}
\end{eqnarray}
Here $\Phi_T$ corresponds to the vacuum expectation value of the thermal moduli fields.
To suppress the thermal inflation moduli at the required level $n_{\Phi_T} / s \lsim 10^{-12}$ we require
\begin{equation}
\lambda \gsim \frac{10^8 \, {\rm GeV}^4}{M^4} \gamma_\chi^2 \,. \label{fentropyb2}
\end{equation}
Again this is a weaker bound on $\lambda$ than those in Fig. \ref{lambda_g}.

%

The Eqs. (\ref{fabundance}) and (\ref{fsupp_abun_2}) give us information about
 the necessary conditions for the suppression of the big-bang and thermal inflation
 moduli, but they are based on the unknown parameters $M$ and
$\lambda$. Since we still need to know if the range $M \lsim 10^{15} \hspace{1mm}
 {\rm GeV}$,
required to obtain a low-energy scale inflation, is not forbidden by the requirements
coming from the solution to the moduli problem, we must find a
$\lambda$-independent
 relation on
$M$. This relation can be found noting that the increment in the entropy due
 to the curvaton
decay [c.f. Eq. (\ref{ds})] can be written in an alternative way:
\begin{equation}
\Delta_\sigma \simeq \left[\frac{g_\ast (T_{\rm dec})}{g_\ast (T_\chi) (1-\Omega_{\rm{dec}})^3}\right]^{1/4} \,,
 \label{new_Delta}
\end{equation}
so the abundance of big-bang moduli after thermal inflation is:
\begin{eqnarray}
\frac{n_\Phi}{s} &\sim& \frac{\Phi^2}{10\,m_P^{3/2}m_\Phi^{1/2}\Delta_{PR}\Delta_\chi\Delta_\sigma}
\sim \frac{10\Phi^2T_\chi T_C^3(1-\Omega_{\rm{dec}})^{3/4}}{\beta V_h m_\Phi^{1/2}m_P^{3/2}} \nonumber \\
&\gsim& 10^{24}{\rm GeV}^3 M^{-3} \gamma_\chi^{1/2} \left(1-\Omega_{\rm{dec}}\right)^{3/4} \left(\frac{\Phi}{m_P}\right)^2 \times \nonumber\\
&& \times \left(\frac{T_C}{m_\Phi}\right)^3 \left(\frac{m_\Phi}{10^3 \, {\rm GeV}}\right)^{1/2} \frac{1}{\beta}
\left(\frac{m_\Phi^2 M^2}{V_h}\right) \,.
\end{eqnarray}
This means that
\begin{equation}
M \gsim 10^{12} \, {\rm GeV} \,, \label{lower_M}
\end{equation}
to satisfy $n_\Phi/s \lsim 10^{-12}$. This is the lower bound on $M$ we have
 used throughout this chapter. A similar treatment to the abundance of thermal inflation moduli [\hbox{c.f. Eq. (\ref{fsupp_abun_2})}] leads to the bound $M \lsim 10^{16}$ GeV, which is weaker than that obtained in Fig. \ref{lambda_g}.

Of course we might have considered the scenario where there are no moduli fields at all. Without the introduction of the moduli problem
Eq. (\ref{lower_M})
becomes unnecessary.
This does not help for the improvement of the required
 range of values for $m_\sigma$ but it does for $\lambda$ as the lower bound
 on $M$ in Eq. (\ref{lower_M}) becomes replaced by $M \gg 10^3$ GeV, which comes from the definition of the flaton fields. In this way the range of values for
 $M$ extends to smaller values well below $10^{12}$ GeV until the coupling constant $\lambda$ eventually reaches the lower bound $10^{-4}$.

The introduction of a period of thermal inflation into our curvaton scenario, sketched in Fig. \ref{figstandard},
has helped us not only to lower the energy scale of the main inflationary epoch, but also to solve the moduli problem \cite{banks,carlos,coughlan,ellislina,ellisnaq} still present after ordinary inflation. The
required parameter space for $\lambda$ has been plotted in Fig.~\ref{lambda_g},
and the vacuum expectation value for the flaton field has been showed to be
 in the range $10^{12} \, {\rm GeV} \lsim M \lsim 10^{15} \,
 {\rm GeV}$.
 Our flaton field, in view of the allowed range of values for $M$, should be one of the gauge singlets present in string theory \cite{polchinski}.
 The upper bound on $m_\sigma$ [\hbox{c.f. Eq. (\ref{whitinbarem})}], $m_\sigma \lsim 10^{-1}$ GeV,
 suggests the
 curvaton field could be a PNGB \cite{chun,dlnr,hofmann}. This is because in the presence
 of supergravity all the scalar fields, whose masses are not protected by a
 global symmetry, acquire soft masses of the order of the gravitino mass if $H \lsim m_{3/2}$, and contributions contributions to the squared mass of order $H^2$ if $H \gsim m_{3/2}$ except during the radiation dominated era \cite{lymo}. The
 smallness of the curvaton mass is in turn because of the very small value for
$H_{\rm{osc}}$. The parameter $H_{\rm{osc}}$ is directly proportional to $M$,
 so the bigger $M$ is, the more possible to obtain a range of values for $m_\sigma$
 compatible with the soft supersymmetric contributions. We, in the next chapter, will look for a mechanism to
 improve the required range of values for the bare mass $m_\sigma$ and the coupling constant $\lambda$ in presence of the moduli problem \cite{yeinzon04}.

\section{Conclusions}

We have presented a different type of curvaton scenario \cite{yeinzon}, in which the scale of inflation can be much lower than \mbox{$H_*\sim 10^7$ GeV}, which 
is the default lower bound for the standard curvaton model \cite{lyth04}. This 
scenario considers a curvaton, whose mass, being appropriately Higgsed, 
is substantially enlarged at a phase transition after the end of inflation 
(`heavy curvaton'). We have shown that this mechanism is indeed 
able to accommodate inflation scales as low as \mbox{$H_*\sim$ 1~TeV} or even 
lower.

We have implemented 
the idea of a thermal inflation epoch, introduced in 
Refs. \cite{lazarides86,lyth95,lyth96} to solve the moduli problem, as a second
inflationary
period necessary to lower the energy scale of the main inflationary stage.
In our model, a flaton field $\chi$ with bare mass coming from soft 
supersymmetric contributions and vacuum expectation value in the range 
$10^{12} \, {\rm GeV}\lsim M\lsim 10^{15} \, {\rm GeV}$ (i.e. one of the gauge singlets
in string theory \cite{polchinski}), is 
held at the origin of the scalar potential by finite-temperature effects. 
These effects are associated to the thermal background created by the main 
reheating epoch. When temperature falls below $V_h$ thermal inflation begins.
This period of thermal inflation lasts around ten e-folds until the temperature
falls below $m_\chi$ liberating the flaton field to roll away towards the 
minimum of the potential. The curvaton field is coupled to the flaton one, through
a coupling constant $\lambda$ in the range $10^{-22} \lsim \lambda \lsim 10^{-10}$, so 
its mass is largely increased at the end of thermal inflation. This increment 
is enough to lower the bound on $H_\ast$ to satisfactory levels, without 
sending the non-gaussianity constraint to the limit. However, the energy scale 
of the thermal inflation epoch is very small, requiring in turn a bare mass for
the curvaton field of at most $10^{-4} - 10^{-1}$ GeV. Taking into account the 
soft supersymmetric contributions to $m_\sigma$, the required smallness of 
$m_\sigma$ points toward using a PNGB curvaton \cite{chun,dlnr,hofmann} to achieve low-scale inflation.
 
The type of mechanism that we presented is not completely compelling.
It suffers from the problem that the mass 
of the curvaton before oscillation, as well as its coupling, have to 
be much smaller than one would expect. In the next chapter \cite{yeinzon04}
it will be shown how this tuning problem can be at least alleviated.

\chapter{Low scale inflation and the immediate heavy curvaton decay} \label{secondpaper}

\section{Introduction}

Low scale inflation {\it is} desirable in order to identify the inflaton field with one of the MSSM flat directions \cite{berkooz,kasuya} or with one of the fields appearing in the SUSY breaking sector, giving the inflaton a much deeper particle physics root. In contrast low scale inflation {\it is not} desirable because it makes very difficult the generation of the adiabatic perturbations by the inflaton, leading to multiple fine-tuning and model-building problems, unless the curvaton mechanism is invoked \cite{lyth03c,lyth02,moroi01b} (see also Refs. \cite{moroi01a,earlier1,earlier2}). With the aim of generating the curvature perturbation that gives origin to the large-scale structure in the observable universe, the curvaton mechanism has appeared as a nice and plausible option and a lot of research has been devoted to its study. Making the curvaton mechanism viable in a low energy inflationary framework would be the ideal situation but, unfortunately, the simplest curvaton model has shown to be incompatible with low enough values for the Hubble parameter during inflation \cite{lyth04}. Some general proposals to make the curvaton paradigm accommodate low scale inflation have recently appeared and specific models have been studied too \cite{yeinzon,matsuda03,postma04}. In the previous chapter, a thermal inflation epoch was attached to the general curvaton mechanism making the curvaton field gain a huge increment in the mass at the end of the thermal inflationary period, triggering this way a period of curvaton oscillations, and lowering the main inflationary scale to satisfactory levels \cite{yeinzon}. However, the parameters of the model required for this effect to take place showed to be extremely small to affect the reliability of the model. The purpose of this chapter is to study the same mechanism but in the case where the increment in the mass is so huge that the decay rate becomes bigger than the Hubble parameter and the curvaton decays immediately \cite{yeinzon04}. The results are very positive, offering a more natural parameter space.

\section{Thermal inflation and the immediate heavy curvaton decay}

The thermal inflation model has been investigated before and found to be a very efficient mechanism to dilute the abundance of some unwanted relics, like the moduli fields, that the main inflationary epoch is not able to get rid of (see Refs. \cite{lyth95,lyth96}). We will constrain the available parameter space for $\lambda$ and $m_\sigma$ in the scalar potential of \hbox{Eq. (\ref{mypotential})} so that enough dilution of the moduli abundance is obtained.
In \hbox{Chapter \ref{firstpaper}} this was done for the case in which the flaton-curvaton coupling term gives a huge contribution to the mass of the curvaton when the flaton field is released and gets its vacuum expectation value $M$. In that case the effective curvaton mass $\tilde{m}_\sigma$ may become bigger than the Hubble parameter giving birth to a period of curvaton oscillations and making the scale of the main inflationary period low enough ($H_\ast \sim m_{3/2} \sim 10^3 \ {\rm GeV}$) to think about the inflaton as a field associated to the SUSY breaking sector \cite{lyth04, matsuda03, postma04}. The evolution of the energy densities associated to the different fluids in that case are sketched in Fig. \ref{figstandard}.

The purpose of this chapter is to analyse the scenario where there are no oscillations of the curvaton field. As it was pointed out in Ref. \cite{postma04} both the curvaton decay rate and the associated lower bound [c.f. Eq. (\ref{decay_rate})]
\begin{equation}
\Gamma_\sigma \geq \frac{\tilde{m}_\sigma^3}{m_P^2} \,,
\end{equation}
are also increased when the flaton field acquires its vacuum expectation value so that, if this increment is big enough for the curvaton decay rate to be bigger than the Hubble parameter, the curvaton field may decay immediately rather than oscillating for some time. Low scale inflation in this case is also possible to be attained \cite{yeinzon,postma04}, but the lower bound on $H_\ast$ changes with respect to the case when the curvaton oscillatory process is triggered. The evolution of the energy densities associated to the different fluids in this case are sketched in Fig. \ref{figmodified}.

\begin{figure}
\begin{center}
\includegraphics[width=15cm,height=20cm]{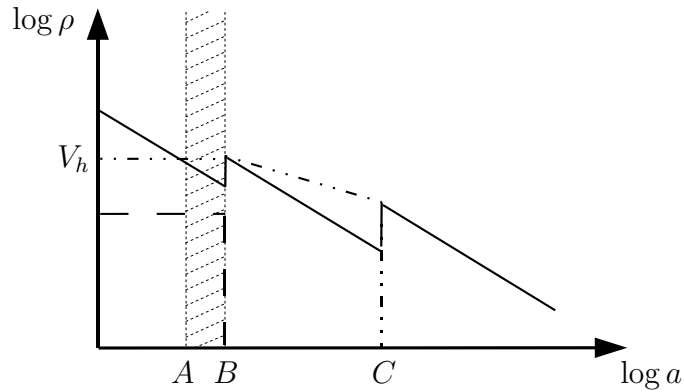}
\put(-340,519){$\log \rho$}
\put(-323,467){$V_h$}
\put(-280,385){$A$}
\put(-264,385){$B$}
\put(-204,385){$C$}
\put(-110,385){$\log a$}
\vspace*{-13.5cm}
\caption{Evolution of the energy densities in the thermal inflation model where the curvaton field $\sigma$ decays immediately at the end of thermal inflation \cite{yeinzon04}. The continuous line corresponds to the global radiation energy density $\rho_{r_0}$, the dashed dotted line corresponds to the global flaton energy density $\rho_{\chi_0}$, and the dashed line corresponds to the global curvaton energy density $\rho_{\sigma_0}$. The horizontal axis represents the expansion parameter $a$. From the left to $A$ radiation dominates the energy density, although it decreases following $\rho_{r_0} \propto a^{-4}$. At this stage the unperturbed components of the flaton and curvaton fields $\chi$ and $\sigma$ are frozen at $\chi_0 = 0$ and $\sigma_0 = \sigma_\ast$ making their energy densities constants. When $\rho_{r_0}$ reaches $V_h$ at $A$, thermal inflation begins. The thermal inflation period lasts until $B$ when the temperature $T$ becomes of the order of the flaton mass $m_\chi$. Thermal inflation is portrade by the dashed region. After thermal inflation ends, the parametric resonance process transforms a substantial fraction of $\rho_\chi$ into $\rho_r$ \cite{wandsreview,kofman,kls,preheat1}. The flaton field is liberated by this time and begins oscillating around the minimum of its potential, behaving then as a matter fluid with $\rho_{\chi_0} \propto a^{-3}$. The curvaton field increments suddenly its mass $m_\sigma$ at $B$ as a result of the oscillations of $\chi$ around the vacuum expectation value $M$. The increment is enough for the decay rate $\Gamma_\sigma$ to overtake $H_{\rm pt}$ (the Hubble parameter at $B$) so that $\sigma$ decays immediately. The curvaton energy density is transfered then completely to $\rho_r$ as it is the curvature perturbation too. By the time $C$, $\chi$ already dominates the energy density before decaying into radiation.
\label{figmodified}}
\end{center}
\end{figure}

In the scenario where curvaton oscillations are allowed, corresponding to \mbox{$\Gamma_\sigma < H_{\rm pt}$}, the lower bound on $H_\ast$ is [\hbox{c.f. Eqs. (\ref{H*bound}), (\ref{Hbound}), and (\ref{Hbound-0})}]
\begin{equation}
H_\ast \geq {\rm max} \{ \ f^{4/5} \times 10^7 \ {\rm GeV}, \sqrt f \delta^2 \times 10^{11} \ {\rm GeV}, 
\delta^4 \times 10^{7} \ {\rm GeV} \ \} \,, \label{H_bound}
\end{equation}
where $f$ and $\delta$, given by Eqs. (\ref{fff}) and (\ref{fdelta}), are less than 1, 
and with $H_{\rm osc} = H_{\rm pt}$ being the Hubble parameter at the end of the thermal inflation period (which also corresponds to the beginning of the curvaton oscillations). In contrast, the lower bound in the scenario where the curvaton field decays immediately, corresponding to $\Gamma_\sigma > H_{\rm pt}$, is [c.f. Eqs. (\ref{H*bound}) and (\ref{Hbound-00})]
\begin{equation}
H_\ast \geq {\rm max} \{ \ f^{4/5} \times 10^7 \ {\rm GeV}, f \times 10^{14} \ {\rm GeV} \ \} \,, \label{H_bound_n}
\end{equation}
which may challenge the WMAP constraint for the curvaton scenario \cite{liber} leading to excessive curvature perturbations from the inflaton field if $f$ is not much smaller than unity.

The lower bound in Eq. (\ref{H_bound}), for $H_\ast \sim 10^3$ GeV, was shown in the previous chapter to be satisfied for very small values for the flaton-curvaton coupling constant, $\lambda \sim 10^{-22} - 10^{-10}$ (see Fig. \ref{lambda_g}), and very small values for the bare mass of the curvaton field, $m_\sigma \lsim 10^{-1} \ {\rm GeV}$ [c.f. Eq. (\ref{whitinbarem})], which suggests that the curvaton field could be a PNGB \cite{chun,dlnr,hofmann}. This is, in any case, a quite negative result due to the required smallness of the parameters $\lambda$ and $m_\sigma$. However, when taking into account the lower bound in \hbox{Eq. (\ref{H_bound_n})}, corresponding to the case when the decay rate $\Gamma_\sigma$ becomes bigger than $H_{\rm pt}$, things change appreciably.

\subsection{The flaton-curvaton coupling constant $\lambda$}
Thermal inflation ends when the thermal energy density is no longer dominant; thus, the Hubble parameter at the end of thermal inflation is associated to the energy density coming from the curvaton and the flaton fields:
\begin{equation}
H_{\rm{pt}}^2 = \frac{\rho_T + V(\chi = 0, \sigma_0 = \sigma_\ast)}{3 m_P^2} \sim \frac{m_\chi^2 M^2}{3 m_P^2} \,,
\end{equation}
so that
\begin{equation}
H_{\rm{pt}} \sim 10^{-16} M \,. \label{tic}
\end{equation}
Since the effective mass of the curvaton field after the end of thermal inflation, i.e., when $\bar \chi = M_\chi$ is the average over oscillations of the flaton field and $\sigma_0 = 0$, is
\begin{equation}
\tilde{m}_\sigma = (m_\sigma^2 + \lambda M^2)^{1/2} \approx \sqrt{\lambda} M \,, \label{cef}
\end{equation}
the parameter $f$ [cf. Eq. (\ref{fff})] becomes
\begin{equation}
f \equiv \frac{H_{\rm pt}}{\tilde{m}_\sigma} \sim 10^{-16} \frac{1}{\sqrt{\lambda}} \,. \label{ff}
\end{equation}
In view of  the Eqs. (\ref{H_bound_n}) and (\ref{ff}) the smallest possible value for $\lambda$, compatible with $H_\ast \sim 10^3 \ {\rm GeV}$, becomes $\lambda \sim 10^{-10}$, which is very good because this already improves the results found in the previous chapter. Moreover, the effective flaton mass during thermal inflation $\tilde{m}_\chi = (m_\chi^2 - \lambda \sigma_\ast^2)^{1/2}$ must be positive to trap the flaton field at the origin of the potential. Thus, $\lambda \sigma_\ast^2 < m_\chi^2$, and the biggest possible value for $\lambda$ becomes [c.f. Eq. (\ref{sosc})]
\begin{equation}
\lambda < \frac{m_\chi^2}{\sigma_\ast^2} \sim \frac{10^{-2} \ {\rm GeV}^2}{\Omega_{\rm dec}^2 H_\ast^2} \lsim 10^{-4} \,, \label{kdlambda2}
\end{equation}
which is already a small value but much bigger and more natural than that found in the case where curvaton oscillations are allowed. The lower bound on $\lambda$ vs $\Omega_{\rm dec}$ is depicted in Fig. \ref{fig}. Note that a small value for $\Omega_{\rm dec}$, which is restricted to be $\Omega_{\rm dec} \geq 0.01$ in order to satisfy the WMAP constraints on non gaussianity \cite{komatsu03, lyth03c}, is desirable to obtain a higher value for $\lambda$, so the biggest possible value $\lambda \sim 10^{-4}$ is at the expense of a high level of non gaussianity. A smaller upper bound for $\lambda$ could be a possibility, according to Eq. (\ref{kdlambda2}) and Fig. \ref{fig}, by increasing $\Omega_{\rm dec}$. Nevertheless $\Omega_{\rm dec}$, in the scenario studied in this chapter, must satisfy $\Omega_{\rm dec} < 1$ to avoid a period of inflation driven by the curvaton field.

\begin{figure}
\begin{center}
\includegraphics[width=6.5cm,height=7.5cm,angle=-90]{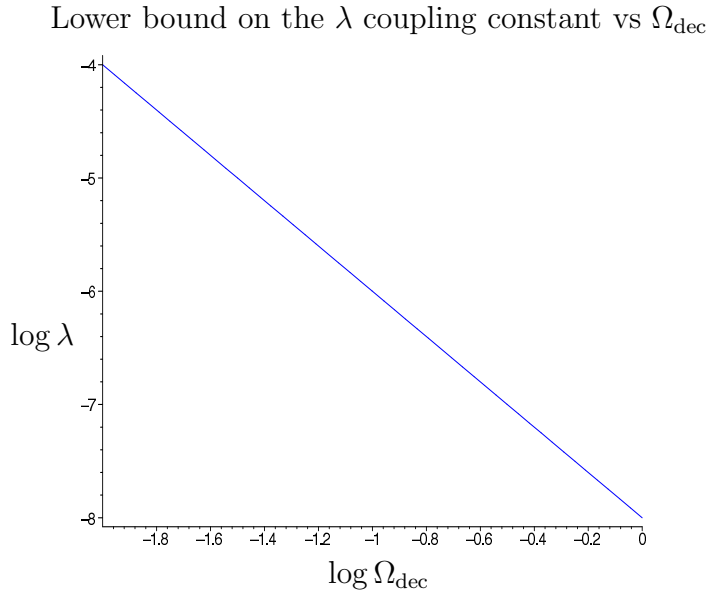}
\put(-225,10){Lower bound on the $\lambda$ coupling constant vs $\Omega_{\rm dec}$}
\put(-240,-110){$\log{\lambda}$}
\put(-120,-200){$\log{\Omega_{\rm dec}}$}
\caption{Lower bound on the flaton-curvaton coupling constant $\lambda$ as a logarithmic plot. A more natural value for $\lambda$ requires a higher level of non gaussianity compatible with the WMAP constraints. \label{fig}}
\end{center}
\end{figure}

Recalling, in the scenario where curvaton oscillations are allowed the coupling constant $\lambda$ is in the range (see Fig. \ref{lambda_g})
\begin{equation}
10^{-22} \lsim \lambda \lsim 10^{-10} \,, \label{klam1}
\end{equation}
whereas in the scenario where the curvaton decays immediately the range is
\begin{equation}
10^{-10} \lsim \lambda \lsim 10^{-4} \,. \label{klam2}
\end{equation}
It is easy to see that the allowed range of values for $\lambda$ in Eq. (\ref{klam2}), valid for the case where the curvaton field decays immediately at the end of the thermal inflation era, is complementary to the allowed range for $\lambda$ in Eq. (\ref{klam1}), valid when the curvaton has some time to oscillate before decaying (see Subsection \ref{rpsk}).

\subsection{The bare curvaton mass $m_\sigma$}
The only bound on $m_\sigma$ is given by the fact that in the heavy curvaton scenario the bare mass must be smaller than the Hubble parameter at the end of the thermal inflation era, so that the sudden increment in the mass and the decay rate leads to the immediate decay of the field avoiding in this case the oscillations. Thus,
\begin{equation}
m_\sigma < H_{\rm pt} \sim 10^{-16} M \,,
\end{equation}
so we need to worry about the possible values for $M$. In the scenario where the curvaton field decays immediately the flaton field is left immersed in a background of radiation, so it must decay before the time of nucleosynthesis in order not to disturb the abundances of the light elements. By setting $\Gamma_\chi \simeq H$ we get the temperature just after the flaton decay
\begin{equation}
T_\chi \simeq 10^{13} \ {\rm GeV}^2 \frac{1}{M} \,, \label{nucl}
\end{equation}
which must be bigger than $1$ MeV to satisfy the nucleosynthesis constraint. Therefore
\begin{equation}
M \lsim 10^{16} \ {\rm GeV} \,, \label{bound_M}
\end{equation}
leading to an upper bound on the bare curvaton mass given by $m_\sigma \lsim 1 \ {\rm GeV}$, which is again a more relaxed constraint than that found in the previous chapter for the case of an oscillating curvaton, but that still reduces the number of possible curvaton candidates, leaving essentially the PNGB \cite{chun,dlnr,hofmann}. Recalling, in the scenario where curvaton oscillations are allowed the bare curvaton mass $m_\sigma$ is in the range [c.f. Eq. (\ref{whitinbarem})]
\begin{equation}
m_\sigma \lsim 10^{-1} \ {\rm GeV} \,,
\end{equation}
whereas in the scenario where the curvaton decays immediately the range is
\begin{equation}
m_\sigma \lsim 1 \ {\rm GeV} \,. \label{bound_m}
\end{equation}

Some important constraints might come from the solution to the moduli problem and could limit the reliability of the Eqs. (\ref{bound_M}) and (\ref{bound_m}). Moduli fields are flaton
fields with a vacuum expectation value $\Phi$ of order the Planck mass. The decay of the flaton field
increments the entropy density $s$, so that the big-bang moduli abundance, defined as that produced before thermal inflation and given by \cite{lyth96}
\begin{equation}
\frac{n_\Phi}{s} \sim \frac{\Phi^2}{10 m_P^{3/2} m_\Phi^{1/2}} \,,
\end{equation}
where $m_\Phi$ is the mass of the moduli fields, gets suppressed by three factors\footnote{Eqs. (\ref{mplfirsts}) and (\ref{mplseconds}) correct a mistake in Ref. \cite{yeinzon04}. However Eqs. (\ref{abundance}) and (\ref{supp_abun_2}) are not affected by that mistake and, therefore, the conclusions in Ref. \cite{yeinzon04} about the lower and upper bounds on $M$ remain unchanged.}. One is
\begin{equation}
\Delta_\sigma \sim \frac{g_\ast(T_\sigma)}{g_\ast(T_C)} \frac{T_\sigma^3}{T_C^3} \,, \label{mplfirsts}
\end{equation} 
due to the curvaton decay, where the $g_\ast$ are the total internal degrees of freedom, $T_\sigma$ is the temperature just after the curvaton decay, and $T_C \sim m_\chi$ is the temperature at the end of thermal inflation; another is
\begin{equation}
\Delta_{PR} \sim \frac{g_\ast(T_{PR})}{g_\ast(T_\sigma)} \frac{T_{PR}^3}{T_\sigma^3} \,, \label{mplseconds}
\end{equation} 
due to the parametric resonance process \cite{wandsreview,kofman,kls,preheat1} following the end of the thermal inflation era, where $T_{PR}$ is the temperature just after the period of preheating; and the other is
\begin{equation}
\Delta_\chi \sim \frac{4\beta V_h / 3T_\chi}{(2 \pi^2 / 45) g_\ast(T_{PR}) T_{PR}^3} \,,
\end{equation} 
due to the flaton decay, where $T_\chi$ is the temperature just after the decay\footnote{This is assuming for simplicity that the flaton has come to dominate the energy density just before decaying (see Fig. \ref{figmodified}).}, and $\beta$ is the fraction of the total
energy density left in the flatons by the parametric resonance process ($\beta \lsim 1$). Thus, the abundance of the big-bang moduli after thermal inflation is:
\begin{eqnarray}
\frac{n_\Phi}{s} & \sim &
\frac{\Phi^2}{10 m_P^{3/2} m_\Phi^{1/2} \Delta_\sigma \Delta_{PR} \Delta_\chi} \sim \frac{10 \Phi^2 T_\chi T_C^3}{\beta V_h m_\Phi^{1/2} m_P^{3/2}} \nonumber \\
& \gsim & 10^6 \ {\rm GeV}^2 M^{-2} \left(\frac{\Phi}{m_P}\right)^2%
\left(\frac{T_\chi}{1 \hspace{1mm} {\rm MeV}}\right) \left(\frac{T_C}{m_\Phi}\right)^3 \times \nonumber \\
&& \times \left(\frac{m_\Phi}{10^3 \hspace{1mm}{\rm GeV}}\right)^{1/2}%
\left(\frac{1}{\beta}\right) \left(\frac{m_\Phi^2 M^2}{V_h}\right) \,. \label{abundance}
\end{eqnarray}
which must be suppressed enough ($n_\Phi / s \lsim 10^{-12}$) so that the nucleosynthesis constraints studied in Ref. \cite{ellis92} are satisfied. This is easily achieved by imposing a lower bound on $M$:
\begin{equation}
M \gsim 10^9 \ {\rm GeV} \,,
\end{equation} 
which does not affect the upper bounds on $M$ and $m_\sigma$ in Eqs. (\ref{bound_M}) and (\ref{bound_m}).

We also have to take care about the abundance of the thermal inflation moduli, defined as that
 produced during the preheating stage following the end of the thermal inflation era: 
\begin{eqnarray}
\frac{n_{\Phi_T}}{s} & \sim  &
\frac{\Phi_T^2 V_h^2 / 10 m_\Phi^3 m_P^4}{(2 \pi^2 / 45) g_\ast(T_{PR}) T_{PR}^3%
\Delta_\chi} \sim \frac{\Phi_T^2 V_h T_\chi}{ 10 \beta m_\Phi^3 m_P^4} \nonumber \\
& \gsim & 10^{-44} \hspace{1mm} {\rm GeV}^{-2} M^2 \left(\frac{\Phi_T}{m_P}\right)^2 \left(\frac{T_\chi}{1 \, {\rm MeV}} \right) \times \nonumber \\
&& \times \left(\frac{1}{\beta}\right) \left(\frac{10^3 \, {\rm GeV}}{m_\Phi}\right) \left(\frac{V_h}{m_\Phi^2 M^2}\right) \,. \label{supp_abun_2}
\end{eqnarray}
Here $\Phi_T$ corresponds to the vacuum expectation value of the thermal moduli fields.
To suppress the thermal inflation moduli at the required level $n_{\Phi_T} / s \lsim 10^{-12}$ we require
\begin{equation}
M \lsim 10^{16} \ {\rm GeV} \,, \label{entropyb2}
\end{equation}
which is precisely the same bound as in Eq. (\ref{bound_M}).  Recalling, the allowed range of values for the vacuum expectation value of the flaton field is
\begin{equation}
10^9 \ {\rm GeV} \lsim M \lsim 10^{16} \ {\rm GeV} \,,
\end{equation}
so that the moduli problem is solved and, in the best case, $m_\sigma \sim 1$ GeV. The latter allowed range for $M$ means that, unlike the case where the curvaton has some time to oscillate before decaying, the flaton field could be the GUT Higgs field studied in Ref. \cite{lyth95}.

\section{Some useful remarks}
Before concluding, we want to stress some points that can help to avoid possible confusion. The parameter space compatible with low scale inflation is a feature of the specific model studied, and we cannot say it is the same for all classes of models in the basis of Eqs. (\ref{H*bound}), (\ref{Hbound}), and (\ref{Hbound-0}), which provide just some general bounds. That is why specific models have been studied (see Refs. \cite{yeinzon,postma04}), even when the general bounds were already known from Refs. \cite{lyth04,matsuda03}.
Although the claim, that the available parameter space is bigger for the immediate curvaton decay,
was given before in Ref. \cite{postma04}, we again cannot say that the available parameter space is the same for all classes of models in the basis of the bounds required to have low energy scale inflation.  For example, from Eq. (\ref{tic}), $H_{\rm pt}$ depends on $M$ so there is no direct bound on it unless we know the bound on $M$ \footnote{Notice that the bounds required to have low energy scale inflation [c.f. Eq. (\ref{H_bound_n})] depend only on the ratio $f = H_{\rm pt}/\tilde{m}_\sigma$, and not exclusively on $H_{\rm pt}$.}. The bound on $M$ comes in turn from the requirement that the flaton decays before nucleosynthesis [\hbox{c.f. Eqs. (\ref{nucl}) and (\ref{bound_M})}] and must be consistent with the adequate suppression of the thermal inflation moduli [c.f. Eqs. (\ref{supp_abun_2}) and (\ref{entropyb2})]. These are, of course, features specific only to the model we are studying, and are therefore not present in Ref. \cite{postma04}.

Naively, one would think that the bounds on $\lambda$ and $m_\sigma$ are found from that on $H_{\rm pt}$ only through a mere change of variables. This is of course not true as the bound on $H_{\rm pt}$ is a very sensitive quantity that has to avoid disturbing the nucleosynthesis process and the adequate moduli abundance suppression. It is worth mentioning that the scenario discussed in this chapter differs appreciably from that studied in Chapter \ref{firstpaper}, due to the immediate curvaton decay, so that the conditions to satisfy the nucleosynthesis and thermal inflation moduli constraints are completely different\footnote{For example, the expressions for the big-bang and thermal inflation moduli abundances in Chapter \ref{firstpaper} [c.f. Eqs. (\ref{fabundance}) and (\ref{fsupp_abun_2})] are different from those in this chapter [c.f. Eqs. (\ref{abundance}) and (\ref{supp_abun_2})].}. 

Finally, the agreement between the bounds found in Ref. \cite{postma04} (which are supposed to be general) and those found in this chapter is apparent and corresponds just to a mere coincidence.  We justify this observation by noting  that Eq. (6) in \hbox{Ref. \cite{postma04}} is essentially the same as our Eq. (\ref{H*1}), the latter being generalized to give Eq. (\ref{H*bound}), except for $\Gamma_\sigma$ which in our Eq. (\ref{H*1}) appears to be $H_{\rm dec}$. The expressions in the previous chapter were carefully derived so that the correct expression is that given there \cite{yeinzon} . In contrast, Eq. (6) in Ref. \cite{postma04} is just valid for the standard case where the curvaton field has some time to oscillate before decaying, so we can identify $\Gamma_\sigma$ with $H_{\rm dec}$. However, for the immediate decay case, $\Gamma_\sigma > H_{\rm dec} = H_{\rm pt}$, which renders Eq. (6) in Ref. \cite{postma04} invalid.
Based on the previous discussion we claim that the bound $H_{\rm pt} < 1$ GeV, as are those on $\lambda$ and $m_\sigma$, is presented in this thesis for the first time {\em in a correct way}.

\section{Conclusions}

In this chapter we have investigated the required parameter space compatible with low scale inflation in the thermal inflation curvaton scenario where there are no oscillations of the curvaton field \cite{yeinzon04}. We have shown that the parameter space is greatly enhanced when the increment in the curvaton decay rate is big enough for the curvaton field to decay immediately at the end of the thermal inflation era. The best case corresponds to a flaton-curvaton coupling constant $\lambda \sim 10^{-4}$ and a bare curvaton mass $m_\sigma \sim 1 \ {\rm GeV}$, which are much bigger and more natural than the ranges $10^{-22} \lsim \lambda \lsim 10^{-10}$ and $m_\sigma \lsim 10^{-1} \ {\rm GeV}$ found previously in \hbox{Chapter \ref{firstpaper}} for the case where the curvaton oscillates for some time before \hbox{decaying \cite{yeinzon}}. In addition we have found $10^9 {\rm GeV} \lsim M \lsim 10^{16} {\rm GeV}$ for the vacuum expectation value $M$ of the flaton field. Therefore, our flaton field as the GUT Higgs field discussed in Ref. \cite{lyth95} is a viable option in this scenario.

\chapter{Non-gaussianity from the second-order cosmological perturbation} \label{fourthpaper}

\section{Introduction}

In chapters \ref{mechanismsiandc}, \ref{firstpaper}, and \ref{secondpaper}, we discussed some of the theoretical aspects of the origin of the large-scale structure in the Universe, emphasising the possibility to achieve low scale inflation in the curvaton scenario. We begin now the discussion of the statistical aspects, specifically the presence of non-gaussianities in the fields responsible for the origin of the curvature perturbation $\zeta$ and/or in $\zeta$ itself. Chapters \ref{fourthpaper} and \ref{fifthpaper} will deal with such an interesting subject.

Cosmological scales
leave the horizon during inflation and re-enter it after Big Bang 
Nucleosynthesis. Throughout the super-horizon era
 it is very useful
to define a primordial cosmological curvature perturbation,
 which is 
conserved if and only if  pressure throughout the Universe is 
a unique function of energy density 
(the adiabatic pressure condition) (see Subsection \ref{acgauge}) \cite{bardeen,bst,lyth84,sasaki1,rs1,sb,llmw}. 
Observation directly constrains the  curvature perturbation
at the very end of the super-horizon era,
a few Hubble times before cosmological scales start to enter the horizon,
when it apparently  sets the initial condition for the subsequent evolution of
all cosmological perturbations. As discussed in Chapter \ref{mechanismsiandc}, the observed curvature perturbation
is almost Gaussian with an almost
scale-invariant spectrum. 

Cosmological perturbation theory expands the exact equations in
 powers of the perturbations
and keeps terms only up to the $n$th order. Since the observed
curvature perturbation 
is  of order $10^{-5}$,
one might think that first-order
perturbation theory will be  adequate for all comparisons with observation. 
That may not be the case however, because 
the PLANCK satellite \cite{planck} and its successors may  be sensitive to non-gaussianity
of the curvature perturbation at the level of second-order perturbation
theory \cite{spergel}. 

Several authors have treated the non-gaussianity of the primordial curvature
perturbation in the context of
second-order perturbation theory. They have adopted different definitions 
of the curvature perturbation and 
obtained  results for a variety of situations.
In this chapter we  revisit  the calculations, using a single
definition of the curvature perturbation
 which we denote by $\zeta$ \cite{lr}. In some cases we disagree with the findings of 
the original authors.

The outline of this chapter is the following:  in Section \ref{definitions} 
we review two definitions of the curvature perturbation found in the literature,
which are valid during and after inflation,
and establish definite relationships between them; in section \ref{thirddef} we review
a third curvature perturbation definition, which applies {\it only} during inflation,
and study it in models of inflation of the slow-roll variety;
in Section \ref{origin} we describe the present framework for thinking about
the origin and evolution of the curvature perturbation;
in Section \ref{gaussianity} we see how non-gaussianity is defined
and constrained by observation;
in Section \ref{finitial} we
study the initial non-gaussianity of the curvature perturbation, 
a few Hubble times after horizon exit; in Section \ref{after}
we study its  subsequent  evolution according to some different models.
 The conclusions
are summarised  in Section \ref{conclusions}. 

We shall denote unperturbed quantities by a subscript $0$,
and generally work with conformal time $\eta$.
Sometimes though we revert to physical time $t$.
We  shall 
adopt the convention that a generic perturbation $g$
is split into a first- and second-order part according to the formula
\be
g =g\one + \frac12 g\two
\label{gsplit}
\,.
\ee

\section{Two definitions of the  curvature perturbation}
\label{definitions}

\subsection{Preliminaries}

Cosmological perturbations describe small departures
of the  actual Universe, away from some perfect homogeneous and isotropic
universe with the line element in \hbox{\eq{unperturbedle}}.
For a  generic perturbation  it is 
convenient to make the Fourier expansion
\be
g(\bfx,\eta) = \frac1{(2\pi)^{3/2}} \int d^3k \ g_{\bf k}(\eta) e^{i\bfk\cdot\bfx}
\,,
\ee
where the spacetime coordinates are those of the unperturbed Universe.
The inverse of the comoving wavenumber, $k^{-1}$, 
is often referred to as the scale. {\em Except where 
otherwise stated, our discussion applies only to the super-horizon regime} ($k \ll a H_{\rm inf}$).

When evaluating an observable quantity only a limited range of
scales  will be involved. The largest scale,
relevant for the low multipoles of the
Cosmic Microwave Background anisotropy,
is $k^{-1}\sim H_{\rm today}^{-1}$ where $H_{\rm today}$ is the present Hubble
parameter. The smallest  scale usually considered
is the one enclosing  matter with mass $\sim 10^6\msun$,
which corresponds to $k\mone\sim 10\mtwo\Mpc \sim 10^{-6}H_{\rm today}\mone$.
The cosmological range of scales therefore extends over only six  
orders of magnitude or so. 

To define cosmological perturbations in general, 
one has to introduce in the perturbed Universe a coordinate
system $(t,x^i)$,  which defines a slicing of spacetime (fixed $t$) and
a threading (fixed $x^i$). To define the curvature perturbation 
it is enough to define the slicing \cite{sasaki1}. 

\subsection{Two definitions of the curvature perturbation} \label{2defz}
 
In this chapter we 
take as our definition of $\zeta$ the following expression for the 
spatial metric \cite{calcagni,singlecreminelli,zc,sasaki1,maldacena,rst,rst2,sb,seery,seery2} 
which applies non-perturbatively: 
\be
g_{ij}  =  a^2(\eta) \tilde\gamma_{ij}  e^{2\zeta}
\,.
\label{mal}
\ee
Here $\tilde\gamma_{ij}$ has unit determinant, and the time-slicing
is one of uniform energy density\footnote
{It is proved in Ref. \cite{sasaki1} that this  
definition of $\zeta$ coincides with that of Lyth and Wands \cite{lyth},
provided that their slices of uniform coordinate expansion are taken
to correspond to those on which the line element has the form \eq{mal}
{\em without the factor $e^{2\zeta}$} (this makes the slices practically
flat
if $\tilde\gamma_{ij} \simeq \delta_{ij}$).}. 

It has been shown under weak assumptions \cite{sasaki1} that this
defines $\zeta$ uniquely, and that $\zeta$ is conserved as long as the 
pressure is a unique function of energy density. 
Also, it has been shown that the uniform density slicing
practically  coincides with the
comoving slicing (orthogonal to the flow of energy),   
and with the uniform Hubble 
slicing (corresponding to  uniform proper expansion, that expansion being
practically  independent of the threading which defines it) \cite{sasaki1}. 
The coincidence of these
slicings  is important since all three have been invoked
by different authors.

Since the matrix $\tilde\gamma$ has unit determinant it can be written
$\tilde \gamma = I e^h$,
where $I$ is the unit matrix and $h$ is traceless \cite{sasaki1}. 
Assuming that the initial condition is set by inflation, $h$ corresponds
to a tensor perturbation (gravitational wave amplitude) which 
will  be negligible unless the scale of inflation is very high. 
As we shall see later (see footnote \ref{teninc} in this chapter), the  results we are going to present
 are valid even if $h$ is not negligible,
but to simplify the presentation we drop $h$ from the equations.
Accordingly, the space part of the metric in the super-horizon
regime is supposed to be well approximated by
\be
g_{ij} = a^2(\eta)\delta_{ij} e^{2\zeta} 
\label{expdef}
\,.
\ee

At first order, \eq{expdef} corresponds to 
\begin{equation}
g_{ij} = a^2(\eta) \delta_{ij} (1+2\zeta)
\label{all}
\,.
\end{equation}
Up to a sign, this is the definition of the first-order
curvature perturbation adopted by all authors [c.f. Eqs. (\ref{perturbedline}) and (\ref{zeta_def2})]. 
There is no universally
agreed  convention for the sign of $\zeta$.
Ours coincides with the convention of most of the papers to which we refer, and
we have checked carefully that the signs in our own set of equations are 
correct.

At  second order we have
\be
g_{ij} = a^2(\eta) \delta_{ij} (1+2\zeta+2\zeta^2)
\,.
\ee
This is our definition of $\zeta$ at second order \cite{lr}.

Malik and Wands \cite{malik} instead  defined 
$\zeta$ by \eq{all} even at second order.
Denoting their definition by a subscript MW,
\be
\zeta\su{MW} = \zeta + \zeta^2
\label{mwdef}
\,,
\ee
or equivalently
\be
\zeta_2\su{MW} = \zeta\two + 2\( \zeta\one \)^2 \label{malmw}
\,,
\ee
where $\zeta\one$ is the first-order quantity whose definition
\eq{all} is agreed by all authors.


To make contact with  calculations of the curvature 
perturbation during inflation, we need some 
gauge-invariant expressions for the curvature
perturbation. As stated in Subsections \ref{zetaninvariant} and \ref{acgauge}, `gauge-invariant' means that the
definition is valid for any choice of the coordinate
system which defines the slicing and threading\footnote
{In the unperturbed limit the slicing has to be the one on which all
quantities are uniform and the the threading has to be orthogonal to it.}.

We shall write gauge-invariant expressions in terms of $\zeta$ and $\zeta\su{MW}$.
First
we consider a quantity $\psi^{\rm MW}$, defined
even at second order by
\begin{equation}
g_{ij} = a^2(\eta) \delta_{ij} (1-2\psi^{\rm MW})
\label{mw}
\,.
\end{equation}
This definition, which is written in analogy to \eq{all}, applies to a generic slicing.
Analogously to \eq{expdef} we can consider a quantity $\psi$, valid also in
a generic slicing, defined by
\be
g_{ij} = a^2(\eta) \delta_{ij} e^{-2\psi}
\label{ourpsi}
\,.
\ee 
On uniform-density slices, $\psi\one= \psi_1^{\rm MW}= -\zeta\one$,
$\psi_2^{\rm MW}=-\zeta_2\su{MW}$, and $\psi_2 = -\zeta_2$. We shall also need the 
energy density perturbation $\delta\rho$, defined on the generic slicing,
as well as the unperturbed energy density $\rho_0$.

At first order, the  gauge-invariant expression for $\zeta$
has the well-known form [\hbox{c.f. Eq. (\ref{zeta_def2})}]
\be
\zeta\one = -\psi\one - \calh \frac{\delta\rho_1}{\rho_0'}
\label{psigi}
\,,
\ee
where $\calh=a'/a$, and the unperturbed energy density satisfies
$\rho_0' = -3\calh (\rho_0 + P_0)$ with $P_0$ being the unperturbed pressure.
This expression 
obviously is correct for the uniform density slicing, and it is 
correct for all slicings because the changes in the first and second terms
induced by a change in the slicing cancel \cite{bardeen,bst,gauget2,lyth84,gauget1}.

At second order, Malik and Wands show that \cite{malik}
\bea
\zeta_2\su{MW} &=& -\psi_2^{\rm MW} - \calh \frac{\delta\rho_2}{\rho_0'}
+2\calh \frac{\delta\rho\one}{\rho_0'} \frac{\delta\rho\one'}{\rho_0'}
+ 2\frac{\delta\rho\one}{\rho_0'} \( \psi\one' + 2\calh \psi\one \) \nonumber \\
&& - \( \calh \frac{\delta\rho_1}{\rho_0'} \)^2
\( \frac{\rho_0''}{\calh \rho_0'} - \frac{\calh'}{\calh^2} - 2 \)
\label{mwgi0}
\,,
\eea
which is, again and for the same reason as before, obviously correct for
all the slices.
Accordingly, from \eq{malmw}, we can write a gauge invariant definition for
our second-order $\zeta$: \footnote{This relation has recently been confirmed
in Ref. \cite{lv} (see also Ref. \cite{extlv}) using a nonlinear coordinate-free approach.}
\be
\zeta_2 = -\psi_2 - \calh \frac{\delta\rho_2}{\rho_0'}
+2\calh \frac{\delta\rho\one}{\rho_0'} \frac{\delta\rho\one'}{\rho_0'}
+ 2\frac{\delta\rho\one}{\rho_0'} \psi\one'
- \( \calh \frac{\delta\rho_1}{\rho_0'} \)^2
\( \frac{\rho_0''}{\calh \rho_0'} - \frac{\calh'}{\calh^2} \)
\label{zetatwo}
\,,
\ee
where the relation
\be
\psi_2^{\rm MW} = \psi_2 - 2(\psi_1)^2 \,,
\ee
coming from Eqs. (\ref{mw}) and (\ref{ourpsi}), has been used.

\section{Slow-roll inflation and a third definition} \label{thirddef}

Now we specialize to the era of slow-roll inflation \cite{albste,book,linde82,treview}.
 We consider
single-component inflation, during which the curvature perturbation 
$\zeta$ is
conserved, and multi-component inflation during which it varies.
After defining  both paradigms, we give a third definition of the 
curvature perturbation which applies only during inflation.

\subsection{Single-component inflation} \label{sicoin}

In a single-component inflation model \cite{book,treview}
the inflaton trajectory is by
definition essentially
unique. 
The inflaton field $\varphi$
parameterises the distance along the inflaton trajectory.
 In terms of the field variation, slow-roll inflation (see Subsection \ref{slow_roll}) is 
characterised by the slow-roll
conditions  \cite{liddle,book,treview}
\bea
\varepsilon &\equiv&  -\left(\frac{\dot{H}_{\rm inf}}{H^2_{\rm inf}}\right)_\ast \ll 1 \,, \label{epsilon} \\
|\eta_\varphi - \varepsilon| &\equiv&  \left|\frac{\ddot{\varphi}_0}{H_{\rm inf} \dot{\varphi}_0}\right| \ll 1 \,. \label{eta}
\eea
The inflaton field 
can be taken to be canonically normalised, in which case 
these definitions
are equivalent  to 
conditions on the potential $V$
\bea
\varepsilon &\equiv& \frac{m_P^2}{2V^2} 
\( \frac{\partial V}{\partial\varphi_0}\)^2 \,,
\label{flat1} \\
\eta_\varphi &\equiv& \frac{m_P^2}{V} \frac{\partial^2V}{\partial\varphi_0^2}
\,,
\label{flat2}
\eea
which, together with the slow-roll approximation,
lead to the slow-roll behaviour
\be
3H_{\rm inf} \dot{\varphi}_0 \approx - \frac{\partial V}{\partial\varphi_0}
\,.
\ee

Even without the slow-roll approximation, slices of uniform $\varphi$
correspond to comoving slices because a spatial gradient of $\varphi$
would give non-vanishing momentum density. 
Since comoving slices coincide with slices of uniform energy density,
the slices of uniform $\varphi$ coincide also with the latter.
 Also, since $\varphi$ is a Lorentz scalar,
its gauge transformation is the same as that of $\rho$. It follows \cite{vernizzi}
that
we can replace $\rho$ by $\varphi$ in the above expressions:
\bea
\zeta\one &=&  -\psi\one - \calh_{\rm inf} \frac{\delta\varphi\one}{\varphi_0'} \,, 
\label{gifirst2}\\
\zeta_2\su{MW} &=& -\psi_2^{\rm MW} - \calh_{\rm inf} \frac{\delta\varphi\two}{\varphi_0'}
+2\calh_{\rm inf} \frac{\delta\varphi\one}{\varphi_0'} \frac{\delta\varphi\one'}{\varphi_0'}
+ 2\frac{\delta\varphi\one}{\varphi_0'} \( \psi\one' + 2\calh_{\rm inf} \psi\one \)
\nonumber \\
&& - \( \calh_{\rm inf} \frac{\delta\varphi\one}{\varphi_0'} \)^2
\( \frac{\varphi_0''}{\calh_{\rm inf} \varphi_0'} - \frac{\calh'_{\rm inf}}{\calh_{\rm inf}^2} - 2 \)
\label{mwgi2}
\,, \\
\zeta_2 &=& -\psi_2 - \calh_{\rm inf} \frac{\delta\varphi_2}{\varphi_0'}
+2\calh_{\rm inf} \frac{\delta\varphi\one}{\varphi_0'} \frac{\delta\varphi\one'}{\varphi_0'}
+ 2\frac{\delta\varphi\one}{\varphi_0'} \psi\one' \nonumber \\
&& - \( \calh_{\rm inf} \frac{\delta\varphi_1}{\varphi_0'} \)^2
\( \frac{\varphi_0''}{\calh_{\rm inf} \varphi_0'} - \frac{\calh'_{\rm inf}}{\calh_{\rm inf}^2} \)
\label{zetagi}
\,.
\eea

\subsection{Multi-component inflation}

Now consider the case of multi-component inflation, where there is a family
of inequivalent inflationary trajectories lying in an $N$-dimensional
manifold of  field space. If the relevant part of the manifold is 
 not too big it will be a good approximation to take the fields to be 
canonically normalised. Then the inequivalent trajectories will be
curved in field space\footnote
{More generally
they will be 
non-geodesics, the 
geodesics being  the trajectories which 
 the background fields
could follow if there was 
 no potential term in the scalar Lagrangian \cite{rigopoulos}.}.
To define the trajectories 
one can choose a fixed basis in field space corresponding to 
fields $\phi_1,\cdots,\phi_N$.

Assuming canonical normalisation, multi-component slow-roll inflation is characterised
by the  conditions
\bea
\frac{m_P^2}{2V^2} \left( \frac{\partial V}{\partial \phi_{n_0}} \right)^2 
& \ll & 1 \,, \label{flat1a} \\
\left|\frac{m_P^2}{V} \frac{\partial^2 V}{\partial \phi_{n_0} \partial\phi_{m_0}}
\right| & \ll & 1 \,, \label{flat2a} \\
3H_{\rm inf} \dot\phi_{n_0} &\approx& - \frac{\partial V}{\partial \phi_{n_0}} \,. \label{svcon}
\eea

The procedure of choosing a fixed basis 
is quite convenient for calculations, but a different
procedure leads to a perhaps simpler theoretical description.
This is to take $\varphi$ to parameterise the distance along the inflaton
trajectories, just as in single-component inflation, but now with the
proviso that uniform $\varphi$ corresponds to uniform field potential
(since we work in the slow-roll approximation, this means that the 
slices in field space of uniform $\varphi$ are orthogonal to 
the trajectories).
Then, in the slow-roll approximation, slices of spacetime with
uniform $\varphi$ will again coincide with slices of uniform density (see Fig.
\ref{fbasis}a).
Since $\varphi$ is a scalar, Eqs. (\ref{gifirst2}) and (\ref{mwgi2}) will then
be
valid. This is the simplest form of the gauge-invariant expression,
though for a practical calculation it may be better to write it in terms
of a fixed basis.

There is a subtlety here. For the first-order case we could
define $\varphi$ in a different way; around  a given point 
on  the unperturbed trajectory we could choose a fixed field
basis, with one of the basis vectors pointing along the trajectory,
and define $\varphi$ as the corresponding field component.
Then we could choose $\varphi$ to be canonically normalised in the 
vicinity of the chosen point in field space. 
That would not work at second order though,  because at that order
it makes a difference whether $\varphi$ is the appropriate parameterisation
of the  distance
along the trajectories (our adopted definition)
or the distance along a tangent vector to the trajectory (the 
alternative definition) (see Fig. \ref{fbasis}b). Only our adopted one will
make \eqs{mwgi2}{zetagi} valid.

\begin{figure}
\begin{center}
\leavevmode
\hspace{2mm}
\hbox{\hspace{-2cm}%
\epsfxsize=2in
\epsffile{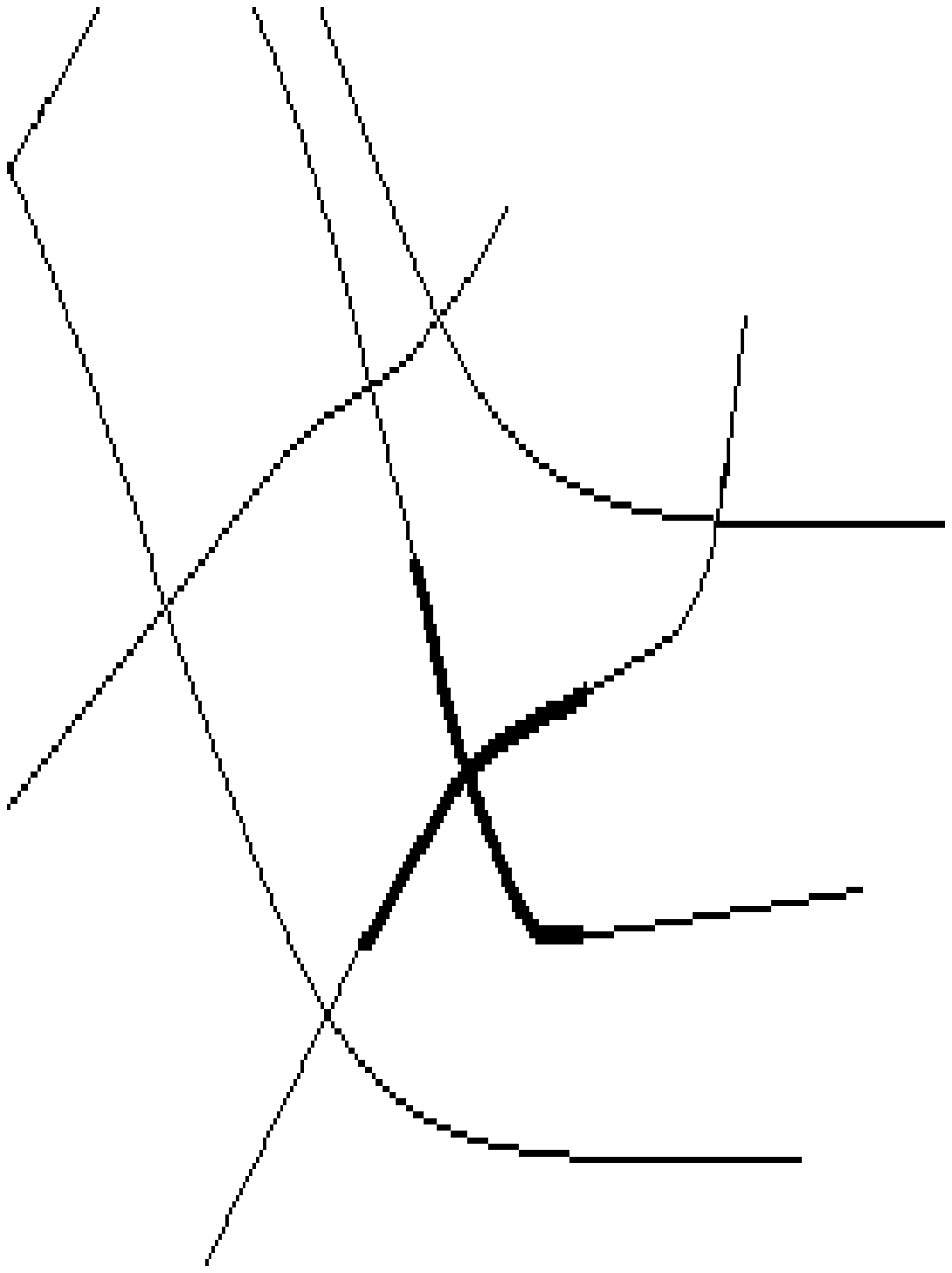}}
\hbox{\hspace{1cm}%
\epsfxsize=2in
\epsffile{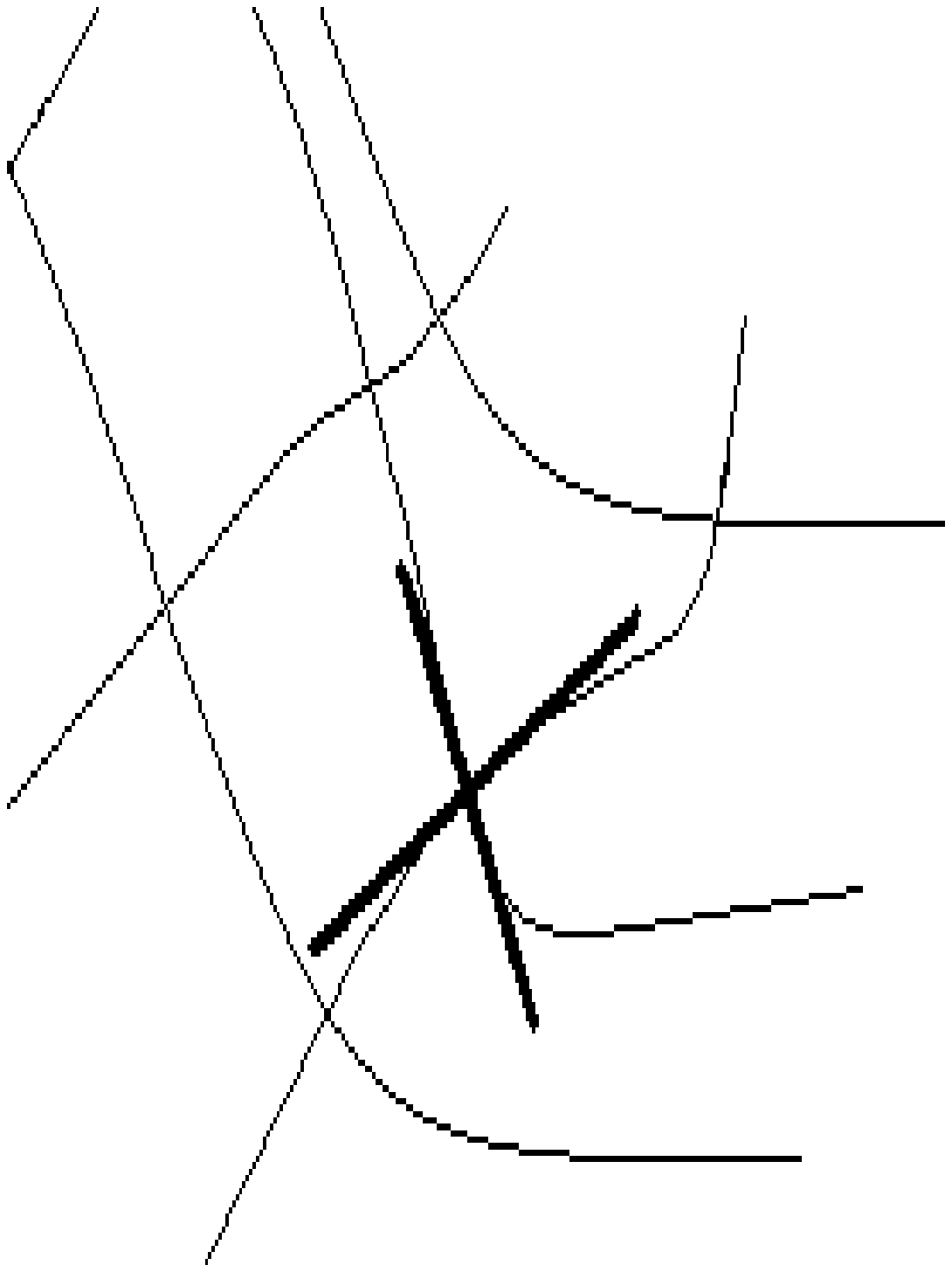}}
\put(-227,140){{\bf $\varphi$}}
\put(-272,163){{\bf $\vartheta$}}
\put(-200,108){$V = {\rm const}$}
\put(-210,75){$V = {\rm const}$}
\put(-217,55){$V = {\rm const}$}
\put(-65,108){{\bf $\varphi$}}
\put(-65,68){{\bf $\vartheta$}}
\put(-25,108){$V = {\rm const}$}
\put(-35,75){$V = {\rm const}$}
\put(-40,55){$V = {\rm const}$}
\put(-250,30){(a)}
\put(-75,30){(b)}
\vspace{-1cm}
\caption{Two different procedures 
for defining the fields in two-component
inflation. The fields are denoted by $\varphi$
and $\vartheta$.
 (a) The field
$\varphi$  parameterises the 
distance along
the inflaton trajectories, with 
 uniform $\varphi$ corresponding to the equipotential
lines.
 The field $\vartheta$ parameterises the 
distance along the equipotentials.
(b) The fields $\varphi$ and $\vartheta$ are the components in
a fixed orthonormal basis, aligned with the inflationary trajectory
 at a certain point in field space. The value of 
$\varphi$ is now  the displacement 
 along the  tangent vector
and the value of  $\vartheta$ is the displacement  along
the orthogonal vector. Working to second order in these
displacements, the equipotentials no longer coincide
with the lines of uniform $\varphi$. \label{fbasis}}
\end{center}
\end{figure}

\subsection{A third definition of the curvature perturbation} \label{3defzetaA}

The third definition in the literature 
applies only during inflation. It  was
given originally by Acquaviva {\it et. al.} \cite{acquaviva} for
the single-component case,
and the generalization to the multi-component case was noted by
Rigopoulos \cite{rigopoulos}. We shall denote this definition by
$\zeta\su{A}$.

The definition of Acquaviva {\it et. al.} and Rigopoulos is
\be
\zeta\su{A} _2 = -\psi_2^{\rm MW} - 
\mathcal{H}_{\rm inf} \frac{\delta\varphi\two}{\varphi_0'}
- \frac{
(\psi_1' + 2\mathcal{H}_{\rm inf}\psi_1 
+ \mathcal{H}_{\rm inf}\delta\varphi\one' / \varphi'_0)^2}{
\mathcal{H}'_{\rm inf} + 2\mathcal{H}^2_{\rm inf} - 
\mathcal{H}_{\rm inf}\varphi''_0 / \varphi'_0
}
\,. \label{abmr}
\ee
This is gauge-invariant by construction, with 
 $\varphi$ defined as in Figure 1(a).
 
It was pointed out by Vernizzi \cite{vernizzi} (actually in the context
of  single-component inflation)  
that comparing this  definition with \eq{mwgi2} gives simply
\be
\zeta_2\su{A} =
 \zeta_2\su{MW}
 - \frac{4 \mathcal{H}_{\rm inf}^2(\zeta_1)^2}{\mathcal{H}'_{\rm inf} 
+ 2 \mathcal{H}_{\rm inf}^2 - \mathcal{H}_{\rm inf}\varphi''_0/\varphi'_0}
\,.
\label{abmr2}
\ee
In the limit of slow-roll the denominator of the last term becomes
just $2\calh^2_{\rm inf}$, and then
\bea
\zeta_2\su{A} &=& \zeta_2
\,.
\eea
In other words, this third definition coincides with our adopted one
in the slow-roll limit.

Making use of the slow-roll parameters in Eqs. (\ref{epsilon}) and (\ref{eta}), the expression
in Eq. (\ref{abmr2}) gives to first-order in the slow-roll approximation
\be
\zeta_2\su{A} = \zeta_2 
 - (2\varepsilon - \eta_\varphi) (\zeta_1)^2 \label{vercom}
\,.
\ee

\section{The  evolution of the curvature perturbation}
\label{origin}

The simplest possibility for the origin of the observed
curvature perturbation  is that it comes from
the vacuum fluctuation of the inflaton field in a single-component model (see Section \ref{section_inflaton}).
More recently other possibilities were recognised and we summarise the 
situation now. Although the  discussion is
usually applied to the magnitude of the curvature perturbation,
it applies equally to the non-gaussianity.

\subsection{Heavy, light and ultra-light fields}

On each scale the  initial epoch, as far as classical perturbations
are concerned, should be taken to be a few Hubble times after horizon
exit during inflation. The reason is that all such perturbations
 are  supposed to originate from
the vacuum fluctuation of one or more light 
scalar  fields, the fluctuation on each scale being 
promoted to a classical perturbation around the time of horizon exit \cite{albrecht,grishchuk,guthpi,lombardo,lyth84}.

Considering a fixed basis with canonical normalisation, a light field
is roughly speaking one  satisfying the 
flatness condition in \eq{flat2a}. 
The terminology is suggested by the important special case
that the effective potential during inflation is quadratic.
 Then, a light field is  roughly speaking 
that whose effective  mass during inflation is less than the value
 $H_*$ of the Hubble parameter. 
More precisely, the condition that the vacuum fluctuation be  promoted to 
a classical perturbation is
\cite{mijic}
\be
m < \frac{3}{2} H_* \label{lightfield}
\,.
\ee
{}From now on we focus on the quadratic potential, and take this as the
 the definition of a light field.
Conversely a heavy field may be  defined as one for which the 
condition in \eq{lightfield} is violated.

 During inflation light fields slowly roll according to \eq{svcon}
(with the vacuum fluctuation superimposed)
 while the heavy fields presumably are pinned down at an instantaneous
minimum of the effective potential. 
As we have seen, multi-component inflation takes place in a subspace of
field space. The fields in this subspace are light, but
their effective masses are  sufficient
to appreciably curve the inflationary trajectories.
In the case of both  multi-component and single-component inflation,
there could also be `ultra-light' fields, which do not appreciably curve
the inflationary trajectory and which therefore
have practically no effect on the dynamics of inflation.

\subsection{The evolution of the curvature perturbation}

To describe  the behaviour of  perturbations during the 
super-horizon era, without making too many detailed assumptions,
 it is useful to invoke the separate universe
hypothesis  \cite{bst,sasaki1,lyth,tanaka,llmw} 
after smoothing on a given comoving scale much bigger than the horizon\footnote{
When considering linear equations, smoothing is equivalent to dropping
short wavelengths fourier components. In the nonlinear situation the smoothing
procedure could be in principle ambiguous. In a given situation one should state
explicitly which quantities are being smoothed.}.
According to this hypothesis the local evolution at each position is
that of some unperturbed universe (separate universe).
 Of course the separate universe hypothesis can and should be 
checked where there is a sufficiently detailed model.
However, it should be correct
on cosmological scales for a very simple reason. The unperturbed Universe
may be  defined as the one around us, smoothed on a scale
a bit bigger than the present Hubble distance.
In other words, the separate universe hypothesis is certainly valid
when applied to that scale. But the whole range of cosmological 
scales spans only a few orders of magnitude.
This means that cosmological scales
are  likely to be huge compared with any
scale that is relevant in the early Universe, and accordingly that the 
separate universe hypothesis should be valid when applied to cosmological
scales even though it might fail on much smaller scales
(this expectation  was verified
in a preheating example \cite{ourreheat}
to which we  return later).

We are concerned with the curvature perturbation, which during the 
 super-horizon era 
 is conserved as long as the pressure is a unique function of the 
energy density (the adiabatic pressure condition) (see Subsection \ref{acgauge}). 
The 
 adiabatic pressure condition 
will be satisfied if and only if the separate universes are identical
(at least as far as the relation between pressure and energy density is
concerned) \footnote
{Of course the identity will only hold after making an 
 appropriate synchronization of the clocks at different positions. Having made that
synchronization, horizon entry will occur at different times
in different positions, which can be regarded as the origin
of the  curvature perturbation.}. 
The condition to have identical universes after a given epoch is  that
the specification of a  {\em single} quantity at that epoch is sufficient
to determine the entire subsequent evolution. 

In the case of single-component inflation,
the initial condition may be supplied by the local  value of the 
inflaton field, at the very 
beginning of the super-horizon era when it first becomes
classical.  Given the separate universe hypothesis, that is the only
possibility if the inflaton is the only light field ever to play a 
significant dynamical role. This means that {\em the curvature perturbation
generated at horizon exit during single-component inflation will be
equal to the one observed at the approach of horizon entry, provided
that the inflaton is the only light field ever to play a dynamical role.}

If inflation is multi-component, more than one field is 
by definition relevant during inflation. Then   the curvature perturbation
cannot be conserved during  inflation. The variation of the curvature 
perturbation during multi-component inflation is caused by the vacuum 
fluctuation orthogonal to the unperturbed inflationary trajectory, which
around the time of horizon exit
kicks the trajectory onto a nearby one so that the local trajectory
becomes position-dependent. After inflation is over, the curvature 
perturbation will be conserved if 
 the local trajectories lead to practically identical  universes.
In other words it will be conserved if the light (and ultra-light) fields,
orthogonal to the trajectory at the end of inflation, do not
affect the subsequent evolution of the Universe.

The curvature perturbation after inflation will vary if some
 light or  ultra-light field, orthogonal to the trajectory at the end of 
inflation, affects the subsequent evolution of the Universe (to be 
precise, affects  the pressure) \cite{wandsreview,gordon}.
As we shall describe in Section \ref{after}, three types of scenario have been proposed for
this post-inflationary variation of the curvature perturbation.

\section{Non-gaussianity} \label{gaussianity}

\subsection{Defining the non-gaussianity} \label{definingng}

A gaussian perturbation is one whose Fourier components are uncorrelated \cite{padmanabhan}.
All of its statistical properties are defined by its spectrum, and the 
spectrum $\calp_g (k) \equiv A_g^2 (k/aH_{\rm inf})^{n_g}$ of generic perturbation is conveniently 
 defined  \cite{book,treview}  by\footnote
{Technically the expectation values in this and the following expressions 
refer to an ensemble of universes but, because the stochastic properties 
of the
perturbations are supposed to be invariant under translations, the 
expectation
values can also be regarded as averages over the location of the observer
who defines the origin of coordinates.}
\be
\langle g_{{\bf k}_1}g_{{\bf k}_2} \rangle = \frac{2\pi^2}{k^3} \delta^3(\bfk_1+\bfk_2) \calp_g(k)
\,,
\ee
the normalisation being chosen so that
\be
\vev{g^2(\bfx)} = \int^\infty_0 \calp_g(k) \frac{dk}{k}
\,.
\ee
On cosmological scales a  few Hubble times before 
horizon entry, observation shows that the
curvature perturbation is almost Gaussian with $|A_\zeta|
\approx 5 \times 10^{-5}$ \cite{observation}. 

The simplest kind of non-gaussianity 
that the  curvature perturbation 
could possess is  of the form
\bea
\zeta(\bfx) = \zeta\sub g(\bfx)  - \frac{3}{5} f\sub{NL} \( \zeta\sub g^2(\bfx)
 - \langle \zeta\sub g^2 \rangle \)
\label{ffnldef}
\,,
\eea
where
$\zeta\sub g$ is Gaussian with $\vev{\zeta\sub g}=0$,
and the non-linearity parameter $f\sub{NL}$ 
is independent of position.  We will call this {\em correlated $\chi^2$ non-gaussianity}.
Note that this definition assumes that $\vev\zeta
=0$, which means that the zero Fourier mode (spatial average) is 
dropped.

Following Maldacena \cite{maldacena}, we have inserted the prefactor
$-(3/5)$ so that in first-order perturbation theory our definition agrees
with that of Komatsu and Spergel \cite{spergel}, which is generally
the definition people use
when comparing theory with observation. Working in first-order
perturbation theory, these authors write  $\Phi(\bfx) = \Phi_g(\bfx) +
\fnl \left(\Phi_g^2(\bfx) - \langle \Phi_g^2 \rangle \right)$, and  their
$\Phi$ is equal to $-3/5$ times our $\zeta$ \footnote{The actual quantity constrained by observational data
is $f_{\rm NL}^T$, which is the non-linearity parameter for the CMB temperature anisotropies:
\be
\frac{\delta T}{T} ({\bf x}) = \left(\frac{\delta T}{T}\right)_g ({\bf x}) + f_{\rm NL}^T \left[\left(\frac{\delta T}{T}\right)_g^2 ({\bf x}) - \left\langle \left(\frac{\delta T}{T}\right)_g^2 \right\rangle \right] \,.
\ee
At first order $f_{\rm NL}^T = 3\fnl$ because $\delta T/T_0 = (-1/5) \zeta$ [c.f. Eqs. (\ref{connections1}) and (\ref{connections2})]. However, to compare adequately the observational data with our $\fnl$, we must calculate $f_{\rm NL}^T$ in terms of $\fnl$ at second order (see e.g. Refs. \cite{bartolo1,bartolo2,bartolotemp,newbartolo,creminelli}). \label{copelandf1}}.

One of the most powerful observational signatures of non-gaussianity
is a nonzero value for the three-point correlator, specified by the 
bispectrum $B$ defined by \cite{review,spergel}
\be
\vev{\zeta_{{\bf k}_1}\zeta_{{\bf k}_2}\zeta_{{\bf k}_3}} = (2\pi)^{-3/2} B({k_1,k_2,k_3}) 
\delta^3(\bfk_1
+\bfk_2 + \bfk_3 ) 
\label{bdef}
\,.
\ee
For correlated $\chi^2$  non-gaussianity (with the gaussian term dominating)
\be
B({k_1,k_2,k_3}) = -\frac{6}{5}f\sub{NL}
\Big[ P_\zeta(k_1) P_\zeta(k_2)
+ {\rm \, cyclic \ permutations \,} \Big]
\label{bchis}
\,,
\ee
where $P_\zeta(k) = 2\pi^2\mathcal{P}_\zeta(k)/k^3$. 
For any  kind of non-Gaussianity
one may use the above expression to define a function $f\sub{NL}
(k_1,k_2,k_3)$.

Given a calculation of
 $f\sub{NL}$ using first-order perturbation
theory, one expects  in general that going to second order will change
$f\sub{NL}$ by an amount of order 1. On this basis, one expects that a 
first-order calculation is  good enough if it yields
$|f\sub{NL}|\gg 1$, but that otherwise a second-order calculation will be
necessary. 

The definition \eq{bchis}  of $f\sub{NL}$ is  made using our adopted definition
of $\zeta$. If $\zeta$ in the definition is   replaced 
by $\zeta\su{MW}$ (with the zero Fourier mode dropped)
then $f\sub{NL}$ should be replaced by
\be
f\sub{NL}\su{MW}\equiv f\sub{NL} - \frac{5}{3} \label{minusone}\label{fnlmw}
\,.
\ee
To obtain this expression we used \eq{mwdef} and dropped terms higher than
second order\footnote{Obviously the parameter $f_{\rm NL}^T$, which is the important one to make comparison with observational data, does not depend on the chosen definition for $\fnl$. \label{copelandf2}}. 

All of this assumes that the non-gaussian component of $\zeta$ is fully 
correlated with the gaussian component. An alternative possibility
\cite{bl}
that will be important for us is if $\zeta$ has the form
\bea
\zeta(\bfx) = \zeta_g (\bfx)  - \frac{3}{5} \tilde f\sub{NL} \( \zeta_\vartheta^2(\bfx)
 - \langle \zeta_\vartheta^2 \rangle \)
\label{fnltildef}
\,,
\eea
where
$\zeta_g$ and $\zeta_\vartheta$ are  uncorrelated  Gaussian perturbations,
normalised to have equal
spectra, and the  parameter $\tilde f\sub{NL}$ 
is independent of position.  We will call this {\em
uncorrelated $\chi^2$ non-gaussianity}.
It can be shown \cite{bl}
that in this case, $f\sub{NL}$ as defined by \eq{bchis} is given by
\be
f\sub{NL} \sim \( \frac{\tilde f\sub{NL}}{653} \)^3
\,.
\ee

\subsection{Observational constraints on the non-gaussianity}

Taking $f\sub{NL}$ to denote the non-linearity parameter at the primordial 
era, let us consider the observational constraints.
Detailed calculations have so far been made  only with
 $f\sub{NL}$ independent of the wavenumbers, and only
by using  first-order perturbation theory for the evolution of the 
cosmological perturbations after horizon entry.
It is found  \cite{komatsu03}
that present observation
 requires
 $|f\sub {NL}|\lsim 10^2$   
making the non-gaussian fraction at most of order $10^{-3}$. 
The use of first-order perturbation theory 
in this context  is amply justified.
Looking to the future though, it is found
that the
PLANCK satellite will either detect non-gaussianity or reduce the bound to
$|f\sub {NL}|\lsim 5$ \cite{review,spergel}, and that
foreseeable  future observations can reach
a level $|f\sub{NL}|\sim 3$ \cite{review,spergel}. 

Although the use of first-order perturbation
theory is not really justified for the latter estimates, we can safely
conclude that it will be  difficult for observation ever to detect
a value $|f\sub{NL}|\ll 1$. That is a pity because,  as we shall see,
such a value is predicted by some theoretical scenarios. On the other 
hand, other scenarios predict $|f\sub{NL}|$ roughly of order 1.
It will  therefore be of great interest to have detailed second-order
 calculations, to establish precisely the level of sensitivity that can
be achieved by future observations. A step in this direction has 
been taken in Ref. \cite{newbartolo} (see also Refs. \cite{bartolo1,bartolo2,bartolotemp,creminelli}),
where a non-linear expression for the large-scale CMB anisotropy
is given in terms of only the curvature perturbation 
(generalizing the first-order Sachs-Wolfe effect \cite{sachs}).

\section{The initial non-gaussianity}

\label{finitial}

\subsection{Single-component inflation} \label{finitialsingle}

At first order, the curvature perturbation during single-component
inflation is Gaussian. The amplitude of its time-independent spectrum is given by 
\cite{book,treview} [c.f. \hbox{Eq. (\ref{Aspectruminf})}]
\be
A_\zeta = - \frac{H_\ast^2}{2\pi \dot{\varphi}_0}
\label{single}
\,,
\ee
and its  spectral index $n_\zeta \equiv d\ln \calp_\zeta(k)/d\ln k$ is given by
\be
n_\zeta = 2\eta_\varphi -6\varepsilon
\,. \label{ngetazeta}
\ee
The squared amplitude of the spectrum $r_{T\zeta}$ of the 
 tensor perturbation, defined 
 as a fraction of $A_\zeta^2$, is also given
in terms of the slow-roll parameter $\varepsilon$ [c.f. Eq. (\ref{randvar})]:
\be
r_{T\zeta} = 16\varepsilon \,.
\ee

{\em If the curvature perturbation does not evolve after single-component
inflation is over} observation constrains $n_\zeta$ and $r_{T\zeta}$, and hence
the slow-roll parameters $\eta_\varphi$ and $\varepsilon$. A current bound
\cite{observation} is $-0.048<n_\zeta<0.016$ and $r_{T\zeta}<0.46$. The second bound
gives $\varepsilon<0.029$, but barring an 
accurate cancellation the first bound gives $\varepsilon \lsim  0.003$.
In most inflation models $\varepsilon$ is completely negligible and then 
the first bound gives $-0.024<\eta_\varphi<0.008$ (irrespective of slow-roll inflation models,
the  upper bound in this expression  holds generally,
and the lower bound is badly violated only if there is an accurate 
cancellation). The bottom line of all this
 is that $\varepsilon$ and $|\eta_\varphi|$ are both constrained to be 
$\lsim 10^{-2}$.

Going to 
second order, Maldacena \cite{maldacena} has calculated the bispectrum
during single-component inflation (see also Refs. \cite{calcagni,singlecreminelli,falk,gangui,rst,sb2,seery}) \footnote{In Ref. \cite{rst} (see also Ref. \cite{rs}) Rigopoulos {\it et. al.} calculated the three-point correlator in single-component slow-roll inflation using a stochastic approach.
Their result agrees with Maldacena's in the squeezed limit (where one of the scales $k^{-1}$ crosses the horizon
much earlier than the other two, $k_1 \ll k_2,k_3$), but disagrees in the limit where the $\vec{k}_i$'s form
an equilateral triangle. Calcagni in Ref. \cite{calcagni} extended this stochastic approach
to calculate the non-gaussianity originated from a Dirac-Born-Infeld tachyonic inflaton and in braneworld
scenarios,
finding results identical to Maldacena's one.
The three-point correlators calculated
for both cases were found identical to that calculated in Ref. \cite{rst}.}.
His result may be written in the form
\be
f\sub{NL} = \frac{5}{12} \[2\eta_\varphi - 6\varepsilon -2 \varepsilon f(k_1,k_2,k_3) \] 
\label{fnlmal}
\,,
\ee
with  $0 \leq f\leq 5/6$.
By virtue of the slow-roll conditions, $|f\sub{NL}|\ll 1$ \footnote
{Near a maximum of the potential `fast-roll' inflation \cite{lotfi,fastroll}
can take place with  $|\eta_\varphi|$ somewhat
bigger than 1. Maldacena's calculation does not apply to that case but,
presumably, it gives initial non-gaussianity $|\fnl|\sim 1$. Although the 
corresponding initial 
spectral index is far from $1$, which means that the initial
curvature perturbation produced by $\varphi$ must be negligible, the
precise initial value of $\fnl$ may in this case be important as long as another
field (like the curvaton) be in charge of generating the observed curvature perturbation.}.
In other words, the curvature perturbation $\zeta$, 
{\em as we have defined it},
is almost Gaussian during single-component inflation.

From \eq{vercom}
$\zeta\su{A}$ is also practically gaussian, but this quantity is defined
only during inflation and therefore could not be considered as a replacement
for $\zeta$. More importantly, $\zeta\su{MW}$ has significant 
non-gaussianity because,  from
\eq{minusone}, it corresponds to   $f\sub{NL}\su{MW} \approx -5/3$.  

One may ask why it is our $\zeta$ and not $\zeta\su{MW}$ which is gaussian
in the slow-roll limit.
 One feature that distinguishes our $\zeta$,
is that any part of it can be absorbed into the scale factor without altering
the rest; indeed
\be
g_{ij}=\delta_{ij}a^2(\eta)e^{2\zeta_1+\zeta_2}
=\delta_{ij}\tilde a^2(\eta)e^{\zeta_2}
\,,
\ee
with $\tilde a = ae^{\zeta_1}$ (if we tried to do that with $\zeta\sub{MW}$,
the part of $\zeta$ not absorbed would have to be re-scaled). This means that
an extremely long-wavelength and possible large part of $\zeta$ has no local
significance. It also means, in the context of perturbation theory, 
  that the first-order part of $\zeta$ can be absorbed into the scale factor
when discussing the second-order part. 
However, the gaussianity of $\zeta$ does not seem to be 
related directly to this feature.
Rather, it has to do with  the gauge
transformation, relating quantities $\psi_A$ and $\psi_B$ defined on 
different slicings.

With our definition \cite{sasaki1}, the gauge transformation is
\be
\psi_A(t,\bfx)-\psi_B(t,\bfx)  = - \Delta N_{AB}(t,\bfx)
\label{psigt}
,
\ee
where $\Delta N_{AB}$ is the number of $e$-folds of  expansion
going from a slice $B$ to a slice $A$, both of them corresponding to 
time $t$ \footnote
{This expression
is valid even when the tensor perturbation is included \cite{sasaki1}. As a result,
 the gauge-invariant expressions mentioned earlier are still valid
in that case, as are the results based on them including the present
discussion. \label{teninc}}.
In writing this expression  we used
 physical time $t$ instead of conformal time,
the two related by $dt=a d\eta$. 
Along a comoving worldline, the number of $e$-folds of expansion is defined as
$N\equiv \int \tilde H d\tau$
where $\tilde H$ is the local Hubble parameter and $d\tau$ is the proper
time interval \cite{sasaki1}.

To understand the relevance of this result, take $\psi_B=0$ 
and $\psi_A=-\zeta$.
The pressure is adiabatic during single-component inflation, 
which means that $dt$ can be identified with the proper time
interval $d\tau$, and the proper expansion rate on slicing $A$
is uniform \cite{sasaki1}. As a result, to second order, 
\bea
\zeta  &=& H_{\rm inf}(t) \Delta t(t,\bfx) + 
\frac12 \dot{H}_{\rm inf}(t) \(\Delta t(t,\bfx) \)^2 \nonumber \\
&\simeq &  H_{\rm inf} \Delta t(t,\bfx) +
\frac12 \frac{\dot{H}_{\rm inf}}{H_{\rm inf}^2} \( H_{\rm inf} \Delta t(t,\bfx) \)^2 \nonumber \\
&\simeq &  H_{\rm inf} \Delta t(t,\bfx) 
\label{gt1}
\,.
\eea
In the last line we made the slow-roll approximation, and from the second
line we can see that the error in $f\sub{NL}$  caused by this approximation
 is precisely $\varepsilon$.

We  also need the gauge transformation for the inflaton
field $\varphi$ in terms of $\Delta t$. 
Since the slices correspond to the same coordinate
time,  the unperturbed inflaton 
field can be taken to be the same on each of them which means that the gauge
transformation for $\delta \varphi$ is
\be
\delta\varphi_A(t,\bfx) - \delta\varphi_B(t,\bfx) = 
\Delta \varphi_{AB} (t,\bfx)
,
\ee
where $\Delta\varphi_{AB}$ is the change in $\varphi$ going from slice
$B$ to slice $A$. But slice $A$ corresponds to uniform $\varphi$,
which means that on slice $B$ to second order
\bea
H_{\rm inf}(t)\frac{\delta\varphi_B(t,\bfx)}{\dot \varphi_0}  &=& -H_{\rm inf}(t) \Delta t (t, \bfx) -
\frac12 H_{\rm inf}(t) \frac{\ddot \varphi_0}{\dot \varphi_0} (\Delta t (t, \bfx))^2  \nonumber \\
& \simeq & -H_{\rm inf} \Delta t (t, \bfx) -
\frac12 \frac{\ddot \varphi_0}{H_{\rm inf} \dot \varphi_0} (H_{\rm inf} \Delta t (t, \bfx))^2 \nonumber \\
& \simeq  &  -H_{\rm inf} \Delta t (t, \bfx)
\label{gt2}
\,,
\eea
where in the last line we used the slow-roll approximation. We can see that the
fractional 
error caused by this approximation is $\ddot\varphi_0/H_{\rm inf}\dot\varphi_0
=\varepsilon -\eta_\varphi$.

Combining \eqs{gt1}{gt2} we have in the slow-roll approximation
\be
\zeta \simeq  -H_{\rm inf}(t) \frac{\delta \varphi_B (t, \bfx)}{\dot \varphi_0}
\,,
\ee
with fractional error of order $\max\{\eta_\varphi,\varepsilon\}$ (this can also be seen directly
from \eqs{gifirst2}{zetagi} evaluated
with $\psi=0$, but we give the above argument because it explains why
the result is valid for $\zeta$ as opposed to $\zeta\su{MW}$).

The final and crucial step is to observe that 
 in the slow-roll approximation $\varphi_B$ is  gaussian,
with again a fractional error of order
 $\max\{\eta_\varphi,\varepsilon\}$. This was demonstrated by Maldacena
\cite{maldacena} but the basic reason is very simple. The  non-gaussianity
of $\varphi$ comes either from  third and higher derivatives of $V$
(through the field equation in unperturbed spacetime) or else through
the back-reaction (the perturbation of spacetime); but
the  first effect is small \cite{book,treview} by virtue of the flatness
requirements on the potential,
and the second effect is small because $\dot\varphi_0/H_{\rm inf}^2$ is small \cite{book}.
This explains why  $\zeta$ with our adopted definition is practically
Gaussian by virtue of the slow-roll approximation.

\subsection{Multi-component inflation}

The flatness and slow-roll conditions \eqss{flat1a}{flat2a}{svcon} 
ensure that 
 the curvature of the inflationary trajectories 
is small during the few Hubble times around horizon exit, during which
the quantum fluctuation is promoted to a classical perturbation.
As a result,  the  {\em initial} curvature perturbation 
in first-order perturbation theory
is still given by the amplitude in \eq{single} and the spectral index in
\hbox{Eq. (\ref{ngetazeta})} in terms of the field $\varphi$
that we defined earlier.

What about the initial non-gaussianity generated at second order?
In the approximation that  the  curvature of the trajectories
around horizon exit is  completely negligible,
we can safely say that the initial 
non-gaussianity corresponds to $|f\sub{NL}|\ll 1$.
Confirming this expectation, Seery and Lidsey \cite{seery2} have calculated the three-point correlator of the perturbations in the fields involved in multi-component slow-roll inflation. Their result is given by
\be
\langle \delta \phi^i_{{\bf k}_1} \delta \phi^j_{{\bf k}_2} \delta \phi^k_{{\bf k}_3} \rangle \equiv \left(\frac{H_\ast}{2\pi}\right)^3(2\pi)^{-3/2}  B_{ijk}(k_1,k_2,k_3) \delta^3 ({\bf k}_1 + {\bf k}_2 + {\bf k}_3) \,, \label{multibi}
\ee
with
\be
B_{ijk}(k_1,k_2,k_3) \equiv -\frac{6}{5} f_{ijk} \left[\frac{4\pi^4}{k_1^3 k_2^3} + {\rm cyclic \ permutations}\right] \,, \label{multifi}
\ee
and\footnote{The cyclic permutations in $k$ and $\phi$ in Eq. (\ref{multimomenta}) must be simultaneous, i.e. when exchanging indices $i$ and $j$, for example, $k_1$ and $k_2$ must also be exchanged. Notice also that the calculation of Seery and Lidsey's is only valid when the magnitudes of the three wavevectors are roughly comparable, so that they exit the horizon at similar epochs.}
\be
f_{ijk} = -\frac{5}{12} \left[\frac{\dot{\phi}_\ast^i}{2\pi m_P^2} \delta_{jk} f_{SL}(k_1,k_2,k_3) + {\rm cyclic \ permutations \ in} \ k \ {\rm and} \ \phi\right] \,. \label{multimomenta}
\ee
As we will discuss in Chapter \ref{fifthpaper}, where the $\delta N$ formalism is used to calculate the stochastic properties of $\zeta$ \cite{lr1}, the contribution $\Delta \fnl$ of the wavevector dependent parameter $f_{ijk}$ to the total $\fnl$ is in any case very small, being $\Delta \fnl$ generically below $(15/24) f_{SL} \sqrt{r_{T\zeta} \varepsilon} \lsim 10^{-2}$ \cite{ignacio} where $f_{SL}$ is in the range $1/3 \leq f_{SL} \leq 11/18$.

\section{The evolution after horizon exit}
\label{after}

\subsection{Single-component inflation and  $\zeta\su{A}_2$}

During single-component inflation the curvature perturbation
$\zeta$, as we have defined it, does not evolve. From its definition
\eq{mwdef}, the same is true of 
 $\zeta\su{MW}$.

 In contrast
$\zeta_2\su{A}$, given by Eq. (\ref{vercom}),
will have the slow variation \cite{vernizzi}
\be
\dot \zeta_2\su{A} \approx -(2\dot\varepsilon - \dot\eta_\varphi) (\zeta\one)^2
\label{dotzetar}
\,.
\ee
This variation has no physical significance, being an artifact of the
definition.

Using a particular   gauge,
Acquaviva {\it et. al.} \cite{acquaviva}
have calculated $\dot\zeta_2\su{A}$ in terms of first-order
quantities $\psi\one$, $\delta\varphi\one$, and their derivatives,
and they have displayed the
result as an indefinite integral
\be
\zeta_2\su{A}(t) = \int^t A(t) dt + B(t)
\label{aeq}
\,.
\ee
Inserting an initial condition, valid a few Hubble times after horizon exit, this becomes
\be
\zeta_2\su{A}(t) = \zeta_2\su{A}(t_{\rm ini}) + \int^t_{t_{\rm ini}} A(t) dt +
\left. B \right|^t_{t_{\rm ini}}
\,. \label{time}
\ee
In view of our discussion, it is clear that these equations will, if
correctly
evaluated, just reproduce the time dependence of \eq{dotzetar}.

The authors of Ref. \cite{acquaviva} also present the respective equation
for $\dot\zeta_2\su{A}$, involving only first-order quantities,
 which is valid also before horizon entry. Contrary to the claim of the 
authors, this   classical
equation  cannot by itself be used to calculate the
initial value (more precisely, the
stochastic properties of the initial value) of $\zeta_2 \su{A}$.
In particular, 
it cannot by itself  reproduce Maldacena's calculation of the 
bispectrum.

It is true of course that in the Heisenberg picture the quantum operators
satisfy the classical field equations. In  first-order perturbation
theory,  where the equations
are linear, this allows one to calculate the curvature perturbation
without going to the trouble of calculating the second-order action
\cite{book} (at the $n$th order of perturbation theory 
the action has to be evaluated to order $n+1$ if it is to be used).
At second order in perturbation theory it remains to be seen whether the 
Heisenberg picture  can provide a useful alternative to Maldacena's 
calculation, who adopted the interaction  picture and calculated the 
action to third order. 

\subsection{Multi-component inflation} \label{hybridev}

During multi-component inflation the curvature perturbation by definition varies
significantly along a generic trajectory, 
which means that non-gaussianity is generated at some level.
So far only a limited range of models
has been investigated 
\cite{multiinfng2,multiinfng4,multiinfng3,enqvistnew,ev,mollerach,multiinfng1}.
To keep the spectral tilt within observational bounds, the unperturbed
trajectory in these  models has to be specially chosen, but the
choice might  be justified by a suitable initial condition.

We shall consider here  a  calculation  by
Enqvist and V$\ddot{{\rm a}}$ihk$\ddot{{\rm o}}$nen in Refs. \cite{ev,lr,antti}.
Following the same line as Acquaviva {\it et. al.} \cite{acquaviva}, 
they study a two-component inflation model, in which the only 
important parts of the potential are 
\be
V(\varphi,\vartheta) = V_h +\frac12 m_\vartheta^2 \vartheta^2 + 
 \frac{1}{2}m_\varphi^2 \varphi^2 
\label{effv}
\,.
\end{equation}
The masses are both supposed to be less than $(3/2)H_*$, so that this is a
two-component inflation model, and the above form of the potential is
supposed to hold for some number $\Delta N$ of $e$-folds after cosmological
scales leave the horizon. They
take the unperturbed inflation trajectory to have $\vartheta_0=0$,
and the idea is to calculate the amount of non-gaussianity generated
after $\Delta N$ $e$-folds. Irrespective of any  later evolution,
this calculated non-gaussianity
will represent the minimal observed one (unless 
non-gaussianity generated later happens to cancel it).

It is supposed  
that the condition $\vartheta_0=0$, as well as the ending of inflation,
will come from a tree-level hybrid potential,
\begin{equation}
V(\varphi,\vartheta) = V_h -\frac{1}{2}m_\vartheta^2 \vartheta^2 + \frac{1}{4}\lambda \vartheta^4 + \frac{1}{2}m_\varphi^2 \varphi^2 + \frac{1}{2}g^2 \vartheta^2 \varphi^2
\label{fullv}
\,.
\end{equation}
Like the original authors though, we shall not investigate the extent to which 
 \eq{fullv} can reproduce \eq{effv} for at least some number of
$e$-folds. We just focus on \eq{effv}, with the assumption $\vartheta_0=0$
for the unperturbed trajectory. 

Because $\vartheta_0=0$, the unperturbed trajectory is straight, and
at first order the curvature perturbation $\zeta$ is conserved.
This is not the case though at second order.
Adopting the definition $\zeta\su{A}$, the authors of  Ref. \cite{ev}
give  an expression for $\zeta\su{A}_2$ similar to that in Eq. (\ref{time}) 
describing the evolution of the second-order curvature perturbation on 
superhorizon scales\footnote
{The fields $\varphi$ and $\vartheta$ in \eq{effv} are supposed to be canonically
normalised, which means that $\varphi$ is {\em not} the field appearing in 
the Rigopoulos definition \eq{abmr} of $\zeta\su{A}$. Instead the authors
of Ref. \cite{ev} give an equivalent definition in terms of the canonically
normalised fields.}.
This equation, 
in the generalized longitudinal gauge,
 reads (from Eq. (67) in Ref. \cite{ev}):
\begin{eqnarray}
\zeta_2\su{A}(t) - \zeta_2\su{A}(t\sub i) &=& 
- \frac{1}{\tilde{\varepsilon} H_{\inf} m_P^2} \Big \{ \int^{t}_{t\sub i} \Big[6H_{\rm inf} \nabla^{-2} 
\partial_i (\delta\dot{\vartheta}\one \partial^i \delta\vartheta\one) 
+ 4 \nabla^{-2} \partial_i (\delta\dot{\vartheta}\one 
\partial^i \delta\vartheta\one)^\cdot \nonumber \\
&& - 2(\delta\dot{\vartheta}\one)^2 + m_\vartheta^2 (\delta\vartheta\one)^2 
+ (\tilde{\varepsilon} - \eta_\varphi)6H_{\rm inf}\nabla^{-4}\partial_i(\partial_k \partial^k 
\delta\vartheta\one \partial^i \delta\vartheta\one)^\cdot \nonumber \\
&&+ (\tilde{\varepsilon} - \eta_\varphi)H_{\rm inf}\nabla^{-4}\partial_i\partial^i(\partial_k \delta
\vartheta\one \partial^k \delta\vartheta\one)^\cdot
- 3\nabla^{-4}\partial_i(\partial_k \partial^k \delta\vartheta\one 
\partial^i \delta\vartheta\one)^{\cdot\cdot} \nonumber \\
&&- \frac12 \nabla^{-4}\partial_i\partial^i(\partial_k \delta\vartheta\one 
\partial^k \delta\vartheta\one)^{\cdot\cdot}\Big] dt 
+ \Big[ - \nabla^{-2}\partial_i(\delta\dot{\vartheta}\one \partial^i 
\delta\vartheta\one) \nonumber \\
&&+ 3\nabla^{-4}\partial_i(\partial_k \partial^k 
\delta\vartheta\one \partial^i \delta\vartheta\one)^\cdot
+ \frac12 \nabla^{-4}\partial_i\partial^i(\partial_k \delta\vartheta\one 
\partial^k \delta\vartheta\one)^\cdot \nonumber \\
&&+ 3\tilde{\varepsilon} H_{\rm inf}\nabla^{-4}\partial_i(\partial_k \partial^k \delta
\vartheta\one \partial^i \delta\vartheta\one)
+ \frac{\tilde{\varepsilon} H_{\rm inf}}{2}\nabla^{-4}\partial_i\partial^i(\partial_k 
\delta\vartheta\one \partial^k \delta\vartheta\one) \Big] \Big|^{t}_{t\sub i} 
\Big \} \,, \nonumber \\
\label{ourev} 
\end{eqnarray}
where $\nabla^{-2}$ is the inverse of the Laplacian operator, $\eta_\varphi \equiv m_\varphi^2/3H_\ast^2$, and $\tilde{\varepsilon}$ is defined by
\be
\tilde{\varepsilon} \equiv \frac{\dot{\varphi}^2_0(t)}{2m_P^2 H_{\rm inf}^2} \,, \label{tildeepsilon}
\ee
which reduces to the $\varepsilon$ parameter in Eq. (\ref{flat1}) for $t = t_\ast$, being $t_\ast$ the time when cosmological scales exit the horizon.

Assuming that this expression is correct, we consider the non-gaussianity
it may generate. Reviewing what it was done in Ref. \cite{antti}, we note first that at $t = t_\ast$
\be
\delta\vartheta\one(t_\ast) \sim \delta\varphi\one(t_\ast) \,, \label{equalityattast}
\ee
which is a good approximation since at that time the amplitude of the spectrum of
perturbations of any light field $\phi$ is $A_{\delta\phi} \approx H_\ast/2\pi$.
Moreover,
assuming slow-roll conditions we obtain
\bea
\delta\vartheta\one(t) &=& \delta\vartheta\one(t_\ast) e^{-\eta_\vartheta N} \,, \label{anttie1} \\
\varphi_0(t) &=& \varphi_0(t_\ast) e^{-\eta_\varphi N} \,, \label{anttie2} \\
\tilde{\varepsilon} &=& \varepsilon e^{-2\eta_\varphi N} \label{anttie3} \,,
\eea
where we have used $N = \int_{t_\ast}^t H_{\rm inf} dt$, $\eta_\vartheta \equiv m_\vartheta^2/3H_\ast^2$, and Eq. (\ref{tildeepsilon}).
A similar expression for the evolution of $\delta\varphi\one$ is obtained by invoking the constancy
of the first-order curvature perturbation $\zeta_1$:
\be
\delta\varphi\one(t) = \delta\varphi\one(t_\ast) e^{-\eta_\varphi N} \,. \label{anttie4}
\ee

Assuming that $H_{\rm inf}$, $\eta_\varphi$, and $\eta_\vartheta$ are almost constants in time, 
we end up with
\bea
\zeta_2\su{A}(t) - \zeta_2\su{A}(t_\ast) &=& - \frac{1}{\tilde{\varepsilon} H_\ast m_P^2} \Big \{
\int^{t}_{t_\ast} \Big[ 2\nabla^{-2} \partial_i (\delta\dot{\vartheta}\one 
\partial^i \delta\vartheta\one)^\cdot
+ 2H_{\rm inf} (\tilde{\varepsilon} - \eta_\varphi) \dot{\gamma}_\vartheta \nonumber \\
&&- (\delta\dot{\vartheta}\one)^2 
- \ddot{\gamma}_\vartheta \Big] dt
+ \Big[- \nabla^{-2} \partial_i (\delta \dot{\vartheta}\one \partial^i \delta\vartheta\one)
+ \dot{\gamma}_\vartheta + \tilde{\varepsilon} H_{\rm inf} \gamma_\vartheta \Big] \Big|^{t}_{t_\ast} \Big \} \nonumber \\
&=& - \frac{1}{\tilde{\varepsilon} H_\ast m_P^2} \Big \{ \int^t_{t_\ast} \Big[ - (\delta \dot{\vartheta}\one)^2
+ 2 H_{\rm inf} \tilde{\varepsilon} \dot{\gamma}_\vartheta \Big] dt
+ \Big[ \nabla^{-2} \partial_i (\delta\dot{\vartheta}\one \partial^i \delta\vartheta\one) \nonumber \\
&&+ H_{\rm inf} (\tilde{\varepsilon} - 2\eta_\varphi) \gamma_\vartheta \Big] \Big|^{t}_{t_\ast} \Big \} \,,
\label{our3}
\eea
where
\be
\gamma_\vartheta \equiv 3 \nabla^{-4} \partial_i (\partial_k \partial^k \delta\vartheta\one \partial^i \delta\vartheta\one) + \frac{1}{2} \nabla^{-2} (\partial_i \delta\vartheta\one \partial^i \delta\vartheta\one) \,,
\ee
and we have used the equation of motion $\delta\ddot{\vartheta}\one + 3H_{\rm inf}\delta\dot{\vartheta}\one + m_\vartheta^2 \delta\vartheta\one = 0$ to go from Eq. (\ref{ourev}) to Eq. (\ref{our3}).

The order of magnitude for $\zeta_2\su{A}(t) - \zeta_2\su{A}(t_\ast)$ is now easily estimated by means of the expressions in Eqs. (\ref{anttie1}) to (\ref{anttie4}), and by neglecting the scale dependence of the non-local terms:
\bea
\zeta_2\su{A}(t) - \zeta_2\su{A}(t_\ast) &\sim& - \frac{1}{\tilde{\varepsilon} H_\ast m_P^2} \Big \{
\int^t_{t_\ast}\Big[ \eta_\vartheta^2 H_{\rm inf}^2 |\delta\vartheta\one|^2 + \tilde{\varepsilon} \eta_\vartheta H_{\rm inf}^2 |\delta\vartheta\one|^2 \Big] dt \nonumber \\
&&+ \Big[ \eta_\vartheta H_{\rm inf} |\delta\vartheta\one|^2 + \tilde{\varepsilon} H_{\rm inf} |\delta\vartheta\one|^2
+ \eta_\varphi H_{\rm inf} |\delta\vartheta\one|^2 \Big] \Big|^{t}_{t_\ast} \Big \} \,,
\eea
so that, using Eq. (\ref{equalityattast}) to write $\delta\vartheta\one$ in terms of $\zeta\one\su{A}$,
\be
\zeta_2\su{A}(t) - \zeta_2\su{A}(t_\ast) \sim \mathcal{O}(\varepsilon,\eta_\varphi,\eta_\vartheta) e^{2N(\eta_\varphi - \eta_\vartheta)} |\zeta\one\su{A}|^2 \,. \label{anttiestimate}
\ee
It is unlikely that the exponential factor on the right hand side provides any significant enhancement to $\zeta\two\su{A}$ if $\varphi$ produces most of the curvature perturbation. Therefore, the overall slow-roll factors give the actual magnitude. We have to remember
that, in this case,  the right hand side
is uncorrelated with the inflaton perturbation $\delta\phi$ which generates 
$\zeta_1\su A$. Thus, \eq{fnltildef} as opposed to \eq{ffnldef} 
applies, and 
the associated $\fnl$ would be $\sim 10^{-9} \mathcal{O}(\varepsilon^3,\eta_\varphi^3,\eta_\vartheta^3)$, which is extremely small. If the observed  $\zeta$ has a non-gaussian part $\zeta\two\su{A}$
equal to Eq. (\ref{anttiestimate}) and a gaussian part generated mostly
{\em after} inflation, one can obtain $|\fnl| >1$ by choosing
$\eta_\varphi > 0.26$, $\eta_\vartheta = \eta_\varphi/2$, $N=70$, and $\zeta\two\su{A} = 10^{-2} \zeta$.


As we will see in the next chapter, where the Enqvist and V$\ddot{{\rm a}}$ihk$\ddot{{\rm o}}$nen model is studied by means of the non perturbative $\delta N$ formalism (see Subsection \ref{ATM2}), the expression in Eq. (\ref{anttiestimate}) disagrees with the one calculated using the $\delta N$ formalism through the appearance of non-local terms \cite{lr1,newlr}, though the order of magnitude is similar \cite{antti}. We point out that the possible source of discrepancy is the use of a set of cosmological perturbation theory equations in Ref. \cite{ev} based on those presented in Ref. \cite{acquaviva}. A calculation made with another set of cosmological perturbation theory equations that do not involve non-local terms \cite{newkarim} reproduces exactly the result found using the $\delta N$ formalism (see Subsection \ref{ATM2}).


\subsection{Preheating}

Now we turn to the possibility that significant non-gaussianity could be
generated during preheating. Preheating is a stage of non-perturbative explosive resonant
decay of
scalar fields which might occur
between the end of inflation and reheating \cite{wandsreview,kofman,kls,preheat1}, 
the latter being taken to 
correspond to the decay of individual particles which leads
to more or less  complete thermalisation of the Universe. Preheating 
typically produces marginally-relativistic particles, which 
decay before reheating.

It was suggested a long time ago \cite{bkm1,bkm2} 
that preheating might cause
the cosmological curvature perturbation to vary at the level of first-order 
perturbation theory, perhaps providing its  main origin.
More recently it has been suggested
\cite{ev1,ev2,enqvistnew}  that
 preheating might
cause the curvature perturbation to vary at second order, 
providing the main source of its
 non-gaussianity. 

If  the separate universe hypothesis is correct, a variation of the 
curvature perturbation during preheating can occur only in models
of preheating which contain a non-inflaton field that  is light during
inflation. This is not the case for the usual  preheating models
that were considered in \cite{ev1,ev2,enqvistnew},
and accordingly one does not expect that significant non-gaussianity will
be generated in those models\footnote
{The preheating model considered in \cite{enqvistnew} contains a field
which may be heavy or light; we refer here to the part of the 
calculation that considers the former case.}.
 This is not in conflict with the findings
of \cite{ev1,ev2,enqvistnew} 
because the  curvature perturbation is not actually
considered there.
Instead the perturbation $\psi^{\rm MW}$ in the longitudinal
gauge is considered, which is only indirectly related to $\zeta$
by Eqs. (\ref{mwdef}), (\ref{psigi}) and (\ref{mwgi0}) \footnote
{The slices
of the longitudinal gauge are orthogonal to the threads of zero shear,  and 
 $\psi^{\rm MW}$ on them is very different from
the curvature perturbation
 $\zeta$.}.
We conjecture that non-gaussianity for 
the curvature perturbation on cosmological scales
is not generated in the usual preheating
models, but that  instead the curvature perturbation  remains constant
on cosmological scales. This should of course be checked, 
in the same spirit that the constancy of the curvature perturbation 
was checked at the first-order level \cite{ourreheat}.

The situation is different for preheating models which 
contain a non-inflaton field that  is light during
inflation. At least three types of models have been proposed 
with that feature \cite{steve2,bdck,steve1,enqvistnew,kr}.
Except for \cite{enqvistnew} only the magnitude of the curvature 
perturbation has been considered, but in all three cases it might be
that  significant non-gaussianity is also generated. 

\subsection{The  curvaton scenario} \label{fnlcurvatons}

In the simplest version of
 the curvaton scenario \cite{lyth03c,lyth02,moroi01b} (see Subsection \ref{zetaandcurvaton}),
the  curvaton field $\sigma$ 
is ultra-light during inflation, weakly coupled, and has no significant
evolution until
 it starts to oscillate during some 
 radiation-dominated era. Until this oscillation gets under way,
 the
curvature perturbation is supposed to be negligible
 (compared with its final observed value).
 The potential during the oscillation
is taken to be quadratic, which will be
a good approximation after a few Hubble times even if it fails initially.
The curvature perturbation 
is generated during the oscillation, and is supposed to be conserved
after the curvaton decays. Here we give a generally-valid formula
for the non-gaussianity in the curvaton scenario, extending somewhat 
the earlier calculations.

The local energy density $\rho_\sigma$ of the  curvaton field
is given by [c.f. Eq. (\ref{sigma_ed})]
\begin{equation}
\rho_\sigma(\eta,{\bf x}) \approx \frac{1}{2} m_\sigma^2 \sigma_a^2(\eta,{\bf x}) \,, \label{ed}
\end{equation} 
where $\sigma_a(\eta,{\bf x})$ represents the amplitude of the oscillations
and $m_\sigma$ is the effective mass.
It is proportional to $a(\eta, {\bf x})^{-3}$ where $a$ is the locally-defined scale
factor. This means that the perturbation $\delta\rho_\sigma/\rho_{\sigma_0}$
is conserved if the slicing is chosen so that the expansion going from
one slicing to the next is uniform \cite{lyth}.
 The flat slicing corresponding
to $\psi^{\rm MW}=0$ has this property \cite{sasaki1,lyth} and accordingly 
$\delta\rho_\sigma$
is defined on that slicing (see Subsection \ref{zetaandcurvaton}).

Assuming that the fractional perturbation is small (which we shall see is
demanded by observation) it is given by
\be
\frac{\delta\rho_\sigma}{\rho_{\sigma_0}} \approx 2\frac{\delta\sigma_a}{\sigma_{a_0}}
+ \( \frac{\delta\sigma_a}{\sigma_{a_0}} \)^2
\label{delrhos}
\,,
\ee
where we have extended to second order the Eq. (\ref{contrastes}).
We first assume that 
$\sigma(\bfx)$ has no evolution between 
inflation and
the onset of oscillation. Then
$\delta\sigma_a/\sigma_{a_0}$ will be equal to its  value just after
horizon exit, which we saw earlier will be practically gaussian.

The total density perturbation is given by
\be
\left(\frac{\delta\rho}{\rho_0}\right)_{\rm total} \approx \Omega \frac{\delta\rho_\sigma}{\rho_{\sigma_0}}
\,,
\ee
where $\Omega \equiv \rho_{\sigma_0}/\rho_{\rm total_0}\propto a$ 
is the fraction of energy density contributed by the curvaton.
Adopting the sudden-decay
approximation, the constant curvature  perturbation obtaining after 
the curvaton decays is given by  Eqs. (\ref{psigi}) and (\ref{zetatwo}), 
evaluated just
before curvaton decay and with $\psi=0$. In performing that calculation,
the exact expression \eq{delrhos} can, without loss of generality, be
identified with the first-order part 
$\delta\rho_{\sigma_1}/\rho_{\sigma_0}$, the second- and higher-order
parts being set at zero.

Adopting the first-order curvature perturbation in \eq{psigi},   one finds
 \cite{lyth03c}  $\chi^2$ non-gaussianity
coming from the second term of \eq{delrhos},  
\be
f\sub{NL}= - \frac{5}{4r} \,,
\ee
with 
\be
r\equiv \frac{3\rho_{\sigma_0}}{4\rho_{r_0} + 3\rho_{\sigma_0}}
\label{rlin}
\,,
\ee
evaluated just before decay.
Going to the second-order expression one finds 
\cite{curvaton}
additional $\chi^2$
non-gaussianity. The final non-linearity parameter
$f\sub{NL}=f\sub{NL}\su{MW}+5/3$ is given by
\begin{equation}
f_{NL} = \frac{5}{3} + \frac{5}{6}r - \frac{5}{4r}
\,.
\end{equation}

If $\Omega$ just before the curvaton decay ($\Omega_{\rm dec}$) is much less than 1 ($\Omega_{\rm dec} \ll 1$) then $f\sub{NL}$ is strongly negative and the 
present bound on it requires $\Omega_{\rm dec} \gsim 0.01$ (combined
with the typical value $|A_\zeta| \approx 5 \times 10^{-5}$, 
this requires  $\delta \rho_\sigma/\rho_{\sigma_0}\ll 1$
as advertised). If instead $\Omega_{\rm dec} =1$ to good accuracy,
then $f\sub{NL}= + 5/4$. Either of these possibilities may
be regarded as generic whereas the intermediate possibility 
($|f\sub{NL}|\sim 1$ but $f\sub{NL}\neq 5/4$) requires a special 
value of $\Omega_{\rm dec}$ just a bit less  than $1$.

Finally, we consider  the case that $\sigma$ evolves 
between horizon exit and the era when the sinusoidal oscillation begins.
If $\sigma_a$ (the amplitude of  oscillation at the latter era)
is some function $g(\sigma_*)$ of the value  a few Hubble times after
horizon exit, then
\be
\delta\sigma_a = g' \delta \sigma_* + \frac12 g'' (\delta\sigma_*)^2
\,,
\ee
where the prime means derivative with respect to $\sigma_*$.
Repeating the above calculation one finds
\be
f\sub{NL} = \frac53 + \frac56 r -\frac5{4r}
\( 1 + \frac{g g''}{g'^2} \)
\,.
\ee
The final term  is the first-order result
(given originally in \cite{lyth04}),
the middle term
is the second-order correction
found in \cite{curvaton}, and the first
term converts from $f\sub{NL}\su{MW}$ to $f\sub{NL}$.

\subsection{The inhomogeneous reheating scenario}

The final scenario  that has been suggested for the origin of the 
curvature perturbation
is its generation during some  spatially inhomogeneous reheating process
\cite{inhomog1,inhomog3,modreh} (see also Refs. \cite{emp,matarrese,maz,mazpo03,inhomog4,inhomog2}).
Before  a reheating process the cosmic fluid is dominated by matter 
(non-relativistic  particles, 
or small scalar field oscillations which are equivalent to particles)
which then decay into thermalised radiation. At least one reheating 
process, presumably, has to occur to give the initial condition for
Big Bang Nucleosynthesis, but there might be more than one.

The inhomogeneous reheating scenario in its simplest form supposes that the curvature 
perturbation is negligible before the relevant reheating process, and
constant afterwards. The inhomogeneous reheating 
corresponds to spatial fluctuations in the decay rate of the inflaton field to
ordinary matter, which lead to fluctuations in the reheating temperature. The coupling
of the inflaton to normal matter is determined by the vacuum expectation values of scalar
fields in the theory. If those fields are light they will fluctuate leading to density perturbations
through the described mechanism. Inhomogeneities in the inflaton decay rate lead to a
a spatially varying value (a perturbation)
of the local Hubble parameter
$H\sub{reh}(\bfx)$ at the decay epoch
(or equivalently of the local energy density).

In contrast with the curvaton scenario, where the form
$\rho_\sigma$ can reasonably be taken as $\rho_\sigma\propto
\sigma^2$,  the inhomogeneous reheating scenario does not suggest any 
particular form for  $H\sub{reh}(\chi)$. Depending on the form,
the inhomogeneous reheating scenario presumably can produce a wide range of 
values for $\fnl$ \cite{inhomog4,inhomog2}.
 
\section{Conclusions} \label{conclusions}

We have examined a number of scenarios for the production of 
a non-gaussian primordial curvature perturbation,
presenting the results with a unified notation \cite{lr}.
These are the single-component inflation, multi-component inflation,
preheating, curvaton, and inhomogeneous reheating scenarios. 
Although the trispectrum may give a competitive observational
signal \cite{bl,trispectrum2,okamoto,trispectrum1}, we have
 focused only on the bispectrum which is characterised by the parameter
$f\sub{NL}$. In all cases our treatment
is based on existing ones, though we do not always agree with the original
authors.

\begin{table}[h]
\begin{center}
\caption{Non-gaussianity according to different scenarios
for the creation of the curvature perturbation. For the simplest curvaton scenario,
 $f\sub{NL}= +5/4$ is a favoured value. \label{tscenarios}}
\vspace{5mm}
\begin{tabular}{lllll}
Scenario & $|f\sub{NL}|\ll 1$ & $|f\sub{NL}|\simeq 1$
&  $f\sub{NL}\ll -1$ &  $f\sub{NL}\gg 1$ \\
\hline
Single-component inflation  & yes & no & no & no \\
Multi-component inflation  & likely & possible  & possible & possible  \\
Simplest curvaton scenario  & unlikely & likely
 & likely & no \\
\hline
\end{tabular}
\end{center}
\end{table}


The preheating and inhomogeneous reheating scenarios   cover
a  range of possibilities, which have not been fully explored but which 
can presumably allow a wide range for $f\sub{NL}$.
The same is true of multi-component inflation, except that 
extremely large values comparable with the current bound $|f\sub{NL}|\lsim
10^2$  seem relatively unlikely. In contrast, the simplest curvaton
scenario can produce a strongly negative value (even violating the 
current bound). However, in the important special case where 
the curvaton dominates the energy density before it decays, it
gives precisely $f\sub{NL}=+5/4$. Finally,
for the single-component inflation case,  Maldacena's calculation
combined with  current constraints on the spectral
tilt show that 
it has magnitude less than $10^{-2}$. These result are summarised in
the Table \ref{tscenarios}.

In the near future, results from WMAP \cite{wmap} or elsewhere
 may detect a value $|f\sub{NL}|\gg1$.
If that does not happen, then PLANCK \cite{planck} or a successor will either detect
a value  $|f\sub{NL}|\sim 1$, or place a bound  $|f\sub{NL}|\lsim 1$.
The precise level at which this will be possible has yet to be determined
because it  would require a second-order calculation of all relevant
 observational signatures.  The example of the simplest curvaton scenario, 
where $f\sub{NL}
=+5/4$ is a favoured value, shows that such a calculation and the 
eventual observations will be well worthwhile.

\chapter{The inflationary prediction for primordial non-gaussianity} \label{fifthpaper}

\section{Introduction}

In this chapter we present for the first time a powerful method to calculate the normalisation $\fnl$ of the bispectrum in slow-roll inflation, by means of the knowledge of the evolution of a family of unperturbed universes \cite{lr1}. The wavevector dependence of $\fnl$ will be, in general, negligible \cite{ignacio,seery2} compared with the (possibly big) contribution coming from the evolution of the unperturbed universes. This method will be applied to selected examples. In particular we will see how the level of non-gaussianity in the curvaton scenario (see Subsection \ref{fnlcurvatons}) and the second order curvature perturbation in the hybrid model of Enqvist and V$\ddot{{\rm a}}$ihk$\ddot{{\rm o}}$nen \cite{newkarim} are successfully reproduced.

The primordial curvature perturbation  of the Universe, 
is already present a few
Hubble times before cosmological scales start to enter the horizon
\cite{book,treview}.
 Its time-independent value at that
stage seems to set the initial condition for the subsequent evolution of
all cosmological perturbations. As a result, observation probes the stochastic
properties of $\zeta$, which
is found to be almost gaussian  with an almost
scale-invariant spectrum.

According to present ideas $\zeta$ is  supposed
to originate from the vacuum fluctuations
 during inflation of one or more light 
scalar fields, which on each scale are promoted to classical perturbations
around the time of horizon exit \cite{albrecht,grishchuk,guthpi,lombardo,lyth84}. 
One takes 
inflation to be  almost exponential (quasi de Sitter spacetime)
corresponding to a practically constant
Hubble parameter $H_{\rm inf}$, and  the effective masses of the fields
to be  much less than $H_*$. This ensures that the fields are almost
massless and live in almost unperturbed quasi de Sitter spacetime, making their 
perturbations indeed almost gaussian and scale invariant (see Subsections \ref{spectrum2} and \ref{dSpec}). 
This automatically makes $\zeta$ almost scale invariant, (see Subsections \ref{zetaandcurvaton} and \ref{pzetainflation}) and can
(though not automatically \cite{lr,lyth03c})  make it also almost gaussian.

All of this is  of intense interest at the present time,
because  observation over  the next few years  will rule out 
most existing scenarios for the generation of $\zeta$, by detecting or
bounding the scale dependence and  non-gaussianity of $\zeta$. We will now
describe a general  procedure for calculating
 the level of non-gaussianity, by means of the
$\delta N$ formalism \cite{sasaki1,ss} (see also Refs. \cite{lee,starobinsky1,starobinsky2}).

\section{Defining the curvature perturbation}
Perturbations 
of the observable Universe are defined with respect to an unperturbed reference
universe, which is homogeneous and isotropic (a FRW universe) (see Section \ref{metricandzeta}).
Its line element may be written as
\be
ds^2 = -dt^2 + a^2(t) \delta_{ij} dx^i dx^j \,,
\ee
defining the unperturbed scale factor $a(t)$, time $t$, and the Cartesian
spatial coordinates $\bfx$.

The curvature perturbation is only of interest after the universe has
been smoothed on some scale $\left(k/a\right)^{-1}$ much bigger than the horizon
$H^{-1}$.
To define it, one
takes the fixed-$t$ slices of spacetime to have uniform energy density, and the fixed-$x$ worldlines
to be comoving. The spatial metric is
\cite{calcagni,singlecreminelli,zc,sasaki1,lr,maldacena,rst,rst2,sb,seery,seery2} [\hbox{c.f. Eq. (\ref{mal})}]
\be
g_{ij} = a^2(t) e^{2\zeta(t,\bfx)} \gamma_{ij}(t,\bfx) =\tilde a^2(t,\bfx)\gamma_{ij}(t,\bfx)
\,.
\label{zetadef}
\ee
In this expression, $\gamma_{ij}(t,\bfx)$  has unit determinant, so that
a  volume of the Universe bounded by fixed comoving spatial coordinates
 is proportional to the locally defined scale
factor  $\tilde{a}^3 (t,\bfx)$. In the inflationary scenario the factor
$\gamma_{ij}$  just accounts for the tensor perturbation, but its form is
irrelevant here (see Subsections \ref{2defz} and \ref{finitialsingle}). 
According to this definition, $\zeta$ is the perturbation
in $\ln\tilde a$.
Only the spatial variation of $\zeta$ is significant, and to make contact
with observation we can work with its Fourier components in a box
a  bit bigger than the observable Universe, setting the zero mode
equal to zero so that $\zeta$ has vanishing spatial
average.

One can also
consider a slicing 
whose metric has the form in Eq. \eqref{zetadef} without the $\zeta$
factor, which we call  the flat slicing.
Starting from any initial flat slice at time $t\sub{ini}$, 
let us define the amount of expansion
\be
N(t,\bfx) \equiv \ln\left[\frac{\tilde a(t)}{a(t\sub{ini})}\right] \,,
\ee
to a final slice of uniform energy density.
 Then \cite{lee,sasaki1,ss,starobinsky1,starobinsky2}
\be
\zeta(t,\bfx) = \delta N \equiv N(t,\bfx) - N_0(t)
\,,
\label{deln1}
\ee
where
\be
N_0(t)\equiv \ln\left[\frac{a(t)}{a(t\sub{ini})}\right] \,,
\ee
is the unperturbed amount of expansion.

To make use of the above  formalism we assume
that in the superhorizon regime ($aH\gg k$), the evolution of the 
Universe at each position (the local evolution)
is  well approximated 
by the evolution of  some unperturbed universe \cite{bst,sasaki1,lyth,tanaka,llmw}.
This `separate universe' assumption
will presumably be correct on cosmological scales
because these scales are so big \cite{lyth}.

By virtue of the separate universe assumption, $N(t,{\bf x})$ is the amount of expansion in some
unperturbed universe, allowing $\zeta$ to be evaluated knowing the evolution of a family of such universes.
For a given content of the Universe it can be checked using the gradient expansion \cite{lv,extlv,sasaki1,rs1,sb} method, but we do not
wish to assume a specific content.

The separate universe assumption leads also to
local energy conservation. Indeed,  using the uniform density
slicing, and remembering that $\tilde a$
determines the expansion,
\be
\dot\rho(t)= -3\tilde H[\rho(t) + P(t,{\bf x})]= -3\left(H+\dot\zeta\right)[\rho(t)+P(t,{\bf x})]
\,,
\label{rhocons}
\ee
where $\tilde H\equiv \dot{\tilde a}/\tilde a$,
$\rho$ is the energy density, and $P$ is the pressure. 
During any era when 
$P$ is a unique function of $\rho$ (the adiabatic condition),
$P$ is uniform on the chosen slicing; then $\dot\zeta$ 
vanishes (because it is uniform and its
spatial average vanishes) so that $\zeta$ is conserved (see Subsection \ref{acgauge}).
This consequence of the separate universe assumption was
 first recognised in
full generality in Refs. \cite{sasaki1,rs1} 
(see also \hbox{Ref. \cite{sb}} for the case of inflation,
Refs. \cite{lyth,llmw} for the case of linear perturbation theory, and
Refs. \cite{lv,extlv} for a coordinate-free treatment).

\section{The inflationary prediction}
The 
evolution of the observable Universe, smoothed on the shortest cosmological
scale, is supposed to be 
determined by the values of one or more light scalar fields when that scale
first emerges from the quantum regime a few 
Hubble times after horizon exit. 
Defined on a flat slicing, each
field $\phi_i$ at this epoch
will be of the form $\phi_i(\bfx) \equiv \phi_{i_0} + \delta\phi_i(\bfx)$.

Because quasi exponential inflation is assumed, and only light fields are considered,
it is a good approximation to take the $\delta\phi_i$ to be almost massless fields
living in unperturbed quasi de Sitter spacetime \cite{book,treview}. In these circumstances
the  perturbations $\delta\phi_i$ generated from the vacuum are 
 almost gaussian,  with an almost
 flat spectrum whose amplitude is \cite{bdbook,bd}
\be
A_{\delta \phi_i} \approx \frac{H_*}{2\pi} \,. \label{amplitude_again}
\ee
 
Now we invoke the separate universe assumption, 
and  choose the homogeneous quantities $\phi_{i_0}$ to correspond to the 
unperturbed universe. Then \eq{deln1} for  
$\zeta$ becomes 
\bea
\zeta(t,\bfx) = N(\rho(t),\phi_1(\bfx),\phi_2(\bfx),\cdots)
- N(\rho(t),\phi_{1_0},\phi_{2_0},\cdots)
\label{master0}
\,.
\eea
In this  expression, the expansion $N$ is evaluated in an unperturbed 
universe, from an epoch when the fields have assigned values to one
when the energy density has an assigned value $\rho$.
This expression \cite{lee,sasaki1,ss,starobinsky1,starobinsky2} allows one to 
propagate forward the  stochastic properties of $\zeta$
to the epoch when it becomes observable, given those of  the initial field 
perturbations.

Since the observed curvature perturbation is almost gaussian,
it must be given to good accuracy by one or more of the linear terms
\be
\zeta(t,\bfx) \simeq  \sum_i  \ni(t)   \delta\phi_i (\bfx)
\,,
\label{linearterm}
\ee
where we use the notation
\bea
\ni &\equiv& \frac{\partial N}{\partial\phi_{i_0}} \,, \\
\nij &\equiv& \frac{\partial^2 N}{\partial\phi_{i_0}\partial\phi_{j_0}} \,,
\eea
with the field
perturbations being almost gaussian.
Here we include for the first time the quadratic terms \cite{lr1}
\bea
\zeta(t,\bfx) = \sum_i  \ni(t)  \delta\phi_i 
+ \frac12\sum_{ij} \nij(t) \delta\phi_i \delta\phi_j
\,.
\label{master1}
\eea
They may be
 entirely  responsible for any observed non-gaussianity
if  the field perturbations are gaussian to sufficient
accuracy\footnote{Here and elsewhere, we  are not
displaying a  homogeneous term needed to make the spatial average
of $\zeta$ vanish.}.

\section{Non-gaussianity}

\subsection{The bispectrum}
The stochastic properties of the perturbations
are specified  through expectation values 
which, according to the inflationary paradigm, are taken with respect to
the time independent (Heisenberg picture) quantum state of the 
Universe
(to be precise, the quantum state of the universe
 before it somehow collapses to give the 
observed Universe).
Focusing on $\zeta$, we consider
 Fourier components, 
\be
\zeta_\bfk\equiv
\int \frac{d^3k}{(2\pi)^{3/2}} \zeta(t,\bfx)\exp(i\bfk\cdot\bfx) \,.
\ee

The stochastic properties of a gaussian perturbation are specified entirely by the spectrum $\calp_\zeta (k) \equiv A_\zeta^2 (k/aH_{\rm inf})^{n_\zeta}$, defined through
\be 
\vev{\zeta_{{\bf k}_1}\zeta_{{\bf k}_2}} = P_\zeta(k) \delta^3({\bf k}_1+{\bf k}_2) \,,
\ee
and
\be
\calp_\zeta(k) \equiv \frac{k^3}{2\pi^2} P_\zeta(k) \,.
\ee
From Eqs. (\ref{amplitude_again}) and (\ref{linearterm})
\be
A_\zeta^2  = \left(\frac{H_*}{2\pi}\right)^2 \sum_i \ni^2
\label{zspec}
\,.
\ee

Non-gaussianity is defined through higher correlations. We consider only the three-point correlation\footnote{The four-point correlation may give a competitive observational signature and can be calculated in a similar fashion \cite{bl,trispectrum2,okamoto,trispectrum1}.}. It defines the  bispectrum $B_\zeta$ through (see Subsection \ref{definingng})
\be
\vev{\zeta_{{\bf k}_1} \zeta_{{\bf k}_2} \zeta_{{\bf k}_3}} = \tp^{-3/2} B_\zeta(k_1,k_2,k_3) 
\delta^3({\bf k}_1+{\bf k}_2+{\bf k}_3) \,. \label{appbdef}
\ee
Its  normalisation is
 specified by
a parameter $\fnl$ according to \cite{review,spergel,maldacena}
\be
B_\zeta \equiv -\frac{6}{5}\fnl(k_1,k_2,k_3) [ P_\zeta(k_1) P_\zeta(k_2) + 
{\rm cyclic \ perturbations}]
\,.
\label{fnldef}
\ee
In first-order cosmological perturbation theory the gauge-invariant gravitational potential $\Phi_g$
during matter domination before horizon entry is $\Phi_g = -(5/3)\zeta$, and our definition of
$\fnl$ coincides with the definition \cite{spergel}
\be
B_{\Phi_g} \equiv 2 \fnl(k_1,k_2,k_3) [ P_{\Phi_g}(k_1) P_{\Phi_g}(k_2) + 
{\rm cyclic \ perturbations}] \,.
\ee
At second-order these definitions of $\fnl$ differ (see for example Refs. \cite{review,bartolo1,curvaton,bartolo2,bartolotemp,newbartolo}. See also footnotes \ref{copelandf1} and \ref{copelandf2} in Chapter \ref{fourthpaper}).

We shall take $\calp_\zeta (k)$ and $\fnl$ to be evaluated when 
cosmological scales approach the horizon and $\zeta$
becomes observable. Observation gives  $|A_\zeta| \approx 5\times 10^{-5}$ \cite{observation},
and $|\fnl|\lsim 100$ \cite{komatsu03}. Absent of a 
 detection, this will eventually come down to roughly $|\fnl|\lsim 1$ \cite{spergel}.

Ignoring any non-gaussianity of the $\delta\phi_i$, our
 formula in Eq. \eqref{master1} 
makes $\fnl$ practically 
independent of the wavenumbers.
Indeed, generalising the result found in Ref. \cite{bl}, we have calculated (see Appendix \ref{prlapp})
\bea
-\frac35 \fnl = 
\frac{\sum_{ij}\ni\nj\nij}{2\[ \sum_i \ni^2 \]^2 }
+ \ln(kL) \frac{A_\zeta^2}  2 
\frac{\sum_{ijk} \nij N_{,jk} N_{,ki}}{\[ \sum_i \ni^2  \]^3}
\,.
\label{master2}
\eea
In deriving  this expression we used the amplitude of the spectrum $A_{\delta \phi_i} \approx H_*/2\pi$
of the field perturbations, and used \eq{zspec} to eliminate
 $H_*$ in favour of 
 $A_\zeta$.
As discussed in Ref. \cite{bl}, the  logarithm 
can be taken to be of order 1, because
it involves the size $k\mone$  of a typical scale under consideration,
relative to the  size $L$ of the region  
within which  the stochastic properties are specified.
Except for the logarithm, {\em $\fnl$ is scale independent if the 
field perturbations are gaussian}.

If only one $\delta\phi_i$ is relevant,
 \eq{master1}  becomes
\be
\zeta(t,\bfx) = \ni \delta\phi_i + \frac12 \nii (\delta\phi_i)^2
\label{simple1}
\,,
\ee
and because the first term dominates, \eq{master2} becomes
\be
-\frac35\fnl = \frac12 \frac{\nii}{\ni^2}
\label{simple2}
\,.
\ee
In this case, $\fnl$ may equivalently be defined  \cite{spergel} 
by writing
\be
\zeta = \zeta\sub g - \frac 35 \fnl \zeta\sub g^2 \,,
\ee
where $\zeta\sub g$ is  gaussian.

To include 
 the possible non-gaussianity of the $\delta\phi_i$, 
we define the bispectra $B_{ijk}$ of the dimensionless field perturbations
$(2\pi/H_*)\delta\phi_i$  and their normalisation $f_{ijk}$, 
in exactly  the same way that we defined $B_\zeta$ and $\fnl$ [c.f. Eqs. (\ref{multibi}) and (\ref{multifi})].
These bispectra add the following contribution to $\fnl$ in Eq. (\ref{master2}) (see Appendix \ref{prlapp}):
\be
\Delta \fnl = \frac{\sum_{ijk} \ni \nj N_{,k} f_{ijk}(k_1,k_2,k_3) }
{\( \sum_i \ni^2 \)^{3/2} } |A_\zeta^{-1}|
\label{delfnl}
\,.
\ee
The   $f_{ijk}$,
 generated directly from the vacuum fluctuation,  will 
depend strongly on the wavenumbers [c.f. Eq. (\ref{multimomenta})].

\subsection{Cosmological perturbation theory}
In the superhorizon regime
the non-linear theory \cite{sasaki1}
that we have used is a complete description.
The basic expression in \eq{master0} is non-perturbative, giving $\zeta(t,\bfx)$ in
terms of the initial fields and the expansion of a family of unperturbed
universes. The  second-order expansion in \eq{master1} is  a matter of convenience.
As  we shall see it seems to be adequate in practice, but \eq{master0} would still
be applicable if the expansion converged slowly or not at all.

Cosmological 
perturbation theory (CPT) is completely different.
It is applicable both inside and outside the horizon, being at each instant 
 a power series in the 
perturbations  of the metric and the stress-energy tensor, together with
whatever variables are needed to completely specify the latter and close
the system of equations. During inflation these variables are the components
of the inflaton, while afterwards they may involve oscillating fields
and a description of the particle content. First-order CPT
 is usually adequate and
can describe non-gaussianity at the level $|\fnl| \gg 1$ which has to be generated by the
second-order term in Eq. (\ref{master1}).
Second-order CPT\footnote{For a full set of CPT equations see e.g. Ref. \cite{newkarim}.} is generally needed only 
to handle non-gaussianity at the level $|\fnl|\sim 1$.

Quantised CPT is needed to calculate the stochastic properties of the
initial field perturbations $\delta\phi_i$, 
which are the input for our calculation.
The slow-roll spectrum in \eq{amplitude_again} comes from the first-order
calculation. The bispectrum is a second-order effect 
and has, in the context of slow-roll inflation, been calculated in \hbox{Refs. \cite{maldacena,seery2}}
[c.f. Eqs. (\ref{multibi}) - (\ref{multimomenta})].
It is shown elsewhere \cite{ignacio} that $\Delta \fnl$ is in this case negligible compared to 1, being generically below $(15/24) f_{SL} \sqrt{r_{T\zeta} \varepsilon} \lsim 10^{-2}$ with \mbox{$1/3 \leq f_{SL} \leq 11/18$} [c.f. Eqs. (\ref{multimomenta}) and (\ref{delfnl})]\footnote{In Ref. \cite{rst2} (see also Refs. \cite{hattori,rst1}) Rigopoulos {\it et. al.} calculated the three-point correlator of $\zeta$ in general multi-component inflationary models using a stochastic approach. Their result is quite puzzling since the wavevector dependence seems to be significant even well after horizon exit. Reconciliation between the approach followed in Ref. \cite{rst2} and ours \cite{lr1} is desirable \cite{rigo}.}.
Higher order correlators have not been calculated yet and would give an additional
contribution to Eq. (\ref{master2}), which presumably is also negligible.
A few single field non slow-roll models \cite{ghost2,singlecreminelli,seery,silverstein} have been investigated
where it is found that $\Delta \fnl$ could be much bigger than 1. From now on we take the $\delta \phi_i$
to be gaussian.

In the regime $aH \gg k$, perturbation theory must be compatible with Eq. (\ref{master1}).
In particular, the non-local terms, present at second order for a generic perturbation,
must be absent for $\zeta$. Finally,
CPT is needed to evolve the perturbations after horizon
entry, but that is not our concern here. 
In the following, we apply our formalism to calculate
$\fnl$ in various cases and  compare it with  the  CPT result where that 
is known.

\section{The $\delta N$ formalism in some multi-component models}

\subsection{A two-component inflation model}
As a first use of \eq{master2} we 
consider the two-component inflation model of Kadota and Stewart \cite{ks1,ks2},
estimating for the first time  the non-gaussianity which it predicts.
The model
 works with a complex field $\Phi$, which is supposed to be a modulus
with a point of enhanced symmetry at the origin. The scalar potential is given by
\be
V = V_h - m^2 |\Phi|^2 + \frac{1}{3} A m^2 [\Phi^3 + \Phi^{\ast 3}] + \frac{1}{2} \nu (\nu + 1) A^2 m^2 |\Phi|^4 \,,
\ee
with $A$ being fixed so that the vacuum energy vanishes at the minimum of the potential, and $\nu = \{1,2,3,...\}$. Writing
\be
\Phi=|\Phi|e^{i\theta} \,,
\ee
the tree level potential has a maximum at
$\Phi_0=0$ and depends on both $|\Phi|$ and $\theta$. A one-loop correction
turns the maximum into a crater and inflation occurs while $\Phi$
is rolling away from the rim of the crater (see Fig. \ref{kadotafig}). The curvature 
perturbation is supposed to be constant after the end of slow-roll 
inflation. For  $\theta_0\ll\theta_c$, with $\theta_c$ being a parameter of the model,
it is found that \cite{ks2}
\be
N\propto \left|\frac{\theta_c}{\theta_0}\right| \,.
\ee
Through the first term of \eq{master2}
\be
\fnl\simeq \left|\frac{\theta_0}{\theta_c}\right| \,,
\ee
which is too small ever to be observed.

\begin{figure}[t]
\begin{center}
\includegraphics[width=9cm,height=9cm]{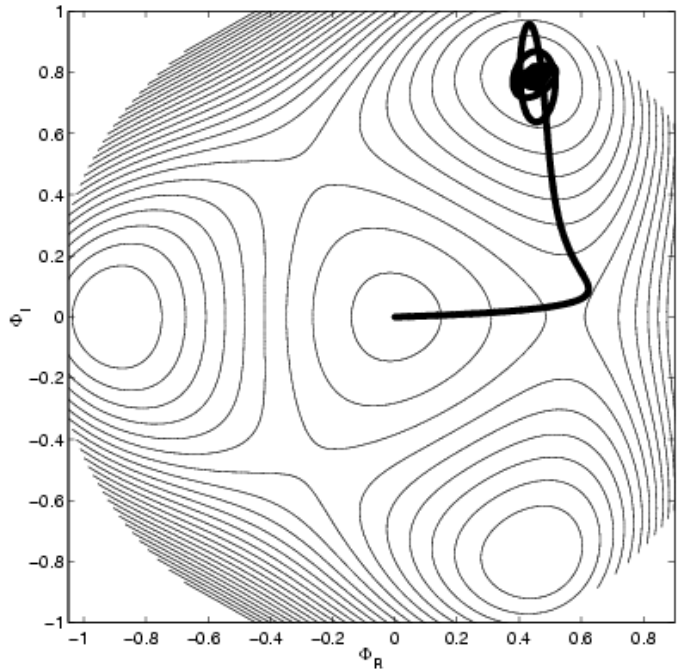}
\caption{Equipotential layers and trajectory of the field $\Phi \equiv (\Phi_R + i \Phi_I)/\sqrt{2}$ in the Kadota and Stewart's model of Refs. \cite{ks1,ks2}. Figure taken from Ref. \cite{ks2}. \label{kadotafig}}
\end{center}
\end{figure}

\subsection{The curvaton model}
In the curvaton model \cite{lyth03c,lyth02,moroi01b} (see also Refs. \cite{moroi01a,earlier1,earlier2} and Subsection \ref{zetaandcurvaton})
the curvature perturbation $\zeta$ grows, from a negligible value in an initially radiation dominated epoch, due to the oscillations of a weakly coupled light field $\sigma$ (the curvaton) around the minimum of its quadratic potential
\be
V_\sigma(t,\bfx) = \frac{1}{2}m_\sigma^2 \sigma^2(t,\bfx) \,,
\ee
where $m_\sigma$ is the curvaton effective mass. Due to the oscillations, the initially negligible curvaton energy density redshifts as
\be
\rho_\sigma (t,\bfx) \approx \frac{1}{2}m_\sigma^2 \sigma_a^2(t,\bfx) \propto a^{-3}(t,\bfx) \,,
\ee
where $\sigma_a$ represents the amplitude of the oscillations. Meanwhile the radiation energy density $\rho_r$ redshifts as $a^{-4}$. Soon after the curvaton decay, the standard Hot Big-Bang is recovered and $\zeta$ is assumed to be conserved until horizon reentry.

To calculate $\fnl$ using Eq. (\ref{master2}) we first realise that $\sigma_*$ (the unperturbed value of $\sigma$ a few Hubble times after horizon exit) is the only relevant quantity since the curvature perturbation produced by the inflaton, and imprinted in the radiation fluid during the reheating process, is supposed to be negligible. Thus, \eq{simple2} applies. Second, we can redefine $N$ as the number of e-folds from the beginning of the sinusoidal oscillations to the curvaton decay. This is because the number of e-folds from the end of inflation to the beginning of the oscillations is completely unperturbed as the radiation energy density dominates during that time. Thus, $N$ is now a function of three variables
\be
N(\rho_{dec},\rho_{osc},\sigma_*)=\frac{1}{3}\ln\left(\frac{\rho_{\sigma_{\rm osc}}}{\rho_{\sigma_{dec}}}\right) = \frac{1}{3}\ln\left[\frac{\frac{1}{2}m_\sigma^2 [g(\sigma_*)]^2}{\rho_{\sigma_{dec}}}\right]
\,,
\ee
where $g \equiv \sigma_{osc}$ is the amplitude at the beginning of the sinusoidal oscillations as a function of
$\sigma_*$. Here the curvaton energy density just before the curvaton decay $\rho_{\sigma_{dec}}$ is expressed in terms of the total energy density $\rho_{dec}$ at that time, the total energy density at the beginning of the sinusoidal oscillations $\rho_{osc}$, and $g$ by
\be
\rho_{\sigma_{dec}} = \frac{1}{2}m_\sigma^2 [g(\sigma_*)]^2 \left(\frac{\rho_{dec} - \rho_{\sigma_{dec}}}{\rho_{osc}}\right)^{3/4}
\,.
\ee
After evaluating $\partial/\partial \sigma_* = g' \partial/\partial g$, at fixed $\rho_{dec}$ and $\rho_{osc}$, we obtain
\be
N_{,\sigma_\ast} = \frac{2}{3} r \frac{g'}{g} \,,
\ee
where
\be
r \equiv \frac{3\rho_{\sigma_{dec}}}{3\rho_{\sigma_{dec}} + 4\rho_{r_{dec}}} \,,
\ee
being $\rho_{r_{dec}}$
the radiation energy density just before the curvaton decay, giving
\be
A_\zeta = \frac{H_\ast}{2\pi} N_{,\sigma_\ast} = \frac{H_\ast r}{3\pi} \frac{g'}{g} \,,
\ee
in agreement with first-order cosmological perturbation theory in the sudden decay approximation \cite{lyth03c,lyth02,moroi01b} (see Subsection \ref{zetaandcurvaton}). Differentiating again we find
from \eq{simple2}
\be
\fnl = -\frac{5}{6} \frac{N_{,\sigma_* \sigma_*}}{N_{,\sigma_*}^2} = \frac{5}{3} + \frac{5}{6} r - \frac{5}{4r} \left(1 + \frac{gg''}{g'^2}\right)
\,,
\ee
which nicely agrees with the already calculated $\fnl$ using first- and second-order perturbation theory (see Refs. \cite{curvaton,lyth04,lr,lyth03c} and Subsection \ref{fnlcurvatons}).

\subsection{Another two-component  model} \label{ATM2}
Finally  we consider
the  two-component inflation model of Ref. \cite{ev} (see also Refs.~\cite{lr,antti} and Subsection \ref{hybridev}). 
For at least some number $N$ of
$e$-folds after cosmological scales leave the horizon, the 
 potential in Eq. (\ref{effv}) is written as
\be
V = V_h \( 1 + \frac12 \eta_\varphi \frac{\varphi^2}{m_P^2} + \frac12 \eta_\vartheta \frac{\vartheta^2}{m_P^2} \) \,,
\ee
with the first term dominating, and $\eta_\varphi$ and $\eta_\vartheta$ being
the slow-roll $\eta$ parameters. The idea is to use \eq{master2} to calculate
the non-gaussianity after the  $N$ e-folds which, barring 
cancellations, will  place  a lower limit on the observed non-gaussianity.

The  slow-roll equations give the background field values $\varphi_0(N)$ and $\vartheta_0(N)$
after $N$ e-folds, 
in terms of
those obtaining just after horizon exit
\bea
\varphi_0(N) &=& \varphi_0 \exp(-N\eta_\varphi) \,, \\
\vartheta_0(N) &=& \vartheta_0 \exp(-N\eta_\vartheta) \,.
\eea
This gives
\be
V_0(N,\varphi_0,\vartheta_0) = V_h \( 1 + \frac12 \eta_\varphi \frac{\varphi_0^2}{m_P^2} e^{-2N\eta_\varphi}  + \frac12 \eta_\vartheta \frac{\vartheta_0^2}{m_P^2} e^{-2N\eta_\vartheta} \) \,,
\ee
and allows us to calculate
the derivatives of $N$ with respect to $\varphi_0$ and $\vartheta_0$ at fixed $V$.
Focusing on the  case $\vartheta_0=0$ considered in Ref. \cite{ev}, we find
\be
\zeta = \frac{\delta\varphi}{\eta_\varphi \varphi_0} - \frac{\eta_\varphi}{2}
\( \frac{\delta\varphi}{\eta_\varphi \varphi_0} \)^2 +  \frac{\eta_\vartheta}{2}
e^{2N(\eta_\varphi - \eta_\vartheta)} \( \frac{\delta\vartheta}{\eta_\varphi \varphi_0} \)^2
\,, \label{goodantti}
\ee
in agreement with the second-order perturbation calculation of Ref. \cite{newkarim}. 
If the observed  $\zeta$ has a non-gaussian part $\zeta_\sigma$
equal to the last term of Eq. (\ref{goodantti}) and a gaussian part generated mostly
{\em after} inflation, one can obtain $|\fnl| >1$ by choosing
$\eta_\varphi > 0.26$, $\eta_\vartheta = \eta_\varphi/2$, $N=70$, and $\zeta_\sigma = 10^{-2} \zeta$.


Our calculated expression for the coefficient of $(\delta \vartheta)^2$ is
in disagreement with the one found in \hbox{Ref. \cite{antti}} [c.f. Eq. (\ref{anttiestimate})]
which uses a set of  CPT equations based on those presented in Ref.
\cite{acquaviva}\footnote{The initial calculations of the structural form of $\zeta$ in this model using CPT \cite{ev,lr} were in gross conflict with \hbox{Eq. (\ref{goodantti})}. This is because the time evolutions of $\varphi$ and $\vartheta$ outside the horizon were not considered. The sources of discrepancy were recognised in Ref. \cite{antti}, showing that the actual order of magnitude is in agreement with Eq. (\ref{goodantti}) except for the presence of non-local terms (see Subsection \ref{hybridev}).}. After converting the variable used there \cite{lr,vernizzi} to our
$\zeta$ (see Subsection \ref{3defzetaA}), these equations give $\dot\zeta$ in terms of first-order quantities,
but they contain non-local terms involving the inverse Laplacian \cite{newlr}. 
Comparison with our non-linear expression in
\eq{master0} shows that such terms must cancel if correctly evaluated.

\section{Conclusions}

The $\delta N$ formalism is a non-perturbative approach to calculate the curvature perturbation $\zeta$ at all orders, in terms only of background quantities that describe the evolution of a family of unperturbed universes. Such a  formalism was originally introduced to calculate the spectrum of $\zeta$ at first order \cite{ss,starobinsky1,starobinsky2} (see also \hbox{Ref. \cite{lee}}) in multi-component inflationary models. Now, with the increasing interest in the non-gaussian features of $\zeta$ in both single- and multi-component inflationary models, the formalism has been extended to calculate the curvature perturbation at second order $\zeta\two$ and the normalisation $\fnl$ of the bispectrum \cite{lr1}. The $\delta N$ formalism relies on the separate universe assumption, which says that on superhorizon scales the Universe behaves locally as if it were unperturbed \cite{bst,sasaki1,lyth,tanaka,llmw}, and on the intrinsic gaussianity of the fields $\phi_i$ involved. The quantities $\zeta\two$ and $\fnl$ are easily given in terms of the first and second derivatives of the unperturbed number of e-folds $N$, from an epoch when the fields have assigned values $\phi_{i_0}$ to one when the energy density has an assigned value $\rho$, with respect to the unperturbed fields $\phi_{i_0}$. The possible intrinsic non-gaussianity of the fields $\phi_i$ would lead to an additional contribution to $\fnl$, highly wavevector dependent \cite{seery2} but in any case negligible compared to 1 \cite{ignacio}.
The $\delta N$ formalism reproduces \cite{lr1,seery2} the well known results in single-component inflation \cite{maldacena}, in the curvaton scenario \cite{curvaton,lyth04,lr,lyth03c}, and in the `hybrid' model of Enqvist and V$\ddot{{\rm a}}$ihk$\ddot{{\rm o}}$nen \cite{newkarim}. In addition, the Kadota and Stewart's modular inflation model \cite{ks1,ks2} has served as an example of the power of this formalism. The $\delta N$ formalism is an interesting alternative to cosmological perturbation theory where, order by order, the relevant expressions to calculate $\zeta\two$ and $\fnl$ tend to be more and more complicated (see for example \hbox{Refs. \cite{acquaviva,ev,newkarim}}). Nevertheless, it is true that the latter is valid on all the scales, while the former is only valid in the superhorizon regime. Fortunately, the subhorizon effects are negligible \cite{ignacio,seery2} making the $\delta N$ formalism reliable and completely independent of cosmological perturbation theory.

\chapter{Conclusions\label{conclusiones}}

Prior to the standard Hot Big-Bang, a period of accelerated expansion seems to have been crucial \cite{guth}. Not only does this period solve the classical problems of the Big-Bang cosmological model, namely the horizon, flatness, and unwanted relics problems, but it also amplifies the fluctuations in the light scalar fields $\phi_i$ living in the Friedmann-Robertson-Walker spacetime \cite{bst,guthpi82,hawking,linde82a,mukhanov,mukhanovrep,riotto,starobinsky1}. This inflationary process serves also for the scalar field fluctuations to become classical soon after horizon exit \cite{albrecht,grishchuk,guthpi,lombardo,lyth84}, giving birth to the primordial perturbations in the energy density that generate the temperature anisotropies in the cosmic microwave background radiation, and, through gravitational collapse, the large-scale structure observed today. Such primordial perturbations can be defined perturbatively on a homogeneous and isotropic background, but the freedom to choose the perturbed coordinate system makes them gauge dependent. To characterize adequately the primordial perturbations, we introduce the gauge-invariant curvature perturbation $\zeta$ \cite{bardeen} which represents the intrinsic spatial curvature on slices of uniform energy density (or slices with zero flow of energy). While the pressure is a unique function of the energy density, $\zeta$ is conserved at all orders \cite{sasaki1,rs1}, which makes this an ideal quantity. In this thesis we have explored some of the theoretical and statistical aspects of the origin of the large-scale structure, such as the two most known scenarios to generate $\zeta$ (the inflaton and the curvaton scenario), the required inflationary energy scale in those scenarios, and the non-gaussianity associated to $\zeta$.

In general $\zeta$ depends on the perturbations in the scalar fields $\phi_i$ during the inflationary period, whose spectra $\mathcal{P}_{\delta \phi_i}(k)$ are generically the same for all kinds of quasi exponential expansion either the respective field dominates the energy density or not. The only appreciable difference is in the way the scale dependence is given in terms of the Hubble parameter $H_\ast$ a few Hubble times after horizon exit and the mass $m_{\phi_i}$ of the respective field. However, the spectrum of $\zeta$, $\mathcal{P}_\zeta(k)$, varies significantly among the different possible scenarios for the origin of the large-scale structure, although in all cases it is almost gaussian and scale invariant. We have studied in \hbox{Chapter \ref{mechanismsiandc}} two different scenarios for the origin of $\zeta$, the inflaton \cite{albste,book,linde82,treview} and the curvaton scenario \cite{lyth03c,lyth02,moroi01b} (see also Refs. \cite{moroi01a,earlier1,earlier2}), pointing out their different signatures and connecting them with the amplitude of gravitational waves produced in each scenario. Both scenarios have their advantages and drawbacks. For example, the single-component inflaton scenario presents a consistency relation that relates the amplitude of $\mathcal{P}_\zeta(k)$ with the amplitude of the gravitational waves spectrum $\mathcal{P}_T(k)$ \cite{abbott,fabbri,liddle,book,rubakov}. The possible detection of gravitational waves is consistent with this scenario and would serve as a smoking gun if the consistency relation is satisfied. The drawback is that the generation of $\zeta$ by the inflaton field $\varphi$ (the field which drives inflation) imposes severe constraints on the theoretical model building \cite{liber,mota,motayo}. In contrast, the curvaton scenario circumvents the latter problem since the generation of $\zeta$ is assigned to a weakly coupled scalar field $\sigma$ (the curvaton) different to the inflaton. The drawbacks are that there is no consistency relation in this case, and that the scenario is inconsistent with a detectable level of gravitational waves \cite{lyth03c}.

There exists several well motivated inflationary models that locate the inflaton field within a particle physics framework \cite{lythlsi,treview}. Most of these models are however unrealistic due to the strong constraints mentioned above. In particular, the generation of adiabatic perturbations is almost inconsistent with the low inflationary energy scale, given by $H_\ast$, required to identify the inflaton with one of the many fields present in supersymmetry. The curvaton scenario comes to rescue these models, allowing for a much lower inflationary energy scale. But how low may this energy scale be in the curvaton scenario?. In Chapter \ref{firstpaper} we have discussed the lower bound on $H_\ast$ in the simplest curvaton setup, showing that it is high enough \cite{lyth04} \mbox{($H_\ast > 10^7$ GeV)} to fail at rescuing the inflationary models whose potentials are generated by some mechanism of gravity-mediated supersymmetry breaking where $H_\ast \sim 10^3$ GeV is required. The general conditions to obtain low scale inflation in the curvaton scenario have been given in that chapter \cite{yeinzon}, in terms of three quantities $\epsilon$, $f$, and $\delta$, that parameterize respectively the evolution of the curvaton field from the time of horizon exit to the beginning of its oscillations, the effective curvaton mass $\tilde{m}_\sigma$ at the end of a phase transition with respect to the Hubble parameter at the same time, and the time of the phase transition with respect to the time of horizon exit. In Chapters \ref{firstpaper} and \ref{secondpaper} we have invoked the `heavy curvaton' picture \cite{lyth04}, defined as the setup where $\sigma$ suddenly increases its mass at the end of a phase transition much later than inflation \cite{matsuda03}. The mass increment is given by the coupling of $\sigma$ (parameterized by the constant $\lambda$) with another field which acquires a large vacuum expectation value at the end of the same period. Thus, the smallness of the parameters $f$ and $\delta$ are exploited to allow for an inflationary scale consistent with gravity-mediated susy breaking\footnote{The parameter $\epsilon$ is in these cases unmodified because it is assumed that the unperturbed component of $\sigma$ is frozen throughout inflation and until oscillations begin.}. In the cases presented in these two chapters the phase transition is associated with the end of a second (thermal) inflationary period, which was originally introduced to dilute the abundances of unwanted relics that the first (main) inflationary period is not able to do (for instance the moduli fields) \cite{lyth95,lyth96}. Thermal inflation is driven by the confinement of a second (flaton) field $\chi$ at the origin of the potential due to the thermal effects from the radiation background left by the inflaton decay \cite{barreiro96}. The eventual rolling of $\chi$, towards the minimum of its potential, ends thermal inflation and triggers an increment in the bare mass $m_\sigma$ of $\sigma$ through the coupling of the latter with $\chi$. Solving the moduli problem while satisfying adequately all the constraints in the first model discussed in Chapter \ref{firstpaper}, where the curvaton oscillates for some time before decaying, restricts the parameter space to a region where the two important parameters, $\lambda$ and $m_\sigma$, are required to be very small ($10^{-22} \lsim \lambda \lsim 10^{-10}$ and $m_\sigma \lsim 10^{-1} \ {\rm GeV}$) \cite{yeinzon}. It is likely then that the curvaton field is a pseudo Nambu-Goldstone boson. In contrast, as discussed in the second model presented in Chapter \ref{secondpaper}, if the increment in the mass of the curvaton field is so high that the decay rate overtakes the Hubble parameter, leading to the immediate decay of $\sigma$, the parameter space is less constrained, with more natural values for $\lambda$ and $m_\sigma$ ($10^{-10} \lsim \lambda \lsim 10^{-4}$ and $m_\sigma \lsim 1 \ {\rm GeV}$) \cite{yeinzon04}.

Although $\zeta$ is found to be almost perfectly gaussian in most of the models that give account of its origin, as required by observation, the possibly present small non-gaussianity is being scrutinized by present experiments like the WMAP satellite \cite{wmap} and will be the focus of future experiments like the PLANCK satellite \cite{planck}. The first statistical significant quantity that gives us information about the level of non-gaussianity is the bispectrum, which corresponds to the three-point correlator of $\zeta$ \cite{spergel}. Its normalisation is given by the parameter $\fnl$, which has been found to be $|\fnl| \lsim 10^2$ \cite{komatsu03} but that will eventually go down to $|\fnl| \sim 1$ in the forthcoming years unless there is an earlier detection \cite{spergel}. In view of the difficulty at discriminating between models by means of only the spectral index and/or consistency relations, the detection of non-gaussianity and a precise determination of $\fnl$ would be useful tools to serve this purpose. For example, in the single-component inflaton scenario the level of non-gaussianity is very small, being $|\fnl|$ or the order of the slow-roll parameters \cite{maldacena}. In contrast, in the curvaton scenario $|\fnl|$ could be much higher, being $\fnl$ negative, or, if $\sigma$ has already dominated the energy density before decaying, the precise value for $\fnl$ would be $\fnl = +5/4$ \cite{curvaton,lr,lyth03c}. In \hbox{Chapters \ref{fourthpaper} and \ref{fifthpaper}} we have addressed this subject, following a perturbative approach in the former, and a non-perturbative one in the latter. In Chapter \ref{fourthpaper} we have presented in an unified way the different definitions of the second-order curvature perturbation $\zeta\two$ present in the literature \cite{acquaviva,lr,lyth,malik,rigopoulos}. The translation rules to go from one definition to another have been explicitly given, and would help to avoid the possible confusion when confronting different papers and results which use different definitions for $\zeta\two$ \cite{lr}. We have examined the predictions for $\fnl$ coming from the single-component inflation model, the multi-component one, and the curvaton scenario, discussing also the respective predictions in preheating and the inhomogeneous reheating scenario. Although multi-component inflation, preheating, and the inhomogeneous reheating scenarios do not predict a definite value or set of values for $\fnl$, being very dependent on the specific model, the single-component inflaton scenario as well as the curvaton one do give definite predictions for $\fnl$ \cite{lr}. Multi-component slow-roll inflation is studied in Chapter \ref{fifthpaper} following the non-perturbative $\delta N$ formalism \cite{lee,sasaki1,ss,starobinsky1,starobinsky2}, that allows us to calculate $\zeta$ at all orders only by knowing the evolution of a family of unperturbed universes. Immediate applications of this formalism are the calculation of the spectral index \cite{ss}, and $\fnl$ \cite{lr1}, in a general multi-component slow-roll inflation model. The formalism relies on the separate universe assumption, which says that the local evolution of the Universe on superhorizon scales is the same as that of an unperturbed universe \cite{bst,sasaki1,lyth,tanaka,llmw}. The normalisation parameter $\fnl$ is easily given in terms of first and second derivatives of the number of e-folds, from an initially flat slice to a final uniform energy density slice, with respect to the fields living in the former. The intrinsic non-gaussianity in the fields $\phi_i$ relevant for the evolution of the family of universes may be also taken into account, but its contribution, which is highly wavevector dependent \cite{maldacena,seery2}, is in any case  negligible compared to 1 \cite{ignacio}.
Comparison of this formalism with cosmological perturbation theory has been made, in the case of the curvaton scenario \cite{lr1} and the `hybrid' \cite{ev} model of Enqvist and V$\ddot{{\rm a}}$ihk$\ddot{{\rm o}}$nen \cite{lr1,newkarim,antti}. According to Ref.~\cite{newkarim} the results for the latter model following the two approaches agree, and moreover they refute the claim about the possible presence of non-local terms in cosmological perturbation theory \cite{acquaviva,review,ev,antti}. 

\appendix

\chapter{$\fnl$ from the $\delta N$ formalism} \label{prlapp}

In this appendix we derive explicitly the expression in Eq. (\ref{master2}) which gives the normalisation $\fnl$ of the bispectrum of $\zeta$ defined in Eqs. (\ref{appbdef}) and (\ref{fnldef}).

We begin by writing $\zeta$ as given by Eq. (\ref{master1}) including the homogeneous term which makes the spatial average of $\zeta$ vanish:
\be
\zeta = \sum_i N_{,i} \delta\phi_i + \frac{1}{2} \sum_{ij} N_{,ij} \delta\phi_i \delta\phi_j - \frac{1}{2} \sum_{ij} N_{,ij} \langle \delta\phi_i \delta\phi_j \rangle \,.
\ee
The corresponding mode function is then written as
\be
\zeta_{\bf k} = \sum_i N_{,i} \delta_{\bf k}\phi_i + \frac{1}{2} \sum_{ij} N_{,ij} \int \frac{d^3p}{(2\pi)^{3/2}} \delta_{{\bf k + p}}\phi_i \delta_{\bf p}\phi^\ast_j - \frac{1}{2} (2\pi)^{3/2} \delta^3({\bf k}) \sum_{ij} N_{,ij} \langle \delta\phi_i \delta\phi_j \rangle \,.
\ee

Making use of the above formula, the product of three mode functions $\zeta_{\bf k}$ is therefore
\bea
\zeta_{{\bf k}_1}\zeta_{{\bf k}_2}\zeta_{{\bf k}_3} &=& \sum_{ijk} N_{,i} N_{,j} N_{,k} \delta_{{\bf k}_1}\phi_i \delta_{{\bf k}_2}\phi_j \delta_{{\bf k}_3}\phi_k \nonumber \\
&+& \frac{1}{2} \sum_{ijkl} N_{,i} N_{,j} N_{,kl} \Big[ \delta_{{\bf k}_1}\phi_i \delta_{{\bf k}_2}\phi_j \int \frac{d^3p}{(2\pi)^{3/2}} \delta_{{\bf k}_3 + {\bf p}}\phi_k \delta_{\bf p}\phi_l^\ast \nonumber \\
&+& \delta_{{\bf k}_1}\phi_i \delta_{{\bf k}_3}\phi_j \int \frac{d^3p}{(2\pi)^{3/2}} \delta_{{\bf k}_2 + {\bf p}}\phi_k \delta_{\bf p}\phi_l^\ast \nonumber \\
&+& \delta_{{\bf k}_2}\phi_i \delta_{{\bf k}_3}\phi_j \int \frac{d^3p}{(2\pi)^{3/2}} \delta_{{\bf k}_1 + {\bf p}}\phi_k \delta_{\bf p}\phi_l^\ast \Big] \nonumber \\
&-& \frac{1}{2} (2\pi)^{3/2} \sum_{ijkl} N_{,i} N_{,j} N_{,kl} \Big[ \delta_{{\bf k}_1}\phi_i \delta_{{\bf k}_2}\phi_j \delta^3({\bf k}_3) \langle \delta\phi_k \delta\phi_l \rangle \nonumber \\
&+& \delta_{{\bf k}_1}\phi_i \delta_{{\bf k}_3}\phi_j \delta^3({\bf k}_2) \langle \delta\phi_k \delta\phi_l \rangle \nonumber \\
&+& \delta_{{\bf k}_2}\phi_i \delta_{{\bf k}_3}\phi_j \delta^3({\bf k}_1) \langle \delta\phi_k \delta\phi_l \rangle \Big] \nonumber \\
&+& \frac{1}{4} \sum_{ijklm} N_{,i} N_{,jk} N_{,lm} \Big[
\int \frac{d^3p}{(2\pi)^{3/2}} \frac{d^3p'}{(2\pi)^{3/2}} \delta_{{\bf k}_1}\phi_i \delta_{{\bf k}_2 + {\bf p}}\phi_j \delta_{\bf p}\phi_k^\ast \delta_{{\bf k}_3 + {\bf p'}}\phi_l \delta_{\bf p'}\phi_m^\ast \nonumber \\
&+& \int \frac{d^3p}{(2\pi)^{3/2}} \frac{d^3p'}{(2\pi)^{3/2}} \delta_{{\bf k}_2}\phi_i \delta_{{\bf k}_1 + {\bf p}}\phi_j \delta_{\bf p}\phi_k^\ast \delta_{{\bf k}_3 + {\bf p'}}\phi_l \delta_{\bf p'}\phi_m^\ast \nonumber \\
&+& \int \frac{d^3p}{(2\pi)^{3/2}} \frac{d^3p'}{(2\pi)^{3/2}} \delta_{{\bf k}_3}\phi_i \delta_{{\bf k}_1 + {\bf p}}\phi_j \delta_{\bf p}\phi_k^\ast \delta_{{\bf k}_2 + {\bf p'}}\phi_l \delta_{\bf p'}\phi_m^\ast \Big] \nonumber \\
&-& \frac{1}{4} (2\pi)^{3/2} \sum_{ijklm} N_{,i} N_{,jk} N_{,lm} \Big[ 
\delta^3({\bf k}_1) \langle \delta \phi_j \delta \phi_k \rangle \int \frac{d^3p}{(2\pi)^{3/2}} \delta_{{\bf k}_2} \phi_i \delta_{{\bf k}_3 + {\bf p}} \phi_l \delta_{\bf p} \phi_m^\ast \nonumber \\
&+&\delta^3({\bf k}_2) \langle \delta \phi_j \delta \phi_k \rangle \int \frac{d^3p}{(2\pi)^{3/2}} \delta_{{\bf k}_1} \phi_i \delta_{{\bf k}_3 + {\bf p}} \phi_l \delta_{\bf p} \phi_m^\ast \nonumber \\
&+&\delta^3({\bf k}_3) \langle \delta \phi_l \delta \phi_m \rangle \int \frac{d^3p}{(2\pi)^{3/2}} \delta_{{\bf k}_1} \phi_i \delta_{{\bf k}_2 + {\bf p}} \phi_j \delta_{\bf p} \phi_k^\ast \nonumber \\
&+&\delta^3({\bf k}_1) \langle \delta \phi_j \delta \phi_k \rangle \int \frac{d^3p}{(2\pi)^{3/2}} \delta_{{\bf k}_3} \phi_i \delta_{{\bf k}_2 + {\bf p}} \phi_l \delta_{\bf p} \phi_m^\ast \nonumber \\
&+&\delta^3({\bf k}_2) \langle \delta \phi_j \delta \phi_k \rangle \int \frac{d^3p}{(2\pi)^{3/2}} \delta_{{\bf k}_3} \phi_i \delta_{{\bf k}_1 + {\bf p}} \phi_l \delta_{\bf p} \phi_m^\ast \nonumber \\
&+&\delta^3({\bf k}_3) \langle \delta \phi_l \delta \phi_m \rangle \int \frac{d^3p}{(2\pi)^{3/2}} \delta_{{\bf k}_2} \phi_i \delta_{{\bf k}_1 + {\bf p}} \phi_j \delta_{\bf p} \phi_k^\ast \Big] \nonumber \\
&+& \frac{1}{4} (2\pi)^3 \sum_{ijklm} N_{,i} N_{,jk} N_{,lm} \Big[
\delta^3({\bf k}_1) \delta^3 ({\bf k}_2) \langle \delta \phi_j \delta \phi_k \rangle \langle \delta \phi_l \delta \phi_m \rangle \delta_{{\bf k}_3} \phi_i \nonumber \\
&+& \delta^3({\bf k}_1) \delta^3 ({\bf k}_3) \langle \delta \phi_j \delta \phi_k \rangle \langle \delta \phi_l \delta \phi_m \rangle \delta_{{\bf k}_2} \phi_i \nonumber \\
&+&\delta^3({\bf k}_2) \delta^3 ({\bf k}_3) \langle \delta \phi_j \delta \phi_k \rangle \langle \delta \phi_l \delta \phi_m \rangle \delta_{{\bf k}_1} \phi_i \Big] \nonumber \\
&+& \frac{1}{8} \sum_{ijklmn} N_{,ij} N_{,kl} N_{,mn} \Big[ \nonumber \\
&&\int \frac{d^3p}{(2\pi)^{3/2}} \delta_{{\bf k}_1 + {\bf p}}\phi_i \delta_{\bf p}\phi_j^\ast \int \frac{d^3p'}{(2\pi)^{3/2}} \delta_{{\bf k}_2 + {\bf p'}}\phi_k \delta_{\bf p'}\phi_l^\ast \int \frac{d^3p''}{(2\pi)^{3/2}} \delta_{{\bf k}_3 + {\bf p''}}\phi_m \delta_{\bf p''}\phi_n^\ast \Big] \nonumber \\
&-& \frac{1}{8} (2\pi)^{3/2} \sum_{ijklmn} N_{,ij} N_{,kl} N_{,mn} \Big[ \nonumber \\
&&\delta^3({\bf k}_1) \langle \delta \phi_i \delta \phi_j \rangle \int \frac{d^3p}{(2\pi)^{3/2}} \delta_{{\bf k}_2 + {\bf p}}\phi_k \delta_{\bf p}\phi_l^\ast \int \frac{d^3p'}{(2\pi)^{3/2}} \delta_{{\bf k}_3 + {\bf p'}}\phi_m \delta_{\bf p'}\phi_n^\ast \nonumber \\
&+&\delta^3({\bf k}_2) \langle \delta \phi_k \delta \phi_l \rangle \int \frac{d^3p}{(2\pi)^{3/2}} \delta_{{\bf k}_1 + {\bf p}}\phi_i \delta_{\bf p}\phi_j^\ast \int \frac{d^3p'}{(2\pi)^{3/2}} \delta_{{\bf k}_3 + {\bf p'}}\phi_m \delta_{\bf p'}\phi_n^\ast \nonumber \\
&+&\delta^3({\bf k}_3) \langle \delta \phi_m \delta \phi_n \rangle \int \frac{d^3p}{(2\pi)^{3/2}} \delta_{{\bf k}_1 + {\bf p}}\phi_i \delta_{\bf p}\phi_j^\ast \int \frac{d^3p'}{(2\pi)^{3/2}} \delta_{{\bf k}_2 + {\bf p'}}\phi_k \delta_{\bf p'}\phi_l^\ast \Big] \nonumber \\
&+& \frac{1}{8} (2\pi)^3 \sum_{ijklmn} N_{,ij} N_{,kl} N_{,mn} \Big[ \nonumber \\
&&\delta^3({\bf k}_1) \delta^3({\bf k}_2) \langle \delta \phi_i \delta \phi_j \rangle \langle \delta \phi_k \delta \phi_l \rangle \int \frac{d^3p}{(2\pi)^{3/2}} \delta_{{\bf k}_3 + {\bf p}}\phi_m \delta_{\bf p}\phi_n^\ast \nonumber \\
&+&\delta^3({\bf k}_1) \delta^3({\bf k}_3) \langle \delta \phi_i \delta \phi_j \rangle \langle \delta \phi_m \delta \phi_n \rangle \int \frac{d^3p}{(2\pi)^{3/2}} \delta_{{\bf k}_2 + {\bf p}}\phi_k \delta_{\bf p}\phi_l^\ast \nonumber \\
&+&\delta^3({\bf k}_2) \delta^3({\bf k}_3) \langle \delta \phi_k \delta \phi_l \rangle \langle \delta \phi_m \delta \phi_n \rangle \int \frac{d^3p}{(2\pi)^{3/2}} \delta_{{\bf k}_1 + {\bf p}}\phi_i \delta_{\bf p}\phi_j^\ast \Big] \nonumber \\
&-& \frac{1}{8} (2\pi)^{9/2} \sum_{ijklmn} N_{,ij} N_{,kl} N_{,mn} \delta^3({\bf k}_1) \delta^3({\bf k}_2) \delta^3({\bf k}_3) \langle \delta \phi_i \delta \phi_j \rangle \langle \delta \phi_k \delta \phi_l \rangle \langle \delta \phi_m \delta \phi_n \rangle \,. \nonumber \\
\label{barethreepoint}
\eea

The next step is to take the average of the latter expression. In doing so, we make use of the following decompositions \cite{review}:
\bea
&&\langle \delta_{{\bf k}_1}\phi_i \delta_{{\bf k}_2}\phi_j \delta_{{\bf k}_3}\phi_k \delta_{{\bf k}_4}\phi_l \rangle = \nonumber \\
&&\langle \delta_{{\bf k}_1}\phi_i \delta_{{\bf k}_2}\phi_j \rangle \langle \delta_{{\bf k}_3}\phi_k \delta_{{\bf k}_4}\phi_l \rangle
+ \langle \delta_{{\bf k}_1}\phi_i \delta_{{\bf k}_3}\phi_k  \rangle \langle \delta_{{\bf k}_2}\phi_j \delta_{{\bf k}_4}\phi_l \rangle
+ \langle \delta_{{\bf k}_1}\phi_i \delta_{{\bf k}_4}\phi_l  \rangle \langle \delta_{{\bf k}_2}\phi_j \delta_{{\bf k}_3}\phi_k  \rangle \,, \nonumber \\
\label{decom4}
\eea

\bea
&&\langle \delta_{{\bf k}_1}\phi_i \delta_{{\bf k}_2}\phi_j \delta_{{\bf k}_3}\phi_k \delta_{{\bf k}_4}\phi_l \delta_{{\bf k}_5}\phi_m \delta_{{\bf k}_6}\phi_n \rangle = \nonumber \\
&&\langle \delta_{{\bf k}_1}\phi_i \delta_{{\bf k}_2}\phi_j \rangle \langle \delta_{{\bf k}_3}\phi_k \delta_{{\bf k}_4}\phi_l \rangle \langle \delta_{{\bf k}_5}\phi_m \delta_{{\bf k}_6}\phi_n \rangle
+ \langle \delta_{{\bf k}_1}\phi_i \delta_{{\bf k}_2}\phi_j \rangle \langle \delta_{{\bf k}_3}\phi_k \delta_{{\bf k}_5}\phi_m \rangle \langle \delta_{{\bf k}_4}\phi_l \delta_{{\bf k}_6}\phi_n \rangle \nonumber \\
&+& \langle \delta_{{\bf k}_1}\phi_i \delta_{{\bf k}_2}\phi_j \rangle \langle \delta_{{\bf k}_3}\phi_k \delta_{{\bf k}_6}\phi_n \rangle \langle \delta_{{\bf k}_4}\phi_l \delta_{{\bf k}_5}\phi_m \rangle
+ \langle \delta_{{\bf k}_1}\phi_i \delta_{{\bf k}_3}\phi_k \rangle \langle \delta_{{\bf k}_2}\phi_j \delta_{{\bf k}_4}\phi_l \rangle \langle \delta_{{\bf k}_5}\phi_m \delta_{{\bf k}_6}\phi_n \rangle \nonumber \\
&+& \langle \delta_{{\bf k}_1}\phi_i \delta_{{\bf k}_3}\phi_k \rangle \langle \delta_{{\bf k}_2}\phi_j \delta_{{\bf k}_5}\phi_m \rangle \langle \delta_{{\bf k}_4}\phi_l \delta_{{\bf k}_6}\phi_n \rangle
+ \langle \delta_{{\bf k}_1}\phi_i \delta_{{\bf k}_3}\phi_k \rangle \langle \delta_{{\bf k}_2}\phi_j \delta_{{\bf k}_6}\phi_n \rangle \langle \delta_{{\bf k}_4}\phi_l \delta_{{\bf k}_5}\phi_m \rangle \nonumber \\
&+& \langle \delta_{{\bf k}_1}\phi_i \delta_{{\bf k}_4}\phi_l \rangle \langle \delta_{{\bf k}_2}\phi_j \delta_{{\bf k}_3}\phi_k \rangle \langle \delta_{{\bf k}_5}\phi_m \delta_{{\bf k}_6}\phi_n \rangle
+ \langle \delta_{{\bf k}_1}\phi_i \delta_{{\bf k}_4}\phi_l \rangle \langle \delta_{{\bf k}_2}\phi_j \delta_{{\bf k}_5}\phi_m \rangle \langle \delta_{{\bf k}_3}\phi_k \delta_{{\bf k}_6}\phi_n \rangle \nonumber \\
&+& \langle \delta_{{\bf k}_1}\phi_i \delta_{{\bf k}_4}\phi_l \rangle \langle \delta_{{\bf k}_2}\phi_j \delta_{{\bf k}_6}\phi_n \rangle \langle \delta_{{\bf k}_3}\phi_k \delta_{{\bf k}_5}\phi_m \rangle
+ \langle \delta_{{\bf k}_1}\phi_i \delta_{{\bf k}_5}\phi_m \rangle \langle \delta_{{\bf k}_2}\phi_j \delta_{{\bf k}_3}\phi_k \rangle \langle \delta_{{\bf k}_4}\phi_l \delta_{{\bf k}_6}\phi_n \rangle \nonumber \\
&+& \langle \delta_{{\bf k}_1}\phi_i \delta_{{\bf k}_5}\phi_m \rangle \langle \delta_{{\bf k}_2}\phi_j \delta_{{\bf k}_4}\phi_l \rangle \langle \delta_{{\bf k}_3}\phi_k \delta_{{\bf k}_6}\phi_n \rangle
+ \langle \delta_{{\bf k}_1}\phi_i \delta_{{\bf k}_5}\phi_m \rangle \langle \delta_{{\bf k}_2}\phi_j \delta_{{\bf k}_6}\phi_n \rangle \langle \delta_{{\bf k}_3}\phi_k \delta_{{\bf k}_4}\phi_l \rangle \nonumber \\
&+& \langle \delta_{{\bf k}_1}\phi_i \delta_{{\bf k}_6}\phi_n \rangle \langle \delta_{{\bf k}_2}\phi_j \delta_{{\bf k}_3}\phi_k \rangle \langle \delta_{{\bf k}_4}\phi_l \delta_{{\bf k}_5}\phi_m  \rangle
+ \langle \delta_{{\bf k}_1}\phi_i \delta_{{\bf k}_6}\phi_n \rangle \langle \delta_{{\bf k}_2}\phi_j \delta_{{\bf k}_4}\phi_l \rangle \langle \delta_{{\bf k}_3}\phi_k \delta_{{\bf k}_5}\phi_m  \rangle \nonumber \\
&+& \langle \delta_{{\bf k}_1}\phi_i \delta_{{\bf k}_6}\phi_n \rangle \langle \delta_{{\bf k}_2}\phi_j \delta_{{\bf k}_5}\phi_m \rangle \langle \delta_{{\bf k}_3}\phi_k \delta_{{\bf k}_4}\phi_l \rangle \,, \nonumber \\
\label{decom6}
\eea
where we have neglected connected $n$-point correlators with $n>3$, which nobody has calculated yet but that are presumably very small \cite{seery2}, and products of $n$-point correlators with $m$-point correlators, where $n \geq 3$ and $m \geq 2$, which we believe give a much smaller contribution than that coming from the three-point correlator \cite{lrz}.

From Eqs. (\ref{barethreepoint}), (\ref{decom4}), and (\ref{decom6}), and after doing some algebra, we obtain the three point correlator function of the $\zeta_{\bf k}$ mode functions:
\bea
&&\langle \zeta_{{\bf k}_1} \zeta_{{\bf k}_2} \zeta_{{\bf k}_3} \rangle = 
\sum_{ijk} N_{,i} N_{,j} N_{,k} \langle \delta_{{\bf k}_1}\phi_i \delta_{{\bf k}_2}\phi_j \delta_{{\bf k}_3}\phi_k \rangle \nonumber \\
&&+ \sum_{ij} N_{,i} N_{,j} N_{,ij} (2\pi)^{-3/2} \delta^3 ({\bf k}_1 + {\bf k}_2 + {\bf k}_3) \left(\frac{H_\ast}{2\pi}\right)^4 \left[\frac{4\pi^4}{k_1^3 k_2^3} + {\rm cyclic} \ {\rm permutations}\right] \nonumber \\
&&+ \frac{1}{2} \sum_{ijk} N_{,ij} N_{,jk} N_{,ki} (2\pi)^{-3/2} \delta^3 ({\bf k}_1 + {\bf k}_2 + {\bf k}_3) \left(\frac{H_\ast}{2\pi}\right)^6 \times \nonumber \\
&& \times \int d^3p \left[\frac{1}{|{\bf k}_1 + {\bf p}|^3 p^3 |{\bf k}_3 - {\bf p}|^3} + \frac{1}{|{\bf k}_1 + {\bf p}|^3 p^3 |{\bf k}_2 - {\bf p}|^3} \right] \,.
\label{quasifapp1}
\eea
The integrals in the last line of the previous expression are calculated following the arguments presented in Ref. \cite{bl}, giving as a result
\be
\int d^3p \left[\frac{1}{|{\bf k}_1 + {\bf p}|^3 p^3 |{\bf k}_3 - {\bf p}|^3} + \frac{1}{|{\bf k}_1 + {\bf p}|^3 p^3 |{\bf k}_2 - {\bf p}|^3} \right] = 8\pi \ln(kL) \left[\frac{1}{k_1^3 k_2^3} + {\rm cyclic} \right] \,,
\label{quasifapp2}
\ee
where $k^{-1}$ represents a typical scale under consideration and $L$ is the size of the region within which the stochastic properties are specified. That is why $\ln(kL)$ can be taken of order 1.

Finally, making use of the $\fnl$ definition in Eqs. (\ref{appbdef}) and (\ref{fnldef}):
\be
\vev{\zeta_{{\bf k}_1} \zeta_{{\bf k}_2} \zeta_{{\bf k}_3}} = \tp^{-3/2} \delta^3({\bf k}_1+{\bf k}_2+{\bf k}_3) \left[ -\frac{6}{5}\fnl(k_1,k_2,k_3) [ P_\zeta(k_1) P_\zeta(k_2) + 
{\rm cyclic}]\right]
 \,,
\ee 
the results found in Eqs. (\ref{quasifapp1}) and (\ref{quasifapp2}), and the expression in Eq. (\ref{zspec}) for the amplitude of the spectrum of $\zeta$ in terms of the first derivatives of $N$:
\be
A_\zeta^2 = \left(\frac{H_\ast}{2\pi}\right)^2 \sum_i N_{,i}^2 \,,
\ee
we arrive to the desired expression
\bea
-\frac35 \fnl = 
\frac{\sum_{ij}\ni\nj\nij}{2\[ \sum_i \ni^2 \]^2 }
+ \ln(kL) \frac{A_\zeta^2}  2 
\frac{\sum_{ijk} \nij N_{,jk} N_{,ki}}{\[ \sum_i \ni^2  \]^3}
\,,
\eea
with an additional contribution
\be
\Delta \fnl = \frac{\sum_{ijk} \ni \nj N_{,k} f_{ijk}(k_1,k_2,k_3) }
{\( \sum_i \ni^2 \)^{3/2} } |A_\zeta^{-1}|
\,.
\ee
The latter expression comes from the first line in Eq. (\ref{quasifapp1}) containing the three-point correlator $\langle \delta_{{\bf k}_1} \phi_i \delta_{{\bf k}_2} \phi_j \delta_{{\bf k}_3} \phi_k \rangle$. This three-point correlator is normalised in such a way that the bispectra $B_{ijk}$ of the dimensionless perturbations $(2\pi/H_\ast)\delta \phi_i$ is written in the same way as in Eqs. (\ref{appbdef}) and (\ref{fnldef}) with $\fnl (k_1,k_2,k_3)$ being replaced by $f_{ijk} (k_1,k_2,k_3)$.

\end{document}